\documentclass[american,english]{article}
\usepackage[T1]{fontenc}
\usepackage[latin9]{inputenc}
\usepackage[a4paper]{geometry}
\usepackage{babel}
\usepackage{array}
\usepackage{verbatim}
\usepackage{refstyle}
\usepackage{float}
\usepackage{units}
\usepackage{textcomp}
\usepackage{multirow}
\usepackage{amsmath}
\usepackage{amssymb}
\usepackage{graphicx}
\usepackage{esint}
\usepackage[unicode=true,pdfusetitle,
 bookmarks=true,bookmarksnumbered=false,bookmarksopen=false,
 breaklinks=false,pdfborder={0 0 1},backref=false,colorlinks=false]
 {hyperref}

\makeatletter

\AtBeginDocument{\providecommand\figref[1]{\ref{fig:#1}}}
\AtBeginDocument{\providecommand\tabref[1]{\ref{tab:#1}}}
\DeclareRobustCommand{\greektext}{%
  \fontencoding{LGR}\selectfont\def\encodingdefault{LGR}}
\DeclareRobustCommand{\textgreek}[1]{\leavevmode{\greektext #1}}
\DeclareFontEncoding{LGR}{}{}
\DeclareTextSymbol{\~}{LGR}{126}
\newcommand{\lyxmathsym}[1]{\ifmmode\begingroup\def\b@ld{bold}
  \text{\ifx\math@version\b@ld\bfseries\fi#1}\endgroup\else#1\fi}

\providecommand{\tabularnewline}{\\}
\RS@ifundefined{subref}
  {\def\RSsubtxt{section~}\newref{sub}{name = \RSsubtxt}}
  {}
\RS@ifundefined{thmref}
  {\def\RSthmtxt{theorem~}\newref{thm}{name = \RSthmtxt}}
  {}
\RS@ifundefined{lemref}
  {\def\RSlemtxt{lemma~}\newref{lem}{name = \RSlemtxt}}
  {}

\@ifundefined{showcaptionsetup}{}{%
 \PassOptionsToPackage{caption=false}{subfig}}
\usepackage{subfig}
\AtBeginDocument{
  
}

\makeatother

\begin{document}

\title{\noindent \textbf{\textit{\emph{CeLAND: search for a 4$^{th}$ light
neutrino state with a 3 PBq $^{144}$Ce-$^{144}$Pr $\bar{\nu_{e}}$-generator
in KamLAND }}}}
\maketitle
\begin{abstract}
The reactor neutrino and gallium anomalies can be tested with a 3-4
PBq (75-100 \foreignlanguage{american}{kilocurie} scale) $\mathrm{^{144}Ce-^{144}Pr}$
antineutrino beta-source deployed at the center or next to a large
low-background liquid scintillator detector. The antineutrino generator
will be produced by the Russian reprocessing plant PA Mayak as early
as 2014, transported to Japan, and deployed in the Kamioka Liquid
Scintillator Anti-Neutrino Detector (KamLAND) as early as 2015. KamLAND's
13 m diameter target volume provides a suitable environment to measure
the energy and position dependence of the detected neutrino flux.
A characteristic oscillation pattern would be visible for a baseline
of about 10 m or less, providing a very clean signal of neutrino disappearance
into a yet-unknown, \textquotedbl{}sterile\textquotedbl{} state. This
will provide a comprehensive test of the electron dissaperance neutrino
anomalies and could lead to the discovery of a 4$^{th}$ neutrino
state for {\normalsize $\Delta m_{new}^{2}\gtrsim0.1$ $eV^{2}$ and
sin$^{2}(2\theta_{new})\gtrsim0.05.$\pagebreak{}}{\normalsize \par}
\end{abstract}
\begin{center}
{\large \vskip 0.5in }A. Gando, Y. Gando, S. Hayashida, H. Ikeda,
K. Inoue, K. Ishidoshiro, H. Ishikawa, M. Koga, R. Matsuda, S. Matsuda,
T. Mitsui,D. Motoki, K. Nakamura, Y. Oki, M. Otani, I. Shimizu, J.
Shirai, F. Suekane, A. Suzuki, Y. Takemoto, K. Tamae, K. Ueshima,
H. Watanabe, B.D. Xu, S. Yamada,Y. Yamauchi, H. Yoshida\\
 \textit{Research Center for Neutrino Science, Tohoku University,
Sendai 980-8578, Japan}
\par\end{center}

\begin{center}
\vskip 0.1in M. Cribier$^{1,2}$, M. Durero$^{2}$, V. Fischer$^{2}$,
J. Gaffiot$^{1}$, N. Jonqueres$^{3}$, A. Kouchner$^{1}$, T. Lasserre$^{1,2,4,5}$,
D. Leterme$^{3}$, A. Letourneau$^{2}$, D. Lhuillier$^{2}$, G. Mention$^{2}$,
G. Rampal$^{3}$, L. Scola$^{2}$, Ch. Veyssi\`ere$^{2}$, M. Vivier$^{2}$,
and P. Yala$^{3}$\\
\textit{$^{1}$}\foreignlanguage{american}{\textit{Astroparticules}}\textit{
et Cosmologie APC, 10 rue Alice Domon et L\'eonie Duquet, 75205 Paris
cedex 13, France}\\
\textit{$^{2}$Commissariat \`a l'\'energie atomique et aux \'energies alternatives,
Centre de Saclay, IRFU, 91191 Gif-sur-Yvette, France}\\
\textit{$^{2}$Commissariat \`a l'\'energie atomique et aux \'energies alternatives,
Centre de Saclay, DEN/DANS \& DEN/DM2S, 91191 Gif-sur-Yvette, France}
\par\end{center}

\begin{center}
\textit{$^{4}$ European Research Council Starting Grant 4th-Nu-Avenue StG-307184}\\
\textit{$^{5}$Corresponding author (thierry.lasserre@cea.fr)}\\
\par\end{center}

\begin{center}
\vskip 0.1in B. E. Berger\\
 \textit{Colorado State University, Fort Collins, CO 80523-1875, USA}
\par\end{center}

\begin{center}
\vskip 0.1in A. Kozlov\\
 \textit{Kavli Institute for the Physics and Mathematics of the Universe
(WPI), University of Tokyo, Kashiwa 277-8583, Japan}
\par\end{center}

\begin{center}
\vskip 0.1in T. Banks$^{2}$, D. Dwyer$^{1}$, B. K. Fujikawa$^{1}$,
K. Han$^{2}$, Yu. G. Kolomensky$^{1,2}$, Y. Mei$^{2}$, T. O'Donnell$^{2}$
\par\end{center}

\begin{center}
\textit{$^{1}$Lawrence Berkeley National Laboratory, Berkeley, CA
94720, USA}\\
 \textit{$^{2}$University of California, Berkeley, CA 94704, USA}
\par\end{center}

\begin{center}
\vskip 0.1in Patrick Decowski \\
 \textit{Nikhef and the University of Amsterdam, Science Park 105
1098 XG, Amsterdam, the Netherlands}
\par\end{center}

\begin{center}
\vskip 0.1in D. M. Markoff\\
 \textit{North Carolina Central University, Durham, NC 27707, USA}
\par\end{center}

\begin{center}
\vskip 0.1in S. Yoshida\\
 \textit{Graduate School of Science, Osaka University, Toyonaka, Osaka
560-0043, Japan}
\par\end{center}

\begin{center}
\vskip 0.1in V.N. Kornoukhov$^{1}$, T. V.M. Gelis$^{3}$, G.V. Tikhomirov$^{2}$,
I.S. Saldikov$^{2}$ 
\par\end{center}

\begin{center}
\textit{$^{1}$ Russian Federation State Scientific Center of Theoretical
and Experimental Physics Institute, 117218 Moscow, Russia}\\
 \textit{$^{2}$Scientific and Research Nuclear University, Moscow
engineering and Physics Institute, Russia}\\
 \textit{$^{3}$ Russian Academy of Sciences A.N. Frumkin Institute
of Physical chemistry and Electrochemistry, Russia}
\par\end{center}

\begin{center}
\vskip 0.1in J. G. Learned, J. Maricic, S. Matsuno, R. Milincic\\
 \textit{University of Hawaii at Manoa, Honolulu, HI 96822, USA}
\par\end{center}

\begin{center}
\vskip 0.1in H. J. Karwowski\\
 \textit{University of North Carolina, Chapel Hill, NC 27599, USA}
\par\end{center}

\begin{center}
\vskip 0.1in Y. Efremenko\\
 \textit{University of Tennessee, Knoxville, TN 37996, USA}
\par\end{center}

\begin{center}
\vskip 0.1in J. A. Detwiler, S. Enomoto\\
 \textit{University of Washington, Seattle, WA 98195, USA} 
\par\end{center}

\begin{center}
\newpage{}\tableofcontents{}
\par\end{center}

\noindent \newpage{}

\noindent \begin{center}
\begin{figure}[h]
\centering{}\includegraphics[scale=0.9]{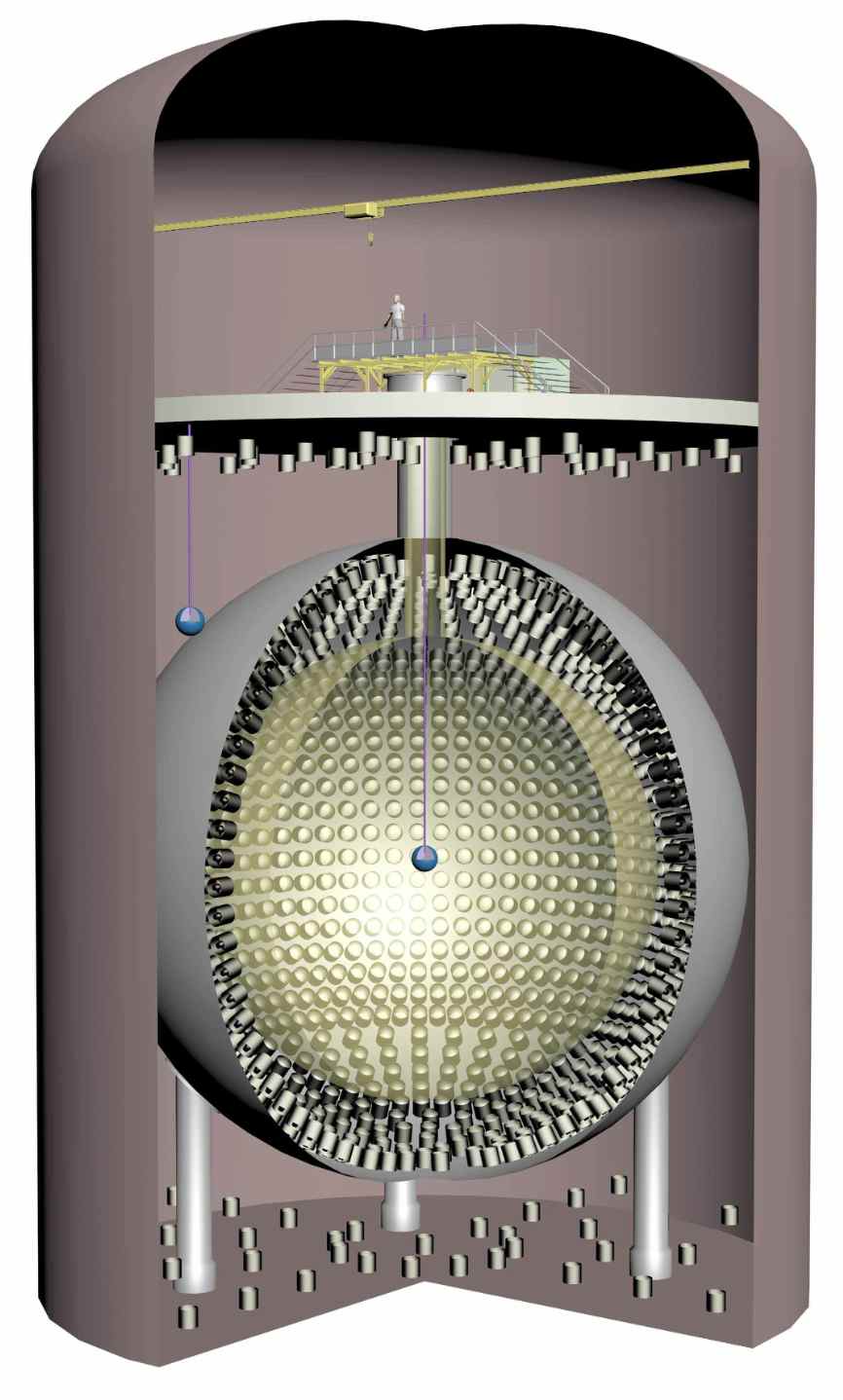}\caption{Schematic view of the CeLAND experiment. In the first phase of the
experiment the collaboration intends to deploy a 75-100 kCi source
in the water Cherenkov outer detector of KamLAND. This run could be
followed by the deployment of the same source (approx. 50 kCi after
6 months) at the center of the KamLAND detector for 1 year (in liquid
scintillator target) in case of detection hint of a fourth neutrino
state.}
\end{figure}

\par\end{center}

\section{\noindent Executive summary}

\noindent Over the last 20 years a standard neutrino oscillation framework
associated with small splittings between the $\mathrm{\nu}$ mass
states has become well established. The three $\mathrm{\nu}$ flavors
($\nu{}_{e}$, $\nu{}_{\mu}$, $\nu{}_{\tau}$) are mixtures of three
massive neutrinos ($\nu{}_{1}$, $\nu{}_{2}$, $\nu{}_{3}$) separated
by squared mass differences of $\Delta m_{21}^{2}=\Delta m_{\mathrm{sol}}^{2}=7.50_{-0.20}^{+0.19}\times10^{-5}\,\mathrm{eV}{}^{2}$
and $\mid\Delta m_{31}^{2}\mid\approx\mid\Delta m_{32}^{2}\mid=\Delta m_{\mathrm{atm}}^{2}=2.32_{-0.08}^{+0.12}\times10^{-3}\,\mathrm{eV}{}^{2}$
\cite{PDG2012}. This is a minimal extension of the Standard Model
that requires a lepton mixing matrix, similarly to the quark sector.

\noindent Beyond this model, indications of oscillations between active
and sterile $\nu{}_{s}$ have been observed in LSND \cite{LSNDAnomaly},
MiniBooNE \cite{MiniBooNEAnomaly1,MiniBooNEAnomaly2}, Gallium \cite{GalliumAnomaly1,GalliumAnomaly2,GalliumAnomaly3}
and more recently by reactor experiments \cite{RAA}. This suggests
the possible existence of a fourth massive $\nu$ with a mass of $\Delta m^{2}\gtrsim0.1\,\mathrm{eV}{}^{2}$
\cite{SterileSuggestion1,SterileSuggestion2,SterileSuggestion3}.
Both the reactor and Gallium anomalies result from the observation
of the disappearance of MeV energy $\nu_{e}$ and $\bar{\nu}_{e}$
in counting experiments. Therefore, to definitivively test the anomalies
one must not only test neutrino disappearance at short baselines,
but also search for an oscillation pattern as a function of $L/E_{\nu}$
energy and baseline-dependent signatures. Probing $\Delta m^{2}$
values on the order of 0.1 to a few eV$^{2}$ implies that an oscillation
search using neutrinos with energies of typical of radioactive decays,
i.e. in the few MeV range, requires a baseline of only several meters.
Assuming CP invariance, both gallium and reactor anomalies can be
unambiguously tested using a neutrino emitter placed inside or next
to a large 10-meter scale liquid scintillator detector, such as Borexino,
KamLAND, SNO+, or Daya Bay. This constitutes an elegant method to
probe the existence of oscillations into a 4$^{th}$ neutrino by using
a well-known neutrino detector.

\noindent There are two options for deploying intense $\nu$ emitters
in large liquid scintillator detectors: monochromatic $\nu_{e}$ emitters,
like $^{51}\mathrm{Cr}$ or $^{37}\mathrm{Ar}$, or $\bar{\nu}_{e}$
emitters with continuous $\beta$-spectra, like $^{144}\mathrm{\mathrm{Ce}}$,
$^{90}\mathrm{Sr}$, $^{42}\mathrm{Ar}$ or $^{106}\mathrm{Ru}$.
In the case of $\nu{}_{e}$ emitters the signature is provided by
$\nu_{e}$ elastic scattering off electrons in the liquid scintillator
molecules. This signature can be mimicked by Compton scattering induced
by radioactivity or cosmogenic background, or by Solar-$\nu$'s. The
constraints of an experiment with $\nu_{e}$ impose the use of a very
high activity source $\gtrsim$ 10 MCi outside of the detector target.
In the case of $\bar{\nu}_{e}$, events are detected via the inverse
beta decay reaction (IBD), which provides a $e^{+}$-n delayed coincidence
that offers an efficient rejection of the mentioned background. For
this reason, we focus our studies on $\bar{\nu}_{e}$ sources in the
current proposal. 

\noindent Based on the concept presented in 2011 in \cite{144CePRL},
the CeLAND collaboration intends to use an intense $^{144}\mathrm{Ce}$
source decaying into the unstable daughter $^{144}\mathrm{Pr}$ which,
in turn, decays into stable $^{144}\mathrm{Nd}$ with a Q-value of
2.996~MeV\@. The $^{144}\mathrm{Pr}$ decay produces antineutrinos
above the 1.8~MeV threshold for the inverse beta decay reaction.
The half-life of $^{144}\mathrm{Ce}$ is 285 days, and that of its
daughter $^{144}\mathrm{Pr}$ is only 17.3 minutes, so that the latter
decay remains in equilibrium at all times. This thus leaves a significant
amount of time for the source production, its transportation to the
detector, and the data taking, without significant decrease in its
initial activity.

\noindent The intense $^{144}\mathrm{Ce}-^{144}\mathrm{Pr}$ antineutrino
generator (ANG hereafter) can be produced at the Federal State Unitary
Enterprise Mayak Production Association (FSUE -Mayak- PA or simply
Mayak) reprocessing plant in Russia as early as 2014. The source will
then be transported within one month to the Kamioka mine, Japan, where
the KamLAND detector is located. The CeLAND collaboration intends
to deploy a 75-100 kCi (2.78-3.73 PBq) antineutrino generator within
13m of the KamLAND detector center, as early as 2015. The plan is
to take data for at least 18 months and to search for a fourth neutrino.
If a hint of oscillation is observed, the collaboration would consider
deploying the source at the center of the detector, or procuring a
second antineutrino generator to increase the statistical power of
the experiment. Such deployment strategy allows a comprehensive test
of the electron disappearance neutrino anomalies and could lead to
the discovery of a 4$^{th}$ neutrino state for $\Delta m_{new}^{2}\gtrsim0.1$
$eV^{2}$ and sin$^{2}(2\theta_{new})\gtrsim0.05.$\clearpage{}

\section{\noindent Physics motivations}

\subsection{\noindent Oscillation overview}

\noindent Neutrino oscillations have been observed in solar, atmospheric,
and long-baseline reactor and accelerator experiments. For a detailed
review, see for example \cite{Giunti_Kim}. The data collected so
far by these experiments are well fitted in the framework of a three-neutrino
mixing approach, in which the three known flavor neutrinos $\nu_{e}$,
$\nu_{\mu}$ and $\nu_{\tau}$ are unitary linear combinations of
three massive neutrinos $\nu_{1}$, $\nu_{2}$ and $\nu_{3}$ with
squared-mass differences and mixing angles \cite{PDG2012}:

\noindent 
\begin{eqnarray}
 & \Delta m_{21}^{2}=\Delta m_{sol}^{2}=7.50_{-0.20}^{+0.19}\times10^{-5}eV^{2},\\
 & sin^{2}(2\theta_{12})=0.857_{-0.025}^{+0.023}\\
 & \mid\Delta m_{31}^{2}\mid\approx\mid\Delta m_{32}^{2}\mid=\Delta m_{atm}^{2}=2.32_{-0.08}^{+0.12}\times10^{-3}eV^{2},\\
 & sin^{2}(2\theta_{23})>0.95\,(90\%\, C.L.),\, sin^{2}(2\theta_{13})=0.098\pm0.013
\end{eqnarray}
where $\Delta m_{jk}^{2}=m_{j}^{2}-m_{k}^{2}$ and m$_{j}$ is the
mass of the neutrino $\nu_{j}$.

\noindent Besides this well-established picture, there are a few anomalies
which could point toward short-baseline (SBL) neutrino oscillations
generated by a larger squared-mass difference: the gallium radioactive
source experiment anomalies \cite{GalliumAnomaly1,GalliumAnomaly2,GalliumAnomaly3},
the recently observed reactor antineutrino anomaly \cite{RAA}, and
the MiniBooNE \cite{MiniBooNEAnomaly1,MiniBooNEAnomaly2} and LSND
\cite{LSNDAnomaly} accelerator neutrino anomalies. Existence of a
fourth yet unobserved neutrino species is not ruled out by cosmological
data obtained by PLANCK \cite{Planck_Cosmo_Results} and WMAP \cite{NeutrinoWhitePaper}.
In this section, we successively review these anomalies and give an
up-to-date interpretation in terms of light sterile neutrinos.

\subsection{\noindent The Gallium anomaly}

\noindent Man-made neutrino sources were originally proposed and designed
to test solar-$\nu$ detection in the GALLEX \cite{Gallex1,Gallex2,Gallex3}
and SAGE \cite{SAGE1,SAGE2,SAGE3} experiments. Intense Mega Curie
(MCi) $\mathrm{^{51}Cr}$ and $\mathrm{^{37}Ar}$ $\nu_{e}$ sources
with precisely measured activities were placed at the center of the
detectors to count for the number of $\nu_{e}$ events, as a tool
to understand the detector responses. Taking into account the uncertainty
of the cross-section of the detection process $\mathrm{\nu_{e}+^{71}Ga\rightarrow^{71}Ge+e^{-}}$,
the ratios R of measured to predicted $\mathrm{^{71}Ge}$ event rates
reported by the experiments

\noindent 
\begin{eqnarray}
 & \mathrm{R_{Cr1}^{GALLEX}=0.95_{-0.12}^{+0.11},}\mathrm{\; R_{Cr2}^{GALLEX}=0.81{}_{-0.11}^{+0.10},}\\
 & \mathrm{R_{Cr}^{SAGE}=0.95_{-0.12}^{+0.12},}\mathrm{\; R_{Ar}^{SAGE}=0.79{}_{-0.10}^{+0.09}}
\end{eqnarray}

\noindent with an averaged ratio of

\noindent 
\begin{equation}
\mathrm{R^{Ga}=0.86_{-0.05}^{+0.05}.}
\end{equation}

\noindent The number of measured events is smaller than the prediction
at the 2.7$\sigma$ level, leading to the so-called Gallium anomaly.
Although the cross-section of the $\mathrm{\nu_{e}+^{71}Ga\rightarrow^{71}Ge+e^{-}}$
detection process might have been overestimated by adding the uncertain
transitions from the ground state of $\mathrm{^{71}Ga}$ to two excited
states of $\mathrm{^{71}Ge}$, the anomaly still remains at the 1.8$\sigma$
level in the absence of such transitions, with a ratio $\mathrm{R^{Ga}=0.90_{-0.05}^{+0.05}}$.
Recently, a re-interpretation of the Gallium data with new measurements
of the Gamow-Teller strengths of the transitions from the ground state
of $\mathrm{^{71}Ga}$ to the two excited states of $\mathrm{^{71}Ge}$
confirmed the anomaly at the 3$\sigma$ level \cite{GalliumAnomaly3}.

\subsection{\noindent The reactor antineutrino anomaly}

\noindent Nuclear reactors emit $\mathrm{\bar{\nu}_{e}}$ through
$\mathrm{\beta^{-}}$ decay of fission products. Any reactor antineutrino
spectrum is mostly the sum of the neutrino spectra of the four fissioning
nuclei ($\mathrm{^{235}U}$, $\mathrm{^{238}U}$, $\mathrm{^{239}Pu}$
and $\mathrm{^{241}Pu}$) weighted by their relative fission rate.
Since thousands of different $\mathrm{\beta}$ branches are available
to the unstable fission products, and many of them are still unknown,
neutrino spectra predictions cannot fully rely on simulations. Neutrino
spectra are then estimated by converting the precise measurement of
the total electron spectrum emitted by nuclear fuel under neutron
irradiation, using energy conservation law. However, some modeling,
such as of the end point distribution of the beta branches, still
remain necessary.

\noindent In 2011, a revised calculation of the rate of $\mathrm{\bar{\nu}_{e}}$
production by nuclear reactors yielded yet another indication of the
hypothetical existence of a fourth neutrino species. The new calculation
relies on detailed knowledge of the decays of thousands of fission
products listed by nuclear databases \cite{NewNeutrinoSpectra1},
while the previous calculation used a phenomenological model based
on 30 effective $\beta$-branches. The revised rate of $\mathrm{\bar{\nu}_{e}}$
production by nuclear reactors is 3.5\% higher than previously thought
and was confirmed independently by another calculation \cite{NewNeutrinoSpectra2}.
This new calculation motivated a re-analysis of 19 past reactor experiments
at reactor distances < 100 m. Figure \ref{fig:RAA-picture} shows
the ratio of measured-to-expected $\mathrm{\bar{\nu}_{e}}$ count
rates in each of these experiments, following the revised calculation
of the rate of $\mathrm{\bar{\nu}_{e}}$ production by nuclear reactors
combined with a new estimation of the inverse $\mathrm{\beta}$ decay
cross-section (driven by the neutron life-time measurement) and off-equilibrium
corrections to the reactor fuel composition. Along with these new
ratios are also displayed the results of middle-baseline experiments
(L > 100 m) such as Chooz \cite{Chooz} or Palo Verde \cite{Palo Verde}.
The dotted line represents a ``classic'' three active neutrinos
mixing scheme. The solid line shows a (3+1) sterile neutrino model.
A constant fit to the 19 SBL experiments' measured-over-expected $\bar{\nu}_{e}$
count rates gives a mean ratio of $\mathrm{R^{R}=0.927\pm0.023}$,
corresponding to a 3$\sigma$ discrepancy called the reactor antineutrino
anomaly (RAA)\cite{RAA}. The SBL reactor data suggests a new $\mathrm{\Delta m_{new}^{2}\sim1\, eV^{2}}$
if interpreted in terms of a new SBL oscillation driven by a $\mathrm{4^{th}}$
neutrino state.

\noindent \begin{center}
\begin{figure}[h]
\begin{centering}
\includegraphics[scale=0.35]{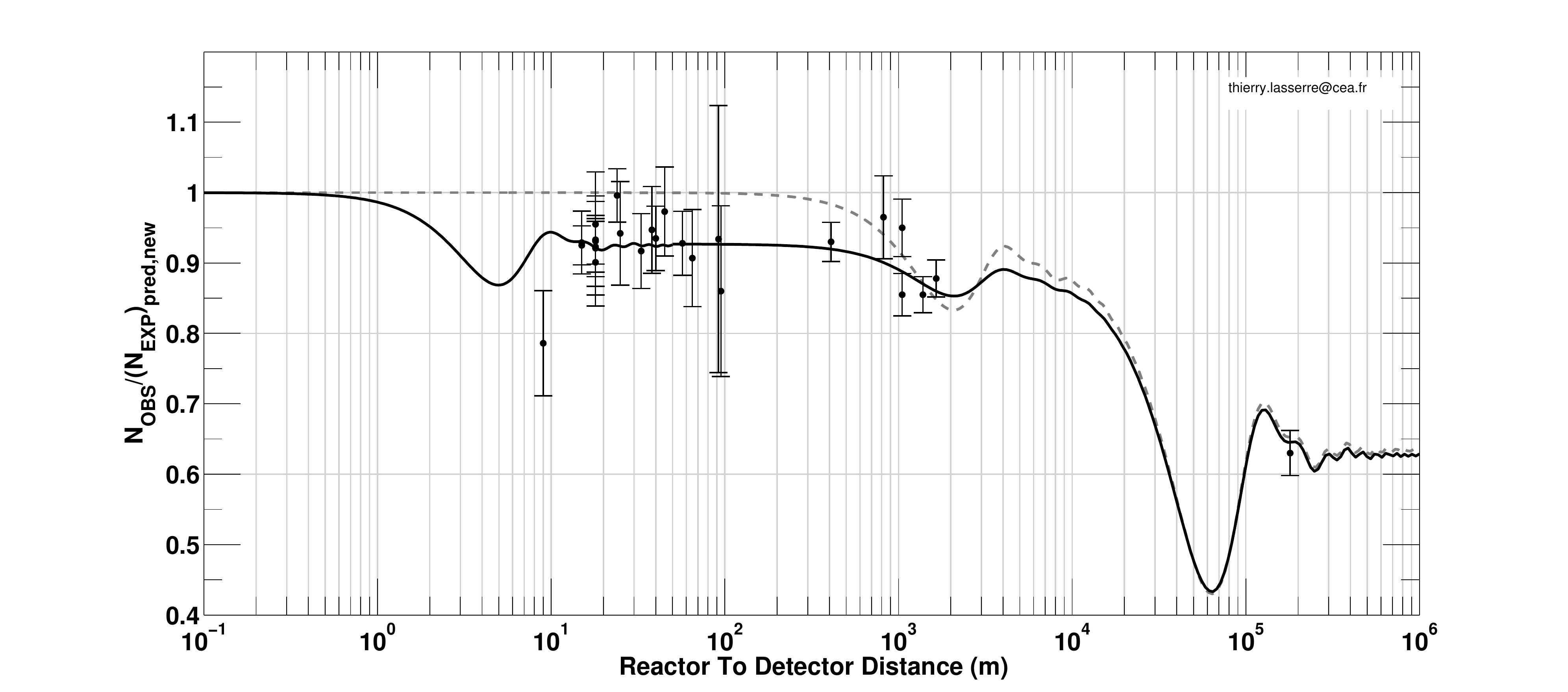}\caption{\label{fig:RAA-picture}Re-analysis of the past reactor antineutrino
experiments, which indicates an anomaly in the expected measured-to-predicted
count rates of $\bar{\nu}_{e}$ at short distances. The dashed line
corresponds to the classic three-neutrino picture with the squared
mass differences quoted above. The solid line corresponds to a 3 active
neutrinos plus one sterile neutrino (3+1) model. }

\par\end{centering}

\end{figure}

\par\end{center}

\subsection{\noindent Accelerator experiment anomalies}

\subsubsection{\noindent The LSND anomaly}

\noindent The LSND experiment was designed to search for $\mathrm{\bar{\nu}_{\mu}\rightarrow\bar{\nu}_{e}}$
oscillations at short distances and ran at the Los Alamos 800 MeV
proton accelerator, producing $\mathrm{\bar{\nu}_{\mu}}$ up to 300
MeV by $\mathrm{\mu^{+}}$ decay at rest \cite{LSND}. The LSND detector
was placed 30 m away from the $\mathrm{\bar{\nu}_{\mu}}$ source and
the detection process of $\mathrm{\bar{\nu}_{e}}$ relied on the inverse
$\mathrm{\beta}$ decay reaction $\mathrm{\bar{\nu}_{e}+p^{+}\rightarrow e^{+}+n}$.
Using the data collected between 1993 and 1998, the collaboration
reported evidence for $\mathrm{\bar{\nu}_{\mu}}$ oscillations at
the 3.8$\sigma$ level, with a total excess of 87.9 $\pm$ 22.4 $\pm$
6.0 events consistent with the $\mathrm{\bar{\nu}_{e}+p^{+}\rightarrow e^{+}+n}$
reaction above the expected background \cite{LSNDAnomaly}. In conjunction
with other known neutrino limits, the LSND data suggest that the observed
neutrino oscillations occur in the 0.2-10 $\mathrm{eV{}^{2}}$ $\mathrm{\Delta m^{2}}$
range, leading to the LSND anomaly.

\subsubsection{\noindent The MiniBooNE anomaly}

\noindent MiniBooNE is an appearance experiment which ran at Fermilab
between 2002 and 2012. It has been searching for a $\mathrm{\nu_{\mu}\rightarrow\nu_{e}}$
and $\bar{\nu}_{\mu}\rightarrow\bar{\nu}_{e}$ oscillation signal
in the LSND $\mathrm{L/E_{\nu}}$ range \cite{MiniBooNE}. The experiment
uses the Fermilab Booster neutrino beam, which produces $\nu_{\mu}$
($\bar{\nu}_{\mu}$) with energies up to 3 GeV with a 8 GeV proton
beam hitting a Beryllium target. The center of the detector is 541
m away from the neutrino source, and the $\nu_{e}$ ($\bar{\nu}_{e}$
) detection process relies on quasi-elastic charged-current scattering
on nucleons $\mathrm{\nu_{e}+C\rightarrow e^{-}+X}$ and $\mathrm{\bar{\nu}_{e}+C\rightarrow e^{+}+X}$
(CCQE). In 2007, the MiniBooNE collaboration released its first results
and showed no evidence of $\nu_{\mu}\rightarrow\nu_{e}$ oscillations
for neutrino energies above 475 MeV \cite{MiniBooNeNoOscillations}.

\noindent However in their second publication \cite{MiniBooNEAnomaly1},
the collaboration reported a sizable excess of 83.7 $\pm$ 15.1 $\pm$
19.3 electron-like events for neutrino energies between 300 MeV and
475 MeV, corresponding to a significance of 3.4$\sigma$. In the antineutrino
mode, MiniBooNE observes an excess of 24.7 $\pm$ 18.0 events in the
475 < $E_{\nu}$ < 3000 MeV energy range, which is consistent with
the antineutrino oscillations suggested by the LSND data \cite{MiniBooNEAnomaly2}.
Finally, a combined analysis of the data collected in the neutrino
and antineutrino mode recently showed an excess of 240.3 $\pm$ 34.5
$\pm$ 52.6 events, updating the significance to 3.8$\sigma$ \cite{MiniBooNEAnomaly3}.
The combined dataset favors $\Delta m^{2}$ in the 0.01-1 eV$^{2}$
range.

\subsection{\noindent Inputs from cosmology}

\noindent Cosmological data can also be used to test the sterile neutrino
hypothesis because the Universe expansion rate is sensitive to the
energy density in relativistic particles during the radiation domination
era \cite{NeutrinoWhitePaper}:

\noindent 
\begin{equation}
\mathrm{H^{2}(t)\approx\frac{8\pi G}{3}(\rho_{\gamma}+\rho_{\nu}),}
\end{equation}

\noindent where $\rho_{\gamma}$ and $\rho_{\nu}$ are the photon
and neutrino energy density, respectively and 

\noindent 
\begin{equation}
\mathrm{\rho_{\nu}=N_{eff}\frac{7\pi^{2}}{120}T_{\nu}^{4}.}
\end{equation}
$\mathrm{N_{eff}}$ could be interpreted as the effective number of
neutrino species. It is equal to 3.046 in the standard model of cosmology
with three generations of fermions. The extra energy density of neutrinos
with respect to $\mathrm{N_{eff}=3}$ comes from QED and non-instantaneous
decoupling effects in the calculation of the neutrino temperature
$\mathrm{T_{\nu}}$. From a particle physics perspective, a sterile
neutrino that mixes with active neutrinos can increase the relativistic
energy density in the early Universe through oscillation-based thermal
production, leading to a $\mathrm{N_{eff}>3.046}$. With $\rho_{\gamma}$
being extremely well determined from measurements of the cosmic microwave
background (CMB) temperature, constraints on H(t) in the early Universe
can be then interpreted as bounds on $\rho_{\nu}$ and hence on $\mathrm{N_{eff}}$. 

\noindent Various cosmological observables can be used to search for
signatures of sterile neutrinos. Those include the light elemental
abundances from big bang nucleosynthesis (BBN), the CMB anisotropies
and the large-scale structure distribution. \figref{CMB-constraints}
summarizes the best constraints on $\mathrm{N_{eff}}$ obtained with
the combination of these different cosmological observables. The recent
Planck result leads to $\mathrm{N_{eff}=3.36}\pm0.66$ at 95\% C.L.
if no additional constraint are being used\cite{Planck_Cosmo_Results}.
However including the astrophysical constraints on the Hubble constant
measurement this results shifts to $\mathrm{N_{eff}=3.52}\pm0.46$
at 95\% C.L. Therefore $\mathrm{N_{eff}>3.046}$ is mildly disfavored
but this cosmological bound is still model dependent. Further laboratory
experiments are thus mandatory to test the hypothetical existence
of a fourth neutrino state.

\noindent \begin{center}
\begin{figure}[h]
\centering{}	\includegraphics[scale=0.35]{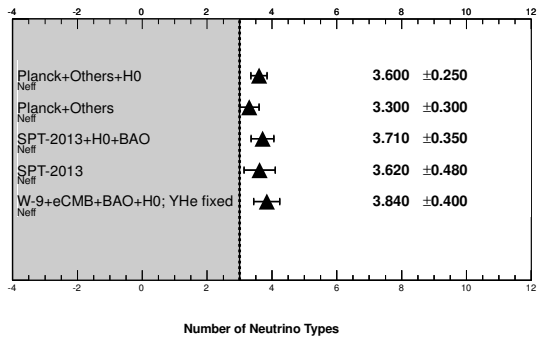}\caption{\label{fig:CMB-constraints}Constraints obtained on the number of
effective neutrino species $\mathrm{N_{eff}}$ from different cosmological
observables, based on WMAP 9-year data, PLANCK, SPT, BAO, and Hubble
constant astrophysical inputs. BAO corresponds to the baryon acoustic
oscillations data. ACT and SPT are the Atacama Cosmology Telescope
and South Pole Telescope, respectively, which measure the CMB damping
tail. Reference \cite{NeutrinoWhitePaper} gives further details about
these experiments.}
\end{figure}

\par\end{center}

\subsection{\noindent Global picture of SBL oscillations}

\noindent Table\ref{tab:tab-anomalies} summarizes the different anomalies
observed so far in past neutrino oscillation experiments. Many studies
have been performed to interpret the results of the different dataset
in terms of oscillations into a fourth neutrino state. Figure \ref{fig:Global-Fits}
shows a comparison of the allowed 95\% C.L. regions in the $\mathrm{sin^{2}(2\theta_{new})-\Delta m_{new}^{2}}$
parameter space obtained from the separate fits of the different anomalies
presented previously in the framework of a (3+1) neutrino mixing model.
A global fit to the Gallium, reactor, solar, and $\mathrm{\nu_{e}C}$
scattering data shows that two regions are preferred (in red contours),
centered around $\mathrm{\Delta m_{new}^{2}=7.6\, eV^{2},\: sin^{2}(2\theta_{new})=0.12}$
and $\mathrm{\Delta m_{new}^{2}\simeq2\, eV^{2},\, sin^{2}(2\theta_{new})\simeq0.1}$
\cite{GalliumAnomaly3}. The second region is preferred by Gallium
and reactor data.

\noindent \begin{center}
\begin{figure}[H]
\centering{}\includegraphics[scale=0.35]{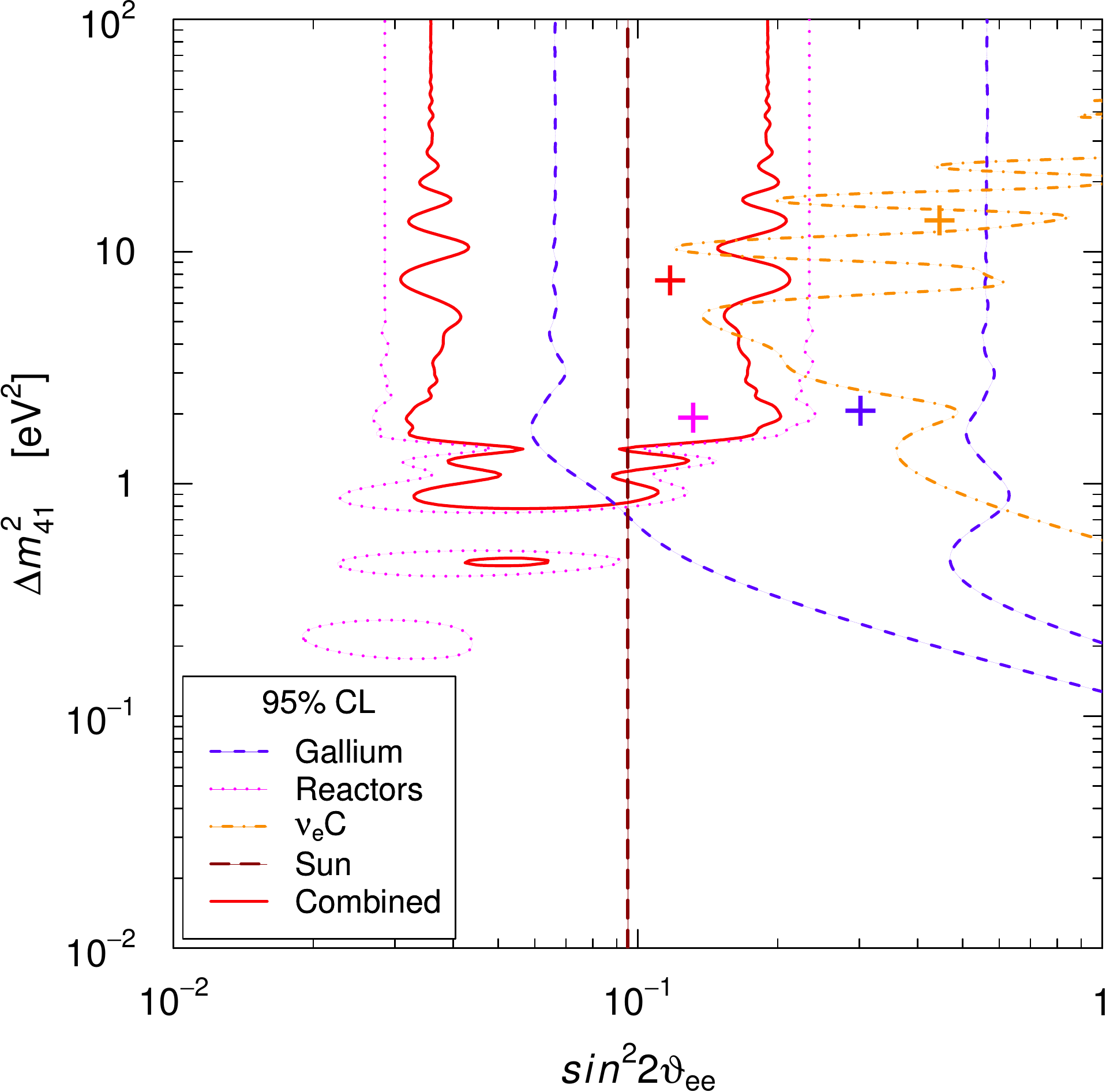}\caption{\label{fig:Global-Fits}Allowed 95\% C.L. regions in the $\mathrm{sin^{2}(2\theta_{new})-\Delta m_{new}^{2}}$
plane obtained from separate fits of Gallium, reactor, solar and $\nu_{e}C$
scattering data and from the combined fits of all data. Crosses indicate
the best-fit points. Figure taken from \cite{GalliumAnomaly3}.}
\end{figure}

\par\end{center}

\noindent \begin{center}
\begin{table}[h]
\begin{centering}
{\footnotesize }%
\begin{tabular}{cccccc}
\hline 
{\footnotesize Experiment} & {\footnotesize $\nu$ source} & {\footnotesize Channel} & {\footnotesize Detection } & {\footnotesize Oscillation signature} & {\footnotesize Significance}\tabularnewline
\hline 
\hline 
{\footnotesize LSND} & {\footnotesize $\mu$ decay at rest} & {\footnotesize $\bar{\nu}_{\mu}\rightarrow\bar{\nu}_{e}$} & {\footnotesize IBD} & {\footnotesize Rate, energy spectrum} & {\footnotesize 3.8$\sigma$}\tabularnewline
{\footnotesize MiniBooNE} & {\footnotesize $\pi$ decay in flight} & {\footnotesize $\nu_{\mu}\rightarrow\nu_{e}$, $\bar{\nu}_{\mu}\rightarrow\bar{\nu}_{e}$} & {\footnotesize CCQE} & {\footnotesize Rate, energy spectrum} & {\footnotesize 3.8$\sigma$}\tabularnewline
{\footnotesize Gallium} & {\footnotesize $^{51}Cr,$$^{37}Ar$} & {\footnotesize $\nu_{e}$ disappearance} & {\footnotesize CCQE} & {\footnotesize Rate} & {\footnotesize $\approx$ 3.0$\sigma$}\tabularnewline
{\footnotesize Reactor} & {\footnotesize $^{235,238}U$, $^{239,241}Pu$} & {\footnotesize $\overline{\nu}_{e}$ disappearance} & {\footnotesize IBD} & {\footnotesize Rate, energy spectrum} & {\footnotesize 3.0$\sigma$}\tabularnewline
{\footnotesize Cosmology} & {\footnotesize CMB, BBN, LSS, ...} & {\footnotesize All} & {\footnotesize -} & {\footnotesize $N_{eff}$} & {\footnotesize $\approx$ 2.0$\sigma$}\tabularnewline
\hline 
\end{tabular}
\par\end{centering}{\footnotesize \par}

\caption{\label{tab:tab-anomalies}Summary of the different anomalies that
possibly indicate the existence of a fourth neutrino state, along
with their corresponding significances.}

\end{table}

\par\end{center}

\noindent The results of the global fits, as well as the results of
Gallium and reactor data give lower limits on $\mathrm{\mathrm{\Delta m_{new}^{2}}}$
but do not give any upper limits. Upper limits on $\mathrm{\Delta m_{new}^{2}}$,
and even upper limits on the absolute mass scale of a new sterile
neutrino $\mathrm{m_{s}}$ can be derived from the effects of heavy
neutrino masses on electron spectrum measured in tritium $\mathrm{\mathrm{\beta}}$-decay
experiments such as Mainz \cite{Mainz1,Mainz2,Mainz3}or KATRIN \cite{KATRIN},
the results of neutrinoless double-$\mathrm{\beta}$ decay experiments
such as KamLAND-Zen \cite{KamLAND-Zen1,KamLAND-Zen3} or EXO \cite{EXO}and
cosmology. The current upper limits at the 95\% C.L. on $\mathrm{\Delta m_{new}^{2}}$
lie at $\mathcal{O}(10^{4}\, eV^{2})$ assuming a normal hierarchy
and come from the Mainz experiment\cite{Mainz3}. For Majorana neutrinos,
the recent combined results of EXO and KamLAND-Zen on the effective
Majorana mass in neutrinoless double-$\mathrm{\beta}$ decay gives
a more stringent constraint between $10^{2}$ and $10^{3}$ $\mathrm{eV^{2}}$
\cite{GalliumAnomaly3}. Finally, cosmological data have also been
used to set a 95\% C.L. upper limit on the mass of a new sterile neutrino
\cite{SterileMassLimitFromCosmo}, which exhibits some tension with
the mixing parameters preferred by the terrestrial experiment data:

\noindent 
\begin{equation}
\mathrm{m_{s}<0.45\, eV.}
\end{equation}

\noindent Nevertheless, the reported anomalies from the various terrestrial
experiments still make a compelling case for a decisive search for
a fourth neutrino state in the larger mass difference region.

\section{\noindent Concept of neutrino generator experiments}

\noindent There are two types of $\nu$-source suitable to search
for light sterile neutrinos: monochromatic $\nu_{e}$ emitters, such
as $^{51}$Cr or $^{37}$Ar and $\bar{\nu}_{e}$ emitters with a continuous
$\beta$-spectrum such as $^{144}$Ce, $^{90}$Sr, $^{42}$Ar or $^{106}$Ru.
Both types provide pure neutrino flux with well-defined energy spectrum
and absolute rate that can be precisely measured, creating an excellent
opportunity to test the fourth neutrino hypothesis using oscillometry.

\subsection{\noindent Oscillometry}

\noindent The unambiguous way to address the existence of a fourth
neutrino state is the determination of the oscillation pattern in
the distance-dependent flux of a pure neutrino flavor source over
the entire oscillation length. A detector-source distance similar
to the expected oscillation length is thus necessary:
\begin{equation}
L\,\sim\,\frac{L_{osc}[m]}{2}=1.24\,\frac{E_{\bar{\nu}_{e}}[MeV]}{\Delta m_{new}^{2}[eV^{2}]}.
\end{equation}

\noindent Electron antineutrinos produced in radioactive decays have
energies up to a few MeV. Assuming a $\Delta m_{{\rm new}}^{2}$$\sim$1
eV$^{2}$, which is preferred by the different anomalies seen in the
past terrestrial neutrino data, the optimum source-detector distance
is $\mathcal{O}(1\, m)$. Therefore, large liquid-scintillator (LS)
detectors such as Borexino \cite{Borexino}, KamLAND \cite{KamLAND},
or SNO+ \cite{SNO+} are able to perform precise neutrino oscillometry
since they are big enough to fully cover the spatial pattern of $\bar{\nu}_{e}$
oscillations into a fourth neutrino state.

\subsection{\noindent Neutrino Emitters}

\noindent In the nineties, intense $^{51}$Cr ($\sim750$ keV, $\mathcal{A_{\mathrm{0}}\,\sim}$
MCi) and $^{37}$Ar (814 keV, $\mathcal{A_{\mathrm{0}}\,\sim}$ 0.4
MCi) sources were used to test the radiochemical experiments Gallex
and Sage. Such radioactive neutrino sources involve either $\beta^{+}$-decay
or electron capture. Electron captures produce mono-energetic $\nu_{e}$'s
allowing for a determination of $L/E_{\nu}$ by only measuring the
interaction vertex position. Production of an $^{37}$Ar source requires
a large fast neutron reactor (which is not easily available), leaving
$^{51}$Cr as the best current candidate for sterile neutrino searches
with $\nu_{e}$ sources. $^{51}$Cr has a half-life of 27.7~days.
It decays 90.1\% of the time to the $^{51}$V ground state and emits
a 751~keV $\nu_{e}$ while 9.9\% of the time it decays to the first
excited state of $^{51}$V and emits a 413~keV $\nu_{e}$ followed
by a 320~keV $\gamma$. The large scale production of $^{51}$Cr
has been  possible thanks to the relatively high $^{50}$Cr neutron
capture cross-section ($\mathrm{\sim}$ 17.9~barns). Natural chromium
is primarily composed of $^{53}$Cr (9.5\%), $^{52}$Cr (83.8\%) and
$^{50}$Cr (4.35\%). The $^{53}$Cr and $^{50}$Cr isotopes have similar
neutron capture cross-section. Under the irradiation of natural chromium,
$^{53}$Cr absorbs 2.5 neutrons every time a neutron is captured on
$^{50}$Cr, thus reducing the $^{51}$Cr yield. Enrichment of natural
chromium with $^{50}$Cr is therefore necessary in order to produce
a $^{51}$Cr source with an activity of several MCi. Such enrichment
is favorable for manufacturing a compact target, which is necessary
for sterile neutrino searches. The material used by the Gallex experiment
was enriched to 38.6\% in $^{50}$Cr while the Sage target was enriched
to 92\%. Because many isotopes have high neutron capture cross-sections,
great care must be taken during the production and handling of the
chromium rods in order to avoid high-energy gamma ray contamination
from chemical impurities.

\noindent In a radiochemical experiment, the interactions of $^{51}$Cr
and $^{37}$Ar neutrinos produce $^{71}$Ge atoms via the $^{71}$Ga($\nu_{e}$,e$^{-}$)$^{71}$Ge
reaction. The $^{71}$Ge atoms are chemically extracted from the detector
and converted to GeH$_{4}$. Ge atoms are then placed in proportional
counters and their number is determined by counting the Auger electrons
released in the transition back to $^{71}$Ga, which occurs with a
half-life of 11.4~days.\\
In liquid scintillator experiments, the $\nu_{e}$ signature is provided
by $\nu_{e}$ elastic scattering off electrons (ES) of the molecules.
The cross section is $\sigma(E_{\nu})\sim0.96\times10^{-44}\times E_{\nu}\,{\rm {cm}^{2}}$,
where $E_{\nu}$ is the neutrino energy in MeV. This signature can
be mimicked by Compton scatterings induced by radioactive and cosmogenic
backgrounds, or by solar-$\nu$ interactions. An experiment running
with a $\nu_{e}$ source then imposes the use of a very high activity
source (5-10 MCi) to provide an interaction rate within the detector
that will exceed the rate from solar $\nu$, and to compensate the
lack of solid angle due to the positioning of a source outside the
detector (since the ES does not provide any specific signature of
$\nu_{e}$ interaction).

\subsection{\noindent Antineutrino Emitters}

\noindent Antineutrino sources are non-monochromatic $\bar{\nu}_{e}$
emitters decaying through $\beta^{-}$-decay. $\bar{\nu}_{e}$ are
detected through the inverse beta decay (IBD) reaction $\bar{\nu}_{e}$
+ p $\rightarrow$ $e^{+}$ + n. The IBD cross-section is $\sigma(E_{e})\sim0.96\,10^{-43}\times p_{e}E_{e}\,{\rm {cm}^{2}}$,
where $p_{e}$ and $E_{e}$ are the momentum and energy (in MeV) of
the detected $e^{+}$, neglecting neutron recoil, weak magnetism effects,
and radiative second order corrections. The IBD cross-section is an
order of magnitude higher than the neutrino scattering off electron
 cross-section at 1 MeV.

\noindent The $\bar{\nu}_{e}$ interaction signature is the time and
space coincidence of two energy depositions in the liquid scintillator.
The first energy deposition consists of $e^{+}$ kinetic energy deposition
closely followed by the energy deposition of two 511~keV $\gamma$-rays
from e$^{+}$ annihilation with an electron of the medium. It is called
the prompt event, releasing a total visible energy of E$_{e}$= E$_{\nu}$-($m_{n}$-$m_{p}$)~MeV.
The second energy deposition originates from the neutron capture on
a free proton, with an averaged capture time of the order of a few
hundreds $\mathrm{\mu s}$, which releases a 2.2 MeV deexcitation
$\gamma$-ray. This is called the delayed event. The time coincidence
between these two energy depositions suppresses any non-source backgrounds
to a negligible level.

\noindent The background free IBD signature plus the relatively high
IBD cross-section allow the source activity to be at the (tens of)
kCi scale for sterile neutrino searches. A suitable $\bar{\nu}_{e}$
source must have a $Q_{\beta}$ greater than 1.8 MeV to exceed the
IBD energy threshold and a long enough lifetime ($\gtrsim$1 month)
not to be limited by production and transportation time to the detector.
Fulfillment of these two requirements is impossible for a single nuclei,
since $Q_{\beta}$ is anti-correlated with lifetime. We then looked
for pairs of nuclei involving a long-lived low-Q nucleus decaying
to a short-lived high-Q nucleus. Since the first isotope, the parent,
has a much longer lifetime than its daughter, both nuclei are in equilibrium
and the activity is driven by the parent. $\bar{\nu}_{e}$ emitted
by the father has not enough energy to undergo an IBD, while $\mathrm{\bar{\nu}_{e}}$
in the tail of the daughter spectrum can:

\noindent 
\begin{equation}
_{Z}^{A}X_{1}\xrightarrow[Q_{\beta}\sim100\,\mathrm{keV}]{\tau_{1/2}\sim\mathrm{year}}\:_{Z+1}^{\phantom{1}\phantom{1}A}X_{2}\xrightarrow[Q_{\beta}>1.8\,\mathrm{MeV}]{\tau_{1/2}\sim\mathrm{hour}}\:_{Z+2}^{\phantom{1}\phantom{1}A}X_{3}
\end{equation}

\noindent Four pairs fulfilling the above mentioned requirements have
been identified in the nuclear databases and are presented in \tabref{Possible-nuclei-couple},
some of them also being reported in \cite{key-2}.

\noindent \begin{center}
\begin{table}[h]
\begin{centering}
\begin{tabular}{ccccccccc}
\hline 
\rule{0pt}{2.25ex} & \multicolumn{2}{c}{$^{144}$Ce-$^{144}$Pr} & \multicolumn{2}{c}{$^{106}$Ru-$^{106}$Rh} & \multicolumn{2}{c}{$^{90}$Sr-$^{90}$Y} & \multicolumn{2}{c}{$^{42}$Ar-$^{42}$K}\tabularnewline
\hline 
\hline 
$\tau_{1/2}$ & 285 d & 7.2 min & 372 d & 30 s & 28.9 y & 64 h & 32.9 y & 12 h\tabularnewline
$Q_{\beta}$ (MeV) & 0.319 & 2.996 & 0.039 & 3.54 & 0.546 & 2.28 & 0.599 & 3.52\tabularnewline
\hline 
\end{tabular}
\par\end{centering}

\caption{\label{tab:Possible-nuclei-couple}Possible nuclei pairs for a $\bar{\nu}_{e}$
source.}
\end{table}

\par\end{center}

\noindent $^{144}$Ce, $^{106}$Ru and $^{90}$Sr are fission products
found within nuclear reactor spent fuels, whereas $^{42}$Ar is too
light to be produced by fission processes. Furthermore, the production
of $^{42}$Ar needs two neutron captures, starting from stable $^{40}$Ar.
Because $^{41}$Ar has a short lifetime of only 110 min and $^{40}$Ar
(n,$\gamma$)$^{41}$Ar reaction has a cross-section of only $\sim$0.4
barn, massive $^{42}$Ar production is very difficult. Therefore,
a solution consisting in the reprocessing and treatment of nuclear
spent fuel rods is more convenient.

\noindent The probability of a nuclei to be created during fission
processes in nuclear reactor fuel is called the fission yield and
is given in percent %
\footnote{The sum of the isotopic fission yield gives 200 \% because of the
normalization with the number of fissions, not with the number of
fissions products.%
}. However, the fission yield of a given nucleus does not take into
account its possible decay to another nucleus, and is therefore not
representative of the abundance of a given nuclei within spent nuclear
fuel. The cumulative fission yield of an isotope is the number of
nuclei of the considered isotope per fission when the reactor is at
equilibrium. This quantity is then more interesting for our purpose
of finding the best pair candidate because with a reasonable approximation
(very long-lived isotopes never reach equilibrium), it represents
a fission product abundance at reactor shutdown. Table \ref{tab:Cumulative-fission-yield}
shows the $^{144}$Ce, $^{106}$Ru and $^{90}$Sr cumulative fission
yields for thermal fission of $^{235}$U and $^{239}$Pu, these two
isotopes being the most fissioned isotopes in nuclear reactors. $^{144}$Ce,
$^{106}$Ru and $^{90}$Sr are clearly abundant fission products,
but $^{106}$Ru suffers from its rather low $^{235}$U cumulative
yield compared to $^{144}$Ce and $^{90}$Sr. Since $^{90}$Sr-$^{90}$Y
suffers from a relatively low $Q_{\beta}$ leading to a small IBD
reaction rate for a given activity, the $\mathrm{^{144}Ce-^{144}Pr}$
pair is therefore the best candidate for a $\bar{\nu}_{e}$ source.

\noindent \begin{center}
\begin{table}[h]
\begin{centering}
\begin{tabular}{cccc}
\hline 
\rule{0pt}{2.25ex} & \multicolumn{3}{c}{Cumulative fission yield (\%)}\tabularnewline
\hline 
\hline 
\rule{0pt}{2.25ex} & $_{58}^{144}$Ce & $_{44}^{106}$Ru & $_{38}^{90}$Sr\tabularnewline
$^{235}$U & 5.50 (4) & 0.401 (6) & 5.78 (6)\tabularnewline
$^{239}$Pu & 3.74 (3) & 4.35 (9) & 2.10 (4)\tabularnewline
\hline 
\end{tabular}
\par\end{centering}

\caption{\label{tab:Cumulative-fission-yield}Cumulative fission yield for
thermal fission of $^{235}$U / $^{239}$Pu for $^{144}$Ce, $^{106}$Ru
and $^{90}$Sr \cite{NNDC}.}
\end{table}

\par\end{center}

\section{\noindent $\mathrm{^{144}Ce-^{144}Pr}$ antineutrino generator}

\noindent The CeLAND experiment will use a $\mathrm{^{144}Ce-^{144}Pr}$
source, which is the best compromise between production feasibility,
intrinsic backgrounds and $\bar{\nu}_{e}$ interaction rate through
the inverse beta decay reaction.

\subsection{\noindent $\mathrm{^{144}Ce-^{144}Pr}$ source features}

\noindent With an atomic number Z=58, cerium belongs to the lanthanide
elements. Natural cerium has an atomic mass of 140.116 g and is a
soft, silvery, ductile metal which easily oxidizes in air. Cerium
has two oxidation states (III and IV) and easily converts to CeO$_{2}$
in contact with oxygen.. Natural cerium contains four stable isotopes
of mass 136 (0.195\%), 138 (0.265\%) and 140 (88.45\%) and 142 (11.1\%),
respectively. Cerium is the most abundant rare earth element, making
up about 0.0046\% of the Earth's crust by weight.

\subsubsection{\noindent Nuclear data}

\noindent There is a considerable interest in the decay of $^{144}$Ce
since it is a prominent fission product. Nuclear data reported below
are taken from the ENSDF database referring to \cite{key-12,key-13}
for $^{144}$Ce and \cite{key-13} for $^{144}$Pr.

\noindent \begin{center}
\begin{table}[h]
\begin{centering}
\begin{tabular}{ccccccc}
\hline 
\rule{0pt}{2.25ex}F. Y. & $\mathrm{t_{1/2}}$ & $\mathrm{1^{st}\,\beta^{-}}$ (keV) & $\mathrm{2^{nd}\,\beta^{-}}$ (keV) & $\mathrm{I_{\gamma>1\, MeV}}$ & $\mathrm{I_{\gamma>2\, MeV}}$ & W/kCi\tabularnewline
\hline 
\hline 
$\mathrm{^{235}U}$: 5.2\% & 284.91d & 318 (76\%) & \rule{0pt}{2.25ex} &  &  & \tabularnewline
 &  & 184 (20\%) & 2996 (99\%) & 1380 (0.007\%) & 2185 & 7.99\tabularnewline
$\mathrm{^{239}Pu}$: 3.7\% &  & 238 (4\%) & 810 (1\%) & 1489 (0.28 \%) & (0.7\%) & \tabularnewline
\hline 
\end{tabular}
\par\end{centering}

\centering{}\caption{\label{tab:sourcefeatures}Summary of the features of $\mathrm{^{144}Ce-^{144}Pr}$
pair.  F.Y. are the fission yields of $\mathrm{^{144}Ce}$, $\mathrm{t_{1/2}}$
the half-lives. $\mathrm{\beta}$ end-points are given for $\mathrm{1^{st}}$
and $\mathrm{2^{nd}}$ nucleus in the pair. The $\mathrm{I_{\gamma}}$'s
are the intensities of the main $\mathrm{\gamma}$-rays per $\mathrm{\beta}$-decay
above 1 and 2 MeV. The last column is the heat production in units
of W/kCi.}
\end{table}

\par\end{center}

\noindent The different cerium isotopes are listed in table \ref{tab:ceisotopes}.
Up to now, 35 radioisotopes have been characterized, the most stable
ones being $^{144}$Ce with a half-life of 284.91 days, $^{139}$Ce
with a half-life of 137.640 days, and $^{141}$Ce with a half-life
of 32.501 days. The remaining radioactive isotopes have half-lives
shorter than 4 days, most of them having half-lives shorter than 10
minutes. Note that the $^{136,138,142}$Ce isotopes are predicted
to undergo double beta decay but this process has never been observed. 

\noindent \begin{center}
\begin{table}[H]
\begin{centering}
{\footnotesize }%
\begin{tabular}{cccccccc}
\hline 
{\footnotesize Parent } & {\footnotesize Z } & {\footnotesize A } & {\footnotesize isotopic mass (u) } & {\footnotesize half-life } & {\footnotesize decay } & {\footnotesize daughter } & {\footnotesize spin/parity}\tabularnewline
\hline 
\hline 
{\footnotesize $^{119}$Ce} & {\footnotesize 58} & {\footnotesize 61} & {\footnotesize 118.95276(64)} & {\footnotesize 200 ms} & {\footnotesize $\beta^{+}$} & {\footnotesize $^{119}$La } & {\footnotesize 5/2+ }\tabularnewline
{\footnotesize $^{120}$Ce} & {\footnotesize 58} & {\footnotesize 62} & {\footnotesize 119.94664(75)} & {\footnotesize 250 ms } & {\footnotesize $\beta^{+}$ } & {\footnotesize $^{120}$La } & {\footnotesize 0+ }\tabularnewline
{\footnotesize $^{121}$Ce} & {\footnotesize 58} & {\footnotesize 63} & {\footnotesize 120.94342(54)} & {\footnotesize 1.1(1) s } & {\footnotesize $\beta^{+}$ } & {\footnotesize $^{121}$La } & {\footnotesize (5/2)(+) }\tabularnewline
{\footnotesize $^{122}$Ce} & {\footnotesize 58} & {\footnotesize 64} & {\footnotesize 121.93791(43)} & {\footnotesize 2 s } & {\footnotesize $\beta^{+}$ } & {\footnotesize $^{121,122}$La } & {\footnotesize 0+}\tabularnewline
{\footnotesize $^{123}$Ce} & {\footnotesize 58} & {\footnotesize 65} & {\footnotesize 122.93540(32)} & {\footnotesize 3.8(2) s } & {\footnotesize $\beta^{+}$,p } & {\footnotesize $^{123}$La, $^{122}$Ba } & {\footnotesize (5/2)(+)}\tabularnewline
{\footnotesize $^{124}$Ce} & {\footnotesize 58} & {\footnotesize 66} & {\footnotesize 123.93041(32)} & {\footnotesize 9.1(12) s } & {\footnotesize $\beta^{+}$ } & {\footnotesize $^{124}$La } & {\footnotesize 0+}\tabularnewline
{\footnotesize $^{125}$Ce} & {\footnotesize 58} & {\footnotesize 67} & {\footnotesize 124.92844(21)} & {\footnotesize 9.3(3) s } & {\footnotesize $\beta^{+}$,p } & {\footnotesize $^{125}$La,$^{124}$Ba } & {\footnotesize (7/2-)}\tabularnewline
{\footnotesize $^{126}$Ce} & {\footnotesize 58} & {\footnotesize 68} & {\footnotesize 125.92397(3)} & {\footnotesize 51.0(3) s } & {\footnotesize $\beta^{+}$ } & {\footnotesize $^{126}$La } & {\footnotesize 0+ }\tabularnewline
{\footnotesize $^{127}$Ce} & {\footnotesize 58} & {\footnotesize 69} & {\footnotesize 126.92273(6)} & {\footnotesize 29(2) s } & {\footnotesize $\beta^{+}$ } & {\footnotesize $^{127}$La } & {\footnotesize 5/2+}\tabularnewline
{\footnotesize $^{128}$Ce} & {\footnotesize 58} & {\footnotesize 70} & {\footnotesize 127.91891(3)} & {\footnotesize 3.93(2) min } & {\footnotesize $\beta^{+}$ } & {\footnotesize $^{128}$La } & {\footnotesize 0+ }\tabularnewline
{\footnotesize $^{129}$Ce} & {\footnotesize 58} & {\footnotesize 71} & {\footnotesize 128.91810(3)} & {\footnotesize 3.5(3) min } & {\footnotesize $\beta^{+}$ } & {\footnotesize $^{129}$La } & {\footnotesize (5/2+) }\tabularnewline
{\footnotesize $^{130}$Ce} & {\footnotesize 58} & {\footnotesize 72} & {\footnotesize 129.91474(3)} & {\footnotesize 22.9(5) min } & {\footnotesize $\beta^{+}$ } & {\footnotesize $^{130}$La } & {\footnotesize 0+}\tabularnewline
{\footnotesize $^{130m}$Ce} & \multicolumn{3}{c}{{\footnotesize 2453.6(3) keV}} & {\footnotesize 100(8) ns } &  &  & {\footnotesize (7-) }\tabularnewline
{\footnotesize $^{131}$Ce} & {\footnotesize 58} & {\footnotesize 73} & {\footnotesize 130.91442(4)} & {\footnotesize 10.2(3) min } & {\footnotesize $\beta^{+}$ } & {\footnotesize $^{131}$La } & {\footnotesize (7/2+)}\tabularnewline
{\footnotesize $^{131m}$Ce} & \multicolumn{3}{c}{{\footnotesize 61.8(1) keV}} & {\footnotesize 5.0(10) min } & {\footnotesize $\beta^{+}$ } & {\footnotesize $^{131}$La } & {\footnotesize (1/2+) }\tabularnewline
{\footnotesize $^{132}$Ce} & {\footnotesize 58} & {\footnotesize 74} & {\footnotesize 131.911460(22)} & {\footnotesize 3.51(11) h } & {\footnotesize $\beta^{+}$ } & {\footnotesize $^{132}$La } & {\footnotesize 0+ }\tabularnewline
{\footnotesize $^{132m}$Ce} & \multicolumn{3}{c}{{\footnotesize 2340.8(5) keV}} & {\footnotesize 9.4(3) ms } & {\footnotesize IT} & {\footnotesize $^{132}$Ce } & {\footnotesize (8-) }\tabularnewline
{\footnotesize $^{133}$Ce} & {\footnotesize 58} & {\footnotesize 75} & {\footnotesize 132.911515(18)} & {\footnotesize 97(4) min } & {\footnotesize $\beta^{+}$ } & {\footnotesize $^{133}$La } & {\footnotesize 1/2+}\tabularnewline
{\footnotesize $^{133m}$Ce} & \multicolumn{3}{c}{{\footnotesize 37.1(8) keV}} & {\footnotesize 4.9(4) h } & {\footnotesize $\beta^{+}$ } & {\footnotesize $^{133}$La } & {\footnotesize 9/2- }\tabularnewline
{\footnotesize $^{134}$Ce} & {\footnotesize 58} & {\footnotesize 76} & {\footnotesize 133.908925(22)} & {\footnotesize 3.16(4) d } & {\footnotesize EC } & {\footnotesize $^{134}$La } & {\footnotesize 0+ }\tabularnewline
{\footnotesize $^{135}$Ce} & {\footnotesize 58} & {\footnotesize 77} & {\footnotesize 134.909151(12)} & {\footnotesize 17.7(3) h } & {\footnotesize $\beta^{+}$ } & {\footnotesize $^{135}$La } & {\footnotesize 1/2(+) }\tabularnewline
{\footnotesize $^{135m}$Ce} & \multicolumn{3}{c}{{\footnotesize 445.8(2) keV}} & {\footnotesize 20(1) s } & {\footnotesize IT } & {\footnotesize $^{135}$Ce} & {\footnotesize (11/2-) }\tabularnewline
{\footnotesize $^{136}$Ce} & {\footnotesize 58} & {\footnotesize 78} & {\footnotesize 135.907172(14)} & \multicolumn{3}{c}{{\footnotesize Observationally Stable}} & {\footnotesize 0+ }\tabularnewline
{\footnotesize $^{136m}$Ce} & \multicolumn{3}{c}{{\footnotesize 3095.5(4) keV}} & {\footnotesize 2.2(2) $\mu$s } &  & {\footnotesize 10+ } & \tabularnewline
{\footnotesize $^{137}$Ce} & {\footnotesize 58} & {\footnotesize 79} & {\footnotesize 136.907806(14)} & {\footnotesize 9.0(3) h } & {\footnotesize $\beta^{+}$ } & {\footnotesize $^{137}$La } & {\footnotesize 3/2+ }\tabularnewline
{\footnotesize $^{137m}$Ce} & \multicolumn{3}{c}{{\footnotesize 254.29(5) keV}} & {\footnotesize 34.4(3) h } & {\footnotesize IT(99.22\%)/$\beta^{+}$(.779\%) } & {\footnotesize $^{137}$Ce/$^{137}$La} & {\footnotesize 11/2- }\tabularnewline
{\footnotesize $^{138}$Ce} & {\footnotesize 58} & {\footnotesize 80} & {\footnotesize 137.905991(11)} & \multicolumn{3}{c}{{\footnotesize Observationally Stable}} & {\footnotesize 0+ }\tabularnewline
{\footnotesize $^{138m}$Ce} & \multicolumn{3}{c}{{\footnotesize 2129.17(12) keV}} & {\footnotesize 8.65(20) ms } & {\footnotesize IT } & {\footnotesize $^{138}$Ce } & {\footnotesize 7- }\tabularnewline
{\footnotesize $^{139}$Ce} & {\footnotesize 58} & {\footnotesize 81} & {\footnotesize 138.906653(8) } & {\footnotesize 137.641(20) d } & {\footnotesize EC } & {\footnotesize $^{139}$9a } & {\footnotesize 3/2+ }\tabularnewline
{\footnotesize $^{139m}$Ce} & \multicolumn{3}{c}{{\footnotesize 754.24(8) keV}} & {\footnotesize 56.54(13) s } & {\footnotesize IT } & {\footnotesize $^{139}$Ce } & {\footnotesize 11/2- }\tabularnewline
{\footnotesize $^{140}$Ce} & {\footnotesize 58} & {\footnotesize 82} & {\footnotesize 139.9054387(26)} & \multicolumn{3}{c}{{\footnotesize Stable}} & {\footnotesize 0+ }\tabularnewline
{\footnotesize $^{140m}$Ce} & \multicolumn{3}{c}{{\footnotesize 2107.85(3) keV }} & {\footnotesize 7.3(15) $\mu$s } &  &  & {\footnotesize 6+ }\tabularnewline
{\footnotesize $^{141}$Ce} & {\footnotesize 58} & {\footnotesize 83} & {\footnotesize 140.9082763(26)} & {\footnotesize 32.508(13) d } & {\footnotesize $\beta^{-}$ } & {\footnotesize $^{141}$Pr } & {\footnotesize 7/2- }\tabularnewline
{\footnotesize $^{142}$Ce} & {\footnotesize 58} & {\footnotesize 84} & {\footnotesize 141.909244(3) } & \multicolumn{3}{c}{{\footnotesize Observationally Stable}} & {\footnotesize 0+ }\tabularnewline
{\footnotesize $^{143}$Ce} & {\footnotesize 58} & {\footnotesize 85} & {\footnotesize 142.912386(3) } & {\footnotesize 33.039(6) h } & {\footnotesize $\beta^{-}$ } & {\footnotesize $^{143}$Pr } & {\footnotesize 3/2- }\tabularnewline
\textbf{\footnotesize $^{144}$Ce} & {\footnotesize 58} & {\footnotesize 86} & {\footnotesize 143.913647(4) } & {\footnotesize 284.91(5) d } & {\footnotesize $\beta^{-}$ } & \textbf{\footnotesize $^{144m}$Pr}{\footnotesize{} } & {\footnotesize 0+ }\tabularnewline
{\footnotesize $^{145}$Ce} & {\footnotesize 58} & {\footnotesize 87} & {\footnotesize 144.91723(4)} & {\footnotesize 3.01(6) min } & {\footnotesize $\beta^{-}$ } & {\footnotesize $^{145}$Pr } & {\footnotesize (3/2-) }\tabularnewline
{\footnotesize $^{146}$Ce} & {\footnotesize 58} & {\footnotesize 88} & {\footnotesize 145.91876(7)} & {\footnotesize 13.52(13) min } & {\footnotesize $\beta^{-}$ } & {\footnotesize $^{146}$Pr } & {\footnotesize 0+ }\tabularnewline
{\footnotesize $^{147}$Ce} & {\footnotesize 58} & {\footnotesize 89} & {\footnotesize 146.92267(3)} & {\footnotesize 56.4(10) } & {\footnotesize $\beta^{-}$ } & {\footnotesize $^{147}$Pr } & {\footnotesize (5/2-) }\tabularnewline
{\footnotesize $^{148}$Ce} & {\footnotesize 58} & {\footnotesize 90} & {\footnotesize 147.92443(3)} & {\footnotesize 56(1) s } & {\footnotesize $\beta^{-}$ } & {\footnotesize $^{148}$Pr } & {\footnotesize 0+ }\tabularnewline
{\footnotesize $^{149}$Ce} & {\footnotesize 58} & {\footnotesize 91} & {\footnotesize 148.9284(1)} & {\footnotesize 5.3(2) s } & {\footnotesize $\beta^{-}$ } & {\footnotesize $^{149}$Pr } & {\footnotesize (3/2-) }\tabularnewline
{\footnotesize $^{150}$Ce} & {\footnotesize 58} & {\footnotesize 92} & {\footnotesize 149.93041(5)} & {\footnotesize 4.0(6) s } & {\footnotesize $\beta^{-}$ } & {\footnotesize $^{150}$Pr } & {\footnotesize 0+ }\tabularnewline
{\footnotesize $^{151}$Ce} & {\footnotesize 58} & {\footnotesize 93} & {\footnotesize 150.93398(11) } & {\footnotesize 1.02(6) s } & {\footnotesize $\beta^{-}$ } & {\footnotesize $^{151}$Pr } & {\footnotesize 3/2- }\tabularnewline
{\footnotesize $^{152}$Ce} & {\footnotesize 58} & {\footnotesize 94} & {\footnotesize 151.93654(21)} & {\footnotesize 1.4(2) s } & {\footnotesize $\beta^{-}$ } & {\footnotesize $^{152}$Pr } & {\footnotesize 0+ }\tabularnewline
{\footnotesize $^{153}$Ce} & {\footnotesize 58} & {\footnotesize 95} & {\footnotesize 152.94058(43)} & {\footnotesize 500 ms } & {\footnotesize $\beta^{-}$ } & {\footnotesize $^{153}$Pr } & {\footnotesize 3/2- }\tabularnewline
{\footnotesize $^{154}$Ce} & {\footnotesize 58} & {\footnotesize 96} & {\footnotesize 153.94342(54)} & {\footnotesize 300 ms } & {\footnotesize $\beta^{-}$ } & {\footnotesize $^{154}$Pr } & {\footnotesize 0+ }\tabularnewline
{\footnotesize $^{155}$Ce} & {\footnotesize 58} & {\footnotesize 97} & {\footnotesize 154.94804(64)} & {\footnotesize 200 ms } & {\footnotesize $\beta^{-}$ } & {\footnotesize $^{155}$Pr } & {\footnotesize 5/2- }\tabularnewline
{\footnotesize $^{156}$Ce} & {\footnotesize 58} & {\footnotesize 98} & {\footnotesize 155.95126(64)} & {\footnotesize 150 ms } & {\footnotesize $\beta^{-}$ } & {\footnotesize $^{156}$Pr } & {\footnotesize 0+ }\tabularnewline
{\footnotesize $^{157}$Ce} & {\footnotesize 58} & {\footnotesize 99} & {\footnotesize 156.95634(75)} & {\footnotesize 50 ms } & {\footnotesize $\beta^{-}$ } & {\footnotesize $^{157}$Pr } & {\footnotesize 7/2+}\tabularnewline
\hline 
\end{tabular}
\par\end{centering}{\footnotesize \par}

\caption{\label{tab:ceisotopes} Cerium isotopes. EC refers to Electron capture
and IT to Isomeric transition. There are 10 meta states.}
\end{table}

\par\end{center}

\noindent The half-life of the $^{144}$Ce, 284.91 days ($\tau$ =
411.04 days), is long enough for uploading of irradiated fuel from
the nuclear reactor, transportation to the reprocessing facility,
extraction of cerium, cerium source packaging and transportation to
the detector site. It is however not too long, minimizing the mass
of reprocessed fuel and source material within the detector. The $^{144}$Ce
is therefore a very good compromise, suitable for a year time scale
experiment. The source compactness requirement imposes the time elapsed
prior to the source manufacturing time. The time period must be shorter
than a few years (typically three) starting from the end of the nuclear
fuel irradiation within the reactor core. 

\noindent \begin{center}
\begin{figure}[h]
\begin{centering}
\includegraphics[scale=0.19]{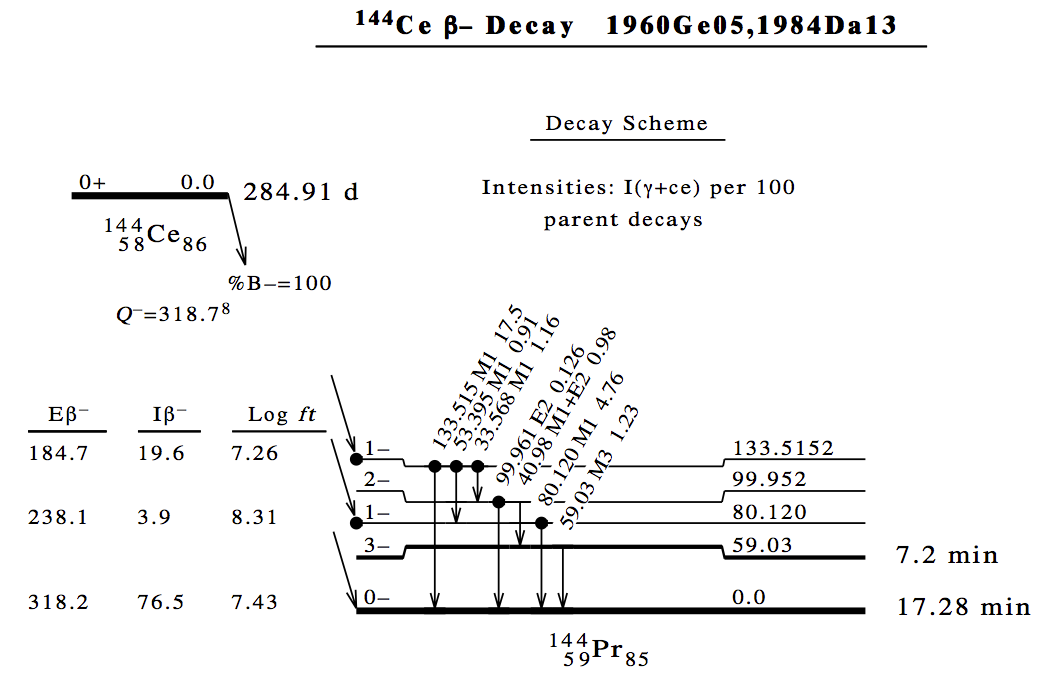}\includegraphics[scale=0.19]{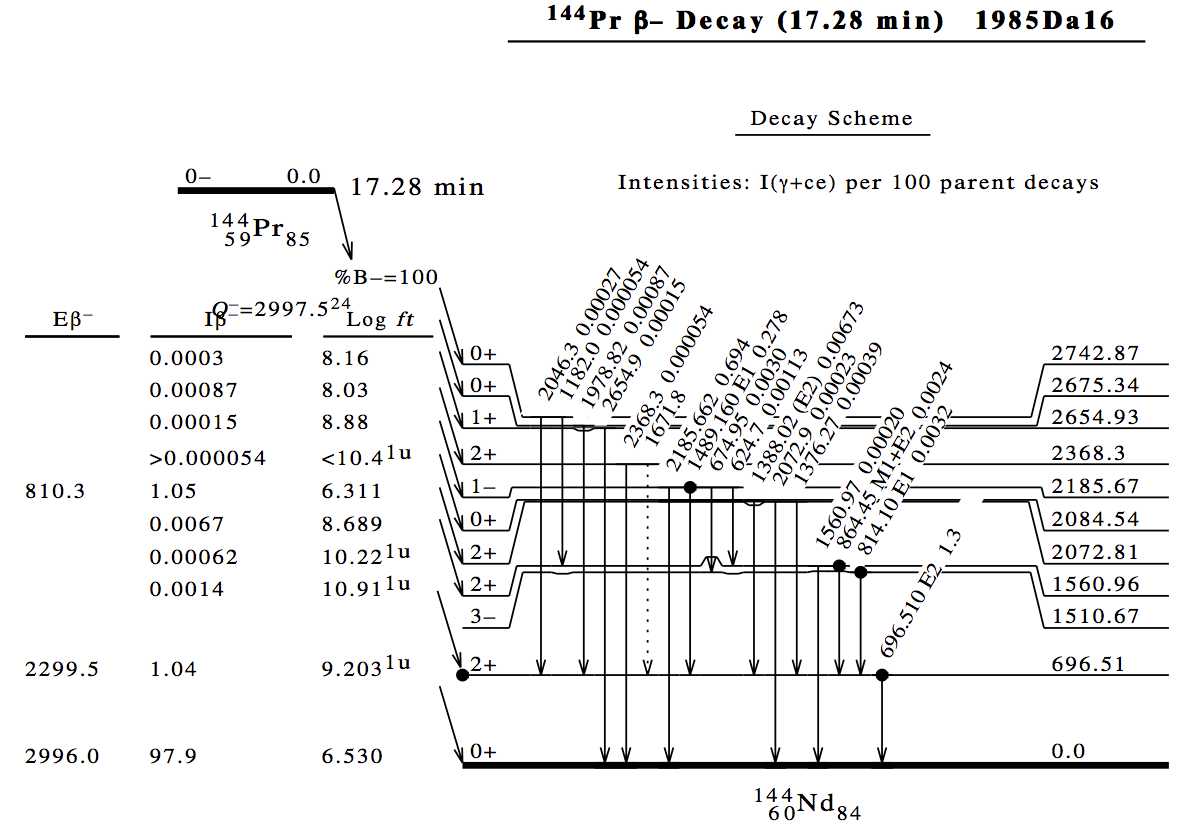}
\par\end{centering}

\caption{(left) \label{fig-144ceds}Decay scheme of $^{144}$Ce. The two intensities
sum to give total transition intensity from each energy level in percent.
Energies of $\gamma$ rays and levels are given in keV. (right) \label{fig-144prds}Decay
schemes of the 17.3-min $^{144}$Pr. Energies of $\gamma$ rays and
levels are given in keV.}

\end{figure}

\par\end{center}

\noindent A major concern regarding an $\bar{\nu}_{e}$ source is
the rate and energy of accompanying $\gamma$-rays because it is a
source of backgrounds for the detection of $\bar{\nu}_{e}$, and requires
substantial shielding to comply with safety rules for anybody working
in the vicinity of the source during production, transportation, and
deployment. For an artificial $\bar{\nu}_{e}$ source, the main channel
of $\gamma$-ray background is the long-lived $\gamma$-emitting isotopes
of the same chemical element simultaneously extracted from irradiated
nuclear fuel. In case of cerium, all isotopes except $^{139}$Ce and
$^{144}$Ce (see table \ref{tab:ceisotopes}) decay fast enough not
to be sources of $\mathrm{\gamma}$ backgrounds for the CeLAND experiment.
$^{139}$Ce cumulative fission yield is only $2.10^{-9}$ for $^{239}$Pu
and $9.10^{-12}$ for $^{235}$U, at least 7 orders of magnitude lower
than $^{144}$Ce. Therefore $^{139}$Ce gamma background yield is
negligible. The main concern then arises from the decay of $^{144}$Ce
and its daughter $^{144}$Pr deexcitation. The decay scheme of $^{144}$Ce
and of $^{144}$Pr ground and exited states are shown in \figref{cepremitspectr}
and \ref{fig-144prds}, respectively. The most dangerous source of
background originates from the energetic 2185 keV $\gamma$ produced
by the decay through excited states of $^{144}$Pr (see also table
\ref{tab:CePrDecayHeat}). This $\gamma$ line has a branching ratio
of 0.7\%. Starting from a $\beta$ activity of 75 kCi, this leads
to a huge $\gamma$ activity of 525 Ci (1.9 10$^{13}$ Bq) that must
be attenuated through a thick high-Z shielding material, such as tungsten
alloy. In addition, it is necessary to guarantee a low level of radio-impurities
containing long-lived $\gamma$ radioisotopes during the cerium chemical
extraction procedure. Specifically, rate of $\gamma$-rays above 1
MeV from impurities must be smaller than 10\% of the 2185 keV $\gamma$
rate, leading to an activity smaller than 50 Ci/75 kCi, ensuring that
this gamma background is negligible.

\noindent \begin{center}
\begin{figure}[h]
\begin{centering}
\includegraphics[scale=0.5]{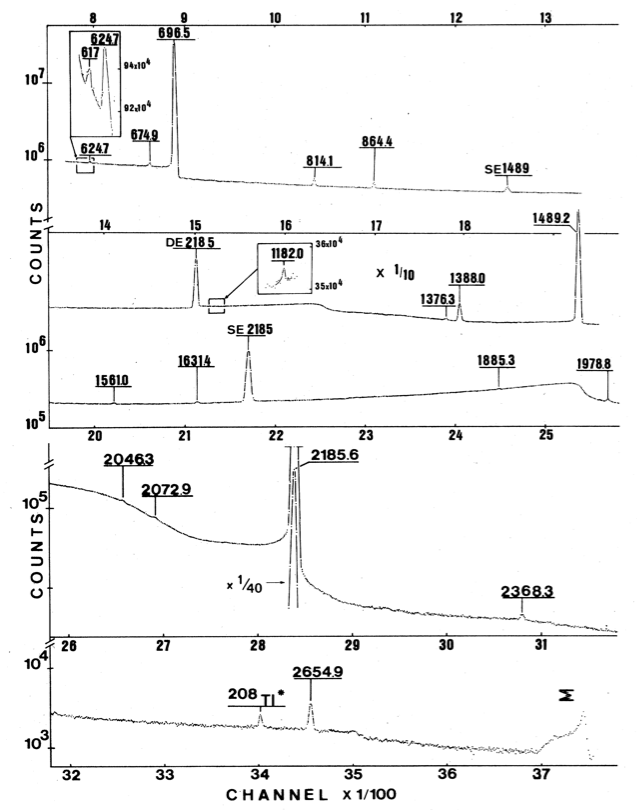}
\par\end{centering}

\caption{$\gamma$ ray spectrum of $^{144m}$Pr isomers in equilibrium with$^{144}$Ce,
measured with a 17\% HPGe detector \cite{key-14}. Energies are in
keV; the energy dispersion is 0.76 keV/channel. SE and DE stand for
single and double escape peaks, respectively. An asterisk denotes
a $\gamma$ ray from the background; the $\Sigma$ symbol indicates
summing a peak from 2185.6 and 696.5 keV $\gamma$ rays.}

\end{figure}

\par\end{center}

\subsubsection{\noindent Decay heat}

\noindent The $\mathrm{^{144}Ce-^{144}Pr}$ pair releases heat through
the collisions of $\beta$ and $\gamma$ radiations with atoms. Details
of the $\mathrm{^{144}Ce-^{144}Pr}$ decay chain and heat production
per decay branch are given in \tabref{CePrDecayHeat}. A $\mathrm{^{144}Ce-^{144}Pr}$
source releases $7.991\pm0.044$ Watt/kCi (0.56 \%) of thermal power
per units of activity, mostly coming from $\mathrm{\beta}$ decay
electrons (95.72 \%) contained in the cerium oxide. 

\noindent A 75 kCi activity therefore corresponds to an overall 599.3
W thermal power. For CeLAND phase 1, the heat will be dissipated within
the 3 kilotons of water circulating in the outer detector. Heat release
as a function of time (and decreasing activity) is given in \tabref{CePrHeatvsTime}.
Starting with a 75 kCi source at the KamLAND site, thermal power decreases
down to 387 W after 6 months (roughly corresponding to a 50 kCi activity).
Two years after the start of the data taking with an initial 75 kCi
source, the heat release drops down to 101 W (corresponding to about
13 kCi). After 10 years of running with an initial 75 kCi source,
the heat release should be roughly 0.1 W, corresponding to an activity
of 10 Ci.

\noindent \begin{center}
\begin{table}[H]
\begin{centering}
{\small }%
\begin{tabular}{lcccc}
\hline 
 & {\small End-point} & {\small Energy} & {\small Intensity} & {\small Dose}\tabularnewline
 & {\small energy (keV)} & {\small (keV)} & (\%) & {\small (keV / Bq)}\tabularnewline
\hline 
\hline 
\multicolumn{5}{l}{{\small $\mathrm{^{144}Ce\,\beta}$ branches}\rule{0pt}{2.25ex}}\tabularnewline
\hline 
 & 318.7 (8) & 91.1 (7) & 76.5 (5) & 69.7 (7)\tabularnewline
 & 238.6 (8) & 66.1 (6) & 3.90 (20) & 2.58 (13)\tabularnewline
 & 185.2 (8) & 50.2 (6) & 19.6 (4) & 9.84 (23)\tabularnewline
{\small Total} &  &  & 100 & 82.1 (11)\tabularnewline
\hline 
\multicolumn{5}{l}{{\small $\mathrm{^{144}Ce\,\gamma}$ lines}\rule{0pt}{2.25ex}}\tabularnewline
\hline 
 &  & 133.515 (2) & 11.09 (19) & 14.8 (3)\tabularnewline
 &  & 99.961 (15) & 0.04 (4) & 0.04 (4)\tabularnewline
 &  & 80.12 (5) & 1.36 (6) & 1.09 (5)\tabularnewline
 &  & 59.03 (3) & 0.0004 (5) & 0.00058 (3)\tabularnewline
 &  & 53.395 (5) & 0.1 (8) & 0.053 (4)\tabularnewline
 &  & 40.98 (10) & 0.257 (16) & 0.105 (7)\tabularnewline
 &  & 33.568 (10) & 0.2 (22) & 0.067 (8)\tabularnewline
{\small Total} &  &  &  & 16.2 (4)\tabularnewline
\hline 
\multicolumn{5}{l}{{\small $\mathrm{^{144}Pr\,\beta}$ branches}\rule{0pt}{2.25ex}}\tabularnewline
\hline 
 & 2997.5 (24) & 1222 (11) & 97.9 (4) & 1196 (5)\tabularnewline
 & 2301 (24) & 895 (11) & 1.04 (20) & 9.31 (18)\tabularnewline
 & 1436.5 (24) & 526.27 (99) & 0.0014 (3) & 0.0074 (16)\tabularnewline
 & 924.7 (24) & 322.85 (92) & 0.00062 (5) & 0.002 (16)\tabularnewline
 & 913 (24) & 306.77 (96) & 0.0067 (10) & 0.0206 (3)\tabularnewline
 & 811.8 (24) & 267.19 (93) & 1.05 (4) & 2.81 (11)\tabularnewline
 & 342.6 (24) & 98.9 (78) & 0.00015 (3) & 0.00015 (3)\tabularnewline
 & 322.2 (24) & 92.33 (77) & 0.00087 (9) & 0.0008 (8)\tabularnewline
 & 254.6 (24) & 71.11 (76) & 0.0003 (10) & 0.00021 (7)\tabularnewline
{\small Total} &  &  & 100 & 1208.2 (53)\tabularnewline
\hline 
\multicolumn{5}{l}{{\small $\mathrm{^{144}Pr\,\gamma}$ lines}\rule{0pt}{2.25ex}}\tabularnewline
\hline 
 &  & 2654.9 (2) & 0.00015 (3) & 0.0039 (7)\tabularnewline
 &  & 2368.3 (3) & $5.4\,10^{-5}$ (13) & 0.0013 (3)\tabularnewline
 &  & 2185.66 (7) & 0.694 (15) & 15.2 (3)\tabularnewline
 &  & 2072.9 (2) & 0.00023 (3) & 0.0047 (6)\tabularnewline
 &  & 2046.3 (2) & 0.00027 (5) & 0.0055 (11)\tabularnewline
 &  & 1978.82 (10) & 0.00087 (8) & 0.0173 (16)\tabularnewline
 &  & 1560.97 (10) & 0.00020 (3) & 0.0031 (4)\tabularnewline
 &  & 1489.17 (5) & 0.278 (5) & 4.14 (7)\tabularnewline
 &  & 1388.02 (10) & 0.00672 (9) & 0.0933 (12)\tabularnewline
 &  & 1376.27 (10) & 0.00039 (4) & 0.0054 (6)\tabularnewline
 &  & 1182 (3) & $5.37\,10^{-5}$ & $6.34\,10^{-4}$\tabularnewline
 &  & 864.45 (10) & 0.0024 (3) & 0.0209 (23)\tabularnewline
 &  & 814.1 (10) & 0.0032 (3) & 0.0262 (22)\tabularnewline
 &  & 696.51 (3) & 1.342 & 9.35\tabularnewline
 &  & 674.95 (10) & 0.003 (3) & 0.0199 (18)\tabularnewline
 &  & 624.7 (1) & 0.00113 (3) & 0.00704 (18)\tabularnewline
{\small Total} &  &  &  & 28.9 (4)\tabularnewline
\hline 
\hline 
{\small Total $\beta$ branches} &  &  &  & 1290.3 (64)\tabularnewline
{\small Total $\gamma$ lines} &  &  &  & 45.1 (8)\tabularnewline
Total X lines &  &  &  & 3.2 (1)\tabularnewline
\multicolumn{3}{l}{Total CE and Auger lines} &  & 9.5 (3)\tabularnewline
{\small Grand total} &  &  &  & 1348.0 (75)\tabularnewline
\hline 
\end{tabular}
\par\end{centering}{\small \par}

\centering{}{\small \caption{\label{tab:CePrDecayHeat}Energy released for branch of $\mathrm{^{144}Ce}$
and $\mathrm{^{144}Pr}$ decays \cite{NNDC}. The $^{144m}\mathrm{Pr}$
is neglected, due to its very low branching ratio (<1\%) and its very
low $\beta$ decay probability (0.7 \%) compared to $\gamma$ isomeric
transition (99.3 \%){\small .}}
}
\end{table}

\par\end{center}

\noindent \begin{center}
\begin{table}[H]
\centering{}%
\begin{tabular}{cccccccccc}
\hline 
Time (months) & 0 & 6 & 12 & 18 & 24 & 36 & 48 & 60 & 120\tabularnewline
\hline 
\hline 
Activity (kCi) & 75.0 & 48.1 & 30.8 & 19.8 & 12.7 & 5.2 & 2.1 & 0.9 & 0.01\tabularnewline
Activity (PBq) & 2.78 & 1.78 & 1.14 & 0.73 & 0.47 & 0.19 & 0.08 & 0.03 & 0.0004\tabularnewline
Heat (W) & 599 & 384 & 246 & 158 & 101 & 41.7 & 17.1 & 7.05 & 0.0829\tabularnewline
\hline 
\end{tabular}\caption{\label{tab:CePrHeatvsTime}$\mathrm{^{144}Ce-^{144}Pr}$ activity
and heat release as a function of time after the production. After
20 years, the activity is only 50 Bq.}
\end{table}

\par\end{center}

\subsubsection{\noindent Electron and antineutrino spectra}

\noindent Electron and antineutrino spectra of a given $\mathrm{\beta}$
decaying radioisotope are simply related by energy conservation at
the $\mathrm{\beta}$-branch level. However, $^{144}$Pr undergoes
several $\mathrm{\beta}$-branch decays with forbidden transitions
and makes the $^{144}$Pr antineutrino spectrum calculation difficult.
Here is presented a direct calculation of the neutrino spectrum using
the Fermi theory \cite{Wilkinson1989}. With $E$ and $p$ the total
energy and momentum of the beta particle, the differential energy
spectrum at the $\mathrm{\beta}$ branch level writes:
\begin{equation}
N(E)dp=Kp^{2}(E-E_{0})^{2}F(Z,E)dp,\label{eq:FermiSpectrum}
\end{equation}
with $E_{0}$ the end point energy. Fermi theory is a simplified description
of $\mathrm{\beta}$ decay and needs to be corrected for various atomic
and nucleus effects, which especially affect low energy electrons
and hence high energy neutrinos. The $^{144}$Pr spectrum is carefully
modeled by including major corrective effects to the Fermi spectrum:

\noindent 
\begin{equation}
N(E)dp=Kp^{2}(E-E_{0})^{2}F(Z,E)L_{0}(E,Z)C(E,Z)R(E,M)G(E,Z)S(Z,E)B(E,Z)dp,
\end{equation}

\noindent where:
\begin{itemize}
\item \noindent $L_{0}(E,Z)$ and $C(E,Z)$ takes respectively into account
the finite size and mass of the nucleus considering both electromagnetic
and weak interactions \cite{Wilkinson1990} (Fermi theory considers
the nucleus to be point-like with an infinite mass)
\item \noindent $R(E,M)$ is the recoil effect on the phase space factor
and the nucleus Coulomb field, with $M$ the nuclear mass \cite{Wilkinson1990}
\item \noindent $G(E,Z)$ corresponds to radiative corrections (i.e. emission
of virtual or real photons in the decay process) \cite{Sirlin2011}
\item \noindent $S(Z,E)$ is the screening of the beta particle \cite{Behrens1982}
by the atomic electrons
\item \noindent $B(E,Z)$ is the weak magnetism correction \cite{Huber2011}
(similar to induced electromagnetic current in a moving Coulomb field,
but for weak interactions). 
\end{itemize}
\noindent All corrective terms have the structure $1+\delta(E,Z,M)$.
These terms represent up to a few percent correction to the Fermi
spectrum (eq. \ref{eq:FermiSpectrum}). The recoil effects turn out
to have a nearly negligible impact (0.013\% correction) on the uncorrected
Fermi spectrum. Figure \ref{fig:Compared-rel-corr} compares the relative
weights of each of the above-mentioned corrections. Any theoretical
uncertainty associated to a correction is propagated in the spectrum.
The dominant source of uncertainty comes from the weak magnetism effect.
Following the prescription of \cite{Huber2011}, a linear correction
is used to estimate the weak magnetism effect. The linear coefficient
is estimated following two approaches. The first one uses a theoretical
calculation based on the so-called impulse approximation \cite{Huber2011}.
The second one assumes that vector currents in a $\mathrm{\beta}$-decay
behave the same as for an electromagnetic decay process (conserved
vector current hypothesis), allowing to make an analogy between a
Fermi type (vector current) $\mathrm{\beta}$-decay and an electromagnetic
M1 $\mathrm{\gamma}$-decay. In such an approximation, the linear
coefficient of the weak magnetism correction can be estimated using
the measured M1 electromagnetic form factors of a reduced (and hardly
representative) sample of 13 nuclei, which present the same nucleus
quantum number change in their respective M1 $\mathrm{\gamma}$-decay
and $\mathrm{\beta}$-decay\cite{Huber2011,Huber2012}. Both methods
give strongly different outputs and the first one is retained as the
default estimation. The second largest uncertainty is associated to
the screening correction term, and comes from the modeling of the
distribution of the atomic electrons. Various models along with their
different results \cite{Behrens1982} are used to conservatively estimate
the uncertainty on the screening correction.

\noindent \begin{center}
\begin{figure}[h]
\begin{centering}
\includegraphics[scale=0.6]{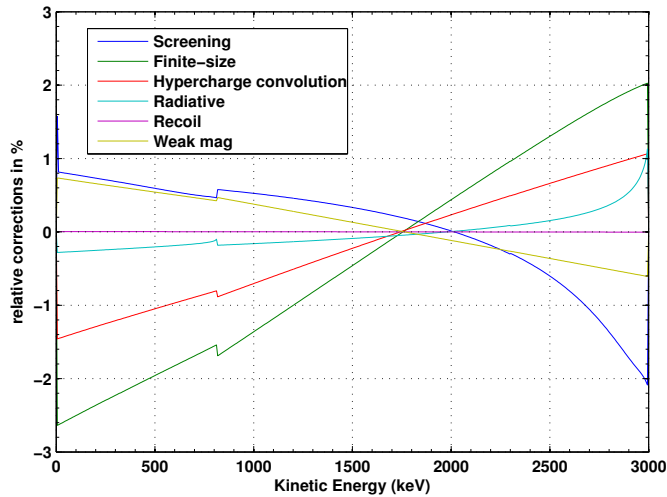}
\par\end{centering}

\caption{\label{fig:Compared-rel-corr}Comparison of the relative amplitudes
for the corrective terms. The weak magnetism correction has been computed
here using the impulse approximation. See text for further details.}
\end{figure}

\par\end{center}

\noindent The $^{144}$Pr neutrino spectrum is modeled combining the
eight most important $\beta$-branches (highest branching ratios)
using data from National Nuclear Data Center. They are summarized
in table \ref{tab:Pr--branch}. The resulting neutrino spectrum along
with errors associated to the uncertainties in the considered corrections
is displayed in \figref{pr144spectra}. As a cross-check, the electron
and antineutrino spectra have been calculated using the Bestiole code
\cite{Muller2010}. Bestiole has been originally developed to predict
antineutrino spectra from nuclear reactors (which comprises thousands
of $\mathrm{\beta}$-branches) using available experimental informations
in different nuclear databases. The agreement with our modeling, once
integrated with the same binning configuration, is very good. The
$^{144}$Pr antineutrino spectrum exhibits the same two discontinuities
at 811.8 keV and 2301 keV, which are the end-point energies of the
$\mathrm{\beta}$-branches of intensity of 1.05\% and 1.04\%, respectively.
The next five branches do not have a strong influence on the total
spectrum and have a negligible contribution. 

\noindent In order to refine our knowledge of the source spectrum,
$\mathrm{\beta}$ spectroscopic measurements using cerium nitrate
samples prepared by PA Mayak are scheduled to begin by the end of
2013. Our $^{144}$Pr $\beta$ spectrum modeling will be adjusted
to the data to reduce uncertainties on the shape of the neutrino spectrum,
and especially to constrain the weak magnetism correction, which dominates
the shape error budget.

\noindent The full $\mathrm{^{144}Ce-^{144}Pr}$ antineutrino spectrum
is displayed in \figref{cepremitspectr}. The $^{144}$Ce lifetime
is much shorter than that of $^{144}$Pr, which enables us to assume
secular equilibrium. The first peak corresponds to the $^{144}$Ce
parent $\beta$ decay with an end-point energy of 318.7 keV. $\bar{\nu}_{e}$
originating from $^{144}$Ce $\beta$ decay are hence not detectable
through the inverse beta decay process. The second broad peak corresponds
to the $^{144}$Pr $\beta$ decay with an end-point energy of 2997.5
keV. The $\mathrm{\bar{\nu}_{e}}$ spectrum which lies above the inverse
beta decay threshold of 1.8 MeV represents about 50\% of the $^{144}$Pr
emitted $\bar{\nu}_{e}$.

\begin{center}
\begin{table}[H]
\begin{centering}
\begin{tabular}{>{\centering}m{2cm}>{\centering}m{4cm}c>{\centering}m{5cm}}
\hline 
Branching ratio (\%) & Quantum 

numbers & $Q_{\beta}$(keV) & type\tabularnewline
\hline 
\hline 
97.9 & $0^{+}$ & 2997.5 & First non unique forbidden\tabularnewline
1.05 & $1^{-}$ & 811.8 & Allowed\tabularnewline
1.04 & $2\text{\textsuperscript{+}}$ & 2301.0 & First unique forbidden\tabularnewline
$\begin{gathered}6.70\cdot10^{-3}\end{gathered}
$ & $0^{+}$ & 913 & First non unique forbidden\tabularnewline
$\begin{gathered}1.4\cdot10^{-3}\end{gathered}
$ & $2\text{\textsuperscript{+}}$ & 1436.5 & First unique forbidden\tabularnewline
$\begin{gathered}8.7\cdot10^{-4}\end{gathered}
$ & $0^{+}$ & 322.2 & First non unique forbidden\tabularnewline
$\begin{gathered}6.2\cdot10^{-4}\end{gathered}
$ & $2\text{\textsuperscript{+}}$ & 924.7 & First unique forbidden\tabularnewline
$\begin{gathered}3.0\cdot10^{-4}\end{gathered}
$ & $0^{+}$ & 254.6 & First non unique forbidden\tabularnewline
\hline 
\end{tabular}
\par\end{centering}

\caption{\label{tab:Pr--branch}$^{144}$Pr $\beta$-branches included in the
analytical model, sorted by decreasing probability}
\end{table}

\par\end{center}

\begin{center}
\begin{figure}[h]
\begin{centering}
\includegraphics[scale=0.6]{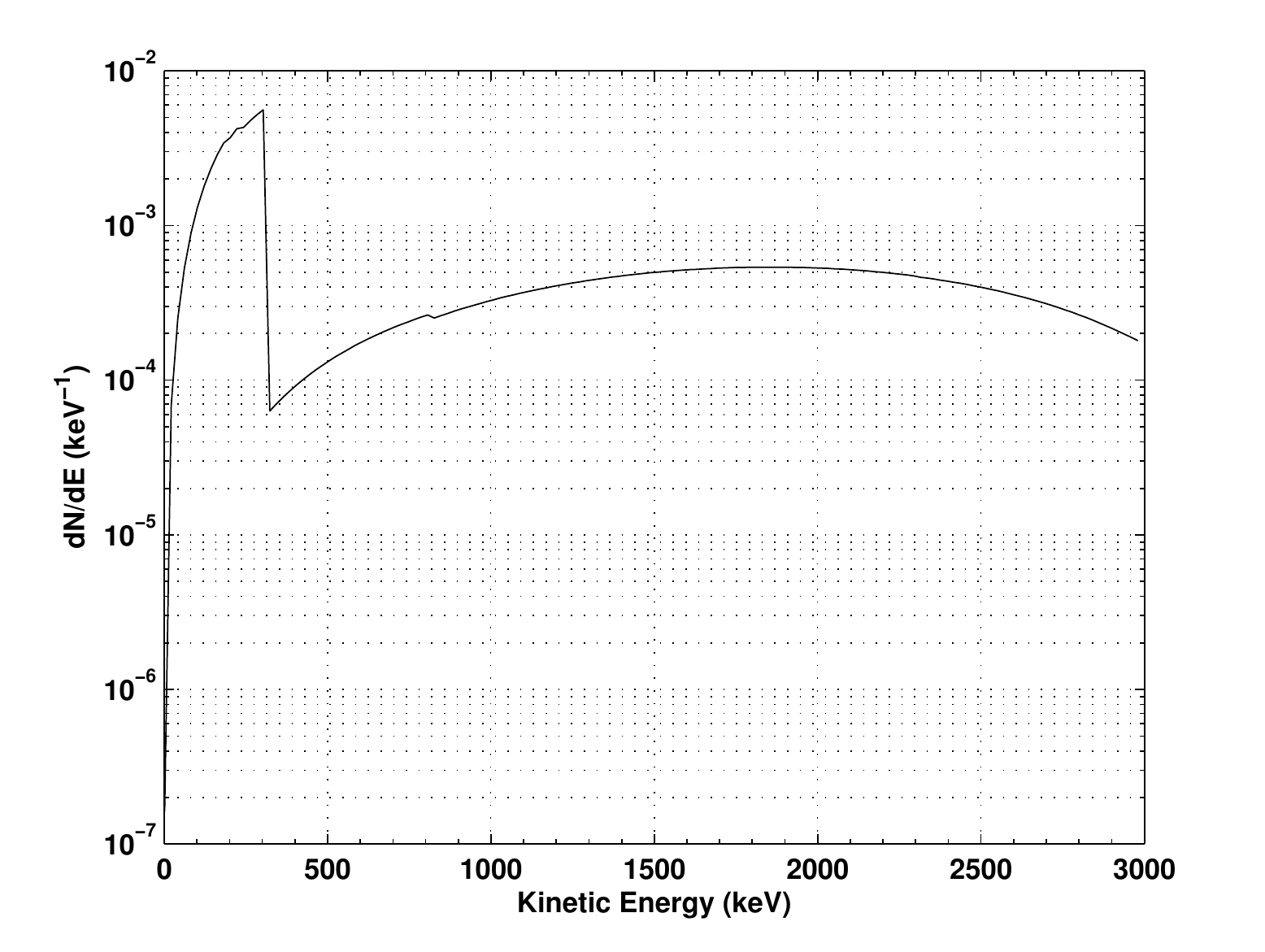}
\par\end{centering}

\caption{\label{fig:cepremitspectr}$\mathrm{^{144}Ce-^{144}Pr}$ antineutrino
emitted spectrum assuming secular equilibrium. }
\end{figure}

\par\end{center}

\begin{center}
\begin{figure}[h]
\begin{centering}
\includegraphics[scale=0.5]{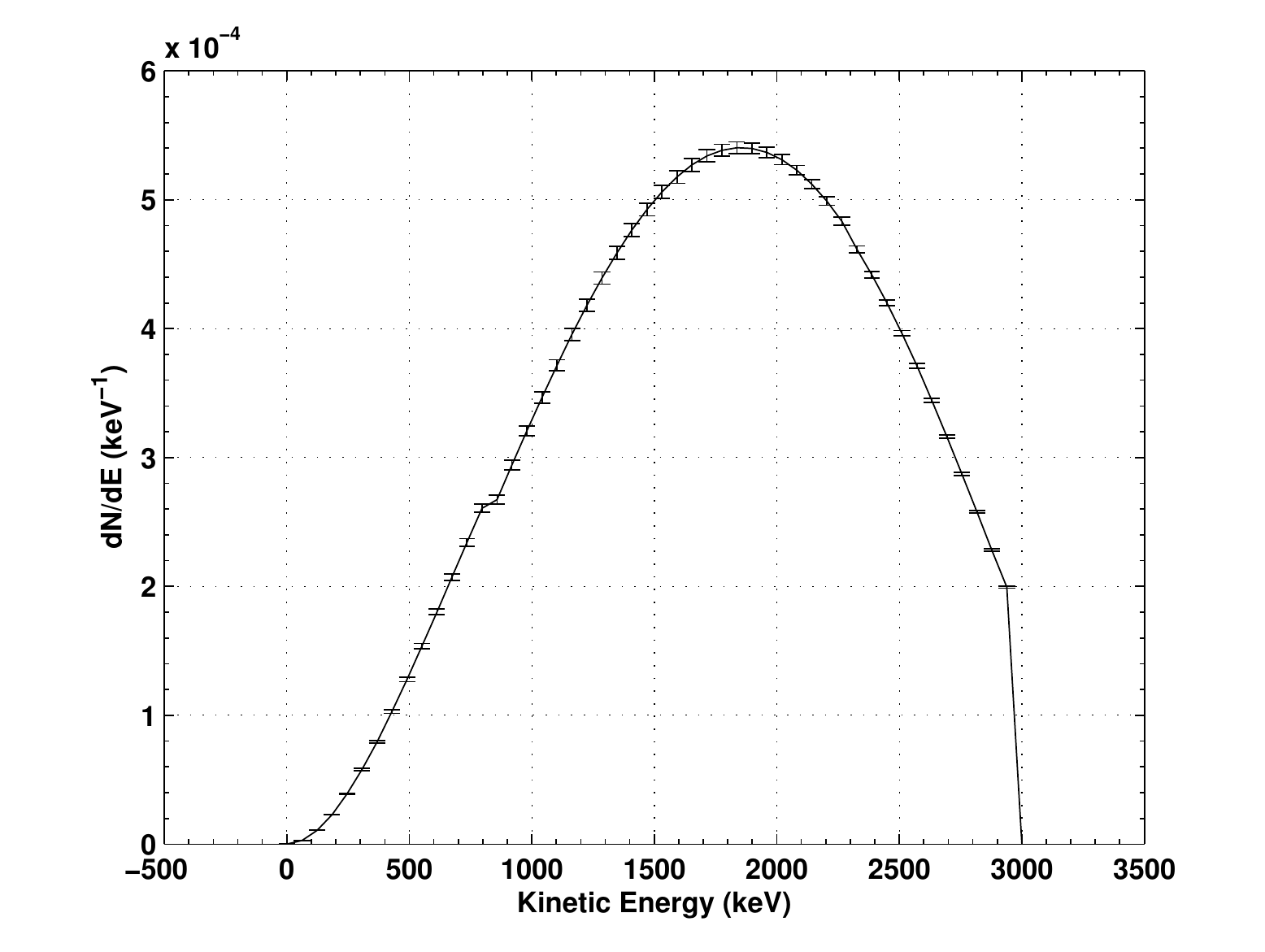}\includegraphics[scale=0.5]{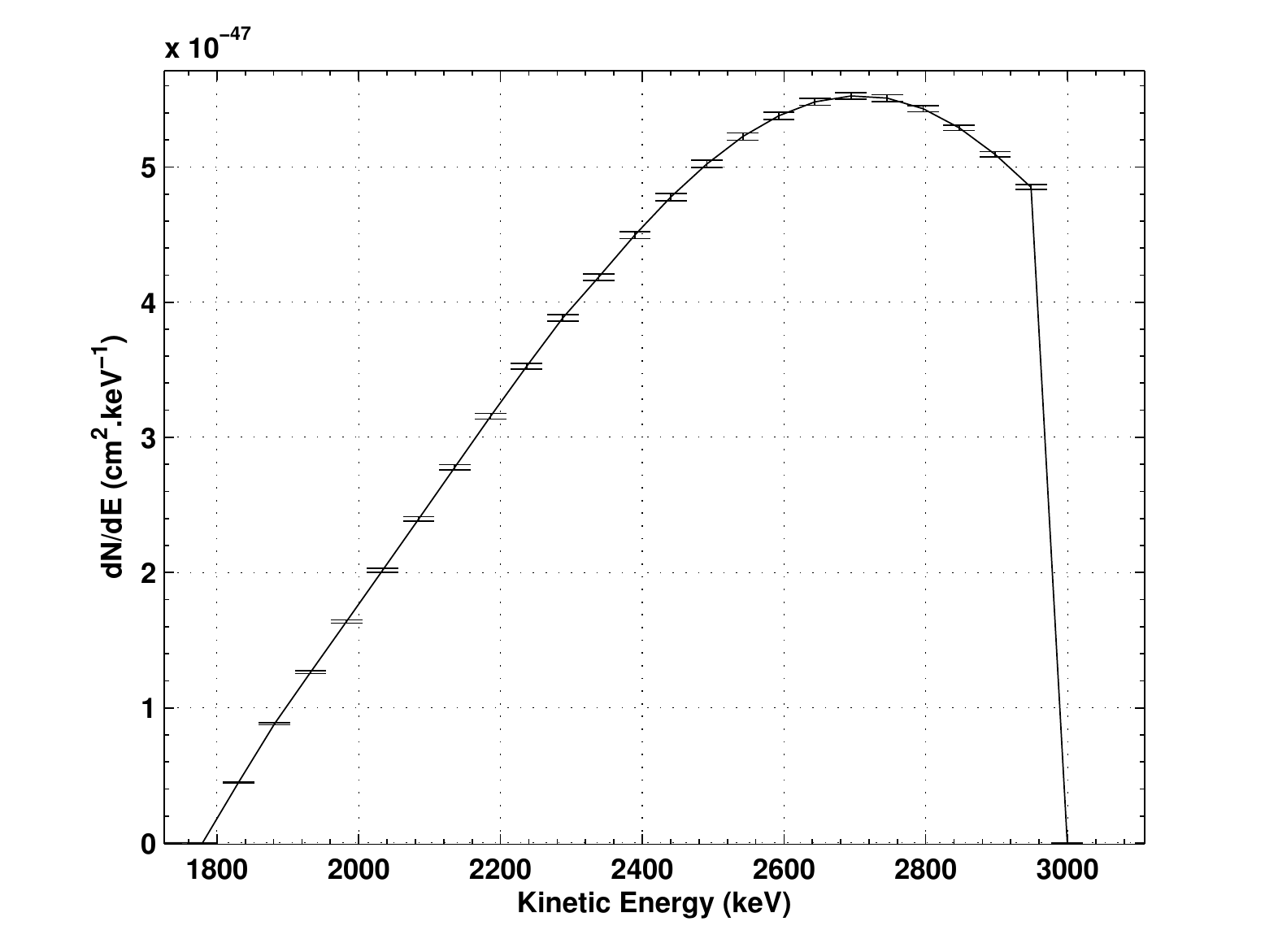}
\par\end{centering}

\caption{\label{fig:pr144spectra}Left panel: Complete neutrino spectrum for
eight first $\beta$-branches of $^{144}$Pr with optimistic weak
magnetism correction. Right Panel: Differential observable neutrino
spectrum from $^{144}$Pr including the IBD cross-section.}
\end{figure}

\par\end{center}

\noindent As a first application, our model was used to predict the
rate of inverse beta decay events that will be detected in KamLAND.
It is also used to estimate the uncertainty coming from the shape
of the neutrino spectrum. The visible energy spectrum in the detector
is obtained by multiplying the IBD cross-section to the $\mathrm{^{144}Ce-^{144}Pr}$
antineutrino spectrum. It is shown on the right panel of \figref{pr144spectra}.
The threshold of IBD is high enough to neglect all influence of Cerium
as its transition end-points are well below 1.8 MeV. The cross-section
is computed using prescriptions from \cite{Fay85}, with a pre-factor
of 9.61 10$^{-48}$$ $ cm$^{2}$$ $/Mev$^{2}$ (which has a negligible
0.1\% uncertainty due to the neutron lifetime uncertainty). Then,
the expected event rate in the KamLAND detector is 
\begin{equation}
N_{\nu}=N_{Ce}\rho_{p}F_{geo}\int_{0}^{\infty}\sigma_{IBD}\left(E_{\nu}\right)S\left(E_{\nu}\right)dE_{\nu},
\end{equation}
where $\rho_{p}=6.62\times10^{22}$ cm$^{-3}$ is the density of protons
in the KamLAND detector and $F_{geo}$ is a geometric factor (with
dimension of a length) including the detector and source geometry.
For a cylindrical source of 13.5~cm diameter and 13.5~cm height,
positioned 9.3~m away from the center of a spherical detector of
6.5~m radius, $F_{geo}=1.10$ m. Considering a pessimistic scenario
where some backgrounds escaping the source (gammas and neutrons) remain
unshielded, a tighter 6 m radius fiducial cut was also considered,
giving $F_{geo}=0.85$ m. $N_{Ce}$ is the number of Cerium decays
corresponding to a data taking period T: 

\noindent 
\begin{equation}
N_{Ce}=\int_{0}^{T}A_{0}e^{\frac{-\mathrm{ln}(2)t}{T_{1/2}}}dt
\end{equation}

\noindent 
\[
=A_{0}\frac{T_{1/2}}{\mathrm{ln}(2)}\left[1-e^{\frac{-\mathrm{ln}(2)T}{T_{1/2}}}\right],
\]

\noindent where $A_{0}$ is the initial source activity. With a 75~kCi
source and a 18-month data taking period, $24165\pm189$ (resp. $18673\pm146$)
antineutrinos are expected to be detected in KamLAND considering a
6.5m radius (resp. 6 m radius) fiducial volume cut. The quoted uncertainties
come from the $\beta$-spectrum modeling only, assuming a weak magnetism
correction factor calculated using the impulse approximation, as done
in \cite{NewNeutrinoSpectra1}.

\subsection{\noindent $\mathrm{^{144}Ce-^{144}Pr}$ source production}

\noindent From an extraction and production perspective, cerium is
a rather abundant element present at the 5.5\% and 3.7\% levels in
the fission products of uranium and plutonium, respectively (see table
\tabref{Cumulative-fission-yield}).

\subsubsection{\noindent Activity}

\noindent As agreed with the PA Mayak, the activity of the cerium
antineutrino generator will be of $85_{-10}^{+15}$ kCi in $^{144}\mathrm{Ce}$
beta activity at delivery in Yekaterinburg. Taking into account the
6-8 \% precision of the activity measurement at Mayak, one guarantee
there is a 90 \% probability to get the activity higher than 75 kCi.
Remaining free space in source (inner) capsule, if any, will be filled
with additional radioactive cerium oxide if available at the end of
the production, up to 100 kCi. Thus, depending on the fuel age and
extraction process there is a possibility that the source activity
may be above 75 kCi at the beginning of the experiment at KamLAND.

\subsubsection{\noindent Geometry}

\noindent The spatial extent of the $\mathrm{CeO_{2}}$ source material
must be small compared to the neutrino oscillation length ($\mathrm{L_{osc}\sim}$
1 m for $\mathrm{\Delta m^{2}\sim1\, eV^{2}}$ neutrino mass splittings
considered above) for the successful oscillometry measurement. Moreover,
the size of the source will directly increase the size and mass of
the shielding, thus requiring a source to be small and compact, with
an extension of a few tens of centimeter.

\noindent But several constraints have to be considered. The first
consideration is that only a mixture of all cerium isotopes produced
during irradiation can be practically separated from spent nuclear
fuel, since isolating the $^{144}$Ce would require a very difficult
isotopic separation. So the $^{144}$Ce content of the cerium oxide
is limited by fission yields and cooling time. Secondly, the $\mathrm{CeO_{2}}$
will be manipulated in hot cells, which requires cylindrically shaped
capsule and limit the achievable density of cerium oxide to $4.0\pm1\ \mathrm{g/cm^{3}}$.

\noindent Finally, the largest capsule that can be handled routinely
in Mayak hot cells is 15 cm in height and diameter, and safety requirements
impose a double container in stainless steel. Mayak will provide the
capsules, 4 mm thick for the outer capsule and 3 mm thick for the
inner capsule. Adding the necessary mechanical gap of 0.5 mm, a maximum
of 13.5 cm in diameter remains available for the $\mathrm{CeO_{2}}$.

\subsubsection{\noindent Spent fuel seeds}

\noindent Preliminary computations have been done to study the amount
of cerium produced within VVER reactors for various condition of irradiation.
The SAS2H (SCALE-4.4) ORIGEN-S software from Oakridge was used by
the MePhi group in Russia. The content of $^{144}$Ce in spent nuclear
fuel (SNF) elements depends on the irradiation history and on the
reactor neutron spectrum. The $^{144}$Ce isotope mostly results from
the decay of $^{144}$Cs and is abundantly produced in the fissions
of $^{235}$U and $^{239}$Pu. The abundance of cerium among produced
rare earth elements is about 22\%. Taking these numbers at their face
values, about 15 SNF elements would be needed to achieve 75 kCi of
$^{144}$Ce, but we will see that technical considerations force us
to consider about five to ten times more SNF elements to reach the
desired activity. 

\noindent To optimize the selection of SNF elements in order to maximize
the $^{144}$Ce content per fuel element units of mass, a dedicated
study was performed by Mayak and MePhi. The general guidelines to
extract the fuel with the highest constant of $^{144}$Ce are the
following:
\begin{enumerate}
\item Select SNF with shortest time of irradiation, since the fission yield
for $^{144}$Ce of $^{239}$Pu is lower than $^{235}$U.
\item Select SNF with highest energy release during the last irradiation,
since the $^{144}$Ce content increases with the fission rate and
therefore power.
\item Select SNF with the lowest burn-up, for the same reason than the first
guideline.
\end{enumerate}
Following these guidelines, a specific activity of $^{144}$Ce in
the range 31 to 35 Ci/g of SNF can be expected, corresponding to an
absolute activity of 8 to 12 kCi per fuel SNF.

\noindent The usual cooling sequence of SNF consists is 3 years of
cooling in the power plant storage pool followed by another 2-year
period of SNF being stored at the reprocessing plant. Delivery of
fresher SNF to the reprocessing plant is possible but not conventional.
This step would maximize the amount of activity within a small volume
and significantly reduce the amount of SNF needed to produce a $^{144}$Ce
source with 75 kCi activity. Each gained month in the source manufacturing
process represents a gain of SNF material of the order of 10\%. To
guarantee that the source is compact enough to fit inside the Mayak
capsule, the possibility of using SNF with a low cooling time has
been considered and approved by Mayak. 

\noindent The transport, by train, will probably use a TUK-6 container
(or equivalent), which is categorized as a B(M) container according
to the international classification and contains typically up to 3.8
tons of UO$_{2}$ (see \figref{TUK-6-Container}). To keep the level
of radiation within the limits for transportation from the reactor
site to the reprocessing plant with a low cooling time fuel, the container
may be not full or filled with a mix of long and low cooling time
fuel.

\noindent \begin{center}
\begin{figure}[H]
\centering{}\includegraphics[scale=0.37]{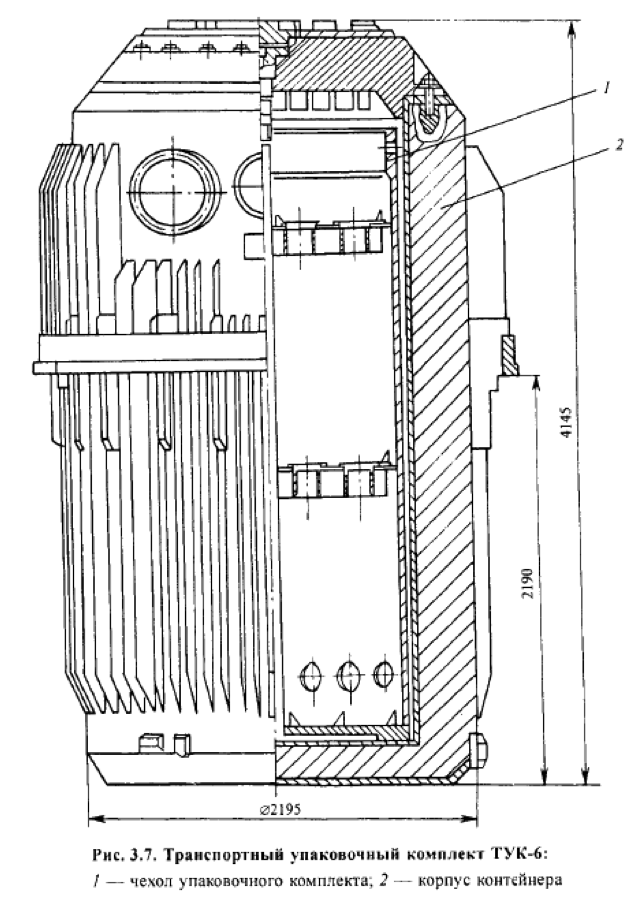}\subfloat{\centering{}\includegraphics[scale=0.43]{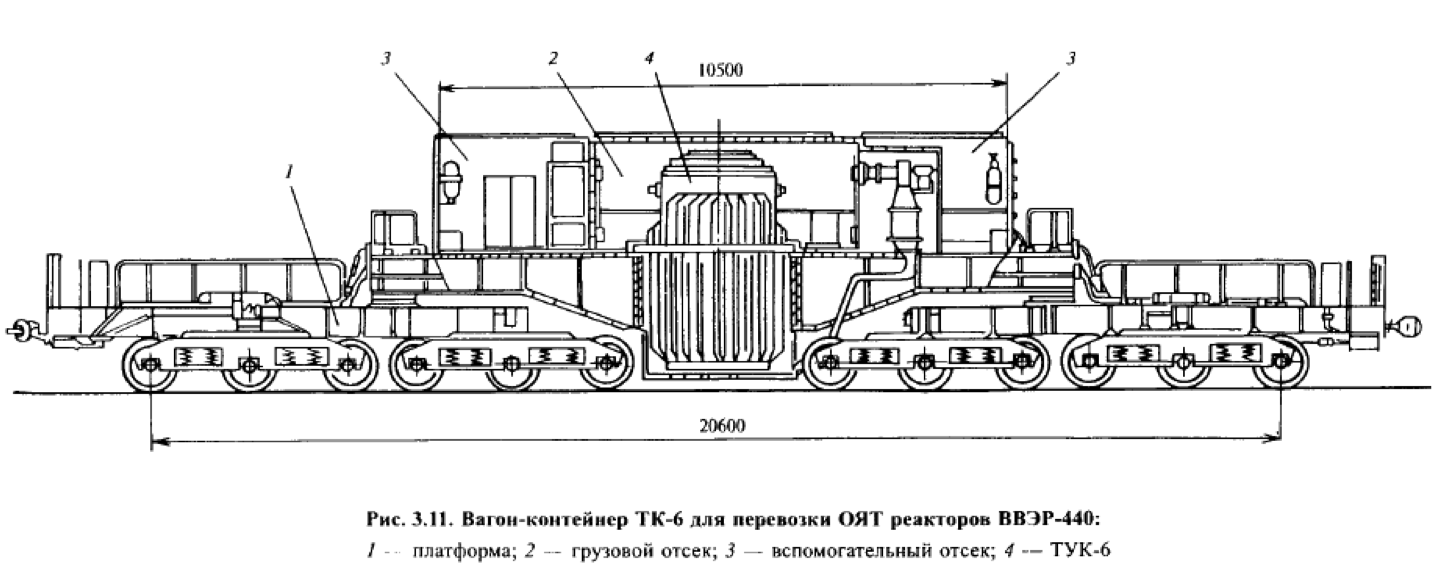}}\caption{\label{fig:TUK-6-Container}TUK-6 container (left) and carriage container
(right) for spent fuel transportation from reactor to reprocessing
plant. Each year about 58 tons of SNF are transported from Cola NPP
to Radiochemistry plant RT-1 at PA Mayak (Ozersk, Chelyabinsk region). }
\end{figure}

\par\end{center}

\noindent An agreement between Rosatom and FSUE Mayak PA has been
reached for delivery of 'fresh' SNF from the Kola Nuclear Power Plant
(VVER-440, 3000 km north of Mayak) with a 1.7 to 2.0 year cooling
period after the end of irradiation. The total mass of this special
batch will be about 15 - 16 tons, including 7 - 8 tons with cooling
time in the range of 580 till 640 days (1.6-1.75 years). The second
part of SNF has cooling time 3.5 years and more. The burnup value
of this \textquotedbl{}fresh\textquotedbl{} fuel will be in the range
of 48 - 51 MW.d/t.

\subsubsection{\noindent Spent fuel cooling time and mass of CeO$_{2}$ containing
75 kCi of $^{144}$Ce}

\noindent \label{sub:Spent-fuel-cooling}

\noindent The ratio of $^{144}$Ce in cerium extracted from SNF as
a function of time $t$, assuming $^{144}\mathrm{Ce}$ is the only
cerium long-lived cerium isotope is, 

\noindent 
\begin{equation}
\chi(t)=\frac{\mathrm{mass}(^{144}\mathrm{Ce},\ t)}{\mathrm{mass}(\mathrm{Ce},\ t)}=\frac{\mathrm{mass}(^{144}\mathrm{Ce},\ t)}{\mathrm{mass}(^{144}\mathrm{Ce},\ t)+\mathrm{mass}(^{\mathrm{stable}}\mathrm{Ce})}=\frac{m_{144}(t)}{m_{144}(t)+m_{\mathrm{stable}}}
\end{equation}

\noindent According to MePhi (private communication during the CeLAND
meeting in Paris, 22 July 2013), the content of $^{144}$Ce in cerium
with $\mathrm{t_{c}}$= 3 years after the end of irradiation is:

\noindent 
\[
\chi(3)=\chi_{3}=0.6{}_{-0.15}^{+0.1}\,\%
\]

\noindent Here $\mathrm{t_{c}}$ is an arbitrary reference cooling
time, used by MePhi for standard simulations. The mass of stable cerium
can be written:

\noindent 
\begin{equation}
m_{\mathrm{stable}}=\frac{m_{144}(3)}{\chi_{3}}-m_{144}(3)=\frac{(1-\chi_{3})\, m_{144}(3)}{\chi_{3}}
\end{equation}
\\
and the mass of $^{144}\mathrm{Ce}$ is:

\noindent 
\[
m_{144}(t)=m_{144}(0)\: e^{-\lambda t}=m_{144}(3)\, e^{-\lambda(t-t_{c})}
\]
\\
with $\lambda=\ln(2)/\tau_{1/2}=2.8158\:10^{-8}$ s$^{-1}$. Then
the $^{144}$Ce mass ratio as a function of time $t$ is: 

\noindent 
\begin{equation}
\mathrm{\chi(t)=\frac{m_{144}(3)\, e^{-\lambda(t-t_{c})}}{m_{144}(3)\, e^{-\lambda(t-t_{c})}+\frac{(1-\chi_{3})\, m_{144}(3)}{\chi_{3}}}=\frac{\chi_{3}\, e^{-\lambda(t-t_{c})}}{1-\chi_{3}+\chi_{3}\, e^{-\lambda(t-t_{c})}}}
\end{equation}

\noindent The $^{144}$Ce mass ratio is plotted on the left panel
of \figref{Ce144CeCeO2-mass-ratio}. The source mass is related to
the $^{144}$Ce activity $\mathrm{\mathcal{A}(t)}$ through:

\noindent 
\begin{equation}
\mathrm{\mathcal{A}(t)=\lambda\ N_{144}(t)=\dfrac{m_{144}(t)\ \mathcal{N}_{a}\ \lambda}{144}\Leftrightarrow m_{144}(t)=\dfrac{144\times\mathcal{A}(t)}{\mathcal{N}_{a}\ \lambda}.}
\end{equation}
\\
with $\mathcal{N}_{a}$ the Avogadro number. Finally, the source mass
is given by:

\noindent 
\begin{equation}
\mathrm{m_{\mathrm{source}}(t)=m_{144}(t)\times\dfrac{176}{144}+m_{\mathrm{stable}}(t)\times\dfrac{174}{142}\simeq\dfrac{\mathcal{A}(t)}{\mathcal{N}_{a}\ \lambda\ \chi(t)}\left(176.5-0.5\chi(t)\right)}
\end{equation}

\noindent \begin{center}
\begin{figure}[h]
\begin{centering}
\includegraphics[scale=0.37]{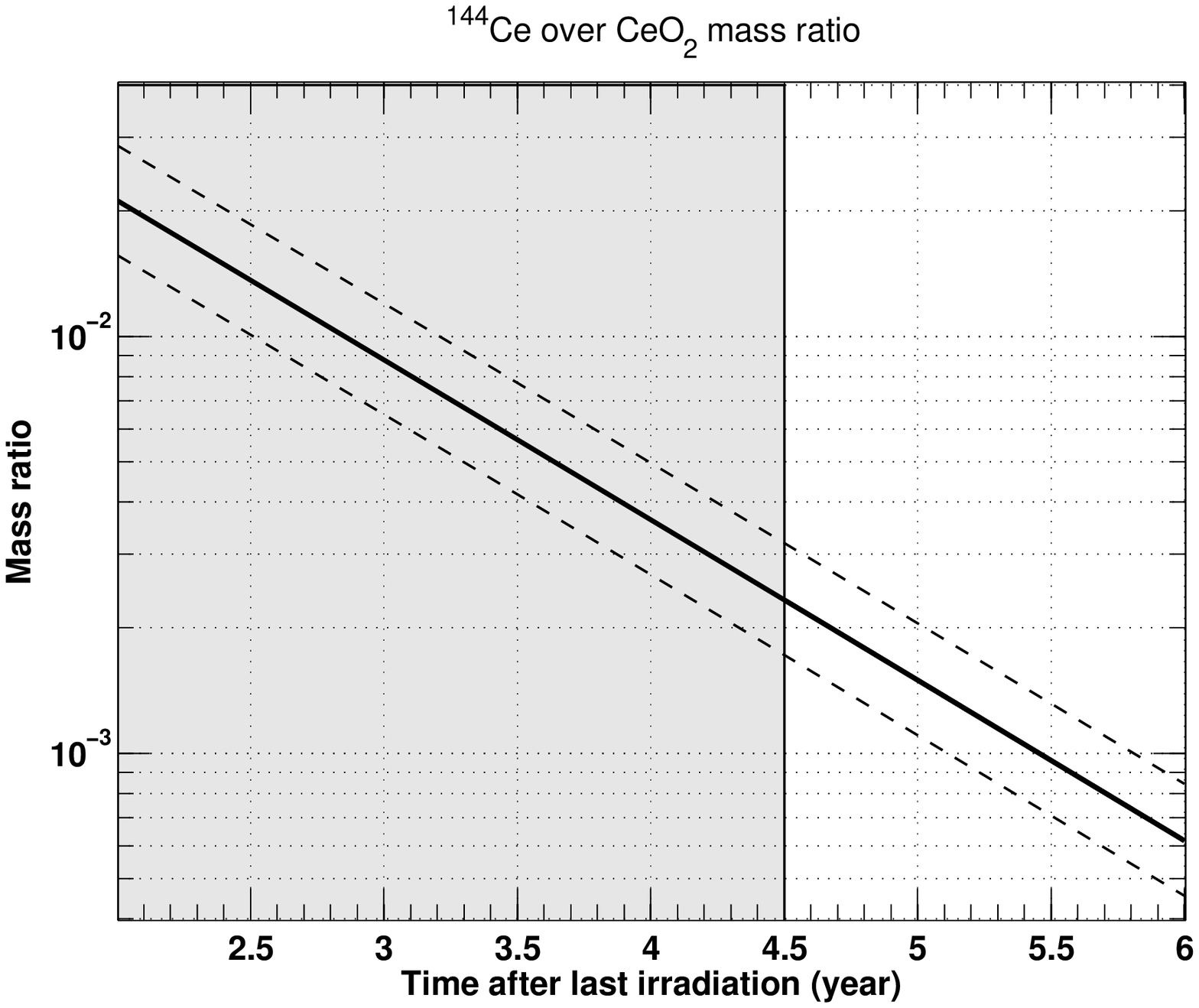}\includegraphics[scale=0.37]{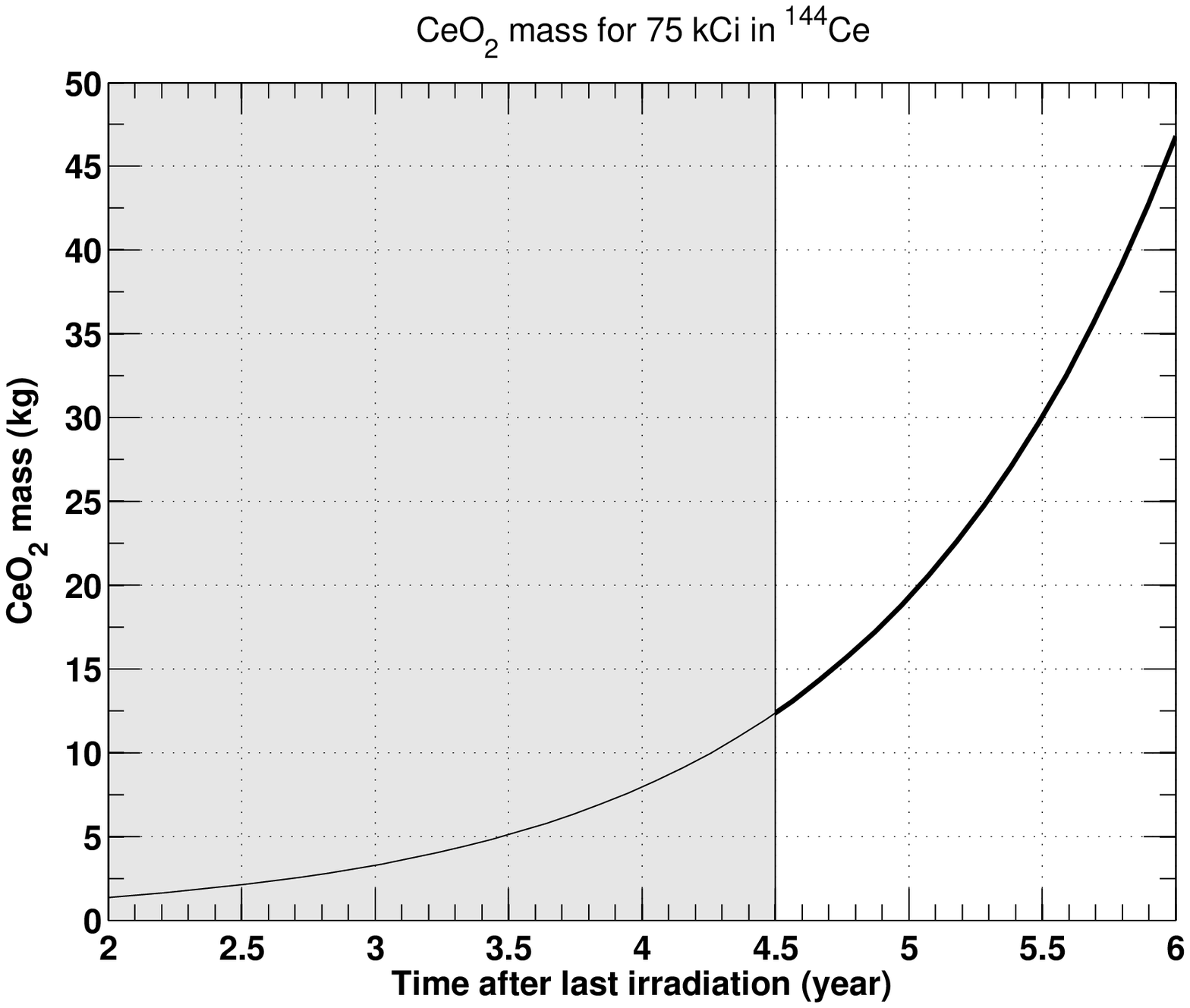}
\par\end{centering}

\centering{}\caption{\label{fig:Ce144CeCeO2-mass-ratio}Left: $^{144}$Ce/CeO$_{2}$ mass
ratio as a function of time after last irradiation at the reactor
(and its uncertainty band computed from reactor core simulations).
Right: Mass of CeO$_{2}$ for a 75 kCi activity in $^{144}$Ce as
a function of the cooling + production time. The antineutrino generator
should be realized with maximum cooling + production time of about
4.5 years in order to fit enough radioactive material inside the inner
capsule (assuming a cerium oxide density of 4 g/cm$^{3}$).}
\end{figure}

\par\end{center}

\noindent The mass of $\mathrm{CeO_{2}}$ required to achieve an activity
of 75 kCi in $^{144}$Ce, is given as a function of time in the right
panel of \figref{Ce144CeCeO2-mass-ratio}. Assuming a 2-year cooling
time and 1 year of production, the source mass included is between
4.8 kg and 6.4 kg. The last case corresponds to a cylinder of 15.4
cm in height and diameter with a minimal density of 3 g/cm$^{3}$.
With all the parameters taken at their baseline values, the source
mass is 5.45 kg, its volume is 1.36 liter corresponding to a cylinder
of 12.0 cm in height and diameter. Considering the uncertainty on
density and $\chi_{3}$, the largest capsule has been validated (15
cm in the external diameter of the outer capsule, 13.5 cm in the internal
diameter of the inner capsule).

\subsubsection{\noindent Cerium production line}

\noindent The Federal State Unitary Enterprise Mayak Production Association
is a Russian spent fuel reprocessing facility located in the city
of Ozersk, which performs the reprocessing of Russian SNF. It can
reprocess spent fuel from various nuclear power plants (VVER, BN-600,
and BN-350) as well as from research reactors, atomic-powered ships
and submarines. 

\noindent Russia reprocesses its SNF with the PUREX process which
extracts uranium and plutonium. A secondary and optional phase can
extract a highly concentrated waste of rare-earth elements (REEs),
and a final step allows to separate through liquid chromatography
a specific element, such as samarium used for powerless lights.

\subsubsection*{Standard reprocessing at the radiochemical plant}

Procedures of spent fuel processing, starting from cutting and digestion
through standard PUREX process to preparation of the required feed
for further individual radionuclides separation are depicted in \figref{CeriumExtraction-1}.

\noindent \begin{center}
\begin{figure}[H]
\centering{}\includegraphics[scale=0.45]{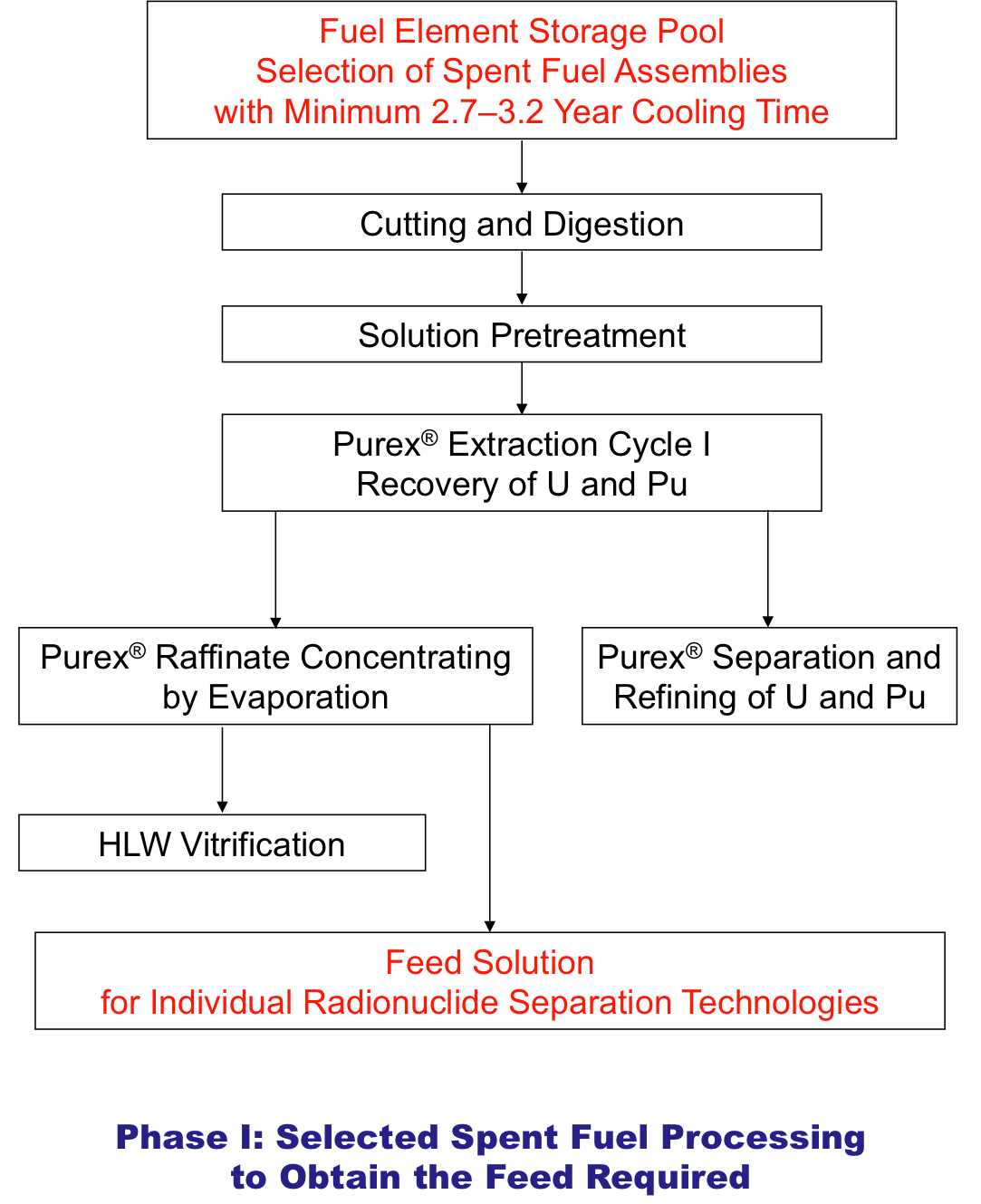}\caption{\label{fig:CeriumExtraction-1}First procedures of PUREX process and
Selected Spent Fuel Processing to obtain the required feed for further
individual radionuclides separation.}
\end{figure}

\par\end{center}

First step is cutting of the ends of fuel assemblies (FA) by electroarc
process. Then fuel part of FA is cut into small pieces and dissolved
by nitric acid. The resulting solution is in form of pulp with superfine
components of graphite, silicone and other elements. These components
can affect the following chemical procedures. This is why one of important
stages of reprocessing is clearing of the pulp by high-molecular organic
flocculating agent and filtration on bulk filter made of stainless
steel metallic spheres and natural minerals. 

Procedure of U, Th and Np recovery (PUREX process) is the extraction
process using multistage mixer settlers with the mixture of tributyl
phosphate and hydrocarbon solvent. A similar process is implemented
by COGEMA (France) and BNFL (UK). After recovery of U, Pu and Np,
PUREX raffinate is transported to another distant workshop of Radiochemical
Plant for its concentrating by evaporation. During the evaporation
procedure a large quantity of nitric acid is recovered from the raffinate
for its reiterative use. Due to big volume of communication pipeline
(the distance of about 1.5 km) the minimum batch of PUREX raffinate
which is transported is about several tens of cubic meters of the
solution. Such a volume of the PUREX raffinate is equivalent to about
10 tons of SNF. It means that minimum quantity of SNF of certain quality
we should select is about 10 tons. 

Most of raffinate after evaporation procedure is directed to vitrification
unit for disposal of high radioactive waste in form of phosphate glasses.
This is one of main task for the Radiochemical Plant RT1 of PA Mayak
in the framework of Russian program of closed fuel cycle. 

Sometimes, accordingly to special request from Customers, a part of
PUREX raffinate is pumped to another distant workshop, Radioisotope
plant, for recovery of radioisotopes for medicine, scientific and
industrial applications. Before pumping the solution is diluted by
water to improve the mobility of this pulp.

\subsubsection*{Additional reprocessing at radioisotope plant}

First operation at Radioisotope plant is recovery of 137Cs isotope
by sorption procedure (see \figref{CeriumExtraction-2}). At the radiochemical
plant the raffinate passed two stages of concentration and purification
of REE and TPE concentrate by double oxalate precipitation. First
and second oxalate procedure consists of oxalate precipitation of
REE and TPE followed by dissolving the precipitate by nitric acid
(digestion of oxalate). Precipitation of oxalates allows to separate
the bulk of impurities present in the raffinate and to obtain relatively
pure REE and TPE group. Main impurities which should be removed are
Cesium ($^{137}$Cs) and Strontium ($^{90}$Sr), and chemical elements
after corrosion (Fe, Ni, Cr). 

\noindent \begin{center}
\begin{figure}[H]
\begin{centering}
\includegraphics[scale=0.45]{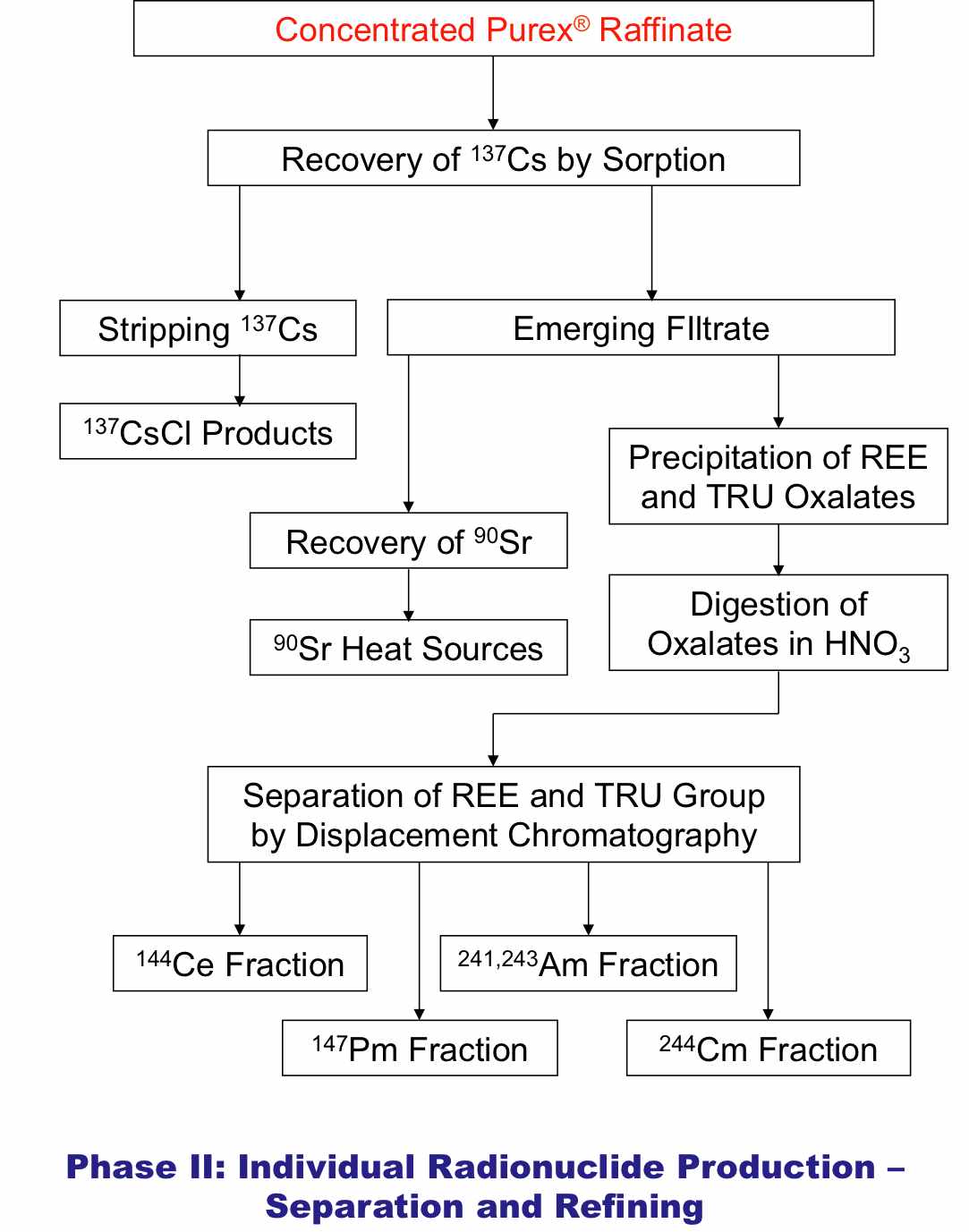}\caption{\label{fig:CeriumExtraction-2}Individual radionuclides separation
and refining.}

\par\end{centering}

\centering{}
\end{figure}

\par\end{center}

\subsubsection*{Cerium Recovery}

\noindent REE and TPE concentrate after correction by means of nitric
acid are transferred by a pipeline to the chromatographic facility
for separation of REE and TPE at the distance of about 500 m. The
pipeline and additional vessels should be purified before this procedure.

\noindent Separation of cerium is performed by complexion displacement
chromatography (CDC) using a series of columns, the first one being
a sorption column and the others being separating columns. REEs are
sorbed as a uniform band in the sorption column and are later separated
using a series of columns packed with a resin containing retaining
ion (Ni$_{2}$$^{+}$, Cu$_{2}$$^{+}$, H$^{+}$, etc.). 

\noindent \begin{center}
\begin{figure}[H]
\centering{}\includegraphics[scale=0.45]{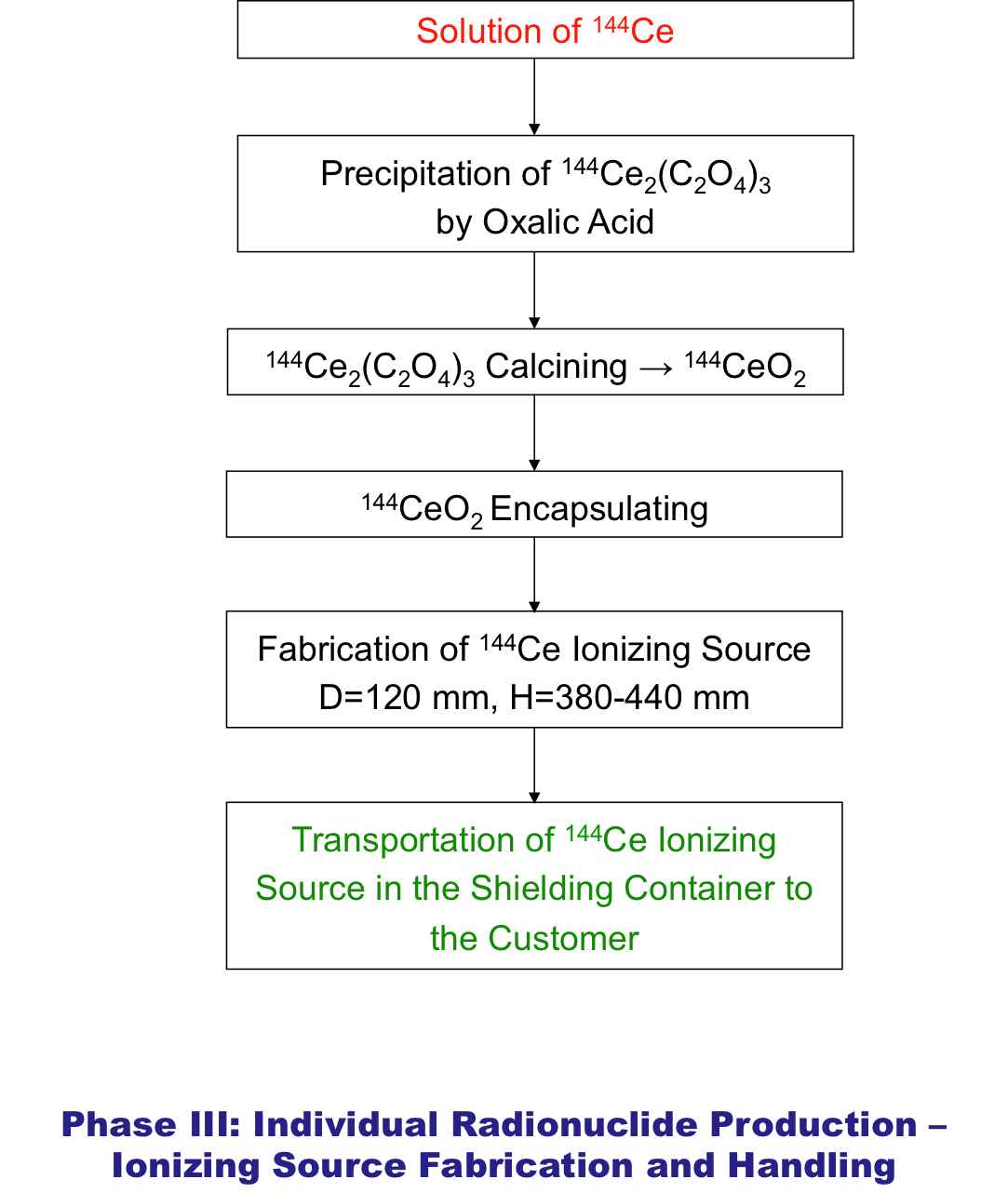}\caption{\label{fig:CeriumExtraction-3}Schematic view of the cerium extraction
process.}
\end{figure}

\par\end{center}

\noindent The main role of the retaining ions is to decompose the
REE complex during the entire separation process and to favor repeated
sorption-desorption cycles. The complexes of the eluent with the retaining
ion should be stronger than those with the ions being separated. The
sorption of the retaining ion on the resin should be weaker than that
of the ions being separated. To separate REEs, a complexing agent
containing displacing ions, NH$_{4}$$^{+}$, or Na$^{+}$, is used.
The sorption of this agent is stronger than that of REEs. When a column
is washed with this agent, NH$_{4}$$^{+}$ cations substitute REE$_{3}$$^{+}$
cations. As a results, the resin is transformed into the NH$_{4}$$^{+}$
form and REE ions form complexes with the eluent (EDTA). These complexes
are decomposed by reaction with the resin in the form of the retaining
ion Ni$_{2}$$^{+}$. The released REE ions are sorbed again and displace
Ni$_{2}$$^{+}$. Thus, separate REE bands are formed in the columns.
The order of these bands corresponds to the strength of the REE complexes
(see \figref{MayakChromatography}). A diagram of the cerium extraction
process is given in \figref{CeriumExtraction-3}. An generic output
of the chromatography is given in \figref{MayakChromatography}.

\noindent \begin{center}
\begin{figure}[H]
\centering{}\includegraphics[scale=0.5]{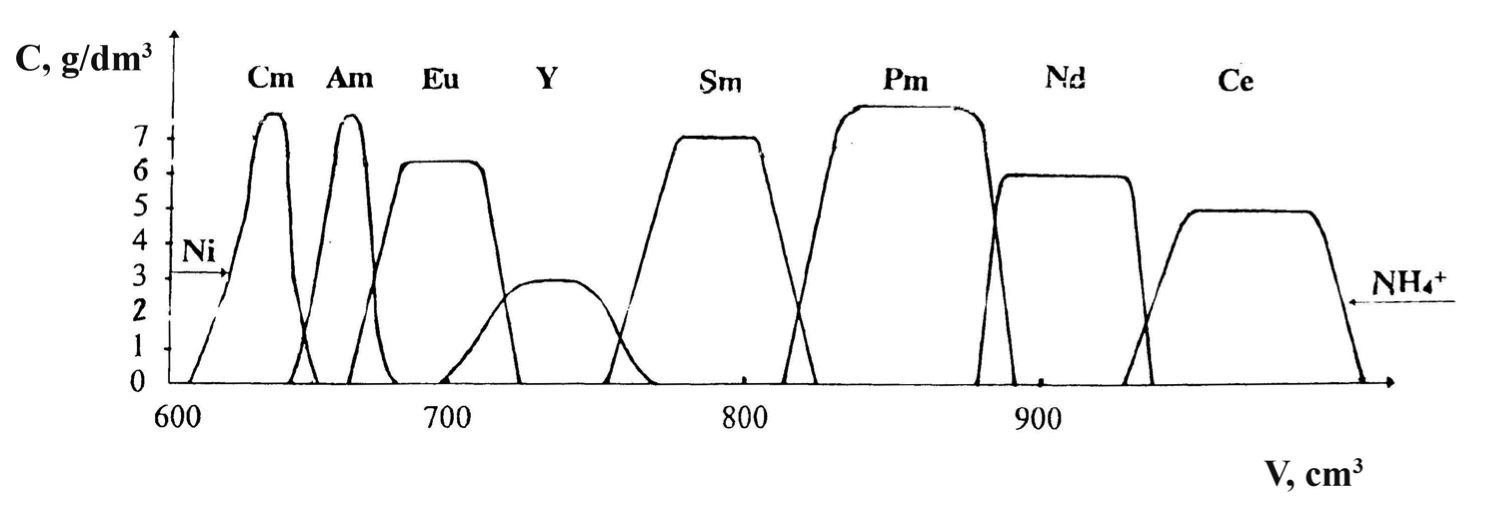}\caption{\label{fig:MayakChromatography}Selection of cerium through displacement
chromatography. We note here that cerium if far away from curium and
americium which guarantee a low concentation of neutron emitter impuritites
in the final product.}
\end{figure}

\par\end{center}

\subsubsection{\noindent $^{144}$Ce source pressing and encapsulation}

\noindent The source will be encapsulated in a cylindrical sealed
double stainless steel container, the outer capsule having a 15 cm
external diameter. The outer capsule will be 4 mm thick, the inner
capsule 3 mm thick, and the mechanical gap will be 0.5 mm (to be confirmed
by Mayak), leading to an inner capsule with a 13.5 cm in internal
diameter (see \figref{75kCiSourceCapsule} for a preliminary design
proposed by CEA before Mayak constraints were known). The final design
is ongoing. The capsules will be welded inside a hot cell (argon-arc
welding). The tightness of the capsules will be tested using the Mayak's
standard helium leakage test. The capsule will be certified by Mayak
according to the ISO 9978-92 (Radiation protection - Sealed radioactive
sources - Leakage test methods) regulation.

\noindent \begin{center}
\begin{figure}[H]
\centering{}\includegraphics[scale=0.25]{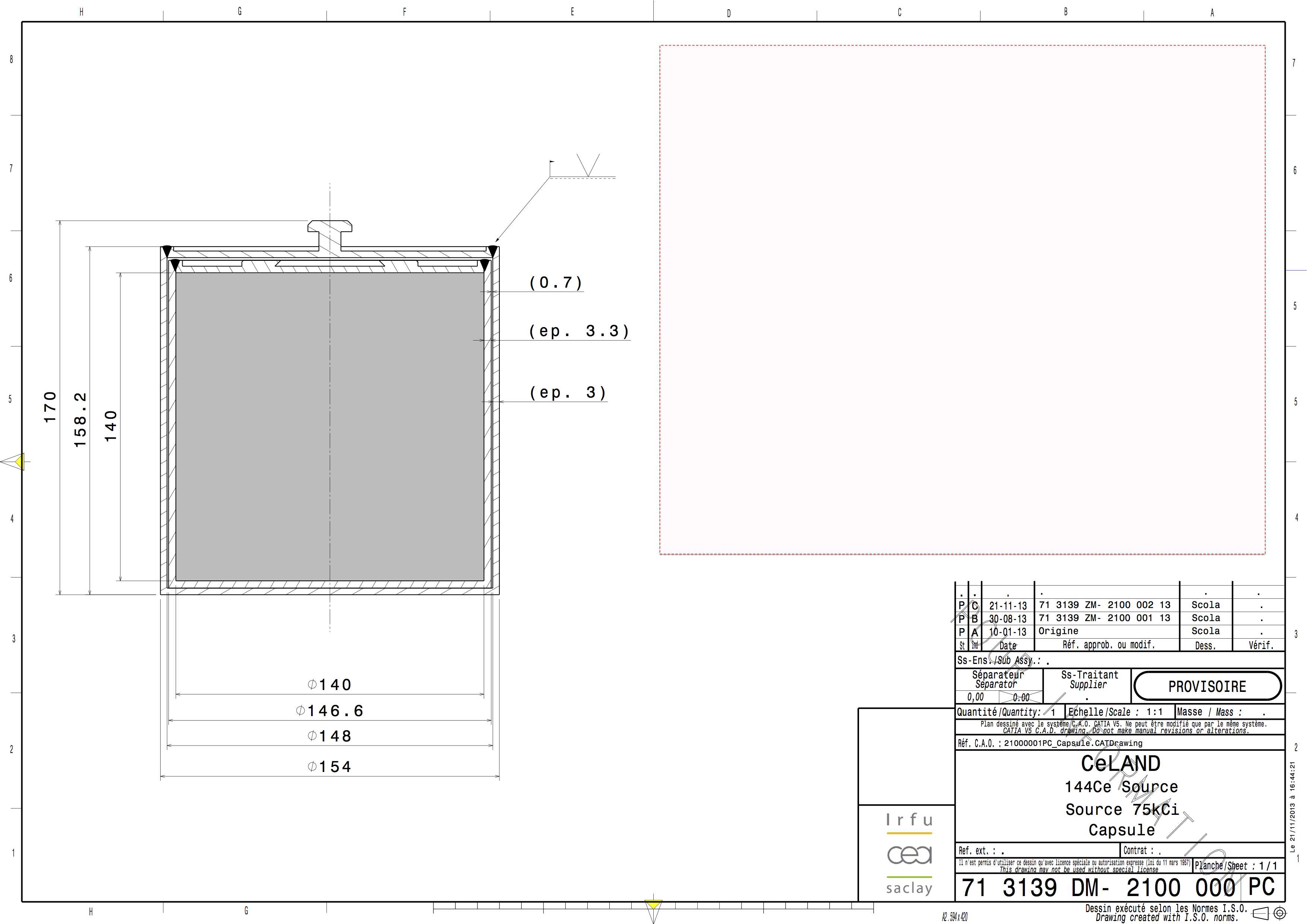}\caption{\label{fig:75kCiSourceCapsule} \emph{Preliminary} 75 kCi $\mathrm{^{144}Ce-^{144}Pr}$
source capsule (cylinder with a diameter of 15 cm height and radius).
Dimensions have to be updated, and a new design of the lid holder
(mushroom-like) is necessary for proper manipulation inside the hot
cells.}
\end{figure}

\par\end{center}

\noindent The cold pressing of the cerium oxide will be done in a
hot cell at PA Mayak. In principle, a density of 5 g/cm$^{3}$ can
be achieved since the $\mathrm{CeO_{2}}$ has a theoretical density
of 7.2 g/cm$^{3}$, but it could be as low as 3 g/cm$^{3}$. This
is being checked with large samples in laboratory. The current density
is expected to be $4\pm1$ g/cm$^{3}$. Prior to insertion, the cerium
oxide will be pressed under the form of large solid pellets, with
a diameter slightly smaller than the capsule.

\noindent Finally, the $\mathrm{CeO_{2}}$ material and source container
will be qualified as SFRM (Special Form of Radioactive Material).
This is a mandatory step prior to the source container insertion within
the tungsten shielding.

\noindent The CeLAND $\mathrm{CeO_{2}}$ source production scenario
assumes a 1-year production time between SNF arrival at the PA Mayak
reprocessing facility and source delivery in Japan. Mayak's schedule
is slightly shorter, about 10 months.

\subsubsection{\noindent Target contamination}

\noindent $^{144}$Ce will be extracted from spent nuclear fuel elements.
Though expected to be very small, contamination of the final $^{144}$Ce
source with other radio-elements is unavoidable. The extraction process
leads to a small contamination from other REEs and minor actinides,
that may be however relevant for realizing a neutrino experiment. 

\noindent The REE contaminants are $\mathrm{\gamma}$ and $\beta$
emitters, which could bias the activity measurement or participate
to $\mathrm{\gamma}$-induced accidental background. The contamination
in REEs must therefore be small enough to provide a negligible contribution
to activity, a requirement easily achieved through the CDC method
used at PA Mayak. Because the source has to be shielded against 2.185
MeV $^{144}$Pr $\mathrm{\gamma}$-rays, any $\mathrm{\gamma}$-rays
from contaminants will be shielded as well. As explained in section
\ref{sec:High-Z-gamma-ray-shielding}, the $\mathrm{\gamma}$ attenuation
length is close to its minimum at 2.2 MeV and will therefore drives
the shielding dimensions. 

\noindent Actinides are produced in a reactor core through successive
neutron captures and beta decays from $^{238}$U, as shown by figure
\ref{fig:actinides}. On one hand, they are $\alpha$, $\beta$ and
$\gamma$ emitters feeding long decay chains that require similar
high-Z shielding than for REE contaminants. On the other hand, the
heavy neutron rich actinides could be more hazardous since they also
undergo Spontaneous Fission (SF). This emission of several $\gamma$-rays
and neutrons is therefore a potential source of correlated background
for the CeLAND experiment.

\noindent \begin{center}
\begin{figure}[h]
\begin{centering}
\includegraphics[scale=0.45]{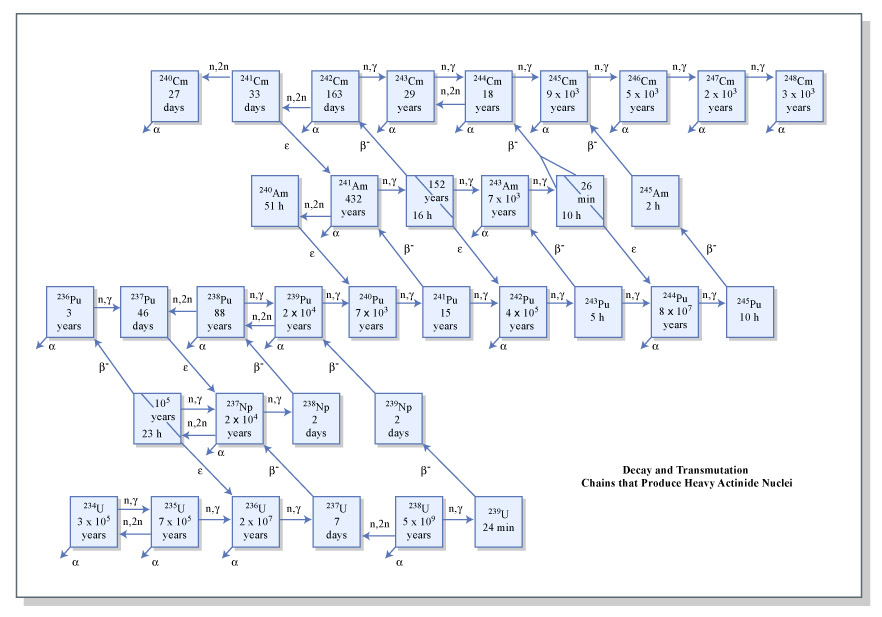}\caption{\label{fig:actinides}Production of minor actinides during irradiation
of a reactor core.}

\par\end{centering}

\end{figure}

\par\end{center}

\noindent Our contamination specifications are consequently the following:
\begin{itemize}
\item \noindent Thermal power of nuclides other than $^{144}$Ce and $^{144}$Pr
should be less than 0.1 \% of the thermal power delivered by $^{144}$Ce
and $^{144}$Pr ($7.991\pm0.044$ W/kCi), in order to guarantee the
precision of the calorimetry measurement;
\item \noindent Gamma activity above 1 MeV should be less than 0.1 \% of
the beta activity of $^{144}$Ce;
\item \noindent Activity of spontaneous fission radio impurities to be defined
by Saclay in collaboration with the MePhi institute, ICP institute
and Isoflex USA.
\end{itemize}
\noindent Since the attenuation corresponding to 2.185 MeV $\mathrm{\gamma}$-rays
is very close to the minimal attenuation length of tungsten, any high
energy $\mathrm{\gamma}$-rays (above 1 MeV) will at least be as attenuated
as the 2.185 MeV $\mathrm{\gamma}$-rays from $^{144}$Pr. The 0.1\%
$^{144}$Ce beta activity specification on $\mathrm{\gamma}$-rays
therefore ensures that the most active $\mathrm{\gamma}$-rays will
come from the $^{144}$Pr 2.185 MeV $\mathrm{\gamma}$ emission (0.7\%
branching ratio). Thus, the gamma radiation shielding dimensions should
only be driven by the activity of the $^{144}$Pr 2.185 MeV $\mathrm{\gamma}$-ray.

\subsubsection{Source induced background}

The CeLAND source could potentially add backgrounds to the signal
measurement through decays of $^{144}$Ce and $^{144}$Pr ($\beta^{-}$
electron, gamma, bremsstrahlung) or through decays of impurities (see
previous section), if deployed too close or inside the KamLAND target
volume. Since the $^{144}$Ce is extracted from spent nuclear fuel,
a particular concern is the presence of heavy minor actinides that
undergo SF with non negligible branching ratios (reaching 74 \% for
$^{250}$Cm), emitting $\gamma$-ray cascades and several fast neutrons.

\paragraph{\noindent Accidental backgrounds}

\noindent First, $\beta^{-}$ decays of $^{144}$Pr emit 2.185 MeV
$\gamma$-rays (0.7\% branching ratio), which lie both within the
prompt and delayed energy windows. The biological shielding should
be able to provide enough attenuation for deployment outside the tank
(an additional plate could be placed between the shielding and the
buffer tank). However, a deployment at the center of the detector
would require much more shielding.

\noindent Second, neutrons from SF can scatter out of the source despite
the high-Z shielding. Escaping neutrons can be captured on the detector
materials, such as hydrogen in water, oil or in the liquid scintillator,
consequently emitting 2.2 MeV $\gamma$-rays both affecting the prompt
and delayed energy windows. As opposed to the 2.185 MeV $\gamma$-rays,
which can be easily attenuated through the high-Z shielding, neutron
induced background might dominate any other source of backgrounds
despite a much lower activity in actinide's SF, which depends on the
$^{144}$Ce contamination with heavy minor actinides.

\paragraph{Correlated backgrounds}

\noindent The only correlated backgrounds induced by the source come
from SF. Fast neutrons can reach the liquid scintillator target and
make a proton recoil faking the prompt signal and slow down until
being captured in the liquid scintillator, with the same coincidence
time than a neutrino signal. Fortunately, the hydrogen-rich buffer
oil separates the source from the target liquid scintillator, and
fast neutrons have great chance to loose energy through scattering
before the target. Moreover, detector response to fast neutron is
affected by quenching and fast neutrons rarely give more than 1.022
MeV of visible energy (IBD energy threshold). This background should
therefore not affect the experiment.

\noindent The multiple neutron background is potentially more hazardous.
Spontaneous fission releases several neutrons in two opposite beams
(neutrons are actually released by the two energetic fission fragments
and are therefore strongly boosted), of which two can be scattered
in the detector direction and then be captured on hydrogen inside
or close to the target. The $\gamma$-rays resulting from the two
captures could fake an IBD signal. Table \ref{tab:Neutron-emission}
shows the half-life, branching ratios to spontaneous fission and specific
neutron activities for all americium, curium, berkelium and californium
isotopes with half-life higher than 180 days. The nuclei of interest
are the curium and californium isotopes with an even mass number ($^{244,246,248,250}$Cm
and $^{248,250,252}$Cf). All americium isotopes and other curium
and californium isotopes with odd mass number have negligible branching
ratios. Berkelium isotopes have very short half-life period, and nuclei
heavier than californium (Es, Fm) can be safely ignored since they
are produced in negligible quantities in a reactor core and have a
very short half-life period. It is worth noting that the mean number
of neutron released per SF $N$ increases with the nucleus mass number,
following roughly $N=0.1094\: A-23.94$ in our mass range. For instance,
$^{252}$Cf releases 50 \% more neutrons per SF than $^{241}$Am (3.6
and 2.4 respectively).

\noindent The specific activities of the relevant heavy minor actinides
have to be weighted by their abundance in spent nuclear fuel to correctly
estimate the level of multiple neutron correlated background in the
KamLAND detector. First, nuclei with odd mass number generally have
large fission cross-sections, leaving radiative capture as a minor
deexcitation process. Second, fresh nuclear fuel does not contain
any isotope heavier than $^{238}$U. All heavy minor actinide isotopes
are then produced through a long chain of successive neutron captures
and $\beta^{-}$ decays, with the fission trap on odd nucleon number
isotopes. The burn-up of spent nuclear fuel has then to be large enough
to start producing heavy elements. Such threshold in burn-up of course
increases with the isotope mass. Moreover, in power plant, reactors
are generally operated at constant power with low enriched fuel. The
neutron flux is then roughly constant, and the mass of $^{238}$U
is so large that it can be considered constant. The production of
heavy actinide isotope with mass number A+1 depends of neutron flux,
cross-section, decays and mass of isotope A. The mass ratio between
an heavy actinide isotope A and the following isotope A+1 is therefore
expected to be constant once the threshold in burn-up of isotope A+1
is reached. Available data show that the total mass of a given curium
isotope with even mass number A is roughly a factor $\mathrm{10^{2}}$
more than the following even isotope A+2. This effect should then
counterbalance the increase of SF branching ratios with mass number.
Detailed simulations are ongoing to precisely estimate heavy isotope
abundance is spent nuclear fuel.

\begin{center}
\begin{table}[h]
\begin{centering}
\begin{tabular}{cc>{\centering}p{2.5cm}>{\centering}p{3.5cm}}
\hline 
Isotope & Half-life & Branching ratio to SF & Specific neutron activity (neutron/g)\tabularnewline
\hline 
\hline 
$^{241}$Am & 432 y & $4\cdot10^{-12}$ & 1.2\rule{0pt}{2.25ex}\tabularnewline
$^{242m}$Am & 141 y & $4.7\cdot10^{-11}$ & 46\tabularnewline
$^{243}$Am & 7370 y & $3.7\cdot10^{-11}$ & 0.72\tabularnewline
$^{243}$Cm & 29 y & $5.3\cdot10^{-11}$ & $2.6\cdot10^{2}$\tabularnewline
$^{244}$Cm & 18 y & $1.4\cdot10^{-6}$ & $1.2\cdot10^{7}$\tabularnewline
$^{245}$Cm & 8500 y & $6.1\cdot10^{-9}$ & $1.1\cdot10^{2}$\tabularnewline
$^{246}$Cm & 4730 y & $3\cdot10^{-4}$ & $1.0\cdot10^{7}$\tabularnewline
$^{248}$Cm & $3.5\cdot10^{4}$ & 0.083 & $4.2\cdot10^{7}$\tabularnewline
$^{250}$Cm & 9000 y & 0.74 & $1.5\cdot10^{10}$\tabularnewline
$^{249}$Bk & 320 d & $4.7\cdot10^{-10}$ & $9.4\cdot10^{4}$\tabularnewline
$^{248}$Cf & 333 d & $2.9\cdot10^{-5}$ & $5.4\cdot10^{9}$\tabularnewline
$^{249}$Cf & 351 y & $5\cdot10^{-9}$ & $2.5\cdot10^{3}$\tabularnewline
$^{250}$Cf & 13 y & $8\cdot10^{-4}$ & $1.1\cdot10^{10}$\tabularnewline
$^{251}$Cf & 898 y & $1\cdot10^{-7}$ & $2.0\cdot10^{4}$\tabularnewline
$^{252}$Cf & 2.65 y & 0.030 & $2.2\cdot10^{12}$\tabularnewline
\hline 
\end{tabular}
\par\end{centering}

\caption{\label{tab:Neutron-emission} Spontaneous fission properties of minor
actinide isotopes with half-life > 0.5 y}

\end{table}

\par\end{center}

\noindent The most abundant curium isotope is $^{244}$Cm, with a
specific neutron activity comparable to $^{246}$Cm and $^{248}$Cm.
According to our preliminary GEANT4 simulations, the correlated background
induced by $^{244}$Cm is 5000 events per year in the KamLAND detector,
with a 6 m radius fiducial volume cut and assuming a contamination
of $10^{-9}$Bq/Bq. This assumption is taken from a limit on the level
of contamination by minor actinides observed in Promethium production.
However, the amount of multiple neutron correlated background in KamLAND
is expected to be lower, because Cerium is more efficiently separated
from actinides than Promethium using the complexion displacement chromatography
method (see figure \ref{fig:MayakChromatography}).

\noindent A neutron shielding solution, using saturated boric water
surrounding the tungsten alloy shielding, is envisaged if the contamination
in heavy minor actinides turns out to be non negligible. Commercial
tungsten alloys are known to resist to boric water at room temperature.
Corrosion test are currently being performed to clearly demonstrate
that point.

\subsubsection{\noindent Summary}

\noindent Below are summarized the specifications of the source in
the standard scenario that we will rely on in estimating the physics
and sensitivity reach of the CeLAND project:
\begin{enumerate}
\item \noindent Spent fuel cooling time less than 2 years prior to transportation
to PA Mayak; 
\item \noindent Source beta activity of 85$_{-10}^{+15}$ kCi of $^{144}$Ce
at delivery ;
\item \noindent Chemical form of cerium is $\mathrm{CeO}_{2}$ with density
$4\pm1$ g/cm$^{3}$;
\item \noindent External diameter of the capsules is 15 cm, internal diameter
is 13.5 cm ;
\item \noindent Samples of cerium from the same production batch will be
provided with the source. 
\end{enumerate}

\section{\noindent High-Z gamma ray shielding\label{sec:High-Z-gamma-ray-shielding}}

\noindent The most serious source of backgrounds is the decay of the
$^{144}$Nd $1^{-}$ excited state, which emits a 2.185 MeV $\mathrm{\gamma}$-ray
with a 0.7\% intensity (see \tabref{sourcefeatures}). The bremsstrahlung
of the $^{144}$Pr $\mathrm{\beta}$ particles might also be another
source of backgrounds. The source must be shielded by a thick high-Z
absorber to attenuate these backgrounds to the required level. With
a theoretical density of 19.3 g/cm$^{3}$, tungsten is widely used
in radiation shielding applications and is among the best suited candidate
for this purpose.

\subsection{\noindent Attenuation of $\mathrm{\gamma}$ rays in CeLAND}

\noindent The total interaction cross-section of $\mathrm{\sim}$MeV
$\gamma$ rays with matter is equal to the sum of three partial cross-sections:
$\sigma=\sigma_{f}+\sigma_{c}+\sigma_{p},$ where $\sigma_{f}$ is
the photoelectric effect cross-section, $\sigma_{c}$ is the Compton
interaction cross-section and $\sigma_{p}$ is the pair creation cross-section.
A beam of gamma radiation passing through a layer of absorber material
of thickness x will be exponentially attenuated: $\mathrm{I(x)=I_{0}e^{-\left(\frac{\mu}{\rho}\right)\rho x}}$,
where $\mathrm{I_{0}}$ is the intensity of the incident beam and
$\mu/\rho$$ $ is the mass attenuation coefficient. This coefficient
is taken from \cite{PhotoAtomicData} for each CeLAND material that
will be later considered in our $\mathrm{\gamma}$ ray attenuation
calculations. The $\mathrm{\mu}$ coefficient definition is $\mu=\sigma\times n$,
where n is the atomic concentration in the material. In the MeV energy
regime, Compton scattering is the dominant interaction process so
that the mass-attenuation coefficient is approximately the same for
absorbers with different Z. This is because the probability of Compton
scattering on atomic electrons is proportional to Z, making $\mathrm{\mu\sim\sigma_{c}Zn}$
and $\mathrm{\mu/\rho\sim\sigma_{c}Z/M_{A}}$ since $\mathrm{\rho=n\times M_{A}}$,
where $\mathrm{M_{A}}$ is the atomic mass. The inverse of the attenuation
coefficient is $\lambda=$1/$\mu$ and has units of length, so that
it can be associated to a mean free path or an attenuation length.
In other words, $\mathrm{\lambda}$ corresponds to the average distance
the $\gamma$ rays can travel in the absorber without interacting.
The attenuation length dependence on the radiation energy is shown
in \figref{gamma-attenuation-length} for typical materials making
the CeLAND source and shielding, as well as the KAMLAND detector liquid
scintillator. A summary of the relevant attenuation lengths used for
the design of the CeLAND phase 1 and 2 shieldings is given in \tabref{SummaryAttLength}.
The design of the shielding of CeLAND phase 2 is still under consideration.

\noindent \begin{center}
\begin{figure}[h]
\begin{centering}
\includegraphics[scale=0.37]{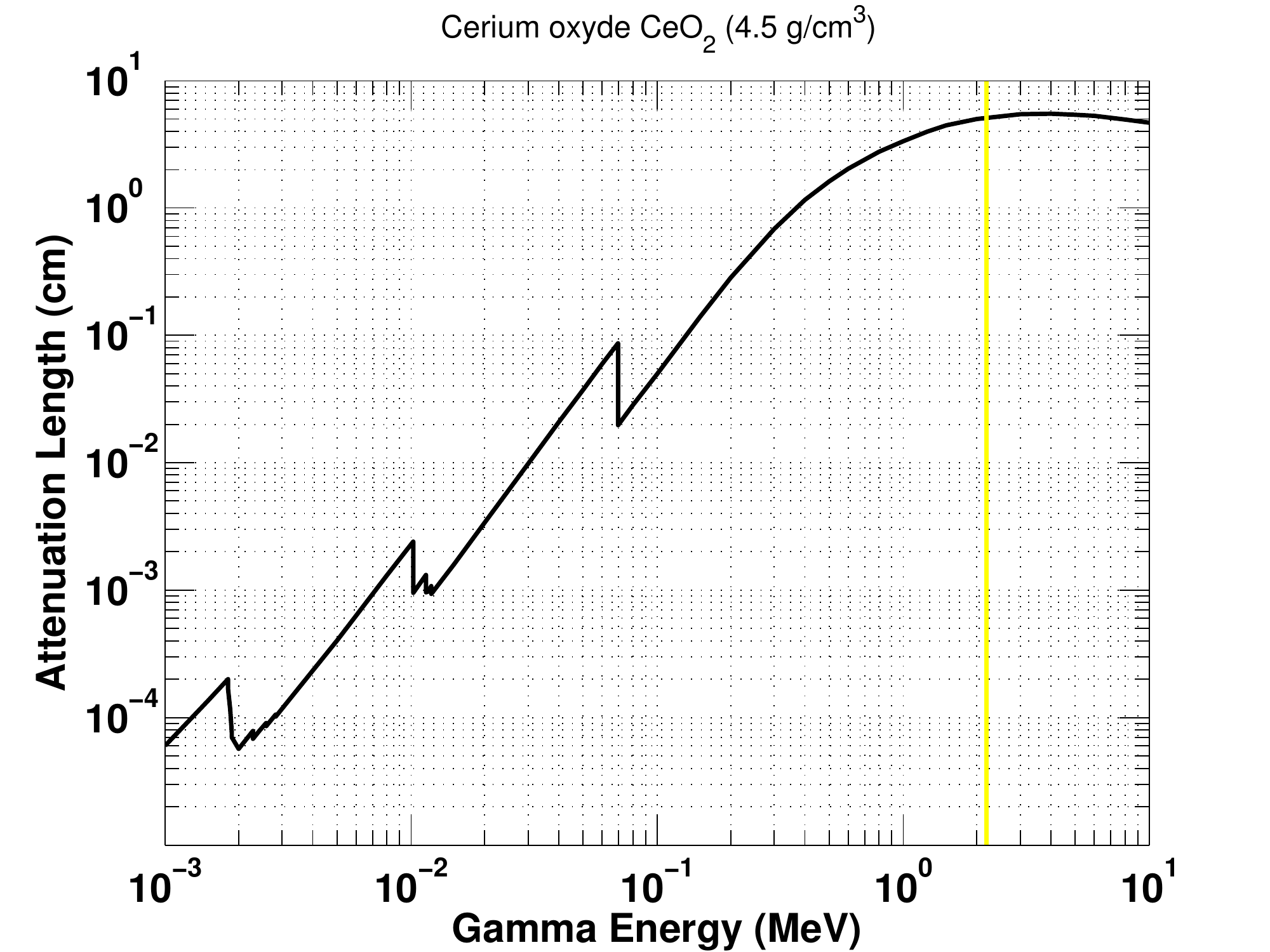}\includegraphics[scale=0.37]{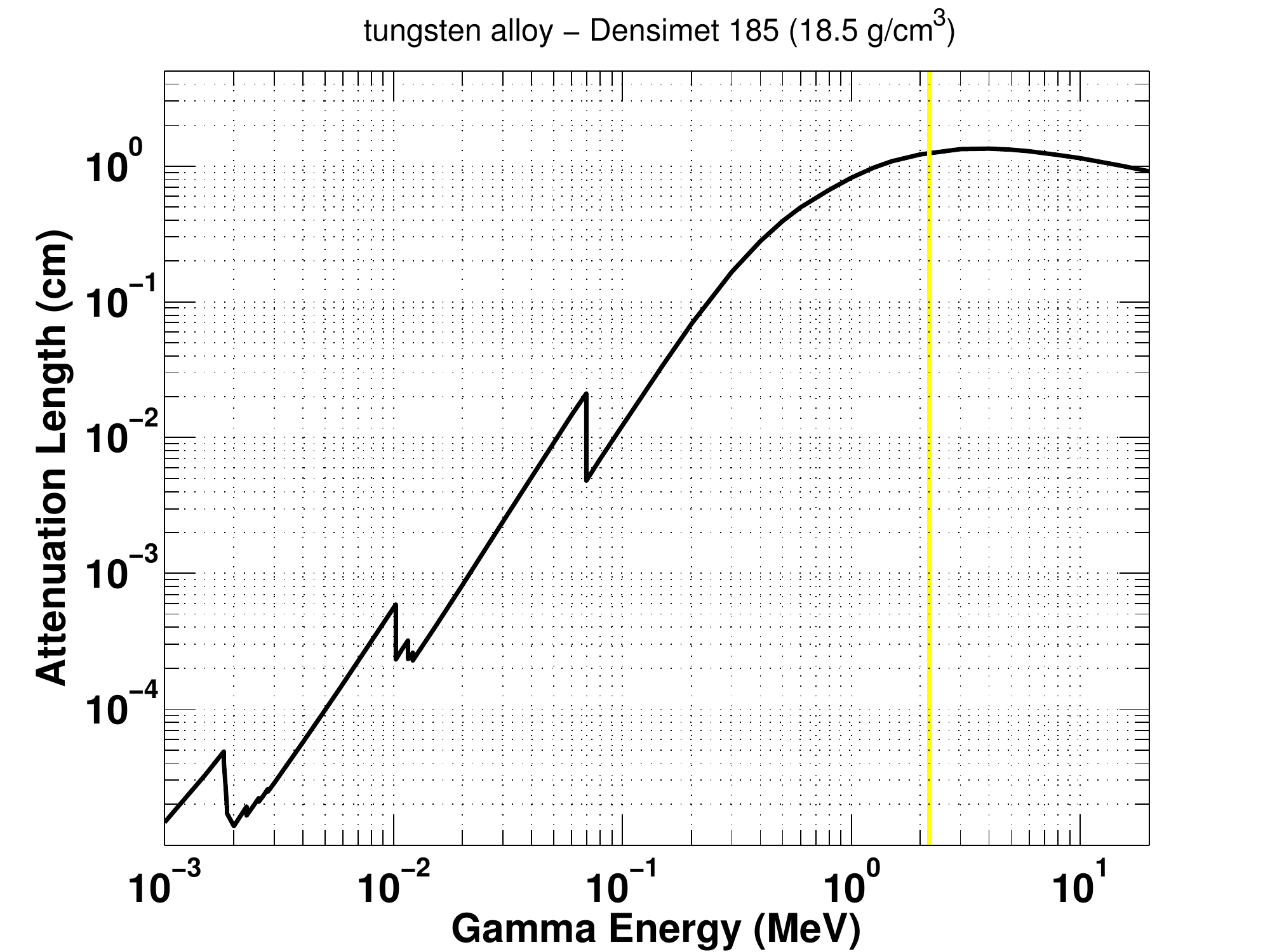}\\
\includegraphics[scale=0.37]{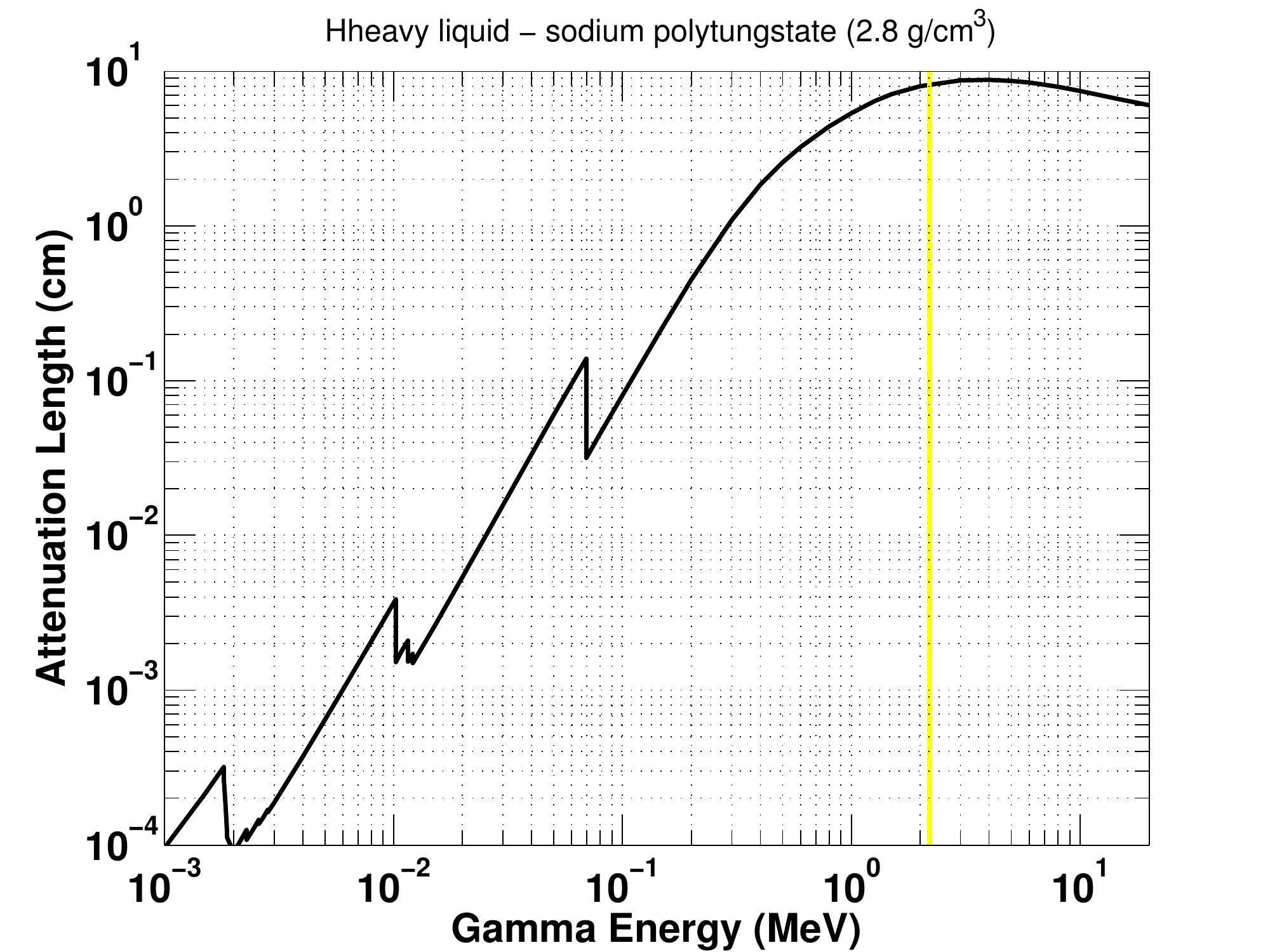}\includegraphics[scale=0.37]{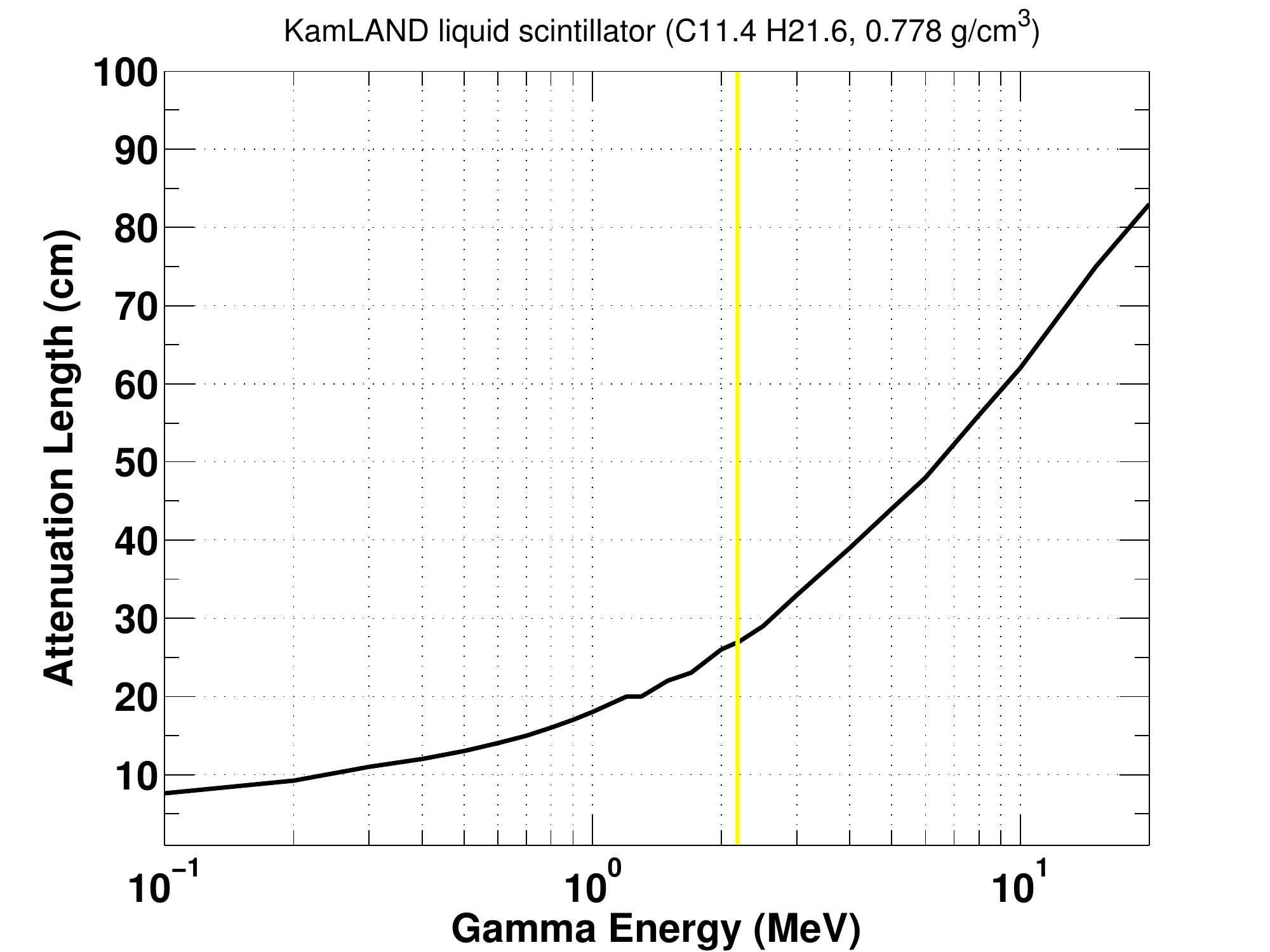}
\par\end{centering}

\centering{}\caption{\label{fig:gamma-attenuation-length}$\mathrm{\gamma}$-ray attenuation
lengths of some CeLAND materials. From left to right and top to bottom:
cerium oxide, tungsten alloy, sodium paratungstate, and KamLAND liquid
scintillator. The yellow line indicates the 2.185 MeV $\gamma$ line
produced by the decay through excited states of $^{144}$Pr, which
set the stronger constraint in the CeLAND source shielding design. }
\end{figure}

\par\end{center}

\noindent \begin{center}
\begin{table}[h]
\begin{centering}
\begin{tabular}{cccccc}
\hline 
$\gamma$ Absorber & Form & density & $\lambda(1$ MeV) & $\lambda$(2.185 MeV) & $\lambda$(2.655 MeV)\tabularnewline
\hline 
\hline 
Tungsten alloy  & solid & 18.5 g/cm$^{3}$ & 0.817 cm & 1.238 cm & 1.288 cm\tabularnewline
Mercury & liquid & 13.5 g/cm$^{3}$ & 1.059 cm & 1.644 cm  & 1.709 cm\tabularnewline
CeO$_{2}$ & solid & 4.5$\pm$0.5 g/cm$^{3}$ & 3.358 cm & 5.089 cm & 5.293 cm\tabularnewline
Sodium paratungstate & liquid & 2.8 g/cm$^{3}$ & 5.358 cm & 8.121 cm & 8.446 cm\tabularnewline
KamLAND scintillator & liquid & 0.778 g/cm$^{3}$ & 18.000 cm & 26.922 cm & 30.132 cm\tabularnewline
KamLAND buffer oil & liquid & 0.762 g/cm$^{3}$ & 18.367 cm & 27.472 cm & 30.747 cm\tabularnewline
\hline 
\end{tabular}
\par\end{centering}

\caption{\label{tab:SummaryAttLength}Summary of the attenuation lengths of
$\gamma$ absorber being considered for CeLAND. The biological shielding
will consist of the source itself (CeO$_{2})$ surrounded by a thick
tungsten alloy cylinder. It will be used as a $\gamma$-ray shielding
for the CeLAND source deployment outside the detector target. In such
a deployment scenario, further attenuation is naturally provided by
the detector buffer oil and liquid scintillator. For deployment inside
the KamLAND sphere, a shielding solution using heavy liquids such
as sodium paratungstate or mercury is being considered.}

\end{table}

\par\end{center}

\subsection{\noindent Biological protection}

\noindent Any exposure to ionizing radiation presents risks of biological
damage. As such, the tungsten alloy shielding must also act as a biological
protection during source transportation and manipulation. After the
source is produced, it will be inserted in the shielding in a hot
cell by the Mayak reprocessing plant. 

\noindent \begin{center}
\begin{table}[H]
\centering{}%
\begin{tabular}{cc}
\hline 
Chemical composition:  & W (97\%), Ni (1.5\%), Fe (1.5\%)\tabularnewline
Density {[}g/cm$^{3}${]} & 18.5\tabularnewline
Tensile strength at 20 \textdegree{}C {[}MPa{]} & 800\tabularnewline
Oxidation resistance & up to 600 \textdegree{}C\tabularnewline
Thermal conductivity at 500 \textdegree{}C {[}W/m�K{]} & 70-90\tabularnewline
Modulus of elasticity  {[}GPa{]} & 385\tabularnewline
Hardness {[}HRC{]} & 43\tabularnewline
Yield strength at 20 \textdegree{}C  {[}MPa{]} & 600\tabularnewline
Breaking elongation at 20 \textdegree{}C  {[}\%{]} & 10\tabularnewline
Coefficient of thermal expansion {[}$10^{-6}$ K$^{-1}${]} & 5.0\tabularnewline
\hline 
\end{tabular}\caption{\label{tab:Densimet-185-properties}Mechanical and thermal properties
of a typical tungsten alloy used for radiation shielding applications.}
\end{table}

\par\end{center}

\subsection{\noindent Equivalent dose}

\noindent In order to assess the tungsten shielding dimensions, and
especially the thickness, a number of factors must be taken into consideration: 
\begin{itemize}
\item \noindent the nature of the ionizing radiation
\item \noindent the strength of the source, which drives the intensity of
the 2.185 MeV escaping $\mathrm{\gamma}$-rays and thus the required
attenuation to suppress this background
\item \noindent the level of exposure of the area surrounding the source
to ionizing radiations, with parameters such as time and distance
from the source to evaluate the impact of escaping radiations on human
bodies
\end{itemize}
\noindent The only ionizing radiation which eventually escapes from
the source are $\mathrm{\gamma}$-rays. As a first estimate, the equivalent
dose (in Sievert) induced by a radioactive point source at the center
of the shielding is computed. The absorbed dose (Gy/s) is given by:

\noindent 
\begin{equation}
\mathrm{D(J/g/s)=\mathcal{A}[Bq]\times\frac{1}{4\pi d^{2}[cm^{2}]}\times\frac{\mu}{\rho}[g/cm^{2}]_{en}\times E[J]}\label{eq:eq-dose-point-source}
\end{equation}

\noindent where d is the source-person distance in cm, $\mathrm{\mathcal{A}}$
is the source $\mathrm{\gamma}$ activity (in Bq) at the surface of
the biological protection (i.e. after shielding), $\mathrm{\mu/\rho}$
is the mass absorption coefficient (the fraction of energy of the
gamma rays transferred to the medium, in $\mathrm{cm^{2}/g}$), and
E the energy of the escaping $\mathrm{\gamma}$-ray (in Joule). Usually,
$\mathrm{\mu/\rho\sim25\, g/cm^{2}}$ for $\mathrm{\gamma}$-ray energies
between 1 and 3 MeV. Furthermore it is almost medium independent,
as explained in the previous section. The equivalent dose (in Sv)
is calculated by multiplying the absorbed dose by the $\mathrm{\gamma}$
radiation weighting factor, w=1. After plugging in the correct numerical
factors in equation \ref{eq:eq-dose-point-source}, the equivalent
dose becomes:

\noindent 
\begin{equation}
\mathrm{D(mSv/h)=4.24\times\mathcal{A}[Ci]\times E[MeV]/d[m]^{2}}\label{eq:dose-boe}
\end{equation}

\noindent For a 75 kCi $\mathrm{^{144}Ce-^{144}Pr}$ source the activity
corresponding to the 2.185 MeV $\gamma$ rays (BR=0.7\%) is 525 Ci.
Considering a 16-cm thick spherical tungsten alloy shielding (d=18.5
$\mathrm{g/cm^{3}}$, attenuation length of 1.26 cm at 2.185 MeV)
the 2.185 MeV $\gamma$ activity is reduced to $\mathrm{1.3\times10^{-3}}$
Ci at the surface of the biological protection. According to equation
\ref{eq:dose-boe}, the absorbed dose received at 1 m from the shielding
is 14 $\mathrm{\mu}$Sv/h. As a comparison, the maximum tolerable
annual dose for people working in environments with ionizing radiations
is given by international regulations to be 20 mSv/y (see table \ref{tab:IAEA-dose-limits}),
corresponding to about 1730 hours/y of exposure for somebody standing
1 m away from the source. 

\noindent \begin{center}
{\small }
\begin{table}[H]
\begin{centering}
\begin{tabular}{cc}
\hline 
\multirow{2}{*}{Criteria} & IAEA dose Limit\tabularnewline
 & for exposed workers\tabularnewline
\hline 
\hline 
Effective dose limit for workers in category A & 20 mSv/y\tabularnewline
Effective dose limit for workers in category B & 6 mSv/y\tabularnewline
Maximum effective dose allowed in any single year & 50 mSv/y\tabularnewline
Total work life (50 y) & 400 mSv\tabularnewline
\hline 
\end{tabular}
\par\end{centering}

\centering{}{\small \caption{\label{tab:IAEA-dose-limits}IAEA radiation dose limits for various
working conditions.}
}
\end{table}

\par\end{center}{\small \par}

\noindent A more accurate computation of the radiation dose using
the GEANT4 simulation package has been performed. We consider here
a 75 kCi $\mathrm{^{144}Ce-^{144}Pr}$ source in the form of $\mathrm{CeO_{2}}$,
encapsulated in a 5-cm radius sphere with a density of 4 g/cm$^{3}$.
A sphere-shaped tungsten alloy shielding is simulated taking a typical
composition for tungsten alloy with a density of 18.5 g/cm$^{3}$
\cite{D185-Properties}. The mechanical and thermal properties of
this material are summarized in table \ref{tab:Densimet-185-properties}.
The simulated sphere is 21 cm diameter, leading to a uniform absorption
thickness of 16 cm along any direction. The computed equivalent dose
corresponding to the 2.185 MeV emission of $^{144}$Pr is 0.930 mSv/h
at the shielding contact. For the 1.489 MeV, the equivalent dose is
0.026 mSv/h at contact. At a distance of 1 m from the shielding surface,
the 2.185 MeV radiation leads to a dose of 0.028 mSv/h. It is two
times higher than the dose obtained with the simple computation of
equation \ref{eq:dose-boe}. This can be explained by the fact that
the simulation also accounts for the degraded $\gamma$-rays that
escape the shielding material. The radiation dose associated to the
1.489 MeV line 1 m away from the biological protection is 0.75 \textmu{}Sv/h.
These results confirm that a tungsten alloy thickness of 16 cm is
sufficient to meet the radiobiological protection requirements.

\noindent Finally a computation of the equivalent dose was performed
by the CEA radio protection division. The two 2.185 MeV and 1.49 MeV
$\gamma$-rays have been simulated through the same shielding geometry
and material, using MCNPX v 2.7.a. Other $\gamma$ rays are neglected,
either because of their low energy or because of their weak intensity.
The source was assumed to be CeO$_{2}$ with a density of 4 g/cm$^{3}$
and contained in a 1-mm thick capsule made of stainless steel with
the same shape and dimensions as previously described. The equivalent
dose rates are summarized in \tabref{CEA-SPR-Dose}. Though slightly
higher, they are similar to our GEANT 4 computation.

\noindent \begin{center}
\begin{table}[H]
\centering{}%
\begin{tabular}{ccc}
\hline 
$\gamma$ rays energy (MeV) & Dose rate at 1 cm (mSv/h) & Dose rate at 1 m (mSv/h)\tabularnewline
\hline 
\hline 
2.185 & 1.5 & 4.2$10^{-2}$\tabularnewline
1.49 & 3.9$10^{-2}$ & <1.5$10^{-3}$\tabularnewline
\hline 
\end{tabular}\caption{\label{tab:CEA-SPR-Dose}CEA - SPR computations of the radiation dose
1 cm and 1 m away from the shielding. }
\end{table}

\par\end{center}

\noindent For completeness, the influence of electromagnetic radiation
produced by the deceleration of the $\beta$-decay electrons when
deflected by the atomic nuclei (Bremsstrahlung) has been studied.
The mean free path of electrons in matter, $\delta$, is taken from
the Katz and Penfold formula \cite{KP-formula}: $\mathrm{\delta(cm)=0.412/\rho E^{n}}$,
where $\rho$ is the material density in g/cm$^{3}$, E is the energy
of the electron in MeV, and the scaling index $n=1.265-0.0954\,\ln(E)$.
The maximum mean free path of electrons of $^{144}$Ce and $^{144}$Pr
is given in \tabref{CEA-SPR-EmeanFreePath} for stainless steel, cerium
oxide, and a tungsten alloy. Bremsstrahlung spectra of $^{144}$Ce
and $^{144}$Pr is computed using MCNPX. The obtained X-ray spectra
are converted into dose rates using the software MERCURIAD v1.04.
Details of the computation can be found in \cite{CEASPR2012-1}. The
dose rate induced by Bremsstrahlung of $\mathrm{\beta}$ electrons
is less than 1 $\mu$Sv/h at 1 cm and can thus be neglected in comparison
with the dose rate directly induced by the 2.185 MeV $\gamma$-ray
of the $^{144}$Pr decay.

\noindent \begin{center}
\begin{table}[h]
\centering{}%
\begin{tabular}{cccc}
\hline 
$\beta$ energy (MeV) & Stainless Steel & CeO$_{2}$ & Tungsten alloy\tabularnewline
\hline 
\hline 
0.32 & 1.1$10^{-2}$ cm & 4.27$10^{-3}$ cm & 2.2$10^{-2}$ cm\tabularnewline
3.0 & 1.9$10^{-1}$ cm & 8.0$10^{-2}$ cm & 3.7$10^{-1}$ cm\tabularnewline
\hline 
\end{tabular}\caption{\label{tab:CEA-SPR-EmeanFreePath}CEA - SPR computations of the mean
free path of electrons in stainless steel, cerium oxide, and a tungsten
alloy. }
\end{table}

\par\end{center}

\noindent \begin{center}
\begin{figure}[h]
\centering{}\subfloat[]{\includegraphics[scale=0.35]{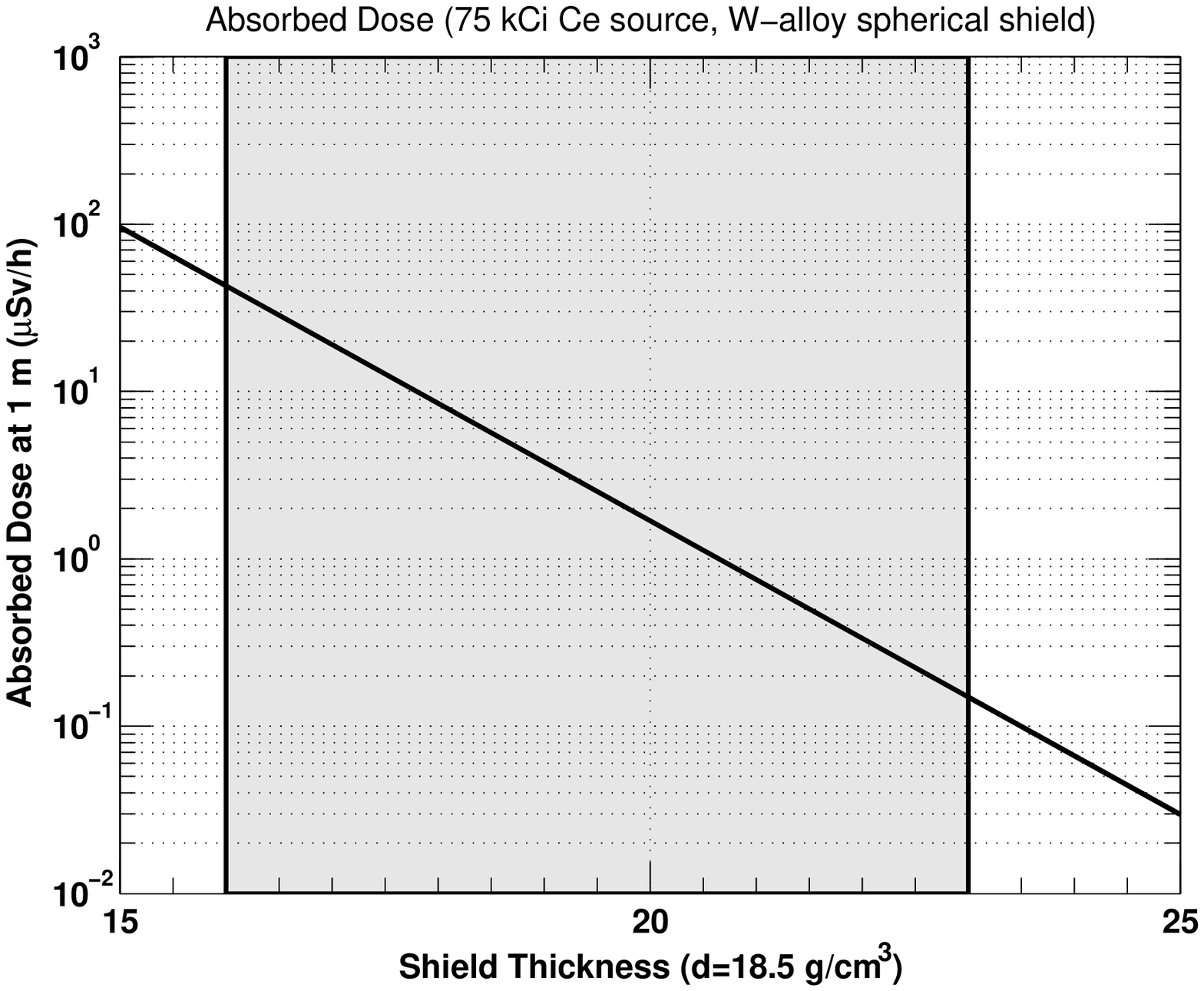}}
\subfloat[]{\includegraphics[scale=0.35]{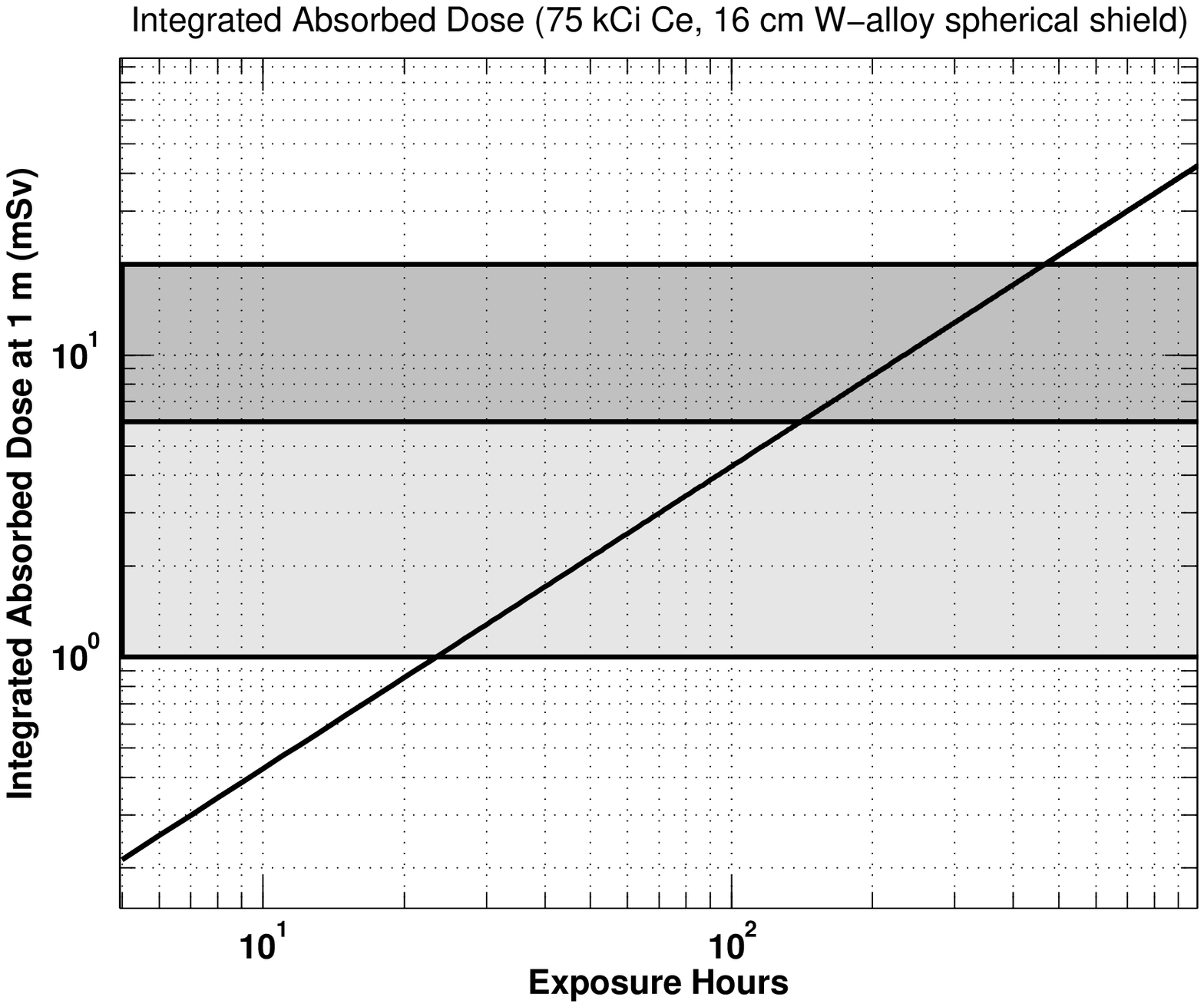}

}\caption{\label{fig:highZshield-Dose}Left panel: absorbed dose received at
1 m from a spherical tungsten alloy shielding as a function of thickness.
The gray area represent the range of thicknesses being considered
for the CeLAND experiment, with a cylinder having a 16 cm minimum
thickness. The maximum dose is 42 $\mathrm{\mu}$Sv/h. Right panel:
Integrated absorbed dose for a 16-cm tungsten alloy shielding as a
function of time, when standing next to the source (initial activity
of 75 kCi). Category A and B equivalent dose limits are represented
by the light and dark gray bands, respectively. A worker could stand
more than 20 hours at 1 meter from the shielding before receiving
an integrated dose of 1 mSv, assuming an activity of 75 kCi.}
\end{figure}

\par\end{center}

\subsection{\noindent Insertion of the $^{144}$Ce source inside the biological
shielding}

\noindent The $^{144}$Ce source will be manipulated and inserted
into the central shielding in a hot cell at the Mayak reprocessing
plant. The hot cell has a 1.1 m high and 1.0 m wide single door and
can contain up to 4 tons of material. The crane available in the area
is certified to lift a maximum weight of 5 tons. The tungsten alloy
shielding will be cleaned at Mayak to avoid surface contamination
of the shielding after being manipulated in the hot cell. The $^{144}$Ce
source must be inserted in the shielding downwards from the top.

\subsection{\noindent High-Z shielding (for the deployment next to KamLAND)}

\noindent The characteristic signature of an inverse beta decay antineutrino
interaction is the time and space coincidence of a prompt energy deposition
($\mathrm{e^{+}}$ energy deposition) followed by a delayed energy
deposition (neutron capture on hydrogen). The main background to the
antineutrino signal are accidental coincidences between prompt-like
(E>0.9 MeV) and delayed-like energy depositions (E>2.0 MeV) occurring
within a time window taken as three neutron capture lifetimes on hydrogen
(equivalent to about $\mathrm{\Delta}$T =772 \textmu{}s), and within
a volume of 10 m$^{3}$. If the prompt and delayed energy deposition
positions can be reconstructed, an additional background rejection
of a factor $\Delta$V =1/100 can be achieved. \\
A dangerous source of accidental coincidences comes from the emission
of energetic 2.185 MeV $\mathrm{\gamma}$-rays from the decay of the
$\mathrm{1^{-}}$ excited state of $^{144}$Nd (Table \ref{tab:sourcefeatures}).
2.185 MeV $\mathrm{\gamma}$-rays can easily mimic the signal neutron
capture on hydrogen at 2.2 MeV and the signal of a positron above
0.9 MeV, generating an accidental coincidence signal. The rate of
accidental coincidences must then stay small enough compared to the
expected antineutrino interaction rate within the KamLAND detector
(16000/y for phase 1). Hence, the reduction of the rate $\mathrm{R_{\gamma}}$
of escaping $\mathrm{\gamma}$-rays is a key parameter in computing
the shielding dimensions. To first order, this requirement translates
into: $\mathrm{R_{\gamma,phase1}^{2}\times\Delta t\times\Delta V\ll16000/y}$
leading to $\mathrm{R{}_{\gamma,phase1}\leq}$ 5 Bq . Starting from
an initial rate of 19.4 TBq for the 2.185 MeV $\mathrm{\gamma}$-ray
line (0.7\% of 2.77 PBq), this requirement on the reduction of R$_{\gamma}$
imposes an attenuation of $\mathrm{\sim10^{-12}}$ .

\subsubsection{\noindent $\gamma$-ray absorption specification}

\noindent In the first phase of the experiment, the $\mathrm{^{144}Ce-^{144}Pr}$
emitter will be placed as close as possible from the steel tank within
the circulating water shielding. The minimum foreseen distance from
the center of the KamLAND detector is 9.6 m. The shielding thickness
is computed taking into account additional attenuation provided by
different media between the source and the liquid scintillator volume:
water plus a 1-cm thick stainless steel vessel ($\lambda$ = 4.15
cm) plus 2.5 m of mineral oil ($\lambda$ = 26.9 cm). As a conservative
assumption, the self-absorption of $\mathrm{\gamma}$-rays within
the CeO$_{2}$ capsule and the solid angle effect are not taken into
account. Achieving a 10$^{-12}$ attenuation in this geometric configuration
finally leads to a 16-cm thick tungsten alloy (18.5 g/cm$^{3}$).

\subsubsection{\noindent Radiopurity specification}

\noindent An important remaining source of background could be the
tungsten shielding itself, especially in the second phase of the experiment
where the shielding will be in contact with the liquid scintillator.
The activity induced by impurities inside the tungsten shielding are
here required to be less than $\sim$1/10 of the activity of the 2.185
MeV $\mathrm{\gamma}$s able to escape the shielding without interaction.
This requirement ensures the absence of a significant background contribution
from the tungsten shielding. For the deployment in the outer veto
or in a room close to the detector, conservative estimates can be
computed as described in the previous section, by imposing a maximum
tolerable background rate of $\mathrm{R{}_{\gamma,phase1}\leq}$ 0.5
Bq. GEANT4 simulations have been run in the CeLAND phase 1 deployment
scenario, with a 16 cm thick spherical shape tungsten alloy shielding
placed 9.6 m away form the KamLAND detector center. Taking into account
the most dangerous $\mathrm{\gamma}$-ray lines emitted in the different
primordial decay chains or nuclei, the maximum tolerable activities
in $^{232}$Th, $^{238}$U, $^{40}$K, $^{60}$Co are the following:
\begin{itemize}
\item \noindent $^{232}$Th (primordial, $^{232}$Th-chain) : $\sim$0.3
Bq/kg
\item \noindent $^{238}$U (primordial, $^{238}$U-chain) : $\sim$3 Bq/kg
\item \noindent $^{40}$K (primordial) : $\sim$120 Bq/kg
\item \noindent $^{60}$Co (anthropogenic) : $\sim$20 Bq/kg
\end{itemize}
\noindent Several tungsten alloy samples from different suppliers
have been analyzed, with activities at the level of ten to hundreds
of mBq/kg for the above-mentioned radioisotopes. They are hence all
suitable for the CeLAND phase 1 deployment scenario of the cerium
antineutrino generator in the KamLAND Outer detector.

\subsubsection{\noindent Preliminary mechanical design}

\noindent The shielding is made up of a hollow tungsten alloy cylinder
with two cavities, each with different inner diameter and located
at different depths: the source fits in the lower cavity at the center
of the shielding and the upper cavity supports the plug to avoid direct
radioactive leaks. The tungsten alloy shielding is 16 cm thick at
minimum and weighs about 1.5 tons. It is represented in \figref{SketchShieldPhase1}. 

\noindent \begin{center}
\begin{figure}[h]
\begin{centering}
\subfloat{\centering{}\includegraphics[scale=0.6]{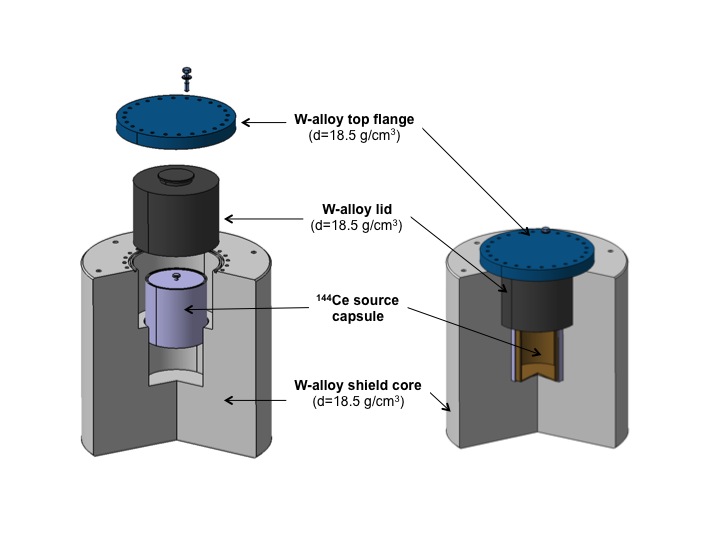}}
\par\end{centering}

\caption{\label{fig:SketchShieldPhase1}Sketch of the tungsten alloy shielding
to be used for radiological protection for the deployment in the outer
detector water (OD) or farther from the target liquid scintillator.}
\end{figure}

\par\end{center}

\noindent The final choice on the tungsten alloy material will mostly
be driven by radiopurity requirements (see previous section). The
upper cavity in the shielding is closed by a tungsten alloy lid. The
tightness of the seal is achieved by a Helicoflex O-ring metallic
gasket with good resistance to radiations and high temperatures. Captive
screws hold the cap in order to ease its mounting in the hot cell.
The final tightening of the screws will require a high torque and
can be done outside the hot cell (the biological protection is ensured
as soon as the cap is plugged on the top of the shielding). A 1 mm
play between the outer capsule and the inner wall of the shielding
lower cavity, as well as a 2 mm chamfer, should ease the insertion
of the source in the shielding. Likewise, there is a 5/10 mm play
between the plug and the inner wall of the shielding upper cavity.
The loading of the capsule in the shielding should be done downwards
from the top. 2 x 6 threaded M16 holes on the top and bottom shielding
sides allow several handling operations: translation and rotation
in the hot cell (with a dedicated tool), insertion of the shielding
in the transportation container, in the calorimeter, handling above
the manhole at KamLAND or inside the detector. They also allow holding
the shielding during measurements in the KamLAND detector or can be
used to fix supports to protect the bottom surface of the shielding.
These threaded holes will be protected by screws fitted with metallic
gaskets to avoid trapping of any dust or any kind of radioactive particles.
A preliminary technical drawing of the CeLAND source shielding is
given in \figref{ShieldPhase1Drawing}.

\noindent \begin{center}
\begin{figure}[h]
\begin{centering}
\includegraphics[scale=0.12]{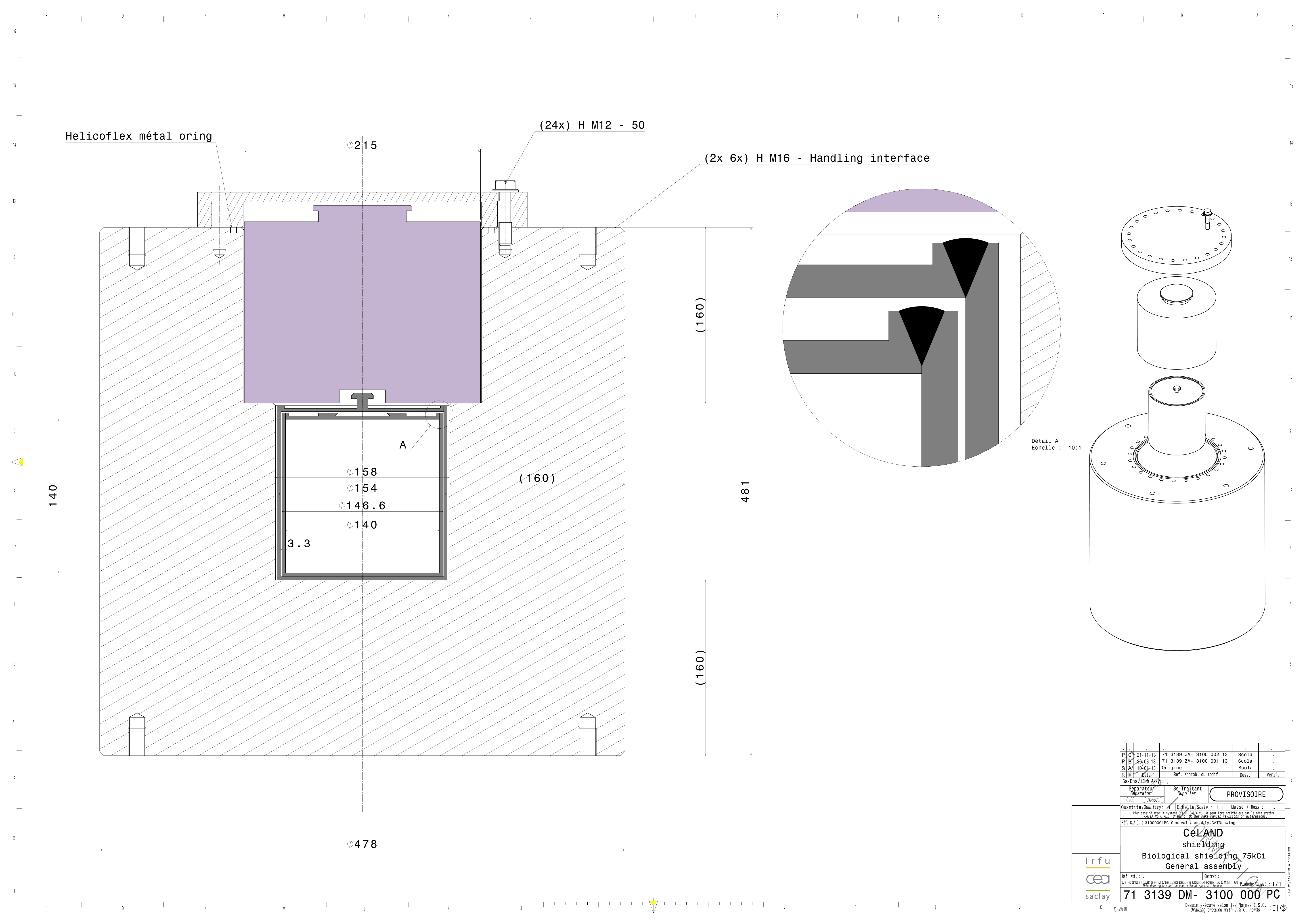}
\par\end{centering}

\caption{\label{fig:ShieldPhase1Drawing}Preliminary drawing of the tungsten
alloy shielding to be used for radiological protection as well as
for the first phase of CeLAND, e.g. the deployment in the outer detector
water (OD).}
\end{figure}

\par\end{center}

\section{\noindent Source and shielding logistics}

\noindent Various heavy elements needed for the experiment will be
shipped between the different manufacturing places to Japan. Here
is a summary of the major heavy shipments.

\subsection{\noindent High-Z shielding transportation }

\noindent Transport involving non radioactive material: 

\noindent \textbullet{} shipment from Saclay to Mayak of the tungsten
radiological protection ;

\noindent \textbullet{} shipment from Saclay to Kamioka of the equipment
to measure its activity prior to deployment and of the tools used
to manipulate the source ; 

\noindent \textbullet{} if the shielding is not manufactured in France,
calibration of the calorimeter will impose a transit through France.

\subsection{\noindent $\mathrm{^{144}Ce-^{144}Pr}$ antineutrino generator transportation }

\noindent The most sensitive part is the transportation of the radioactive
antineutrino generator between Mayak and Japan. Transport of radioactive
materials must follow requirements stated in the reference document
: Regulations for the Safe Transport of Radioactive Material (2009
Edition) issued by IAEA {[}IAEA09{]}. This document describe the type,
characteristics and performance of containers used for these transports,
the different forms of the radioactive material and the radiation
level and temperature outside the transport container.\\

\noindent In our case the parcel (or consignment) which will be transported
is composed of the radioactive Cerium oxide pressed to reach a density
around 4 g/cm$^{3}$ and placed inside a double sealed stainless steel
container ; this object will be inserted in the tungsten shielding
at Mayak and will be tightly closed by a top part in tungsten. Each
capsule will be tested for leaks (Helium leak test) and according
to ISO 9978-92 certification procedure (Radiation protection - Sealed
radioactive sources - Leakage test methods).

\subsubsection{\noindent Transport of radioactive materials }

\noindent The physical characteristics of the $\mathrm{^{144}Ce-^{144}Pr}$
and its encapsulation leads to the classification of the source as
a Special form of radioactive material according to the IAEA terminology.

\subsubsection{\noindent Transport with a B(U) type container }

\noindent During the transport the $\mathrm{^{144}Ce-^{144}Pr}$ source
contained in its biological shielding will be inserted and secured
in an IAEA approved transport container of type B(U) whose characteristics
are being defined. The plan is to use an existing transport container.
As mentioned above the limit on radiation level will be satisfied
thanks to the important tungsten shielding: 

\noindent \textbullet{} 2 mSv/h in contact with the packaging (or
the vehicle);

\noindent \textbullet{} 0.1 mSv/h at two meters from the vehicle.
\\

\noindent These limits will be checked at each transit points. Vehicles
carrying radioactive materials by definition are in motion; the public
exposure times are very short (of the order of a few seconds to minutes)
during transport. The workers involved in the handling of the container
will receive appropriate training on radiological protection. If a
member of the CeLAND collaboration attends to the transportation he/she
will be classified in the category B (6 mSv/y) and therefore has no
impact on his/her health. The limit related to temperature concerns
the external temperature of the transport container and not the temperature
of the shielding itself.\\

\noindent The transport container of type B(U) must fulfill several
requirements as explained above, but must received an agreement from
the safety authorities of the countries involved in the transport.
To simplify the explanations, a US built and certified container in
United Stated needs to be rectified if the transport goes (or transits)
through Japan. For this procedure Russia and Europe have mutual agreements
to accept containers certified in one of the countries. Agreements
of an existing foreign container in Japan can be a long procedure
(several months). The ideal case would be to reuse a container which
has been already used to transport radioactive material from Russia
to Japan.\\

\noindent Two potential transport containers have been identified
:

\subsubsection*{\noindent TN TM \textendash{} MTR }
\begin{itemize}
\item \noindent Engineered by AREVA; used to transport package for irradiated
MTR (Material Test Reactor) and TRIGA (Training, Research, and Isotope
production, General Atomics ) fuel elements Type B(U)F IAEA 1996 
\item \noindent Criticality Safety Index : CSI : 0 
\item \noindent Certified in Europe, Australia and USA; need a certification
in Japan 
\item \noindent External dimensions : H = 2008 mm, D = 2080 mm , weight
= 23 400 kg (max. loaded) 
\item \noindent Useful dimensions: Cavity height = 1080 mm; useful cavity
diameter = 960 mm
\end{itemize}
\noindent \begin{center}
\begin{figure}[h]
\begin{centering}
\subfloat{\centering{}\includegraphics[scale=0.5]{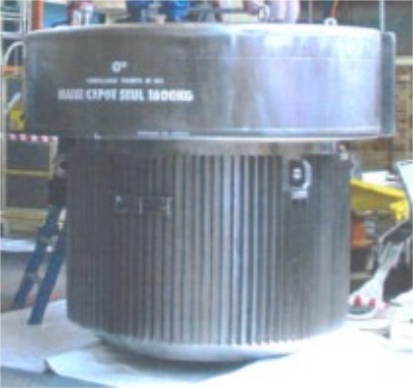}}\subfloat{\includegraphics[scale=0.26]{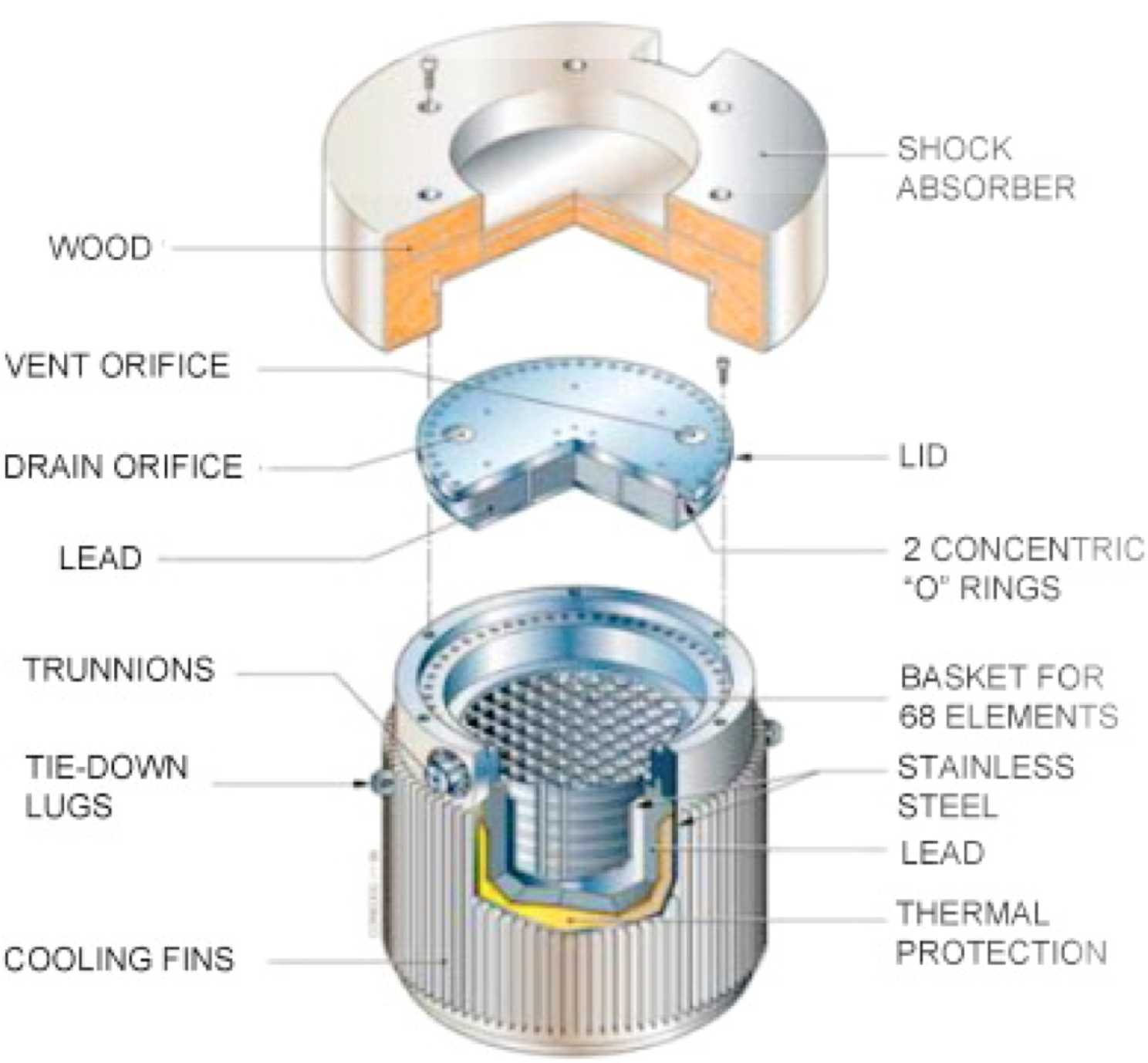}}
\par\end{centering}

\centering{}\caption{TN TM \textendash{} MTR transport container.}
\end{figure}

\par\end{center}

\subsubsection*{\noindent JMS-87Y-18.5T }
\begin{itemize}
\item \noindent Engineered by Kimura Chemical Plant; used to transport package
for irradiated MTR (Material Test Reactor) and TRIGA (Training, Research,
and Isotope production, General Atomics) fuel elements 
\item \noindent Certified in Japan and in USA; needs a certification in
Russia and possibly a new certification in Japan by the new safety
authority (NRA). 
\item \noindent External dimensions : H 1900 mm \textendash{} D 2000 mm
\textendash{} Weight : 18 500 kg (max. loaded) 
\item \noindent Useful dimensions: Cavity height: 820 mm -  cavity diameter:
660 mm
\end{itemize}
\noindent \begin{center}
\begin{figure}[h]
\begin{centering}
\subfloat{\centering{}\includegraphics[scale=0.45]{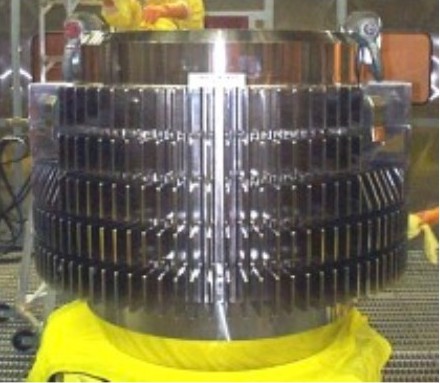}}\subfloat{\includegraphics[scale=0.35]{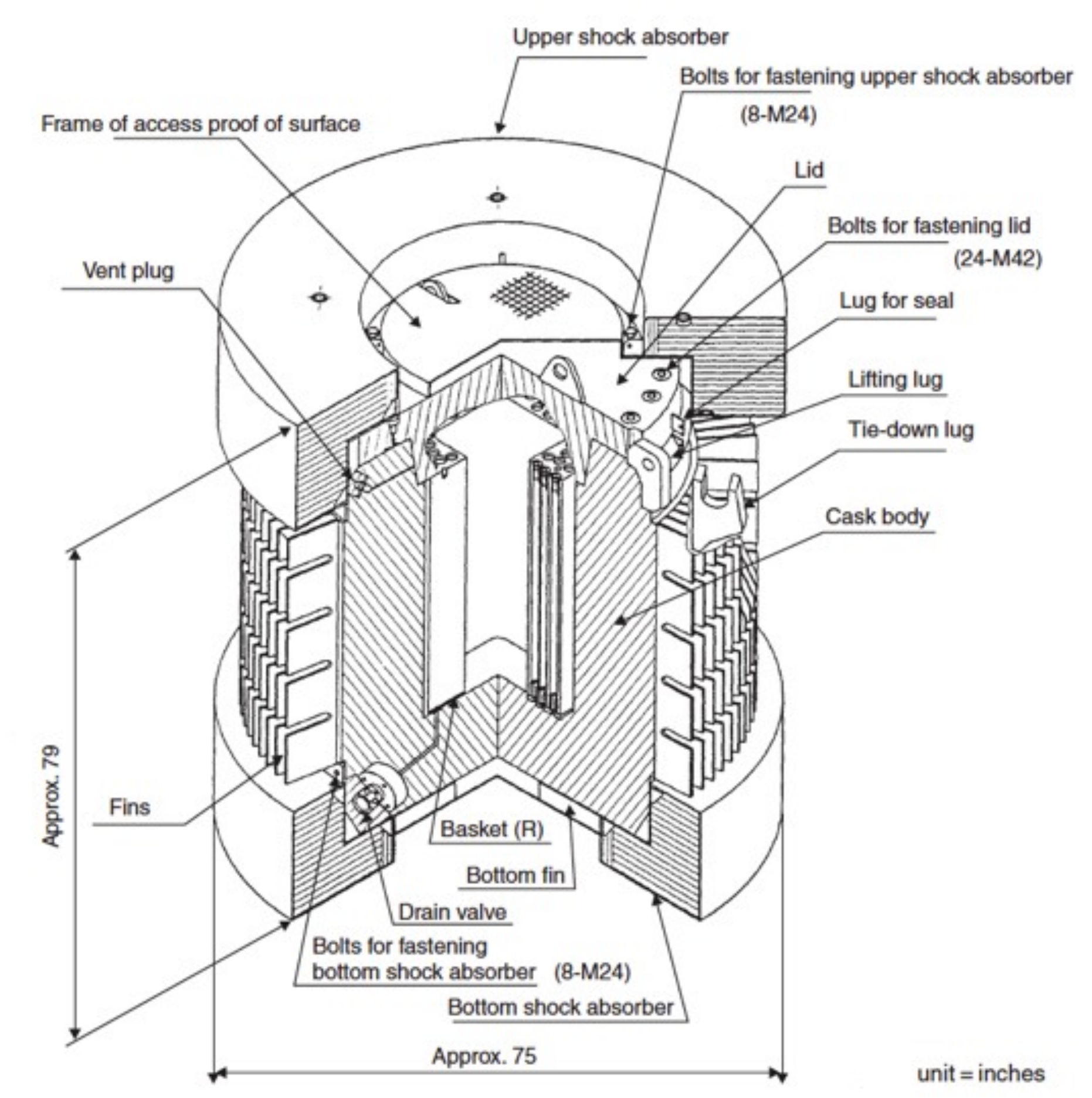}}
\par\end{centering}

\caption{JMS-87Y-18.5T transport container.}
\end{figure}

\par\end{center}

\noindent The antineutrino generator transportation is being studied
with Areva TN International, Hitachi, and PA Mayak that offer a complete
range of transport and storage solutions for radioactive materials
throughout the entire nuclear fuel cycle. From several meetings with
experts we gathered a wealth of useful informations concerning the
transport of our radioactive source. There is a very limited number
of harbors in the world authorized to receive radioactive materials
(so called class 7). Until recently, only the Saint Petersburg and
Mourmansk harbors were certified to receive the class 7 radioactive
materials in Russia. In particular on the east coast, on the sea of
Japan, up to very recently no harbor were approved as class 7. Only
Yokohama and Tokyo harbors have the necessary certifications in Japan.
Areva TNI transported several shipments of radioactive material towards
(and from) Japan, such as fuel elements to be treated at la Hague
plant and Plutonium and MoX fuel elements. Similarly they have organized
the transport from Mayak to western destinations. 

\noindent \begin{center}
\begin{figure}[h]
\begin{centering}
\includegraphics[scale=0.7]{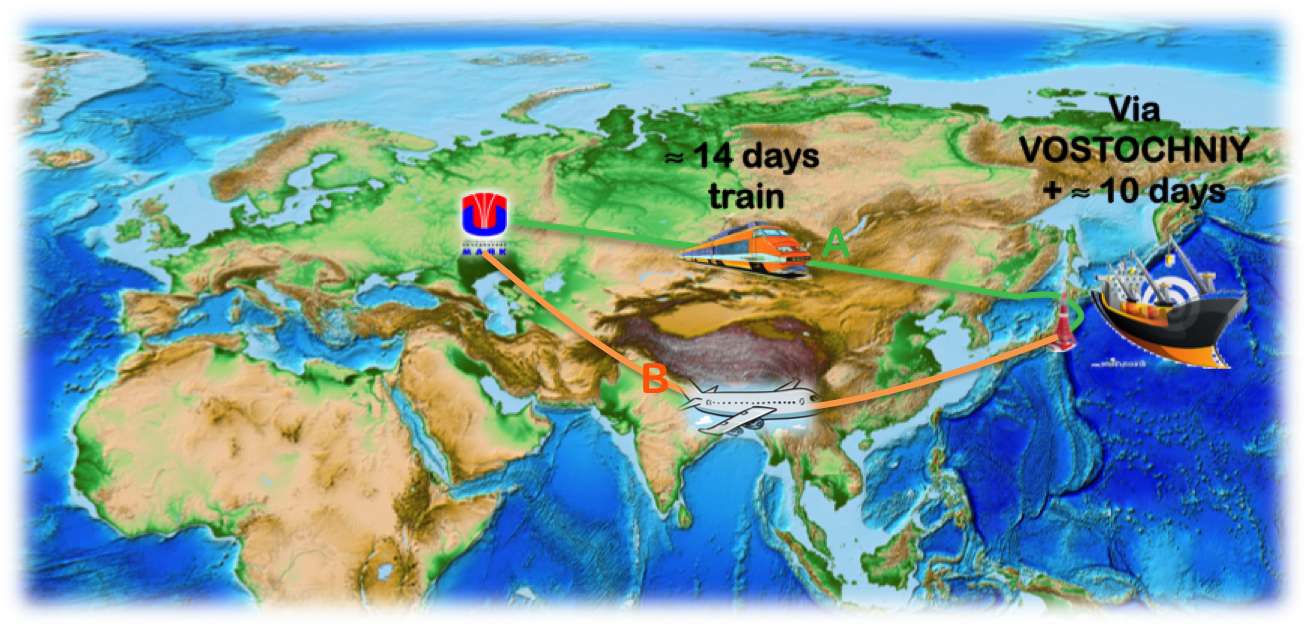}\caption{\label{fig:CeLAND-trasnportations-route}CeLAND transportation routes,
via train + boat (A), and airplane (B).}

\par\end{centering}

\end{figure}

\par\end{center}

\noindent Last fall a transport of enriched uranium took place between
Russia and Japan. It transited through the harbor of Nakhodka on the
east coast of Siberia. This possibility will reduce the time (and
the cost) of the transport by a dedicated ship. IT is shown on \ref{fig:CeLAND-trasnportations-route}.
Indeed, the boat used for this transport is a fully dedicated boat
which means that the cost will include the renting of the boat, the
salary of the crew, the renting of the container and, last but not
least, the cost of the insurance.\\

\noindent Air transport IAEA regulations allow transport of radioactive
material by air, but, in this case there is a limit in the activity
of the material inside a single container. For the case of $^{144}$Ce
the limit is 600 TBq (16.2 kCi), but for $^{51}$Cr it is 90 PBq and
for $^{60}$Co 1200 TBq. Nevertheless the same regulation does not
limit the number of container in a single plane, but each radioactive
source has to be enclosed in a specific container. Thus we studied
the possibility to divide our 75kCi $^{144}$Ce source into 5 sources
of 15 kCi. Even if the individual activity is reduced, we still need
an important shielding to be able to manipulate these sources ; the
estimated mass cannot be less than 500 kg. Existing transport containers
dedicated for air shipments are used to transport alpha or beta sources
with no penetrating radiations. In our case, an extra biological shielding
in tungsten is mandatory to be below the radiation limit, but the
maximum load for the certification is largely exceeded; therefore
they are not suitable for our purpose. \\

\subsubsection*{\noindent Special arrangements }

\noindent Having in mind the constraints on transportation following
the usual standard IAEA regulations, it is wise to look if the regime
of � Special arrangements � as described in paragraph 310 is more
appropriate for our case. It seems useful to quote the text describing
this regime:

\noindent � 310. Consignments for which conformity with the other
provisions of these Regulations is impracticable shall not be transported
except under special arrangement. Provided the competent authority
is satisfied that conformity with the other provisions of these Regulations
is impracticable and that the requisite standards of safety established
by these Regulations have been demonstrated through means alternative
to the other provisions, the competent authority may approve special
arrangement transport operations for single or a planned series of
multiple consignments. The overall level of safety in transport shall
be at least equivalent to that which would be provided if all the
applicable requirements had been met. For consignments of this type,
multilateral approval shall be required. �

\subsubsection{\noindent Transport with a C type container }

The IAEA and the United States Department of Energy (DOE) fabricated
a series of Czech-design SKODa VPVr/M transport overpacks, developed
specifically for spent fuel storage and transport inside the standard
twenty-foot ISO-containers. Sixteen such SKODa VPVr/M overpacks were
produced and subsequently certified first in the Czech republic and
then in Russia, where they are now labelled as TUK145/C. This container
has now performed spent fuel deliveries by automobile, rail, river
and sea transport. Moreover it is the first world\textquoteright{}s
first Type C package certified for the air transport of SNF in compliance
with the classification given in IAEA TS-R-1 Regulations, without
any restriction on the ANG activity.

\subsubsection*{\noindent TUK-145/C }
\begin{itemize}
\item \noindent Engineered by Skoda in Czech Republic; used to transport
package for LEU SNF. 
\item \noindent Certified in Russia for SNF; needs an extension of certification
in Russia for containing cerium, only (not considered to be an obstacle)
and an extension of the certification in Japan by the new safety authority
(NRA). 
\item \noindent External dimensions : H 2155 mm \textendash{} D 1500 mm
\textendash{} Weight : up to 35 000 kg (max. loaded) 
\item \noindent Useful dimensions: Cavity height: 820 mm -  cavity diameter:
660 mm
\end{itemize}
\noindent \begin{center}
\begin{figure}[h]
\begin{centering}
\subfloat{\centering{}\includegraphics[scale=0.25]{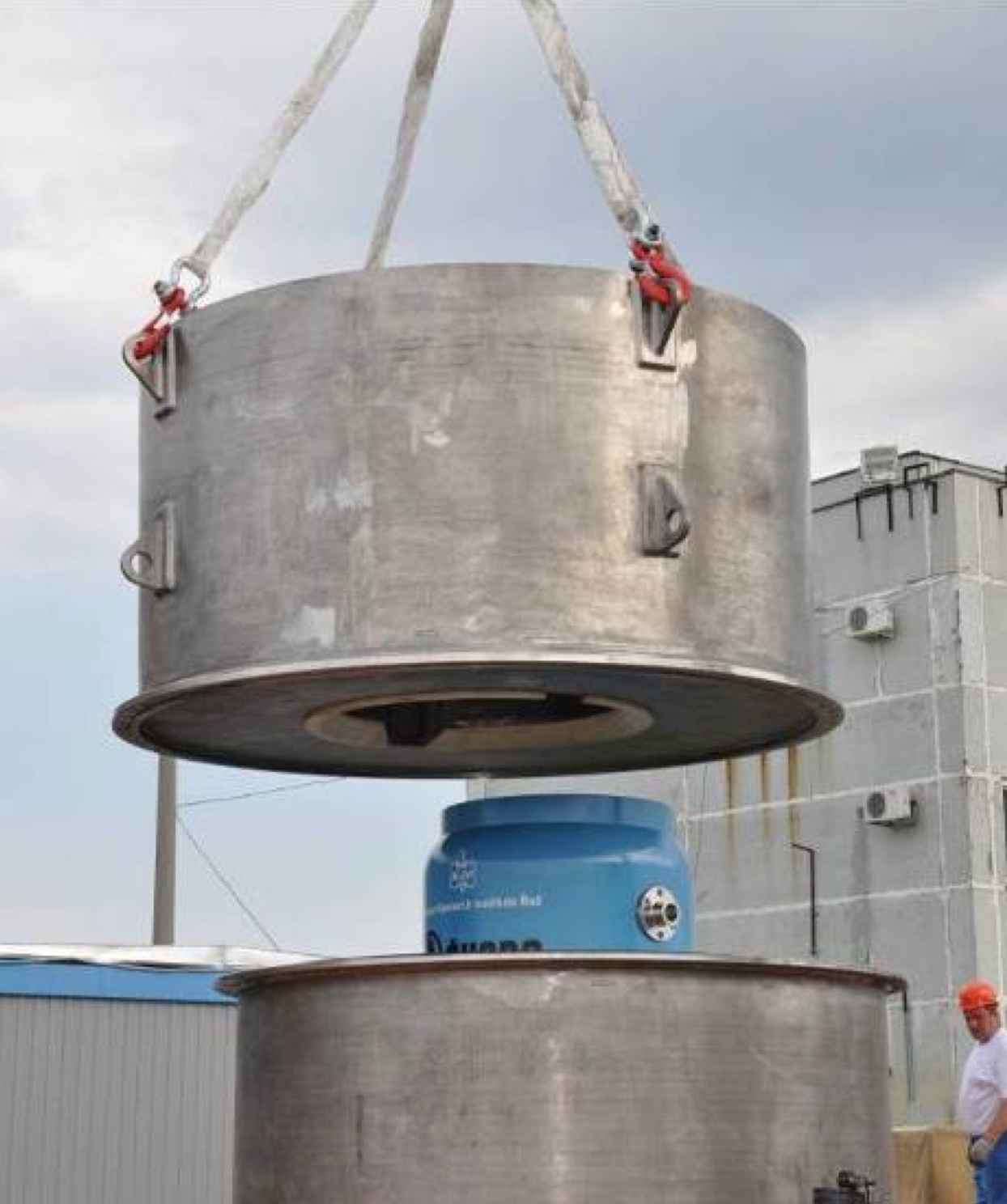}}\subfloat{\includegraphics[scale=0.35]{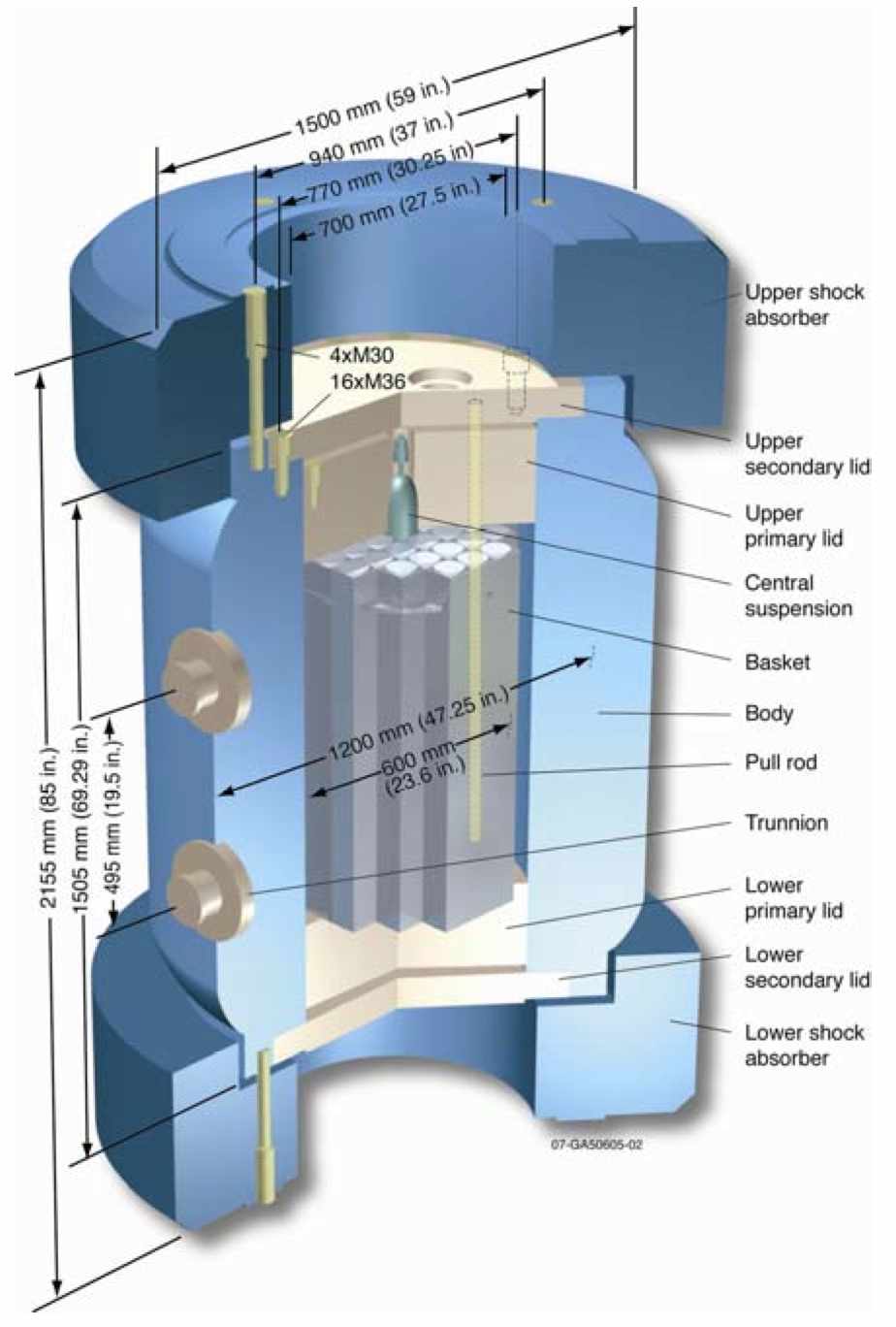}}
\par\end{centering}

\caption{TUK145/C transport container.}
\end{figure}

\par\end{center}

The TUK145/C cask is of a unique construction that allows it to be
loaded and unloaded from below and from the top. The cask body is
made of cast iron. The cask inner walls are plated with aluminum.
Leak-tight containment is provided by the metallic Helicoflex seals.
The inner cavity of the cask is backfilled with helium. The standard
basket would have to be replaced by a custom basket to fit the cerium
antineutrino generator with a reduced shielding not exceeding 450
kg. This corresponds to a thickness of approx. 8.5 cm, leading to
a dose rate at 1 m of 10 mSv/h. The basket could be designed by the
Sosny Science and Production Firm and fabricated by the OZNO plant
in Ozersk, russia. 

The TUK145/C container could be used for the transportation of a 75
kCi $^{144}$Ce ANG with a reduced, 8.5 cm thick W-alloy shielding.
It would be loaded on a truck at PA Mayak, and be transferred to Yekaterinburg
airport and loaded into an aircraft (AN-124-100), including the truck
that would finally drive from the Japanese airport to the Kamioka
mine for unloading. This would necessitate a special operation at
the Kamioka site to unload the $^{144}$Ce ANG with its reduced shielding
from the TUK-145/C and to immediately load it into the additional
shielding cylinder, 8 cm thick as well in order to reach the overall
16 cm thickness.

\section{\noindent Activity measurements}

\subsection{\noindent Requirements}

\noindent The source activity is a key normalization parameter, directly
linked to the prediction of the number of expected events and to the
sensitivity of the experiment, particularly at high $\Delta m^{2}$.
So the minimal uncertainty on activity is desirable. To ensure the
competitiveness of the experiment, a maximum of 2 \% uncertainty on
activity is considered, 1 \% being the goal.

\noindent The main technique considered to measure the activity is
the calorimetry, which is a global technique relatively easy to interpret
and calibrate (compared to radiation measurement). According to nuclear
database, the specific activity of the $\mathrm{^{144}Ce-^{144}Pr}$
couple is $7.991\pm0.044$ W/kCi, so a 0.56 \% systematic uncertainty
will in any case limit the calorimetry sensitivity. In any case, the
source would be inside its tungsten biological protection to contain
gammas, so the calorimeter must be able to support about 1.5 ton and
a cylinder of roughly 60 cm in height and diameter.

\noindent Other techniques rely on radiation measurement ($\beta$
or $\gamma$) on a sample of the source, with extrapolation to the
source by weighing.

\subsection{\noindent Calorimetry}

\subsubsection{\noindent Principle}

\noindent The proposed calorimetric measurement of the source is based
on the fact that emitted $\mathrm{\beta}$, $\gamma$ , X and Auger
radiation will heat up the source and the tungsten shield leading
to a well-defined power rate. The temperature increase of the object
is described by:

\noindent 
\begin{equation}
\mathrm{\frac{dT}{dt}=\frac{P}{C}-\frac{T-T_{ext}}{\tau}}\label{eq:calo}
\end{equation}

\noindent where T is the temperature at a given time t (in K), P the
power released by the source (in W), C the heat capacity (in J.K$^{-1}$)
of the system measured and $\mathrm{\tau}$ (in s) describes the thermal
losses of the calorimeter toward the exterior.

\noindent Several different techniques can be used: 
\begin{itemize}
\item \noindent the temperature increase ($\frac{dT}{dt}$) in a calorimeter
maximally insulated from the environment can be measured (such as
in Gallex \cite{key-4}), which implies negligible heat losses during
the whole measurement and to calibrate the heat capacity;
\item \noindent the equilibrium temperature (when $\frac{dT}{dt}=0$) can
be measured and compared to a calibration phase;
\item \noindent the heat can be transferred to another system and then measured,
typically to a fluid such as water.
\end{itemize}
\noindent $\tau$ can be minimized down to be negligible, and C and
$\mathrm{\tau}$ (if any remaining) can be evaluated precisely in
a preliminary calibration phase of the system with known heat sources.

\subsubsection{\noindent Data}
\begin{itemize}
\item \noindent $^{144}$Ce half life: 284.91 days
\item \noindent $^{144}$Pr half life: 17.28 min
\item \noindent Unitary power: $7.991\pm0.044$ W/kCi
\item \noindent Density of tungsten alloy: 18.5 g/cm$^{3}$
\item \noindent Heat capacity of tungsten: 130 J/kg
\item \noindent Thermal conductivity of tungsten: approx. 70 W.m$^{-1}$.K$^{-1}$
\end{itemize}
\noindent The thermal power delivered by the source is therefore between
600 and 800 W, for respectively 75 and 100 kCi.

\subsubsection{\noindent Requirements}
\begin{enumerate}
\item \noindent Uncertainty on thermal power below 1 \% (as long as reasonably
achievable).
\item \noindent Measurement time below 3 days, and preferentially at the
hour scale.
\item \noindent Capsule temperature below 400\textdegree{}C.
\item \noindent Reproducibility of the measurement, which implies a careful
and limited instrumentation of the shielding to ensure reproducibility
before and after hot cell manipulations and transportation.
\end{enumerate}

\subsubsection{\noindent Current Design}

\noindent The current design (see \figref{Calorimeter}) relies on
the measurement of heat flow \foreignlanguage{american}{transferred}
to water, by measuring the water massive flow rate and the temperature
of the water before and after its circulation at the contact with
the shielding. The flow rate can be very precisely measured with Coriolis
flow meter, which directly gives the massive flow rate with uncertainty
at the 0.1 \% level or even better with dedicated calibration. The
temperature measurement with thermocouple has an uncertainty of 0.1
to 0.2 K, leading to the requirement for the inflow/outflow water
temperature difference of 40 K. This leads to a flow rate of 4 g/s
for 85 kCi.

\noindent \begin{center}
\begin{figure}[h]
\begin{centering}
\includegraphics[scale=0.35]{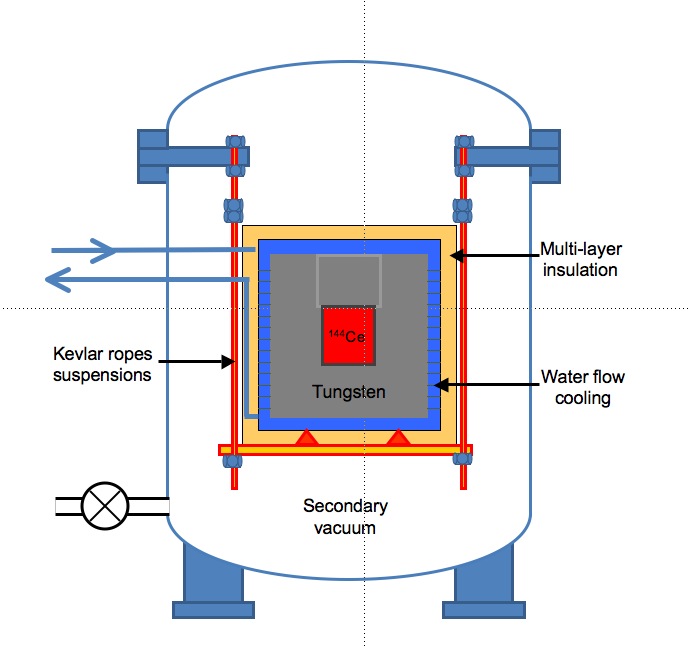}
\par\end{centering}

\caption{\label{fig:Calorimeter}Preliminary design of the calorimeter. The
multi-layer radiation shield, which will take place in the vacuum
between the two tanks, is missing. The shielded source is supported
by 3 pillars, while final design will use 3 strings to hold a board
supporting the shielded source. Thermocouples and the Coriolis flow
meter are not shown either.}
\end{figure}

\par\end{center}

\noindent The source power is measured by the power evacuated by water.
The source has to be insulated at the watt level to reach 1 \% of
uncertainty. This can be achieved by a combination of vacuum (probably
secondary), suspension of the source (probably with Kevlar strings),
multi-layer radiation shield and temperature control of the external
tank.

\noindent This design ensures reproducibility and allows a preliminary
study and calibration phase with electric source power, while being
also an absolute measurement since water heat capacity is well known.
The time needed for the measurement will probably be dominated by
the time needed to reach the target vacuum, since the shielding heating
will be low and consequently thermal equilibrium reached in about
1 hour. As soon as both target vacuum and thermal equilibrium are
reached, the measurement is over. Finally, the capsule temperature
will be far from 400\textdegree{}C since the mean shielding surface
temperature will be several tens of degrees Celsius.

\subsection{\noindent Spectroscopy of samples and weighing of the source}

\noindent Under the assumption of CeO$_{2}$ uniformity (which should
be ensured by the mixing of all the cerium solution in a single tank,
after chromatography at Mayak), one can sample the CeO$_{2}$ and
dilute it, determine the activity of the sample by radiation spectroscopy
and deduce the source activity by weighing both the source and the
sample. Mass measurement can easily achieve excellent accuracy, thus
the main source of uncertainty of this technique is the spectroscopy.

\noindent $\beta$ spectroscopy relies on the measurement of the electron
spectrum emitted by the sample. This measurement is typically realized
with a silicon detector, but the interpretation which relies on simulation
(diffusion in sample, back scattering, efficiency\ldots{}), calibration
and background control can be difficult to control at the target level

\noindent $\gamma$ spectroscopy relies on the measurement of a characteristic
gamma ray of the $\mathrm{^{144}Ce}$-$\mathrm{^{144}Pr}$ couple,
such as the 2.185 MeV ray of the $^{144}$Pr which is intense and
far from common natural background (above $^{40}$K 1.4 MeV ray but
still below the $^{208}$Tl 2.6 MeV ray). This measurement is typically
realized with a High Purity Germanium (HPGe) detector, but the normalization
which also relies on simulation (solid angle, auto absorption, efficiency\ldots{}),
calibration and background control can also be difficult to control
at the target level.

\section{\noindent $\mathrm{^{144}Ce-^{144}Pr}$ source deployment and data
taking}

\subsection{\noindent The KamLAND detector}

\noindent 
\label{sec:kamland}

\noindent The KamLAND detector is located under the peak of Ikenoyama
(Ike Mountain, $36.42^{\circ}$N, $137.31^{\circ}$E) in the Kamioka
mine, see \figref{Kamioka-mine-location.}, with a vertical rock overburden
of approximately 2700 meters water equivalent (m.w.e.). 

\noindent \begin{center}
\begin{figure}[h]
\centering{}\includegraphics[clip,width=1\columnwidth]{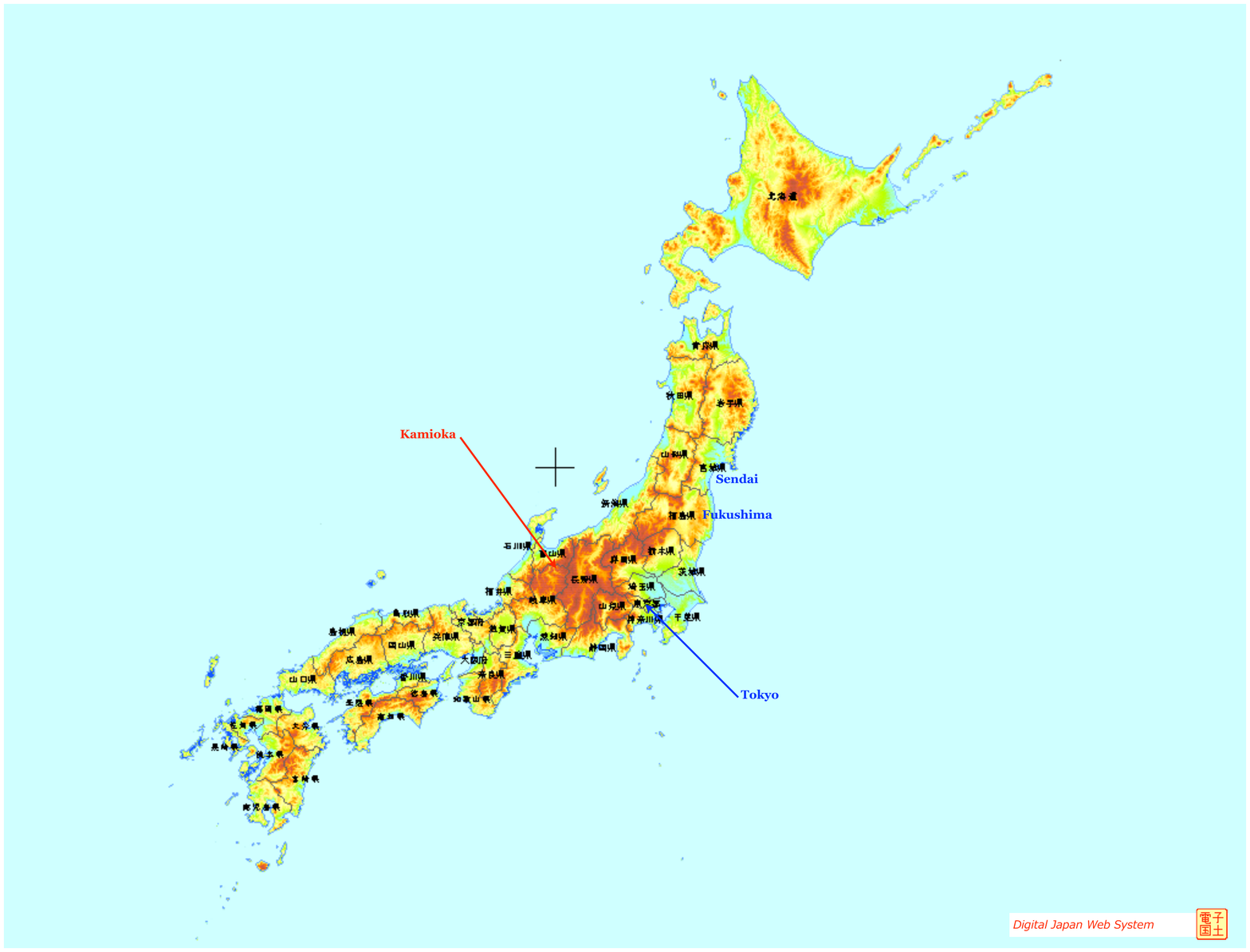}
\caption{\label{fig:Kamioka-mine-location.}Kamioka mine location.}
\end{figure}

\par\end{center}

\noindent A schematic diagram of KamLAND is shown in~\figref{kamlanddetector}.
KamLAND consists of an active detector region of $\sim$1\,kton of
ultra-pure LS contained in a 13-m-diameter spherical balloon made
of 135\,$\mu$m thick transparent nylon/EVOH (ethylene vinyl alcohol
copolymer) composite film and supported by a network of Kevlar ropes.
In addition to providing containment for the LS, the balloon protects
the LS against the diffusion of ambient radon from the surrounding
components. The LS comprises of 80\% dodecane, 20\% pseudocumene (1,2,4-Trimethylbenzene)
by volume, and \mbox{$1.36\pm0.03$}\,g/liter of the fluor PPO
(2,5-Diphenyloxazole). The density of the LS is 0.780\,g/cm$^{3}$
at $11.5^{\circ}$C.

\noindent A buffer comprising of 57\% isoparaffin and 43\% dodecane
oils by volume fills the region between the balloon and the surrounding
18-m-diameter spherical stainless-steel outer vessel to shield the
LS from external radiation.

\noindent An array of photomultiplier tubes (PMTs), 1325 specially
developed fast PMTs masked to 17-inch diameter and 554 older 20-inch
diameter PMTs reused from the Kamiokande experiment are mounted on
the inner surface of the outer containment vessel, providing 34\%
photo-cathode coverage. A 3\,mm thick acrylic barrier at 16.6-m-diameter
helps prevent radon emanating from the PMT glass from entering the
BO. The inner detector (ID), consisting of the LS and BO regions,
is surrounded by a 3.2\,kton water-Cherenkov detector instrumented
with 225 20-inch PMTs. This outer detector (OD) absorbs $\gamma$-rays
and neutrons from the surrounding rock and enables tagging of cosmic-ray
muons.

\noindent The KamLAND front-end electronics (FEE) system is based
on the Analog Transient Waveform Digitizer (ATWD) which captures PMT
signals in 128 10-bit digital samples at intervals of 1.5\,ns. Each
ATWD captures three gain levels of a PMT signal to obtain a dynamic
range from one photoelectron (p.e.) to 1000 p.e. Each ATWD takes 27\,$\mu$s
to read out, so two are attached to each PMT channel to reduce dead
time. The FEE system contains discriminators set at 0.15\,p.e.\ ($\sim$0.3\,mV)\
threshold which send a 125\,ns long logic signal to the trigger electronics.
The trigger electronics counts the number of ID and OD PMTs above
the discriminator threshold with a sampling rate of 40\,MHz and initiates
readout when the number of 17-inch ID PMTs above the discriminator
threshold exceeds the number corresponding to $\sim$0.8\,MeV deposited
energy. The trigger system also issues independent readout commands
when the number of OD PMTs above threshold exceeds a preset number.

\noindent \begin{center}
\begin{figure}[h]
\begin{centering}
\includegraphics[width=0.75\columnwidth]{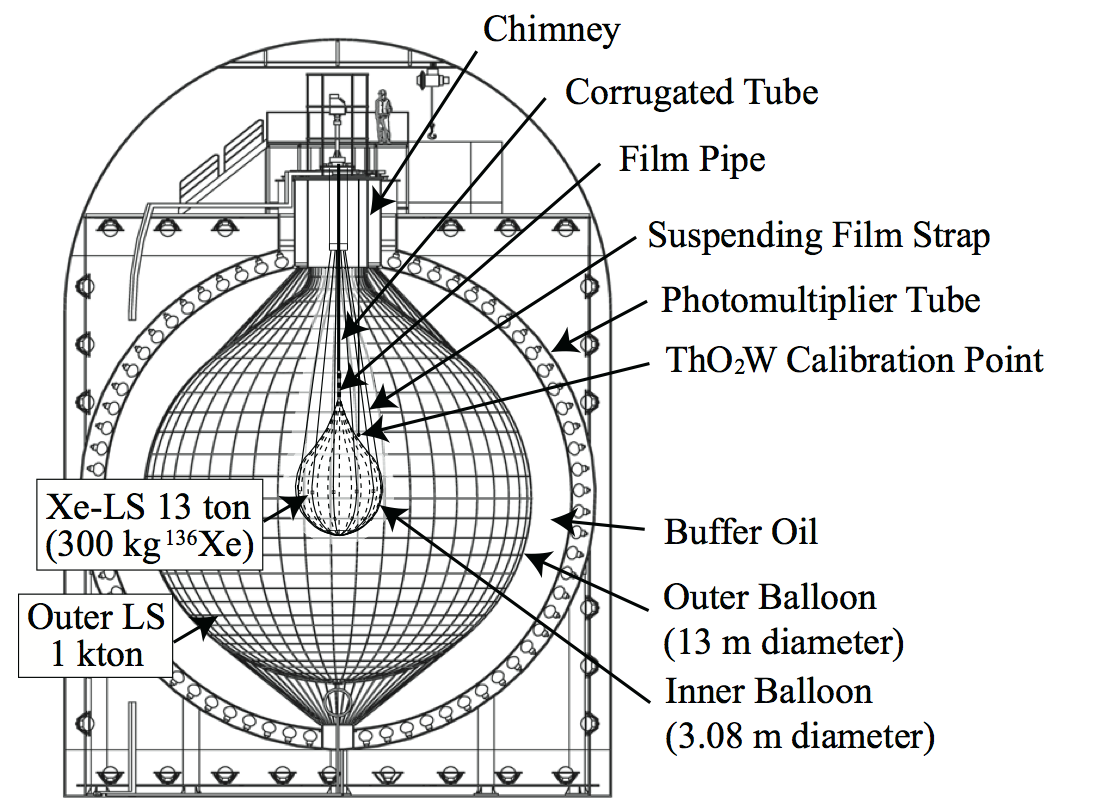} \caption{Schematic diagram of the KamLAND detector at the Kamioka underground
laboratory, including the KamLAND-Zen balloon.\label{fig:kamlanddetector}}

\par\end{centering}

\label{fig:detector} 
\end{figure}

\par\end{center}

\noindent The offline analysis takes full advantage of the information
stored in the digitized PMT signals by identifying individual PMT
pulses in the waveform information that is read out. The time and
charge are computed from the individual pulses. For each PMT, the
average charge corresponding to a single p.e.\ is determined from
\mbox{single-pulse} waveforms observed in low occupancy events.
The ID PMT timing is calibrated with light pulses from a dye laser,
injected at the center of the detector through an optical fiber. The
vertices of spatially localized low-energy ($<$30\,MeV) events are
estimated by comparing calculated time-of-flights of optical photons
from the hypothetical vertex to the measured arrival times at the
PMTs in KamLAND.

\noindent The reconstructed energies of events were calibrated with
$\gamma$\ sources in past deployments: $^{203}$Hg, $^{68}$Ge,
$^{65}$Zn, and $^{60}$Co; and with $n+\gamma$\ sources: \mbox{$^{241}$Am+$^{9}$Be}
and \mbox{$^{210}$Po+$^{13}$C}. These were deployed at various
positions along the vertical axis of the detector and occasionally
off the vertical axis within 5.5 m from the detector center. The energy
calibration is aided with studies of background contaminants $^{40}$K
and $^{208}$Tl, \mbox{$^{212}$Bi-$^{212}$Po} and \mbox{$^{214}$Bi-$^{214}$Po}
sequential decays, $^{12}$B and $^{12}$N spallation products, and
$\gamma$'s from thermal neutron captures on $^{1}$H and $^{12}$C.

\noindent The visible energy ($E_{vis}$) of an event is computed
from the measured light yield. Specifically, $E_{vis}$\ is the number
of detected p.e.\ after corrections for PMT variation, dark noise,
solid angle, shadowing by suspension ropes, optical transparencies,
and scattering properties in the LS. The relationship between $E_{vis}$\ and
the deposited energy ($E_{dep}$) of $\gamma$'s, $e^{\pm}$'s, protons,
and $\alpha$'s is non-linear and modeled as a combination of Birks-quenched
scintillation and Cherenkov radiation. The scale is adjusted so that
$E_{vis}$\ is equal to $E_{dep}$\ for the 2.225\,MeV $\gamma$-ray
from neutron capture on $^{1}$H. The combined 17-inch and 20-inch
PMT energy resolution is \mbox{$\sim$6.5\%$/\sqrt{E_{vis}(\mathrm{MeV})}$}.

\noindent The calibration sources are also used to determine systematic
deviations in position reconstruction by comparison with the source's
known position. This comparison gives an average position reconstruction
uncertainty of less than 3\,cm for events with energies in the range
0.28 to 6.1\,MeV.

\noindent The KamLAND experiment started operation in March 2002 and
has run continuously since then. The experiment has gone through various
phases, the latest phase is a modification to the detector where an
additional balloon at the center of the detector is filled with a
mixture of liquid scintillator and $^{136}$Xe to study neutrinoless
double beta decay.

\subsubsection{\noindent Intrinsic detector backgrounds}

\noindent A number of backgrounds could mask the oscillation signal
from the proposed source measurement. However we demonstrate in what
follows that detector background are absolutely not an issue for CeLAND.
We perform a simulation including all backgrounds measured or the
last 10 years in KamLAND as shown in \figref{KamLANDBkg}. Our background
generator takes into account:
\begin{itemize}
\item \noindent accidental events, usually from natural radioactivity. 
\item \noindent cosmogenics, $^{9}$Li and $^{8}$He.
\item \noindent ($\alpha$,n), an $\alpha$ from Polonium (prompt signal)
gets captured by a carbon atom deexciting into oxygen and a neutron
(delayed signal from capture). 
\item \noindent geoneutrinos, from U and Th inside Earth's crust and mantle.
\item \noindent reactor neutrinos, from the surrounding Japanese and Korean
nuclear reactors. This background will depend on the nuclear activity
in Japan at the time of the experiment, but cannot create significant
background contribution even if all reactors in Japan are running
at the full power level.
\end{itemize}
\noindent \begin{center}
\begin{figure}[h]
\centering{}\includegraphics[scale=0.75]{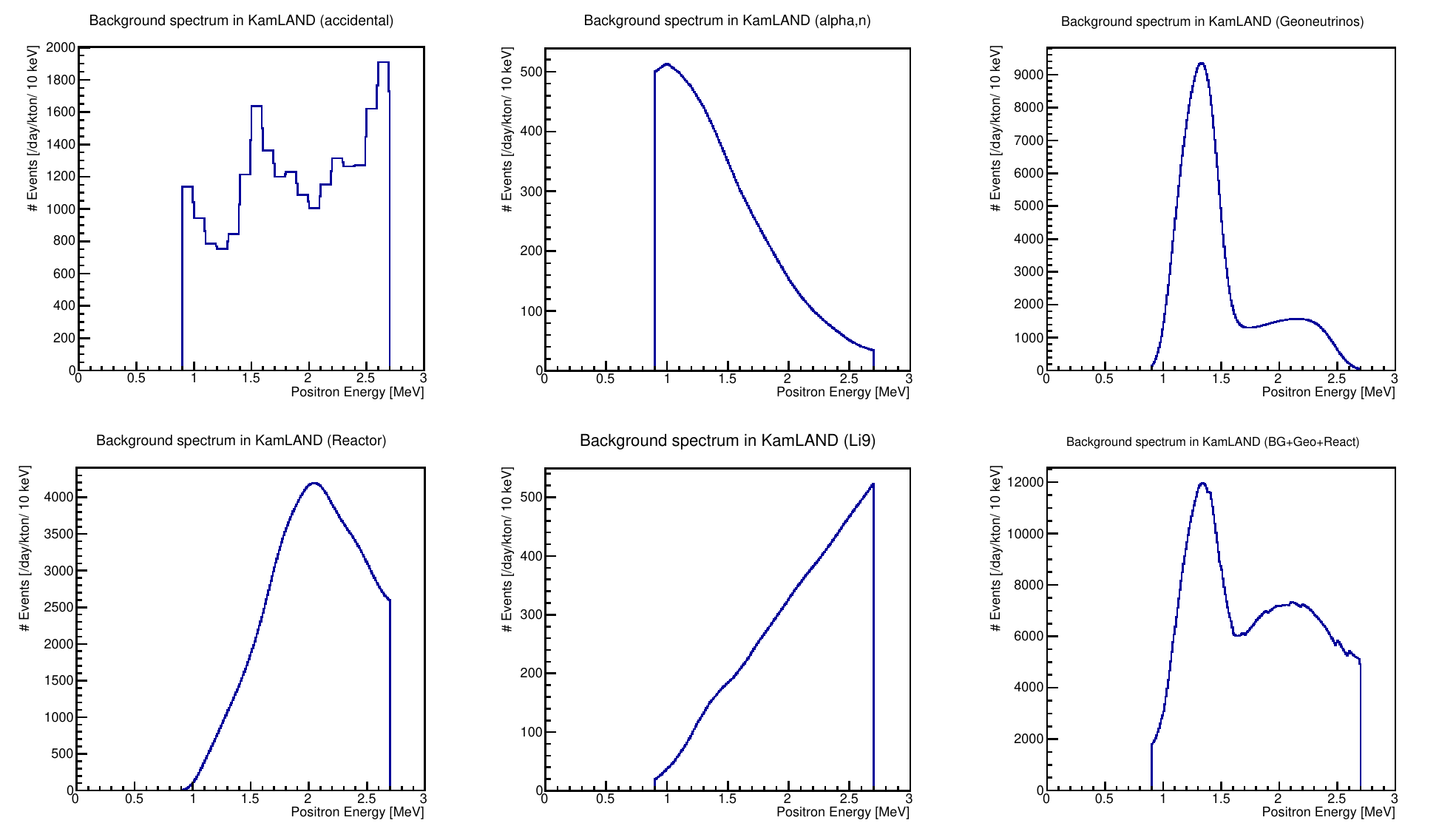}\caption{Energy distributions of the KamLAND intrinsic backgrounds measured
in-situ. From left to right and top to bottom: accidentals,($\alpha$,n),
geoneutrinos, reactor neutrinos, $\mathrm{^{9}Li}$, Sum of all the
previous backgrounds. \label{fig:KamLANDBkg}}
\end{figure}

\par\end{center}

\subsection{\noindent KamLAND infrastructure for CeLAND}

\noindent 
This section discusses the transportation of the Ce-ANG antineutrino
generator in Japan. There are three logical steps to transport the
Ce-ANG generator from the port of entry to the KamLAND deployment
area: 
\begin{enumerate}
\item \noindent Transportation from a port to the mine entrance; 
\item \noindent Transportation from the mine entrance to KamLAND entrance; 
\item \noindent Transportation from the KamLAND entrance to the deployment
area. 
\end{enumerate}

\subsubsection{\noindent Transportation of the antineutrino generator to the mine
entrance}

\noindent The Ce-ANG antineutrino generator will arrive at a port
on the Pacific Coast (i.e., East-side) of Japan. The generator will
be contained in a BU type container with a weight of 10-25 tons. The
generator in the BU container will have to be transported from the
port to the Kamioka mine entrance.

\noindent The national highways are well maintained in Japan and can
be used for transportation. A licensed shipping company can transport
Ce-ANG in the BU container across the mountains to Kamioka. There
are several shipping companies that have transported BU containers
in Japan. The CeLAND Japanese group is currently discussing the transportation
plan with one such company, HITACHI Transport Systems (HTS). From
these discussions it is clear that the technically most challenging
segment of the transportation is the last 500\,m, just before the
Atotsu entrance to the Kamioka mine. The road here follows a narrow
path along the Atotsu river and includes a small bridge that may not
be able to support the weight of the truck and BU container.

\subsubsection{\noindent Transportation of the anti neutrino generator to KamLAND
entrance}

\noindent \begin{center}
\begin{figure}[h]
\begin{centering}
\includegraphics[clip,width=1\columnwidth]{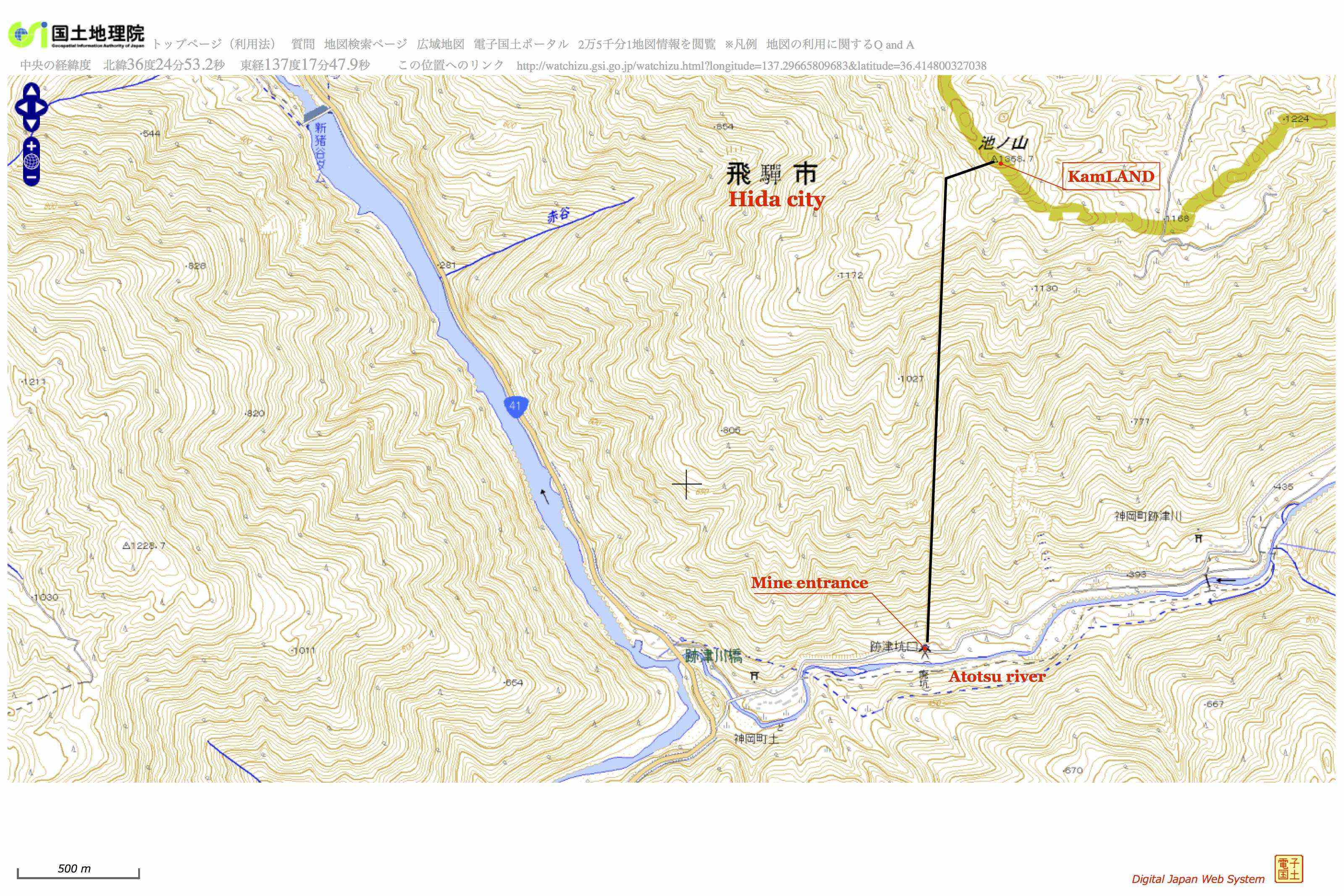} \caption{\label{fig:Access-to-Kamioka}Access to Kamioka mine entrance from
National highway 41.}

\par\end{centering}

\label{fig:SiteMap} 
\end{figure}

\par\end{center}

\noindent The KamLAND detector is located 2.5\,km from the Atotsu
entrance of the Kamioka mine, see \figref{Access-to-Kamioka}. Most
of the tunnel from the mine entrance to the detector is horizontal
and should not be a problem. However, there is a short, 20\,m long
section of the tunnel with a slope that will be challenging for transportation.

\noindent Once the antineutrino generator arrives at the mine entrance,
there are two scenarios:
\begin{itemize}
\item \noindent The most favorable is the case that the antineutrino generator,
contained inside the 1.5-ton $\gamma$-ray shield, can be removed
from the BU container for further transportation. The mine company
has equipment that can lift the 1.5 ton container and transport the
container to the KamLAND entrance. This scenario will need detailed
discussion with the mine company regarding regulation, in particular
whether we will be allowed to lift Ce-ANG at the entrance of the mine
without the BU container.
\item \noindent In case the generator cannot be removed from the BU container,
the BU container will have to be transported to the KamLAND entrance
on the same truck as was used for transportation from the port. The
truck will have to have the capability to clear the slope on the way
towards the KamLAND entrance.
\end{itemize}

\subsubsection{\noindent Transportation from the KamLAND entrance to the KamLAND
dome area}

\noindent \begin{center}
\begin{figure}[h]
\centering{}\includegraphics[clip,width=0.6\columnwidth]{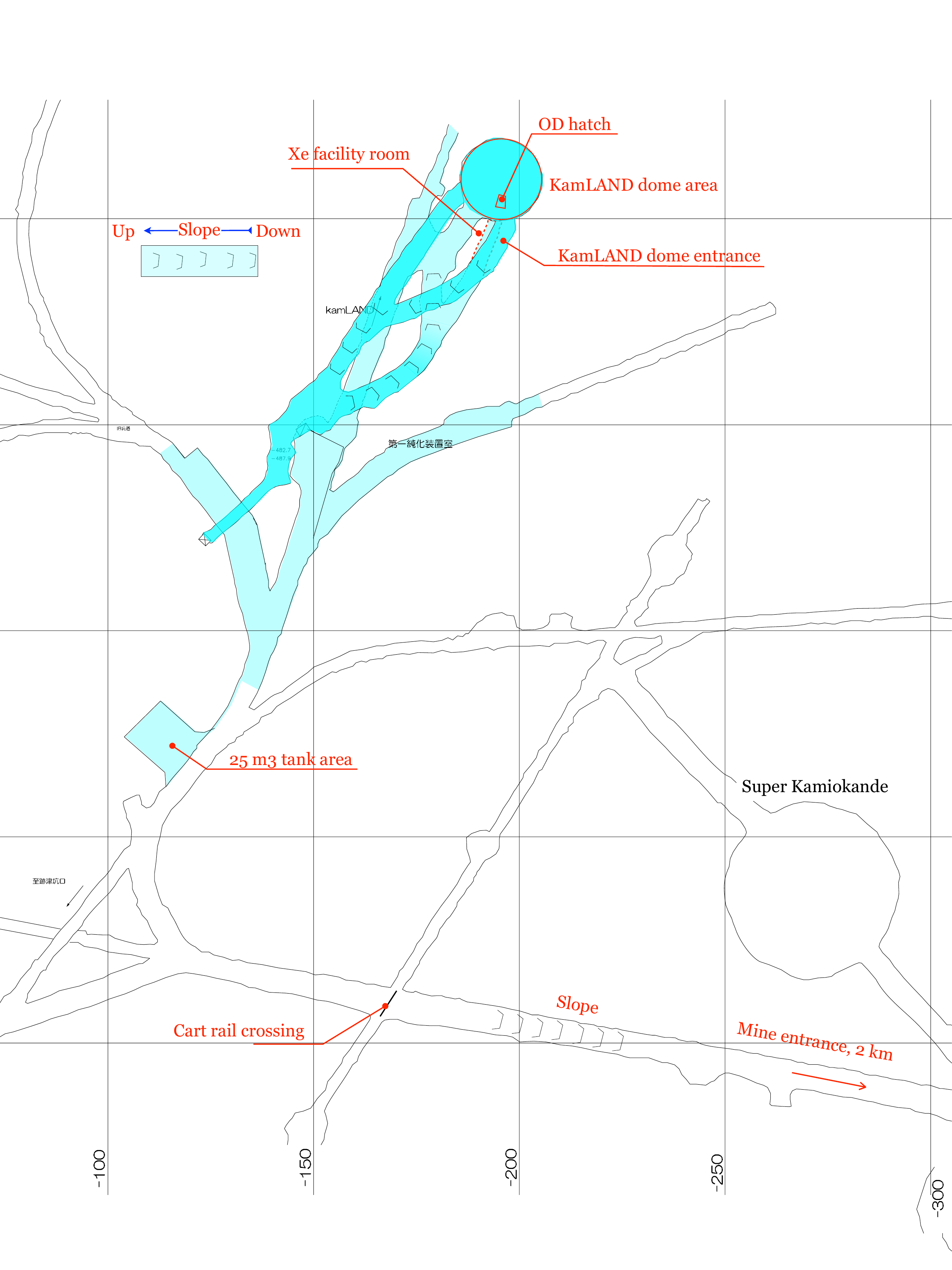}
\caption{\label{fig:KamLAND-area}KamLAND area}
\end{figure}

\par\end{center}

\noindent It will not be feasible to transport the generator inside
the heavy BU container towards the KamLAND dome area. The generator,
still inside the 1.5-ton $\gamma$-ray shield, will have to be removed
from the BU container at this point at the latest. The 25\,m$^{3}$
tank region, at the entrance of KamLAND-controlled area (see \figref{KamLAND-area}),
can be used to remove the generator and $\gamma$-shield from the
BU container. We will have to install lifting fixtures in this area
to accomplish this. Once Ce-ANG is removed, it can be lifted by existing
equipment operated by the Kamioka mine company to the entrance of
the KamLAND dome area. The maximum lifting capacity of this equipment
is 4\,tons and is therefore sufficient for this task.

\noindent No heavy machinery can enter the KamLAND dome area. This
means that personnel will have to transport the 3-ton shield with
the generator over the last $\sim$15\,m. For this we intend to use
cranes, chains or other tools.

\subsection{\noindent Deployment into the outer detector (water veto) and data
taking }

\noindent The existing water Cherenkov detector, the so-called KamLAND
OD, provides a straightforward location for the first phase of source
deployment. 

\begin{center}
\begin{table}[h]
\begin{centering}
\begin{tabular}{cc}
\hline 
Source/detector properties & \tabularnewline
\hline 
\hline 
Source initial activity & 75-100 kCi\tabularnewline
Source location & Water veto\tabularnewline
Source-center distance & 9.3 m\tabularnewline
Fiducial volume radius & between 6.0 and 6.5 m\tabularnewline
Interactions after 6 months & 8930\tabularnewline
Shielding composition & Tungsten alloy, d=18.5 g/cm$^{3}$\tabularnewline
Shielding dimensions & 51 cm height, 54 cm diameter cylinder - TBC\tabularnewline
Shielding mass & 1.5 tons\tabularnewline
\hline 
\end{tabular}
\par\end{centering}

\caption{\label{tab:CeLAND-Phase-I} CeLAND Phase I deployment data}
\end{table}

\par\end{center}

\begin{center}
\begin{figure}[h]
\begin{centering}
\includegraphics[scale=0.35]{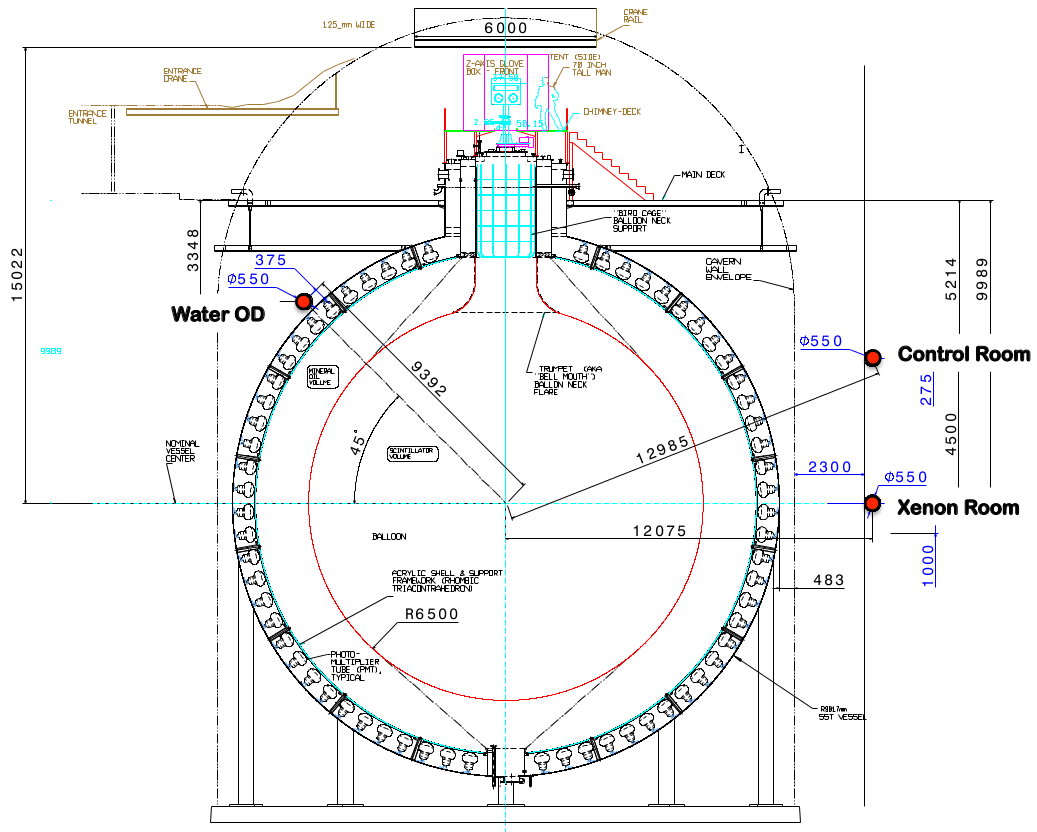}
\par\end{centering}

\caption{\label{fig:3locations-Phase1}Sketch of the 3 different locations
under consideration.}
\end{figure}

\par\end{center}

\subsubsection{\noindent Detector issues and source deployment}

\noindent The recirculating water provides a thermal bath for the
hot source. Buffer oil provides a free shielding of more than 2.5
m with a density of 0.75. Finally spatial constraints for source location
are easier to handle with respect to the detector center. In addition,
the placement of the source in the OD avoids complicated issues of
cleanliness and W-shielding radiopurity. Nevertheless, a dedicated
infrastructure for the transport and lifting of source on the KamLAND
detector upper platform has to be designed and realized. In order
to prevent the unlikely scenario of a radioactive material contamination
in the detector, the source and its biological shielding have to be
isolated from the OD. It is thus planned to put it inside a ``sock''
as sketched in \figref{Ce-ANG-is-contained}. This sock, most likely
made of stainless steel, will act as an additional shielding for the
source intrinsic $\gamma$'s and will be filled with water or any
other liquid that will act as a temperature and neutron flux moderator. 

\begin{center}
\begin{figure}[h]
\centering{}\fbox{ \includegraphics[clip,width=0.7\columnwidth]{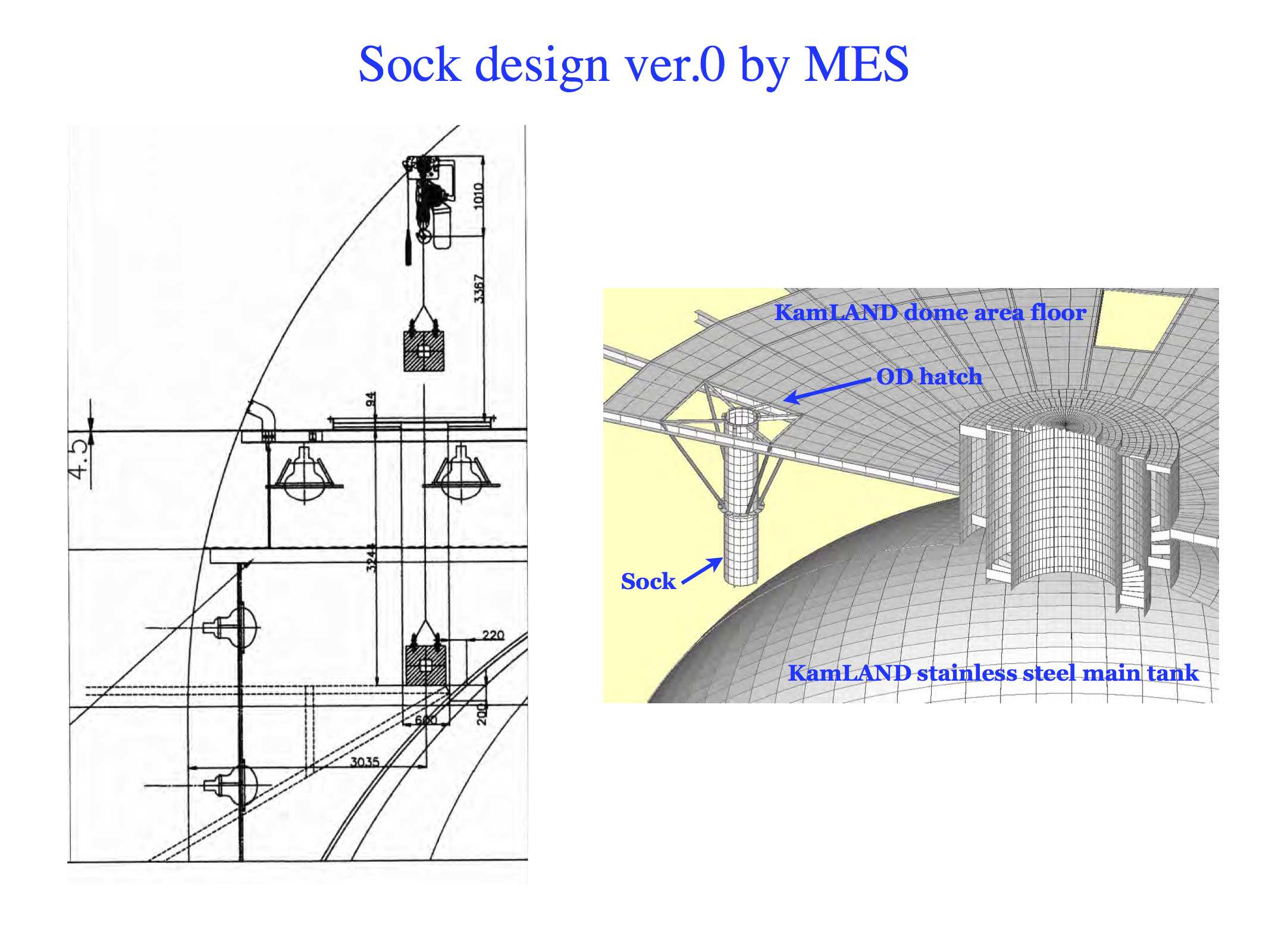}
} \caption{\label{fig:Ce-ANG-is-contained}Ce-ANG is contained inside a stainless
steel enclosure (``sock'') and suspended into the Outer Detector.}
\end{figure}

\par\end{center}

\begin{center}
\begin{figure}[h]
\begin{centering}
\includegraphics[scale=0.27]{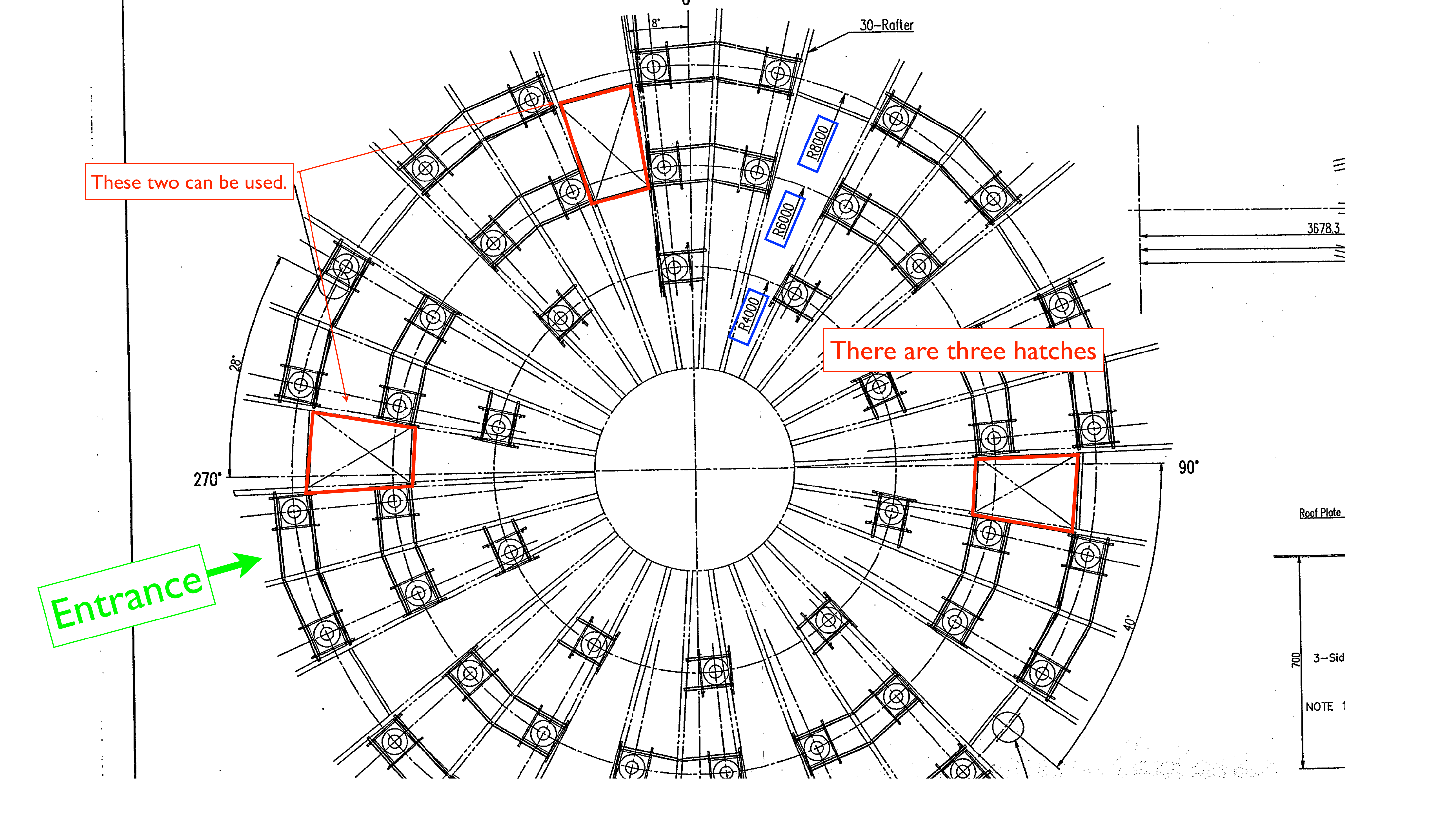}
\par\end{centering}

\centering{}\caption{KamLAND top platform and OD hatch to be used for the source insertion
inside the OD.}
\end{figure}

\par\end{center}

\subsubsection{\noindent Signal and backgrounds}

\noindent Assuming a moderator thickness of 10 cm, the source will
be located at about 9.3 m from the detector center, separated from
the active liquid scintillator volume by at least 2.8 m. CeLAND data
are given in Table \ref{tab:CeLAND-Phase-I}. In the case of no-oscillation
we estimate that 14650 neutrino events in one year are expected within
the detector fiducial volume (taken as R<6.5 m inside the balloon
containing the liquid scintillator). The expected number of interactions
is given in Table \ref{tab:CeLAND-Phase-I}. 

\begin{center}
\begin{figure}[h]
\centering{}\subfloat[]{\centering{}\includegraphics[scale=0.4]{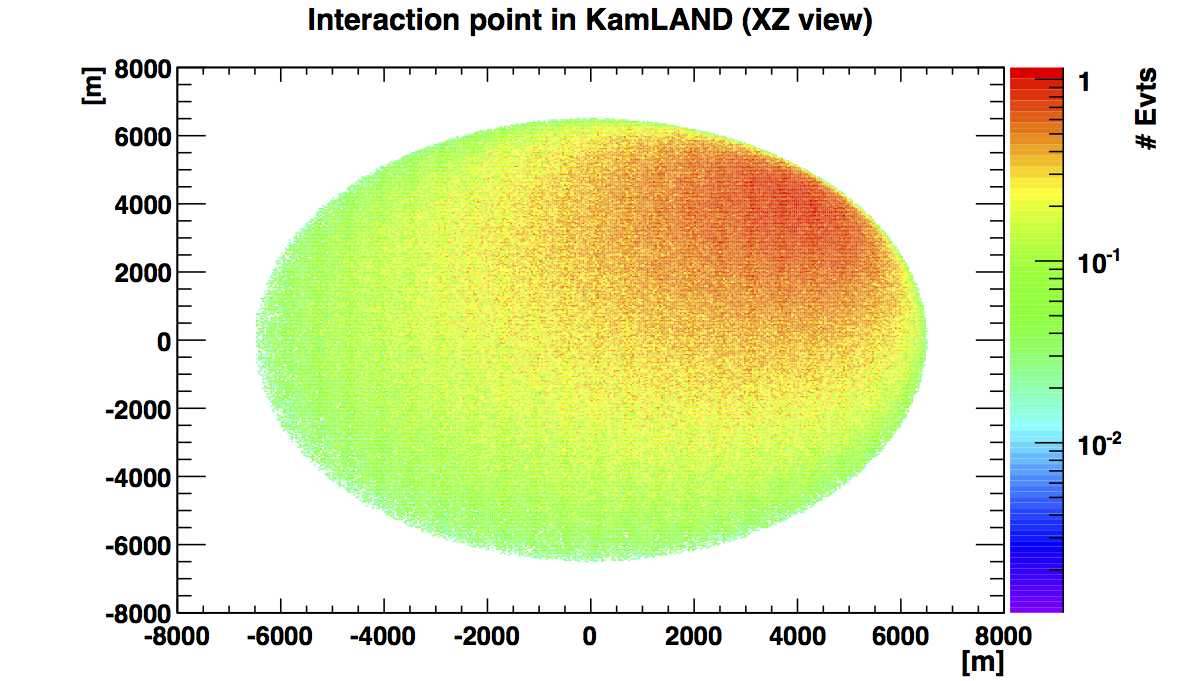}}\subfloat[]{\begin{centering}
\includegraphics[scale=0.4]{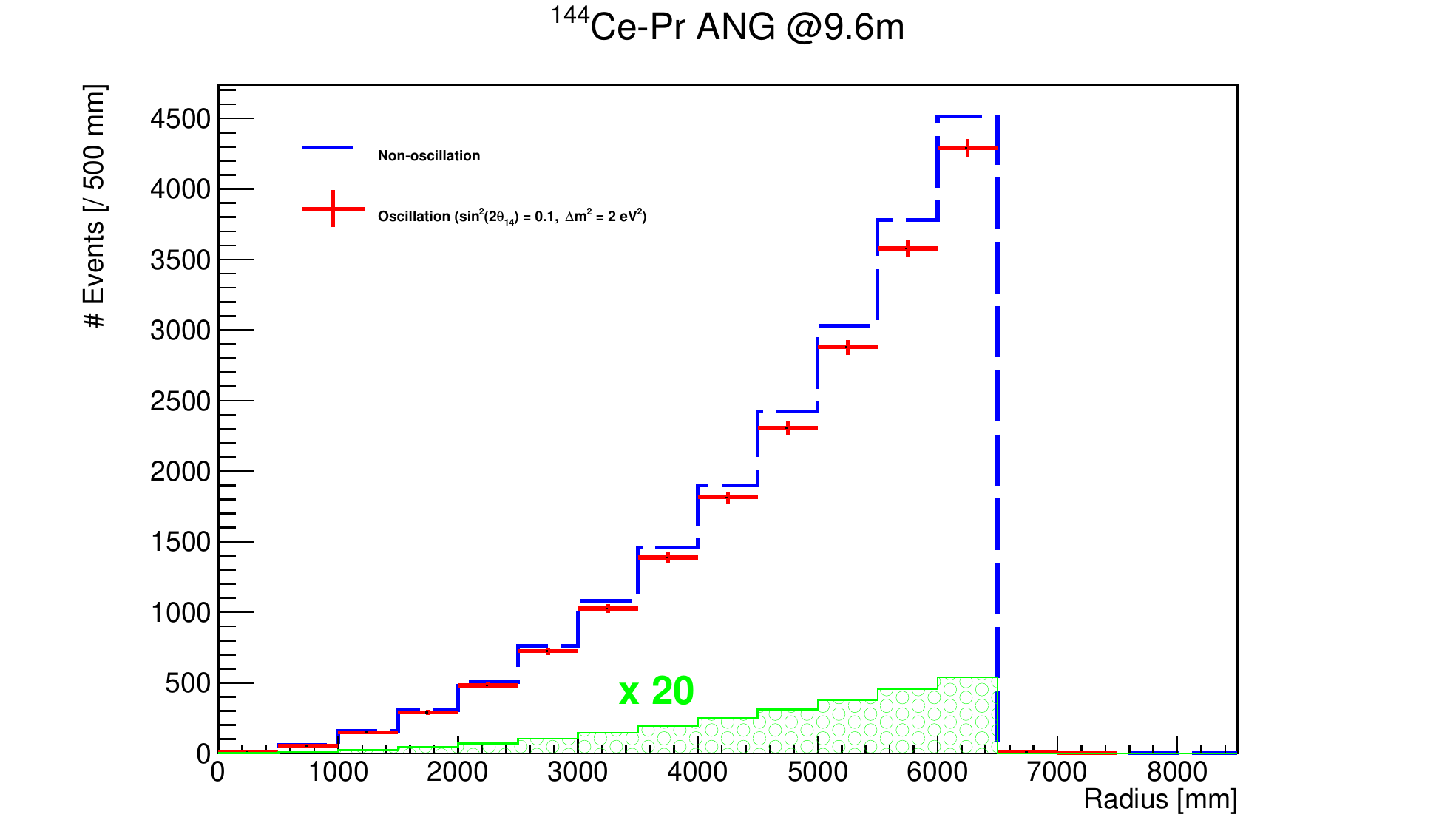}
\par\end{centering}

}\caption{\label{fig:Left-panel:-simulation} Left panel: Simulation of signal
events in the non-oscillation case, for a source located at $\mathrm{Y\sim6.8\,m}$
and $\mathrm{Z\sim6.8\,m}$ (here, 9.3 m away from the detector center).
Right panel: Radial distribution of the signal (blue without oscillation,
red with oscillation) and background events (green). Backgrounds are
enhanced by a factor 20 to be visible on the plot. The origin of the
radius is taken at the center of the detector.}
\end{figure}

\par\end{center}

\noindent The energy and distance traveled determine the oscillation
probability of the $\overline{\nu}_{e}$ flux. Using the GEANT4 KamLAND
Monte-Carlo simulation we simulated 10 million events. However results
that follows are renormalized to 20000 to match the expectation for
a one year run with a 75 kCi source located 9.3 m away from the detector
center. 

\begin{center}
\begin{figure}[h]
\begin{centering}
\subfloat[]{\begin{centering}
\includegraphics[scale=0.39]{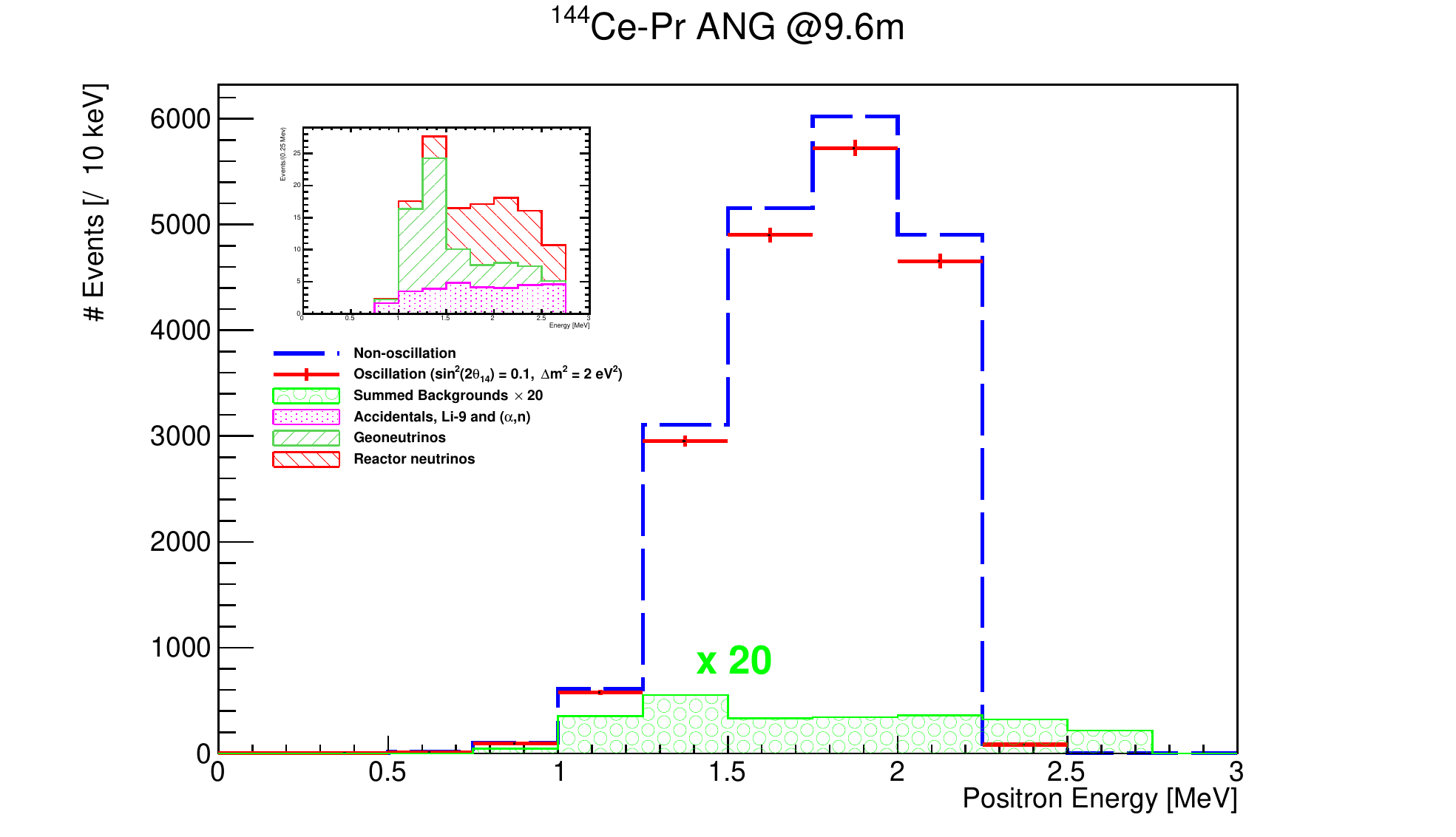}
\par\end{centering}

\centering{}}\subfloat[]{\centering{}\includegraphics[scale=0.39]{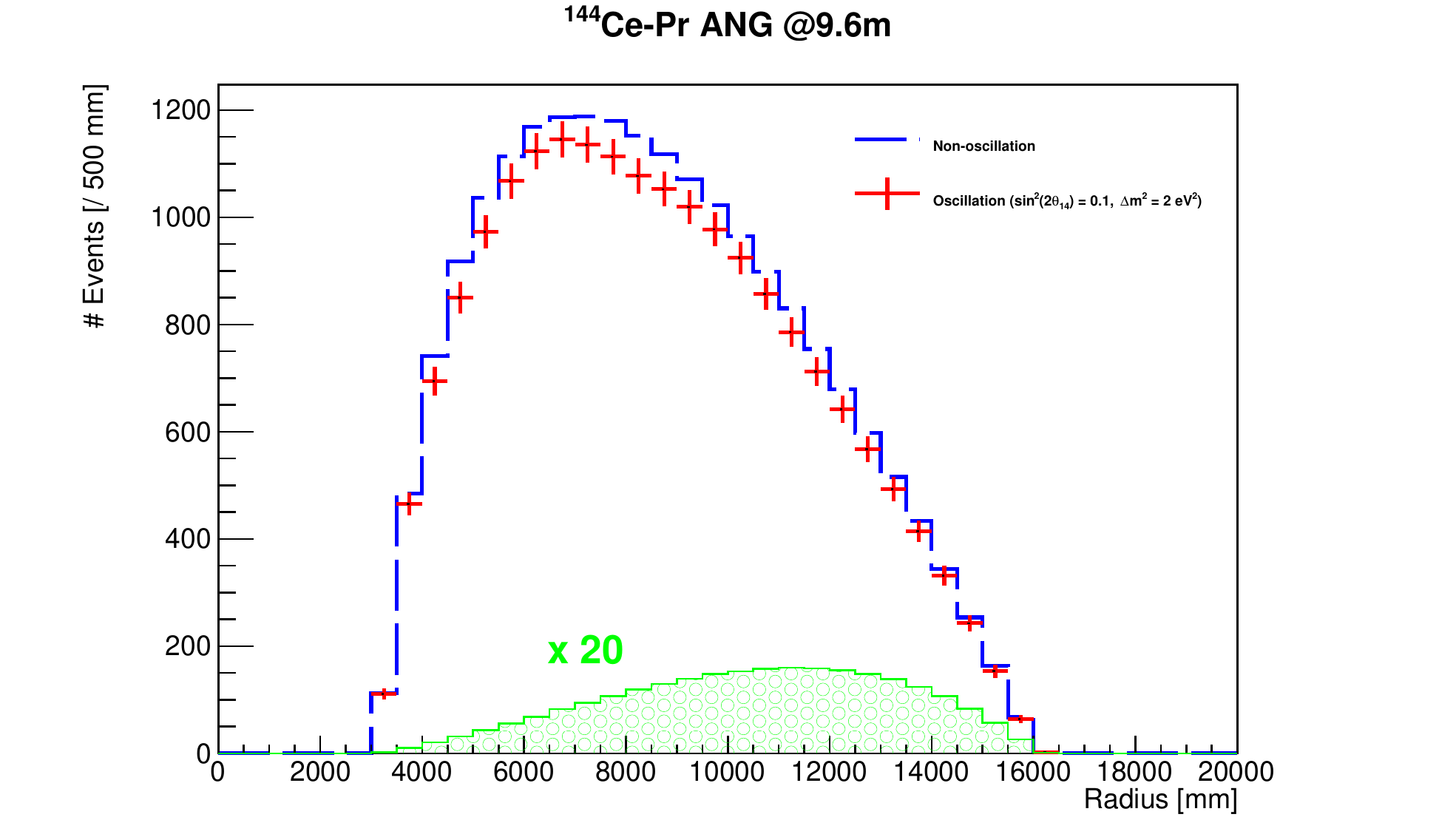}}
\par\end{centering}

\centering{}\caption{Left panel: Energy distribution of the signal in case of non-oscillation
(blue) and oscillation (red) and the background (green, enhanced by
a factor 20). Various background components are stacked in the inset.
Right panel: Radial distribution of the signal and background with
the same color codes. Only the sum of all backgrounds is represented,
dominated by the geoneutrinos. The origin of the radius is taken at
the center of the source, 9.3 m away from the detector center.}
\end{figure}

\par\end{center}

\noindent Figure \ref{fig:Left-panel:-simulation} and shows the effect
of $\overline{\nu}_{e}\rightarrow\nu_{s}$ oscillation as a function
of energy and distance from the source. It is displayed as a function
of R/E in \figref{SignalLoE}. It also illustrates the fractional
oscillation effect normalized to the expected, unoscillated event
rate. In what follows, to be conservative, signal is being considered
within a radius of 6 m with respect to the detector center to limit
the contribution of the accidental background. This fiducial cut could
probably be relaxed to use the whole liquid scintillator contained
in the balloon. As an illustration we considered, as a baseline, the
following neutrino oscillation parameters: $\mathrm{\Delta m^{2}_{14} = 2\,eV^{2}}$, $\mathrm{\sin^{2}(2\,\theta_{14})=0.1}$ . 

\noindent %

\begin{center}
\begin{figure}[h]
\subfloat[]{\begin{centering}
\includegraphics[scale=0.39]{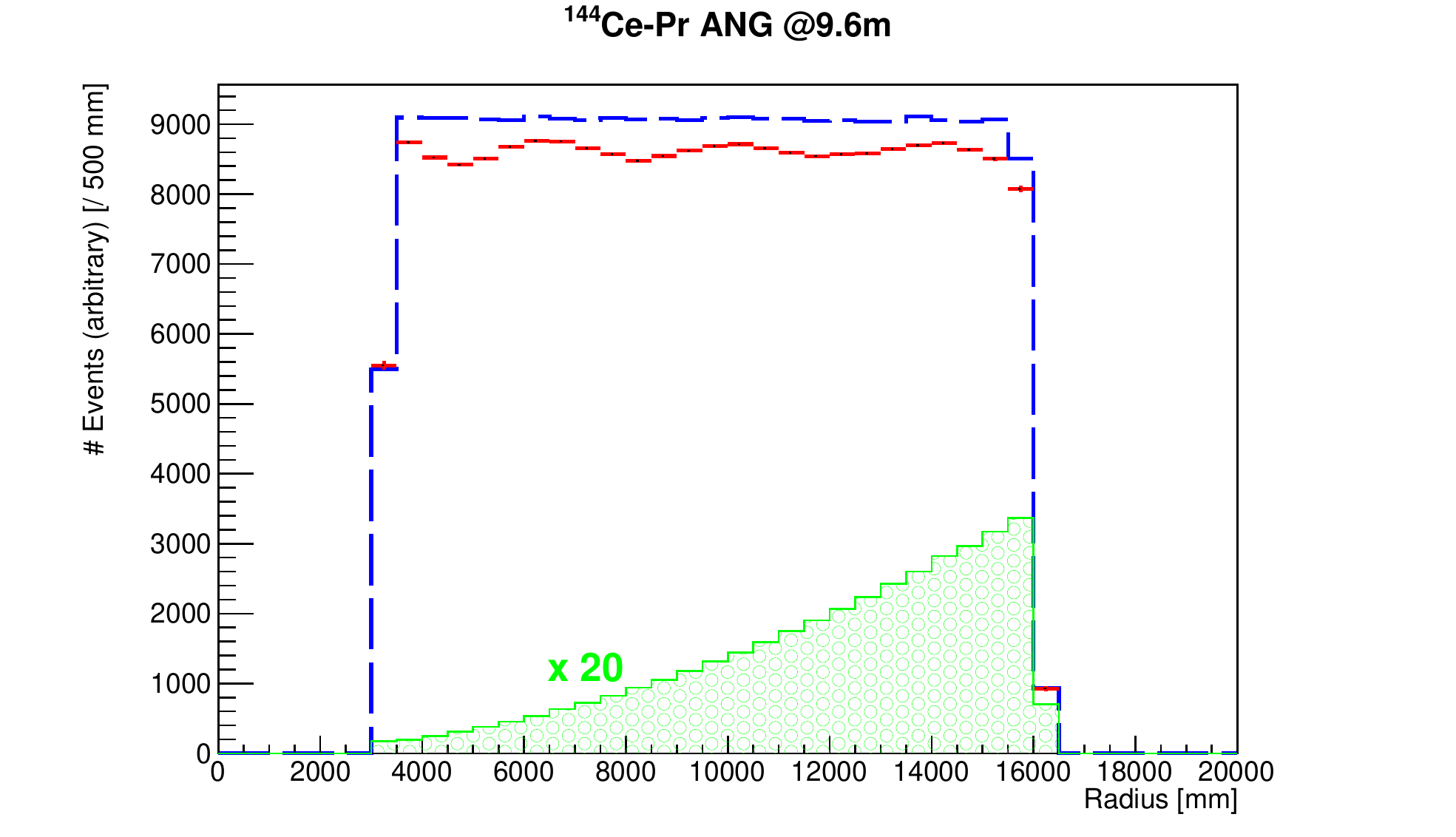}
\par\end{centering}

}\subfloat[]{\begin{centering}
\includegraphics[scale=0.39]{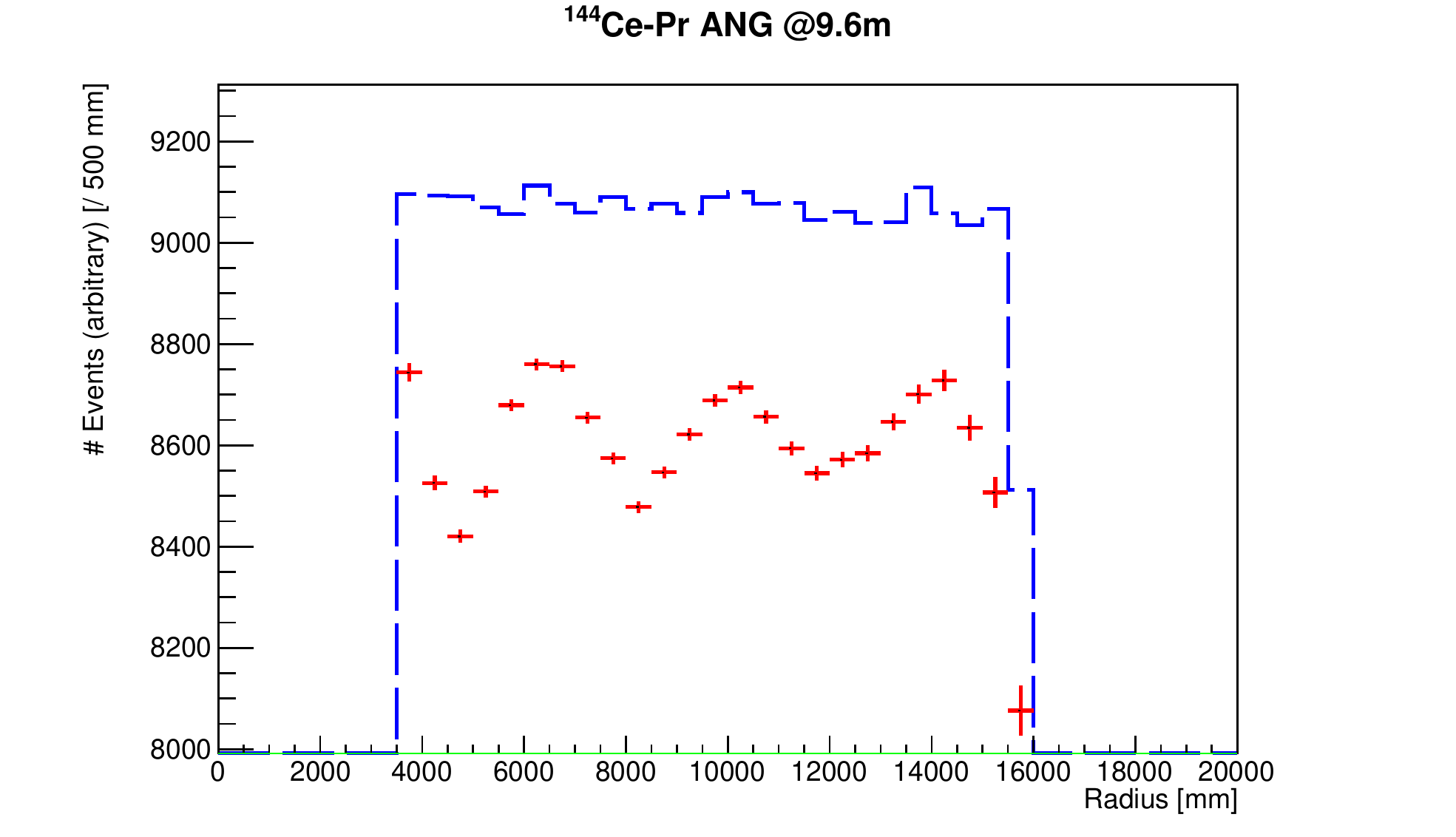}
\par\end{centering}

}

\caption{Left panel: normalized radial distribution of the signal and background
in case of non-oscillation (blue) and oscillation (red) and the background
(green, enhanced by a factor 20). Each bin content at equidistance
to the source is divided by the total scintillator volume at the corresponding
distance. Right panel: zoom on the radial normalized distributions.
Oscillation parameters are $\mathrm{\Delta m^{2}_{14} = 2\,eV^{2}}$, $\mathrm{\sin^{2}(2\,\theta_{14})=0.1}$ 
in the 3 active + 1 sterile neutrino hypothesis.}
\end{figure}

\par\end{center}

\begin{center}
\begin{table}[h]
\begin{centering}
\begin{tabular}{cccccccccc}
\hline 
Months & 3 & 6 & 9 & 12 & 18 & 24 & 30 & 36 & 48\tabularnewline
\hline 
R<6 m  & 4950 & 8930 & 12110 & 14650 & 18330 & 20710 & 22170 & 23160 & 24180\tabularnewline
R<6.5 m & 6425 & 11560 & 15680 & 18970 & 23740 & 26800 & 28710 & 29970 & 31330\tabularnewline
\hline 
\end{tabular}
\par\end{centering}

\caption{Expected $\overline{\nu}_{e}$ interactions inside KamLAND with a
75 kCi $^{144}$Ce antineutrino generator at 9.3 m from the detector
center as a function of the exposure for 2 definitions of the fiducial
volume, R<6 m (current KamLAND analysis) and R<6.5 m (extension of
the fiducial volume to the KamLAND balloon surface).}

\end{table}

\par\end{center}

\begin{center}
\begin{figure}[h]
\centering{}\includegraphics[scale=0.37]{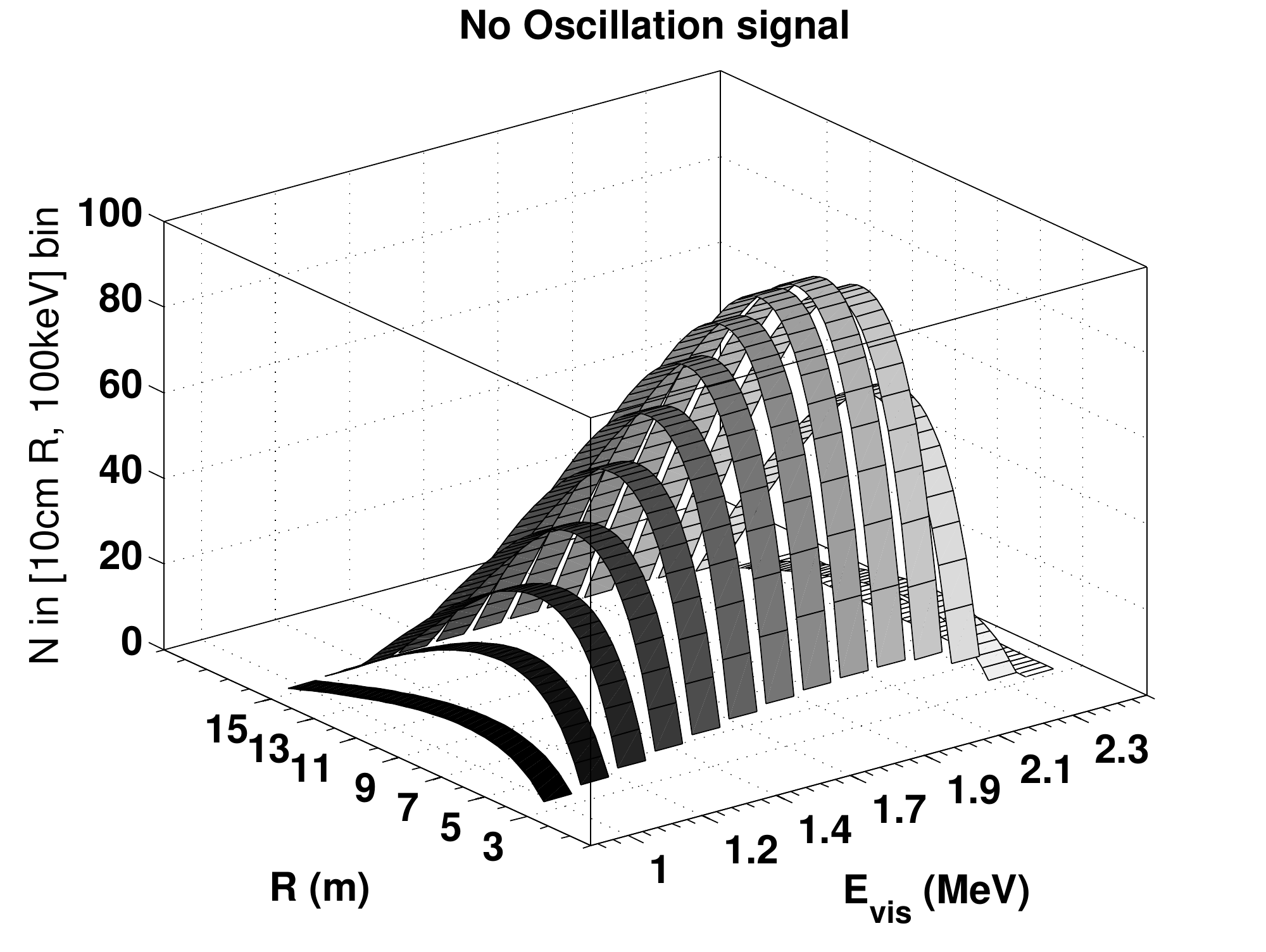}\includegraphics[scale=0.37]{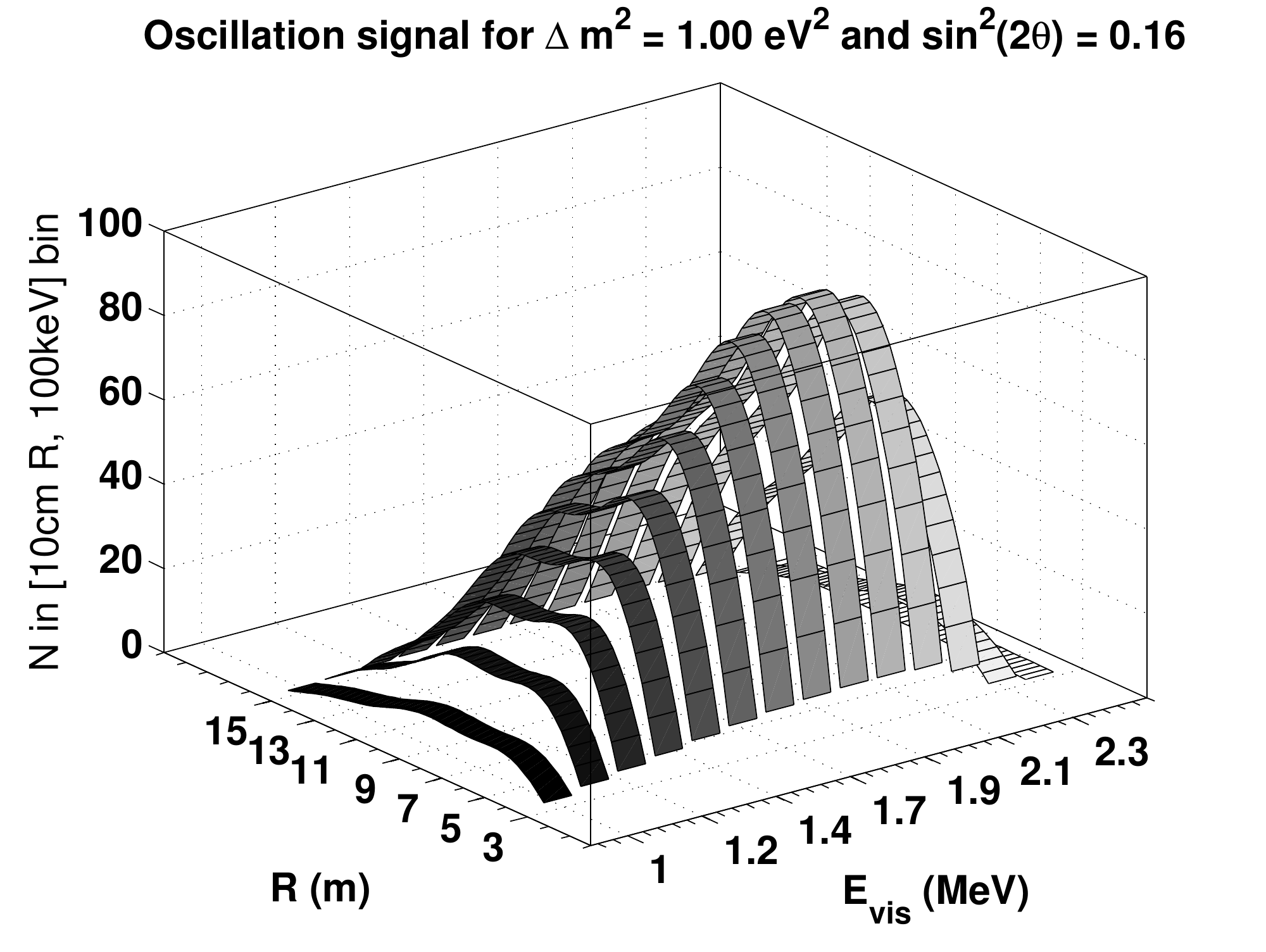}\\
\includegraphics[scale=0.37]{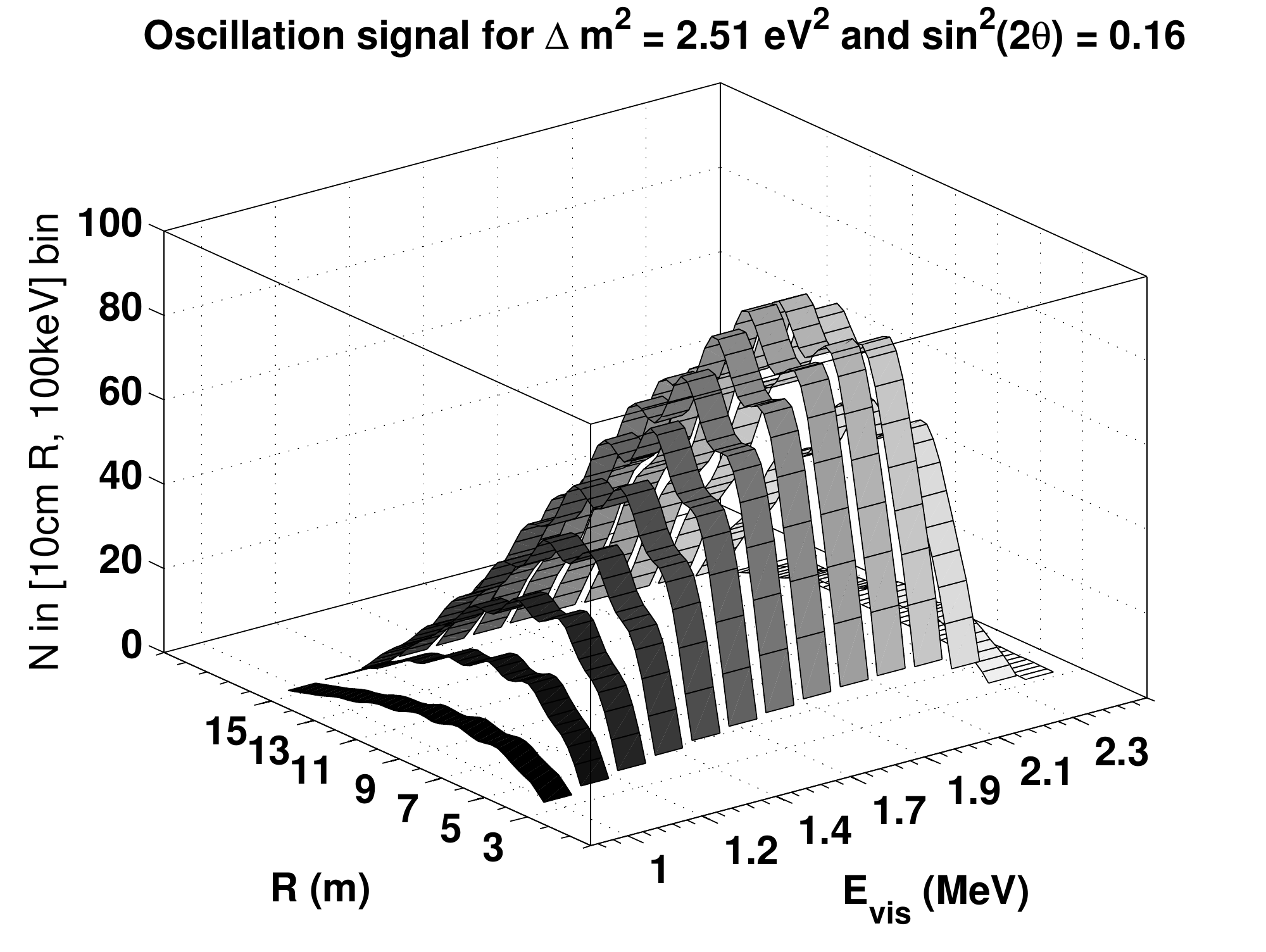}\includegraphics[scale=0.37]{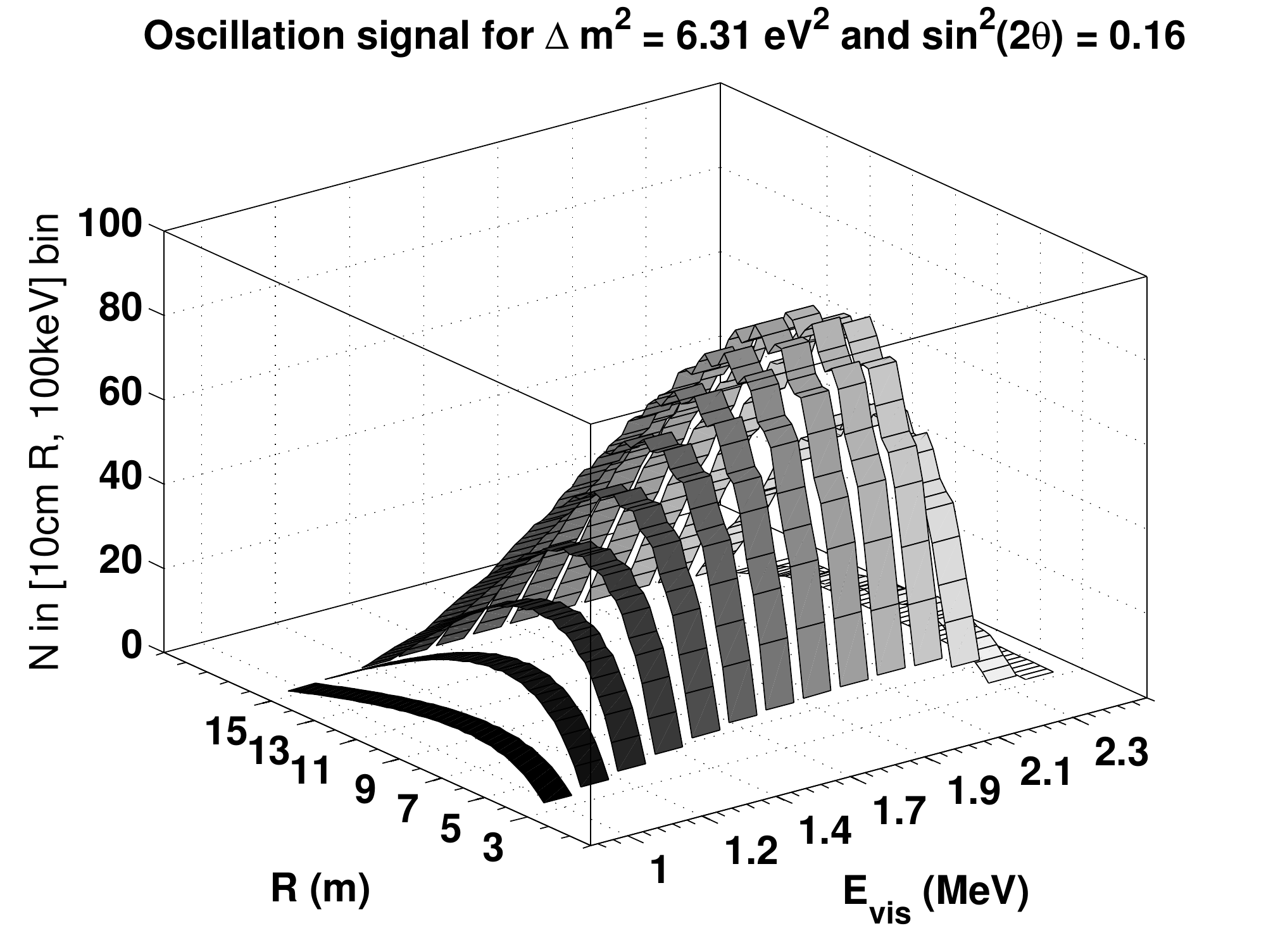}\caption{Advantage of $ $$ $$\bar{\nu}_{e}$ sources providing both R and
E$_{vis}$ oscillation patterns. IBD rate for a 75 kCi $^{144}$Ce
source deployed in the OD of KamLAND, in 10 cm radius bins and 100
keV bins of visible energy, E$_{vis}$=E$_{e}$+2m$_{e}$. In 1.5
year, 20,000 $ $$\bar{\nu}_{e}$ interact within the detector (with
a fiducial volume define as R<6m), for the no-oscillation scenario
(top-left panel). The other panels show the expected oscillation signal
for $\Delta m_{{\rm new}}^{2}$=1, 2.5, 6.3 eV$^{2}$ and $\sin^{2}(2\theta_{{\rm new}})$=0.16.}
\end{figure}

\par\end{center}

\begin{center}
\begin{figure}[h]
\centering{}\includegraphics[scale=0.35]{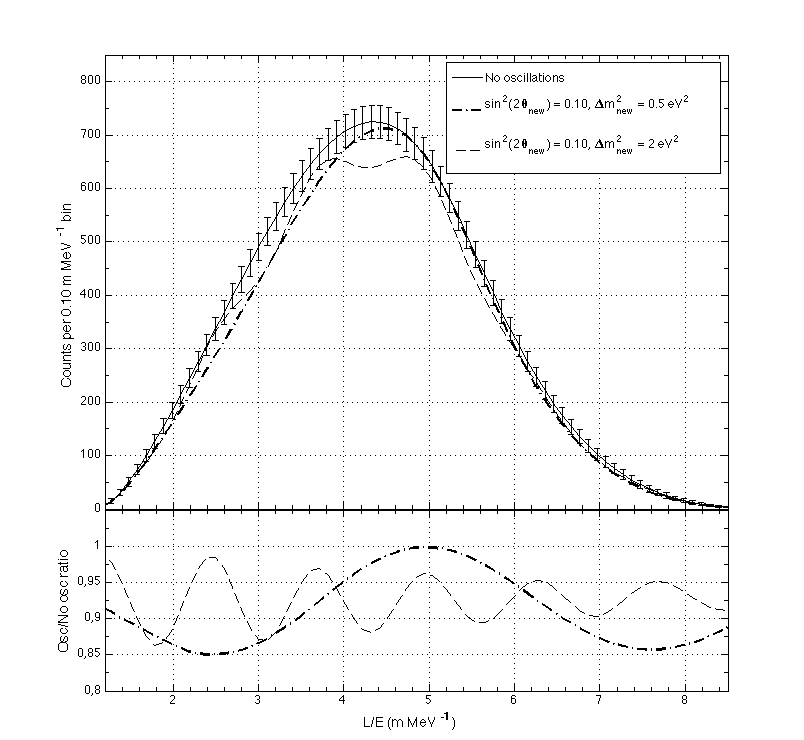}\caption{\label{fig:SignalLoE}Signal displayed as a function of the variable
R/E (L/E on the plot). Statistical error bars are included assuming
18 months of data taking and a fiducial volume defined by R<6.5 m.
The shape uncertainty on the current knowledge of the praseodymium
neutrino spectra is included, as well as an 1.5\% uncertainty on the
source activity. Two oscillation scenarios are given, for $\Delta m_{{\rm new}}^{2}$=0.5
and 2.0 eV$^{2}$ and $\sin^{2}(2\theta_{{\rm new}})$=0.1, and compared
with the no-oscillation case.}
\end{figure}

\par\end{center}

\subsubsection{\noindent Expected sensitivity}

\noindent This section illustrate the expected sensitivity of the
CeLAND experiment for an antineutrino generator deployed at 9.3 m
from the target detector center. 

\begin{center}
\begin{figure}[h]
\begin{centering}
\includegraphics[scale=0.37]{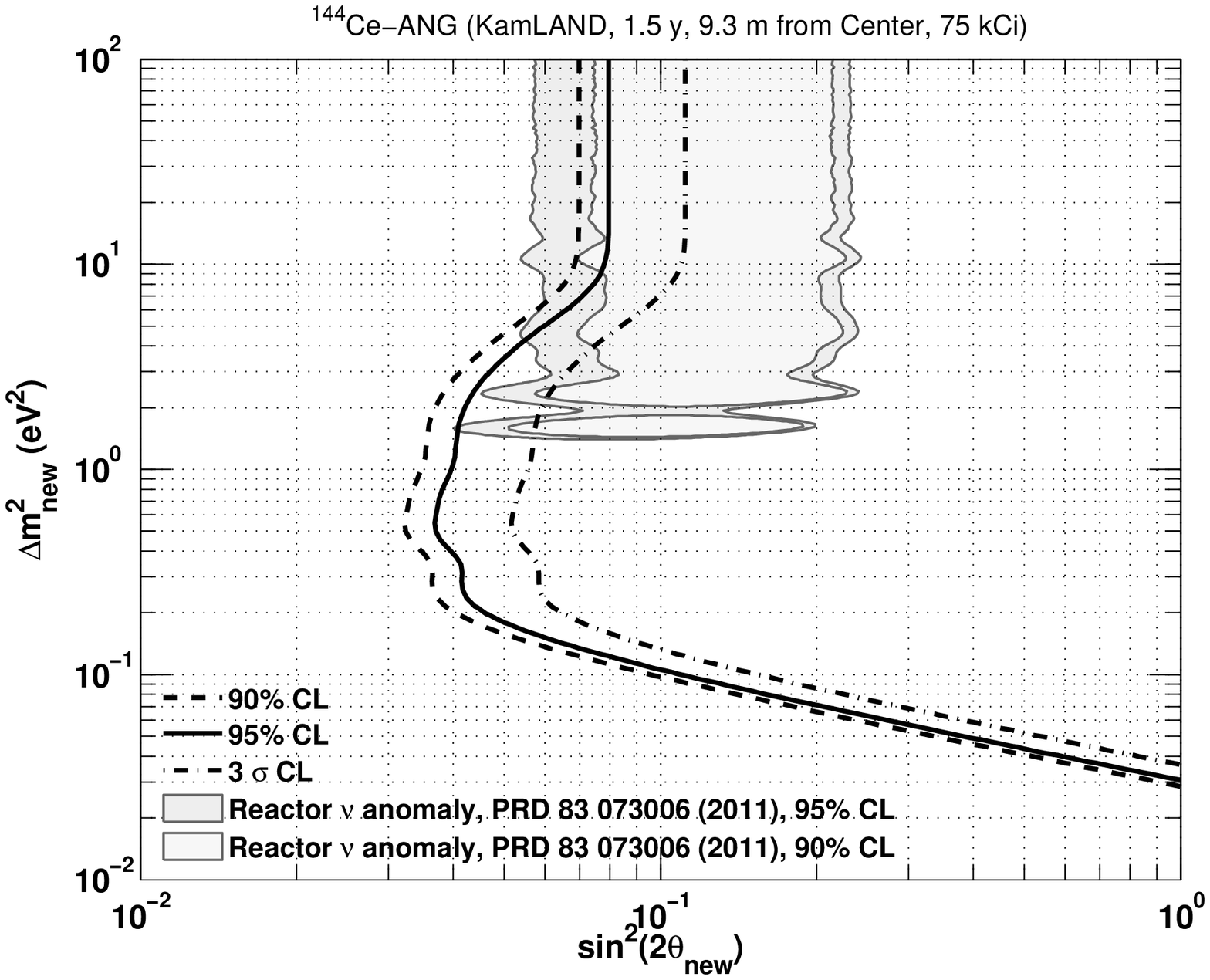}\includegraphics[scale=0.37]{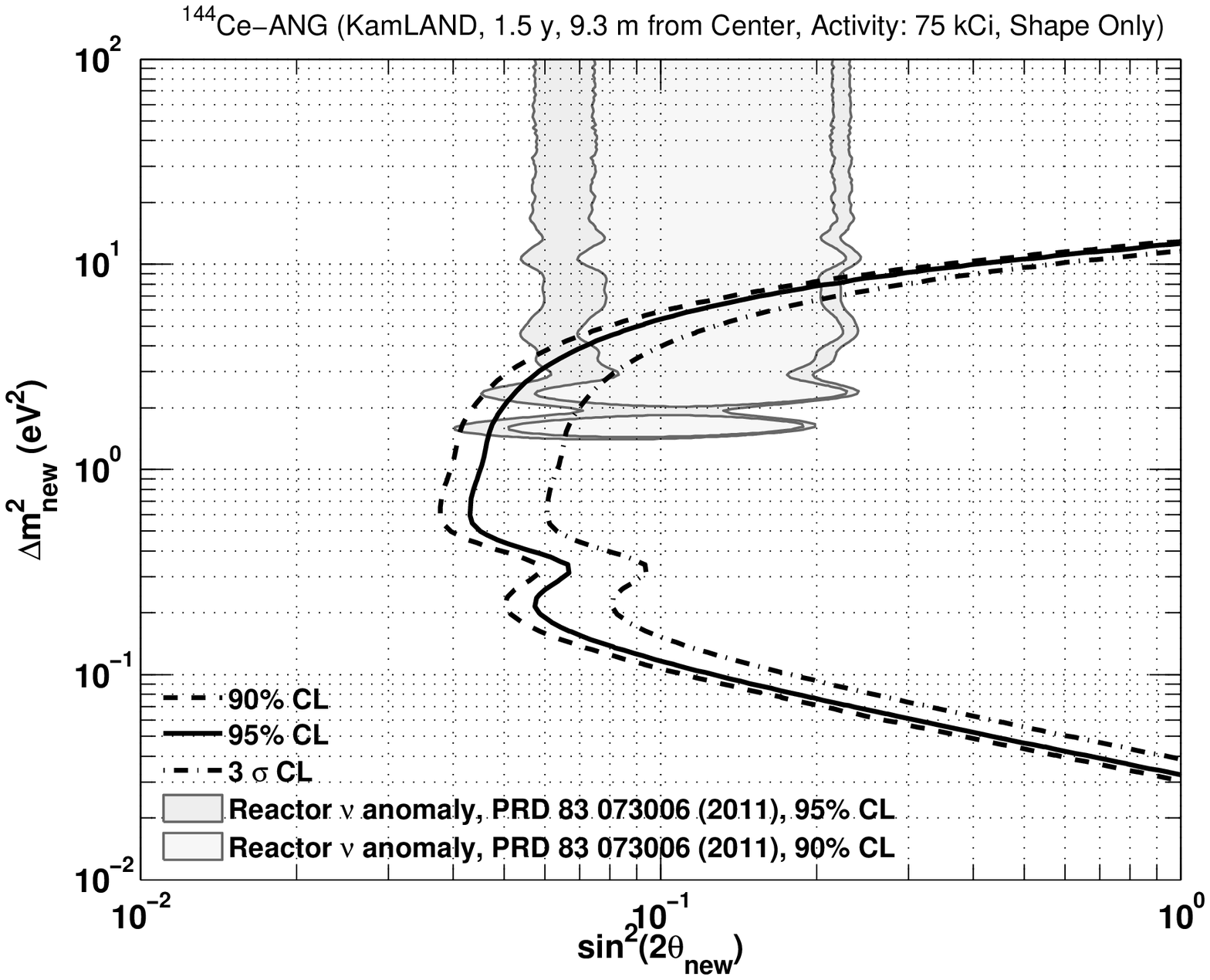}\caption{Expected sensitivity to exclude the non-oscillation. We assume an
initial activity of 75 kCi with a source 9.3 m from the center. The
fiducial volume is starting from the detector center and extends to
a radius of 6.5 m. Results are given for 1.5 year of data taking for
a rate+shape analysis (left panel), and for a shape only analysis
(right panel). The normalization uncertainty is taken at 1.5\%.}

\par\end{centering}

\end{figure}

\par\end{center}

\begin{center}
\begin{figure}[h]
\centering{}\includegraphics[scale=0.37]{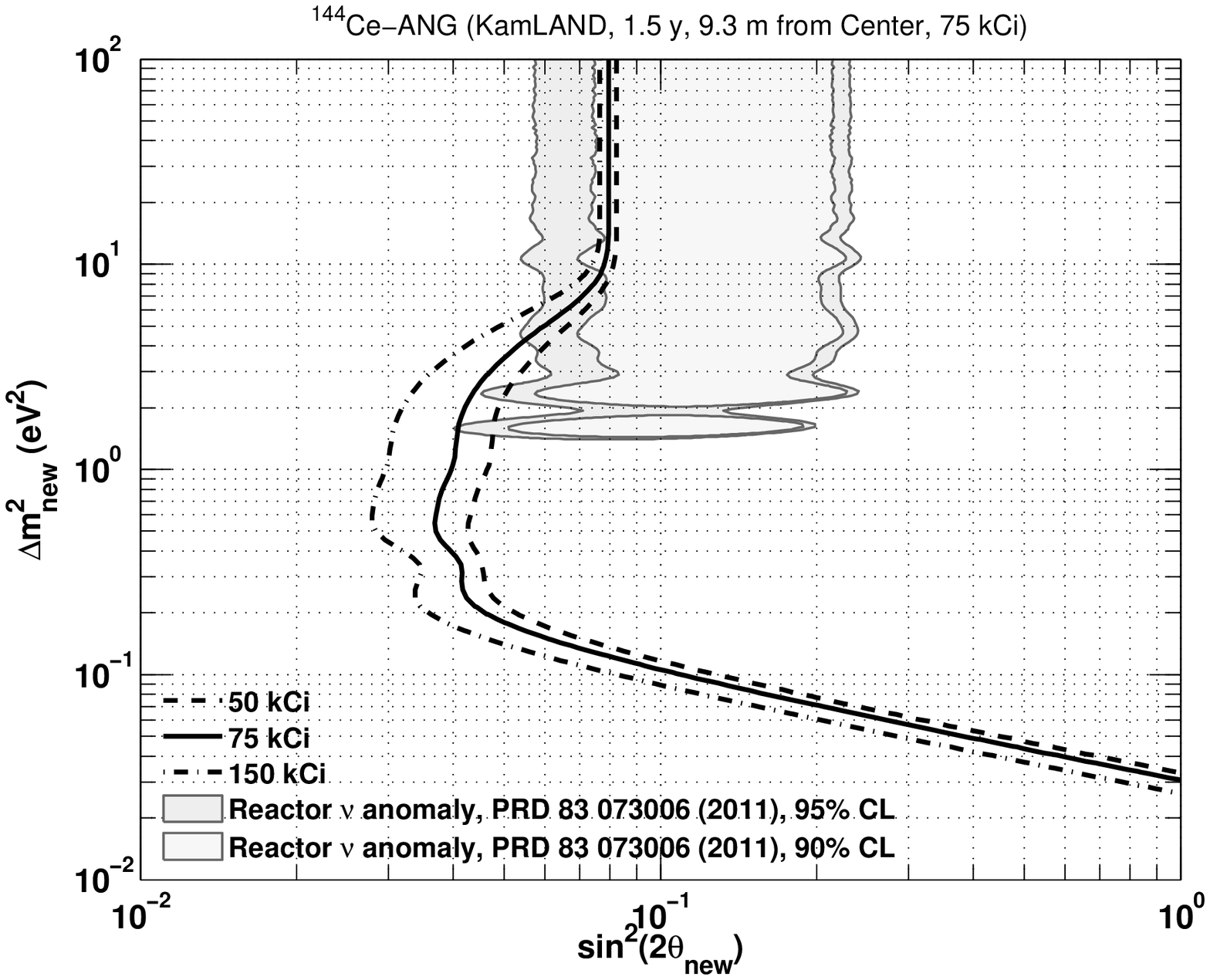}\includegraphics[scale=0.37]{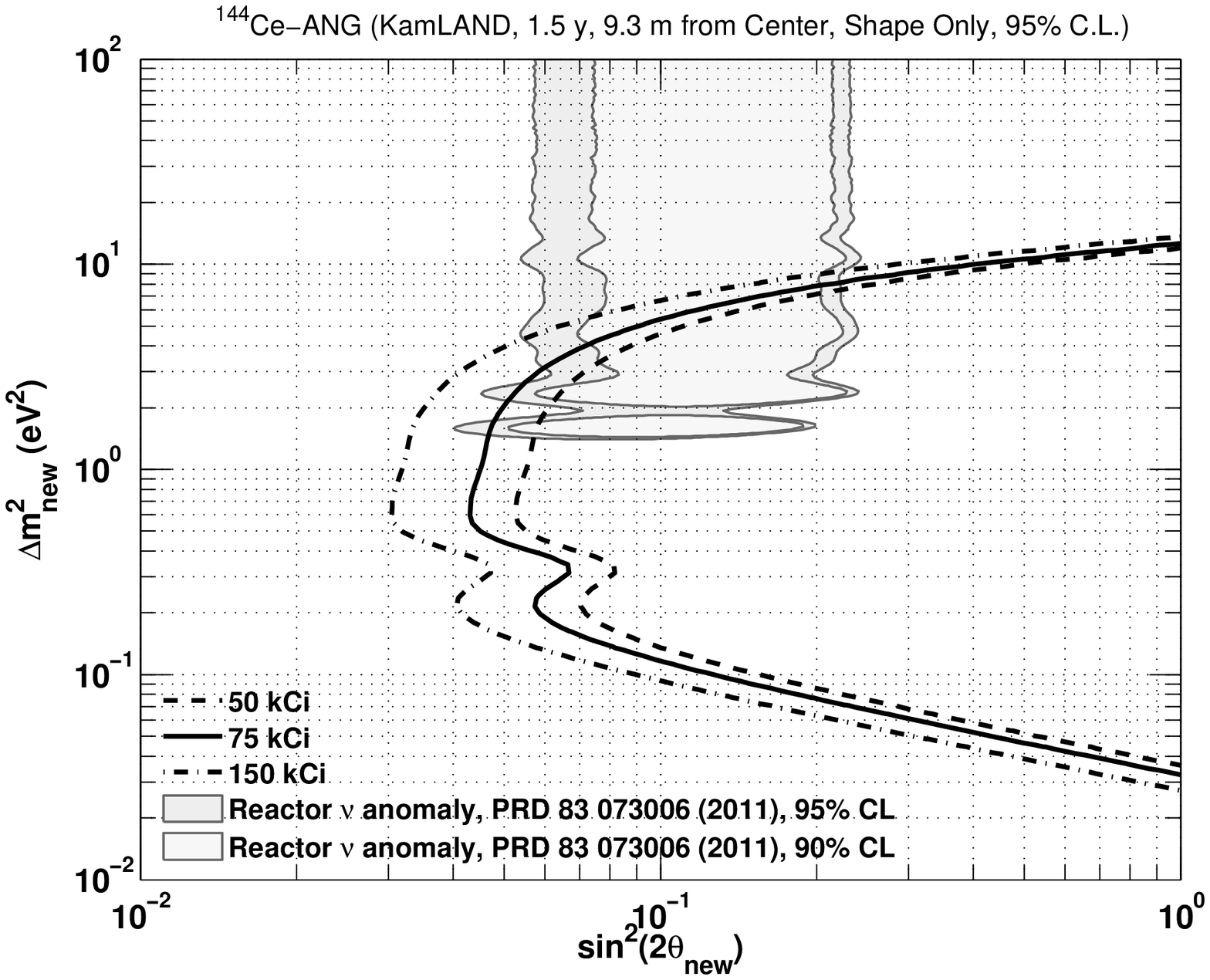}\caption{95\% C.L. expected sensitivity to exclude the non-oscillation, as
a function of the activity of the source. We assume the antineutrino
generator to be places at 9.3 m from the detector center. The fiducial
volume is starting from the detector center and extends to a radius
of 6.5 m. Results are given for 1.5 year of data taking for a rate+shape
analysis (left panel), and for a shape only analysis (right panel).
The normalization uncertainty is taken at 1.5\%.}
\end{figure}

\par\end{center}

\begin{center}
\begin{figure}[h]
\centering{}\includegraphics[scale=0.37]{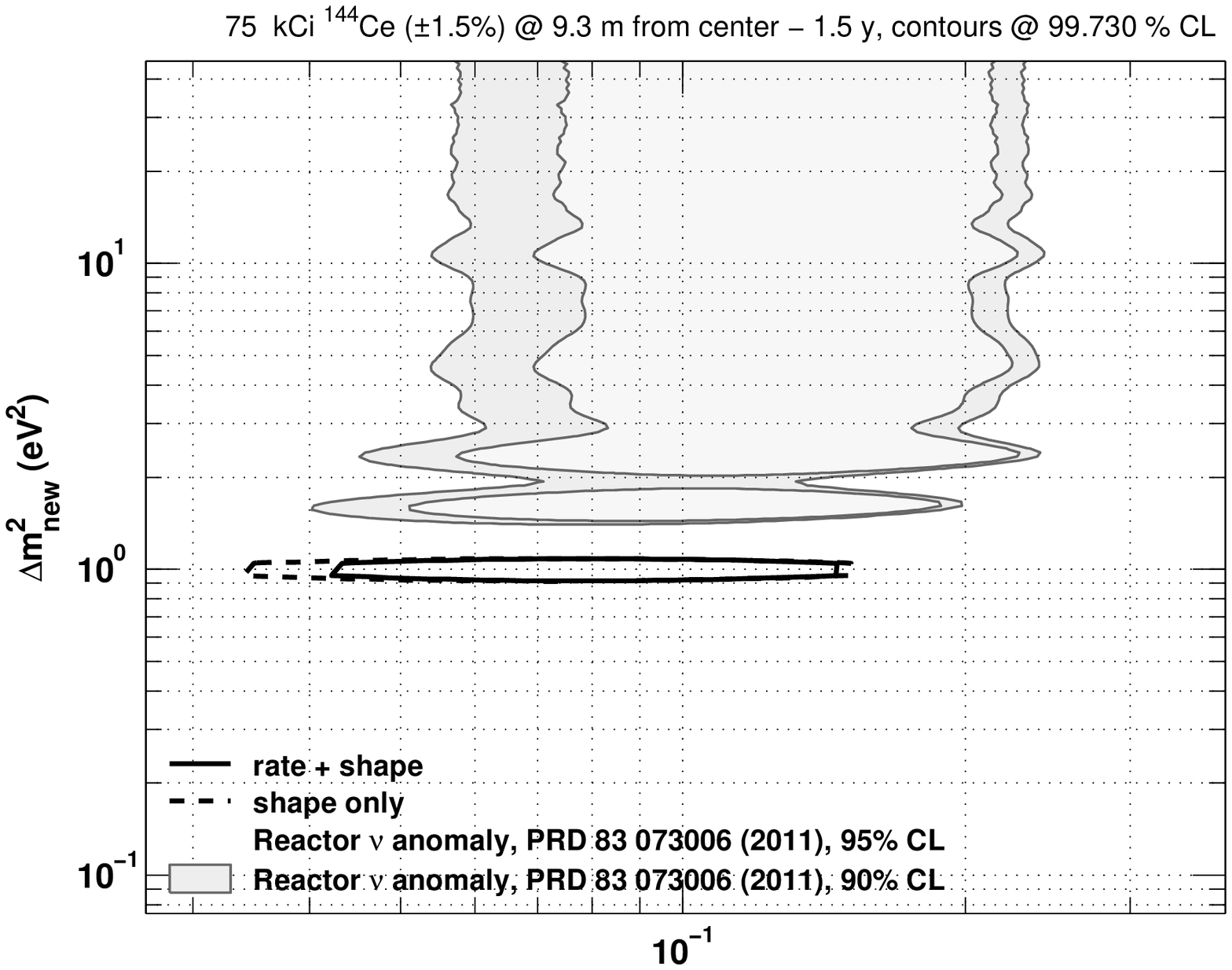}\includegraphics[scale=0.37]{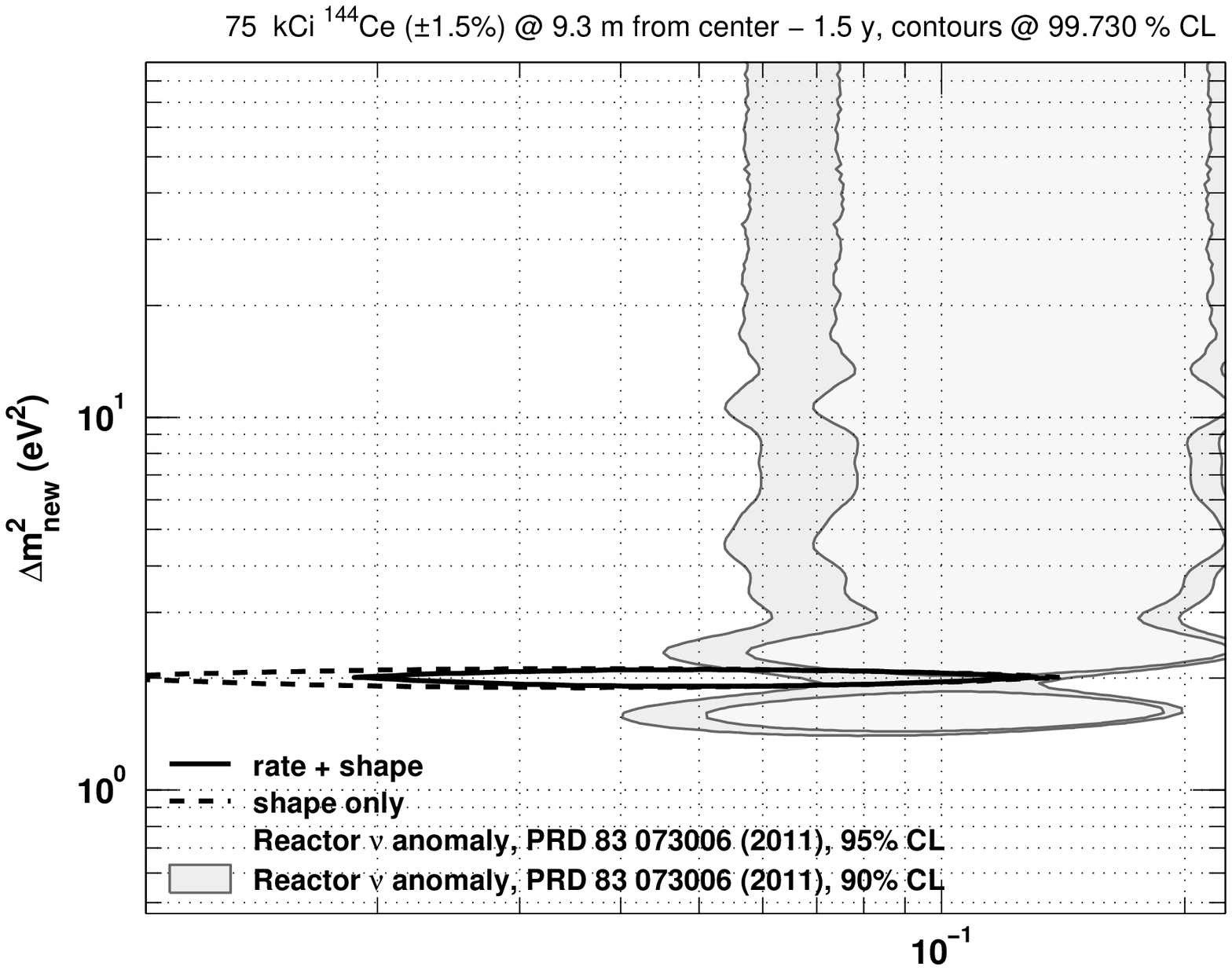}\caption{3$\sigma$ discovery potential illustrated for two examples: $\mathrm{\Delta m_{14}^{2}=1\, eV^{2}}\ensuremath{,}\mathrm{\sin^{2}(2\,\theta_{14})=0.1}$
(left) and $\mathbf{\mathrm{\Delta m_{14}^{2}=2\, eV^{2}}\ensuremath{,}\mathrm{\sin^{2}(2\,\theta_{14})=0.075}}$
(right) . We assume the antineutrino generator to be places at 9.3
m from the detector center. The fiducial volume is starting from the
detector center and extends to a radius of 6.5 m. Results are given
for 1.5 year of data taking. The normalization uncertainty is taken
at 1.5\%.}
\end{figure}

\par\end{center}

\subsection{\noindent Alternative deployment into the Xenon Room and data taking }

\noindent Given the technical issues raised by a deployment in the
OD, another option has been studied: the possibility of placing the
source outside the detector volume. Two locations are being considered:
the Xenon room and the Control room as shown in\figref{3locations-Phase1}.
We will focus on the Xenon room which is currently being used to store
and purify the Xenon used for the KamLAND-Zen experiment.

\subsubsection{\noindent Detector issues and source deployment}

\noindent An alternative option for the deployment of Ce-ANG is the
xenon facility room (Xe-LS room) that is used for the KamLAND-Zen
phase of the experiment. There are significant challenges here as
the area is completely filled with equipment to process the xenon
and load it into the liquid scintillator. In this case, Ce-ANG would
have to be lifted through a narrow corridor (600\,mm width) to the
back of the area, see \figref{Xe-LS-facility-room}. An additional
80\,mm of tungsten shielding would be necessary to allow people to
occupy the area without making it into a controlled radiation area.

\noindent \begin{center}
\begin{figure}[h]
\centering{}\includegraphics[clip,width=0.8\columnwidth]{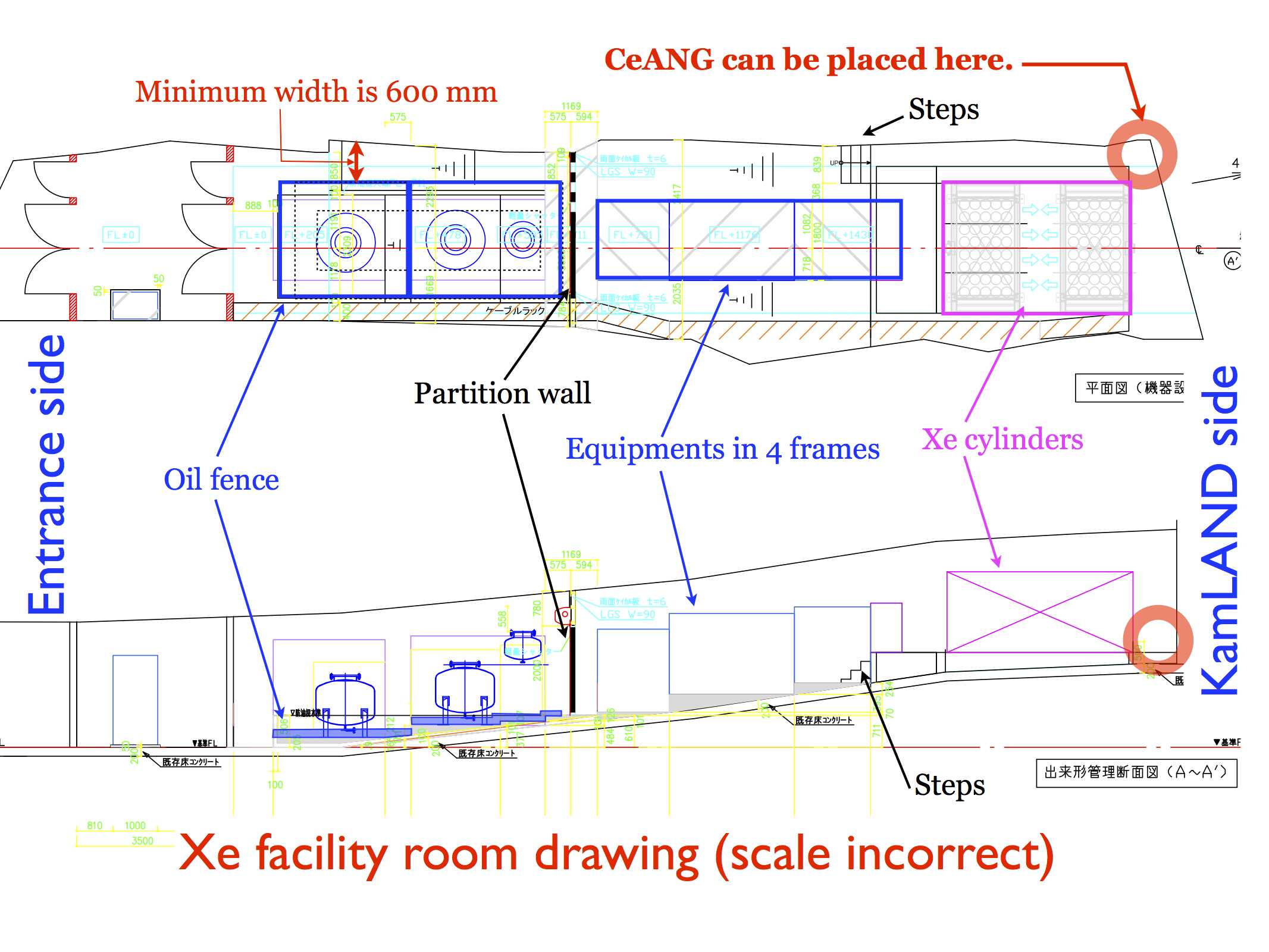}
\caption{\label{fig:Xe-LS-facility-room}Xe-LS facility room, an alternative
deployment area for Ce-ANG.}
\end{figure}

\par\end{center}

\noindent The most optimal spots in the Xenon room and the Control
room are respectively located 12 m and 13 m from the detector center.
Given the relative easiness of integration in both cases, the shorter
baseline of the Xenon room is advantageous. In addition we are considering
the option of performing the calorimetric activity measurement in
the Xenon room as well; this would allow start of the data taking
while the source is still in the calorimeter, therefore increasing
the number of detected neutrinos.

\subsubsection{\noindent Signal and backgrounds and sensitivity}

\noindent Even though the signal quality is lowered by the increase
of baseline at these two locations, it would allow CeLAND to basically
become a background-free experiment. Indeed, at 12 m, the gamma and
neutron activities of the source are no longer an issue. Besides,
if needed, it is much easier to put additional shielding in the Xenon
room than in the OD. The sensitivity is shown in \figref{SensitivityXRCR}.

\begin{center}
\begin{table}[h]
\begin{centering}
\begin{tabular}{cccccccccc}
\hline 
Months & 3 & 6 & 9 & 12 & 18 & 24 & 30 & 36 & 48\tabularnewline
\hline 
\hline 
R<6 m  & 3060 & 5520 & 7480 & 9050 & 11320 & 12770 & 13710 & 14300 & 14940\tabularnewline
R<6.5 m & 3930 & 7080 & 9610 & 11630 & 14530 & 16410 & 17620 & 18380 & 19190\tabularnewline
\hline 
\end{tabular}
\par\end{centering}

\caption{CeLAND in Xenon room: Expected $\overline{\nu}_{e}$ interactions
inside KamLAND with a 75 kCi $^{144}$Ce antineutrino generator at
12 m from the detector center as a function of the exposure for 2
definitions of the fiducial volume, R<6 m (current KamLAND analysis)
and R<6.5 m (extension of the fiducial volume to the KamLAND balloon
surface).}
\end{table}

\par\end{center}

\begin{center}
\begin{table}[h]
\begin{centering}
\begin{tabular}{cccccccccc}
\hline 
Months & 3 & 6 & 9 & 12 & 18 & 24 & 30 & 36 & 48\tabularnewline
\hline 
\hline 
R<6 m  & 2580 & 4660 & 6320 & 7640 & 9560 & 10780 & 11560 & 12080 & 12630\tabularnewline
R<6.5 m & 3320 & 5980 & 8100 & 9800 & 12250 & 13830 & 14850 & 15500 & 16180\tabularnewline
\hline 
\end{tabular}
\par\end{centering}

\caption{CeLAND in Control room: Expected $\overline{\nu}_{e}$ interactions
inside KamLAND with a 75 kCi $^{144}$Ce antineutrino generator at
13 m from the detector center as a function of the exposure for 2
definitions of the fiducial volume, R<6 m (current KamLAND analysis)
and R<6.5 m (extension of the fiducial volume to the second nylon
vessel).}

\end{table}

\par\end{center}

\noindent \begin{center}
\begin{figure}[h]
\begin{centering}
\includegraphics[scale=0.37]{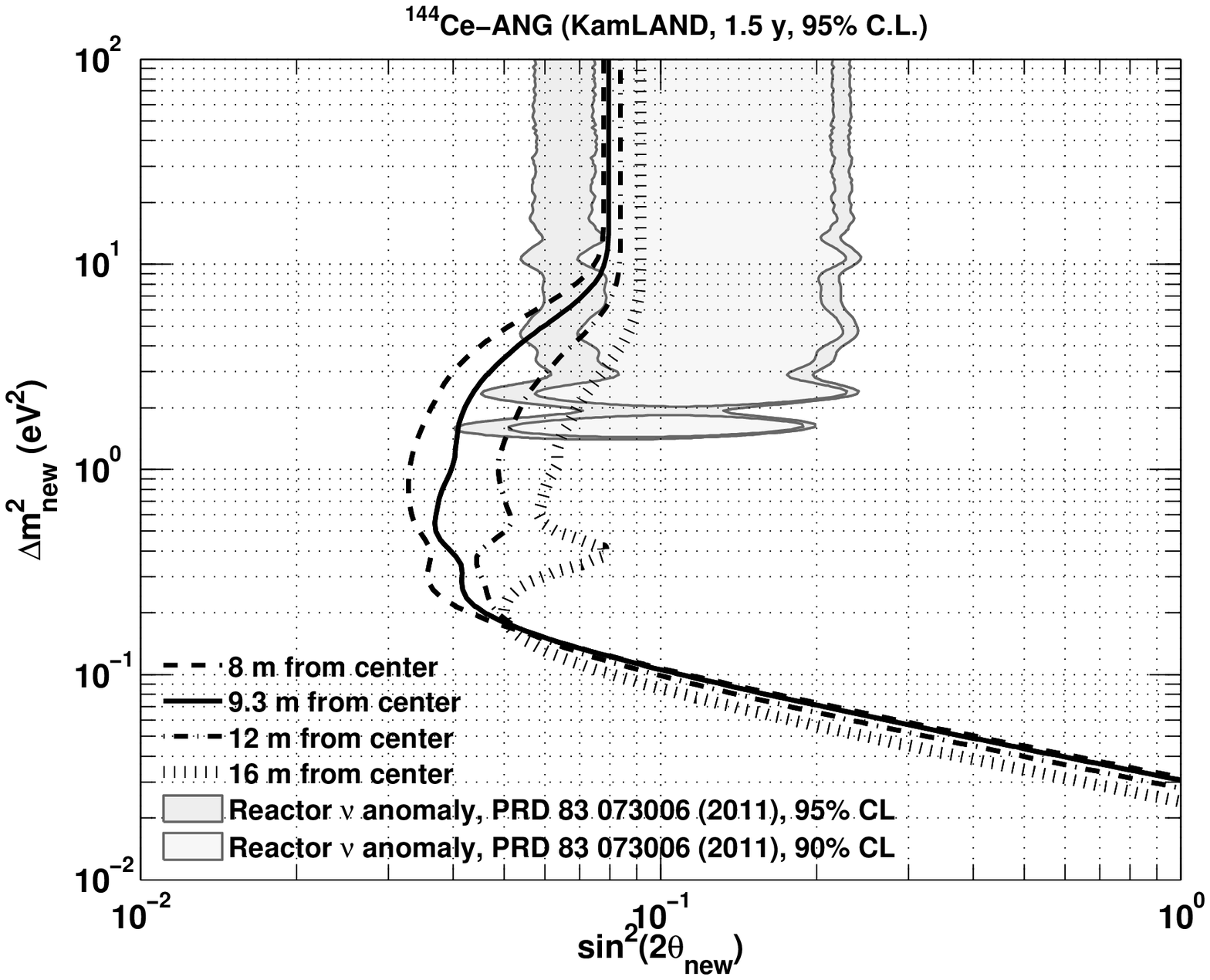}\includegraphics[scale=0.37]{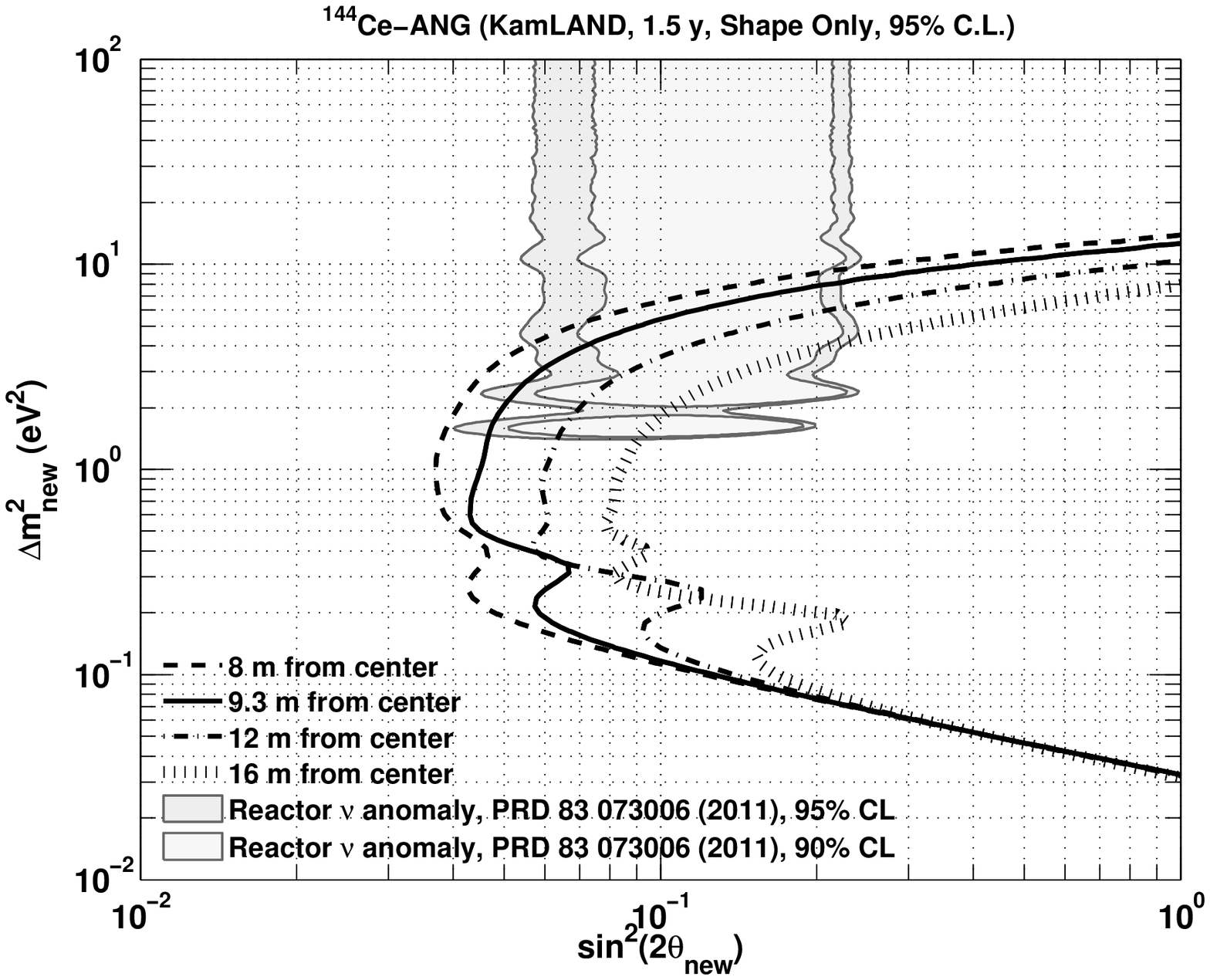}
\par\end{centering}

\caption{\label{fig:SensitivityXRCR}Sensitivity to exclude non-oscillation
with a 75 kCi antineutrino generator as a function of the distance
between the emitter and the detector center. The xenon/control room
sites are located 12 m and 13 m away, respectively. The left panel
shows the rate+shape sensitivity assuming a normalization uncertainty
of 1.5\%, whereas the right panel shows the shape only sensitivity,
which is more affected by a longer baseline. Exclusion contours are
given at the 95\% CL for a fiducial volume is taken as 0<R<6.5 m.}
\end{figure}

\par\end{center}

\subsection{\noindent Running in parallel with KamLAND-Zen}

\noindent The KamLAND-Zen experiment, dedicated to the search of the
rare neutrinoless double beta decay process ($0\nu2\beta$) with $^{136}$Xe,
has been installed in the center of the KamLAND LS balloon in 2011\cite{Zen-XE decay}.
It consists of a 3.08 m nylon-based transparent balloon (IB) filled
with 13 tons of Xe-loaded liquid scintillator as shown in \figref{kamlanddetector}.

\noindent A double beta decay process consists of the simultaneous
emission of two electrons with ($2\nu2\beta$) or without $(0\nu2\beta$)
neutrino emission. In the case of $^{136}$Xe, the $2\nu2\beta$ energy
spectrum consists of a continuum up to the nucleus Q-value of 2.458
MeV while the $0\nu2\beta$ spectrum should be a Dirac delta distribution
at this Q-value. Taking into account the KamLAND energy resolution
of $6.6\%/\sqrt{E(MeV)}$ leads to an overlap of this $0\nu2\beta$
signal with the neutron capture signal at 2.2 MeV and the signals
from several radioactive isotopes such as the $^{214}$Bi from natural
Uranium. These backgrounds have been thoroughly studied by the KamLAND-Zen
collaboration and, in order to suppress the antineutrino background
from both prompt and delayed events of the IBD process, a 1 ms coincidence
cut is applied to the KamLAND-Zen data\cite{Zen-Limit-FirstClaim}.
Furthermore, by loosing the time and space criteria we will achieve
a \textasciitilde{}99.97\% rejection efficiency, similar to $^{214}$Bi-Po.
The remaining number of 2.2 MeV $\gamma$ capture will be then much
lower than any other in-situ backgrounds, at the level of a few events
per year.

\noindent The CeLAND experiment will not be an additional source of
background for the already installed KamLAND-Zen experiment. However,
the background of Zen for CeLAND might be worth looking at. 

\noindent \begin{center}
\begin{figure}[h]
\begin{centering}
\includegraphics[scale=0.23]{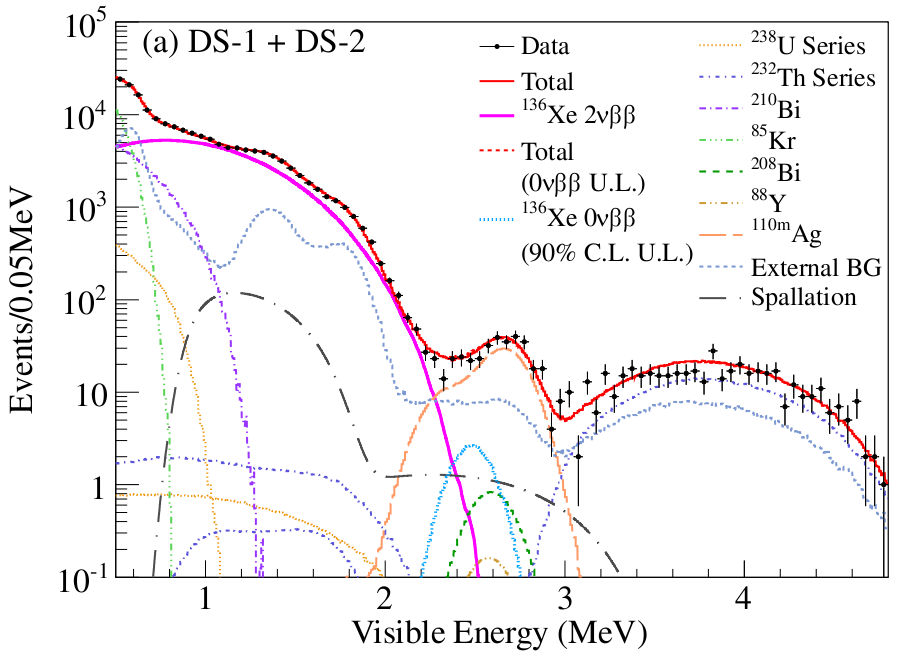}\includegraphics[scale=0.23]{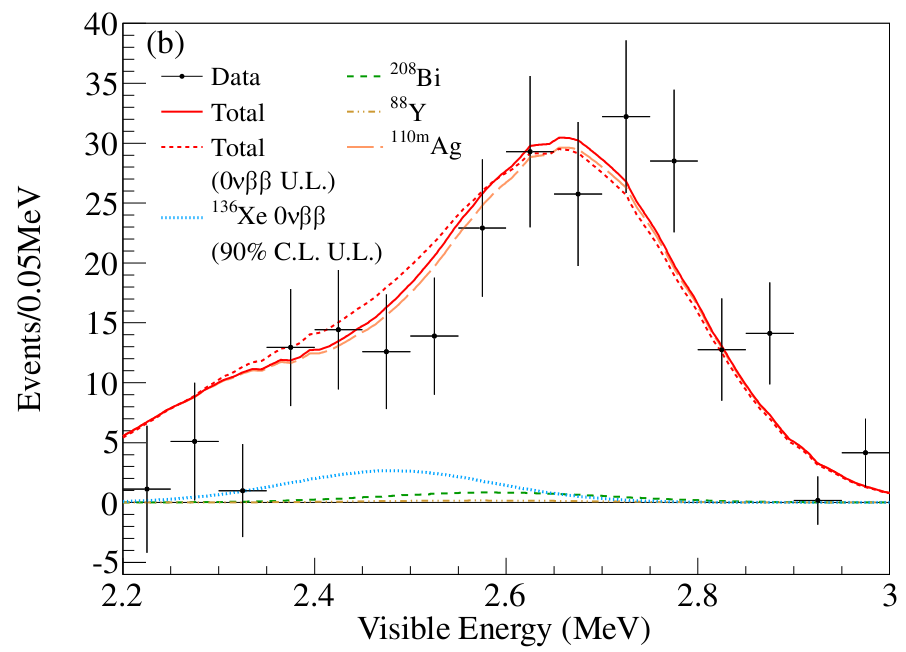}
\par\end{centering}

\caption{\label{fig:kamland-zen-Espec} (a) Energy spectrum of the KamLAND-Zen
candidate events with the best-fit backgrounds, $2\nu2\beta$ decays
and 90 \% CL upper limit on $0\nu2\beta$ decays on the $0.5\,<\, E\,<\,4.8$
MeV range. (b) Close-up on the energy window of interest for $0\nu2\beta$
after known-background subtraction.}
\end{figure}

\par\end{center}

\noindent The energy window of interest for $0\nu2\beta$ in $^{136}$Xe
is located between 2.2 and 3.0 MeV which is the end-point of the $^{144}$Pr
antineutrino prompt spectrum (\figref{kamland-zen-Espec}). Besides,
with a 90 \% CL upper limit of $\mathrm{0.034\: evts.ton^{-1}.day^{-1},}$
the $0\nu2\beta$ background is be considered negligible for both
prompt and delayed events. 

\noindent However with an event rate of $\mathrm{80.2\:\pm\:1.8(stat)\:\pm\:3.4(syst)\; evts.ton^{-1}.day^{-1}}$and
its continuous spectrum up to the $^{136}$Xe Q-value, the $2\nu2\beta$
could be considered a non-negligible background for the detector central
region. In the case of CeLAND, this region is located at the center
of the layer {[}7.76 ; 10.84{]} m from the source and contains about
30 \% of the events (6700 events for an overall number of 20000 events).
A quick computation using the $2\nu2\beta$ event rate and the KamLAND
IBD time window of $620$ \textmu{}s gives a total number of coincidences
induced by the double beta signal of $2.57\,\times\,10^{-3}$ for
a data taking period of 1.5 year. Even though the $2\nu2\beta$ event
rate can be considered negligible, an additional spatial cut is still
needed around the KamLAND-Zen mini-balloon and especially its piping
line since it induces a high background rate due to natural radioactivity.
This spatial cut only affects the delayed IBD events with the following:
R$_{d}$ > 2.5m and $\rho_{d}$ > 2.5m, Z$_{d}$ > 0m (vertical central
cylinder cut at the upper hemisphere)\cite{On-Off-KamLAND}.

\noindent In conclusion KamLAND-Zen is a negligible source of background
for CeLAND as well and both experiments can run simultaneously during
CeLAND run.

\subsection{\noindent Additional systematic studies}

\noindent In this section we evaluate the impact on the experimental
sensitivity of: the finite extension of the CeANG, the effect of the
finite energy and vertex resolutions, and the knowledge of the source
activity. We found out that only only the latter could significantly
affect the experiment.

\subsubsection{\noindent Finite source size effect}

\noindent Since the cerium material will be contained in a cylinder
the $\bar{\nu}_{e}$ emission cannot be exactly taken as a point like
source. We evaluated the impact of the slight extension of the source
through Monte-Carlo simulations and found out that a 15 cm-scale CeANG
provides a sensitivity as good as the point like source case (that
will indeed be used for all other computations). Results are displayed
in \figref{ImpactSourceSize}.

\noindent \begin{center}
\begin{figure}[h]
\begin{centering}
\includegraphics[scale=0.37]{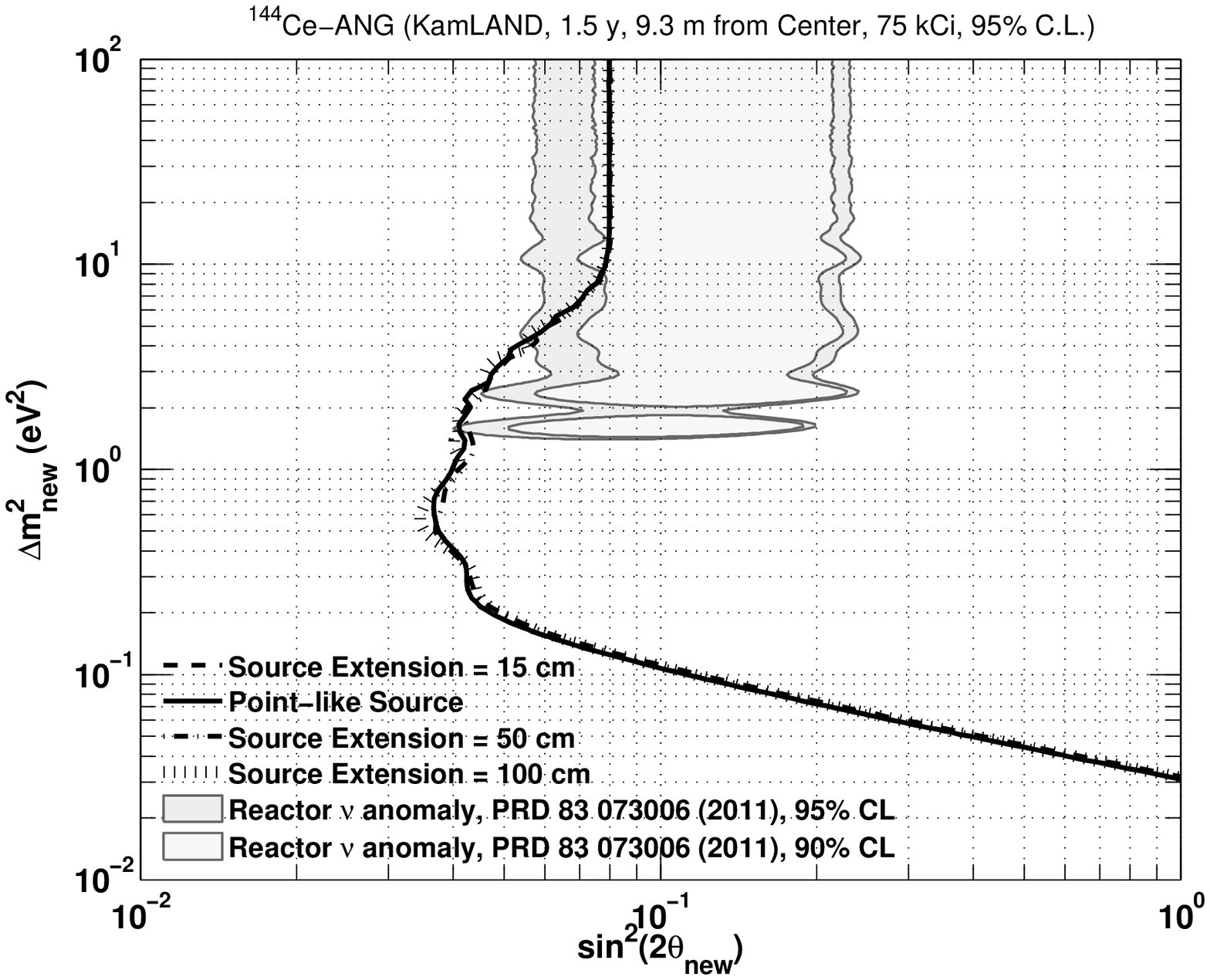}\includegraphics[scale=0.37]{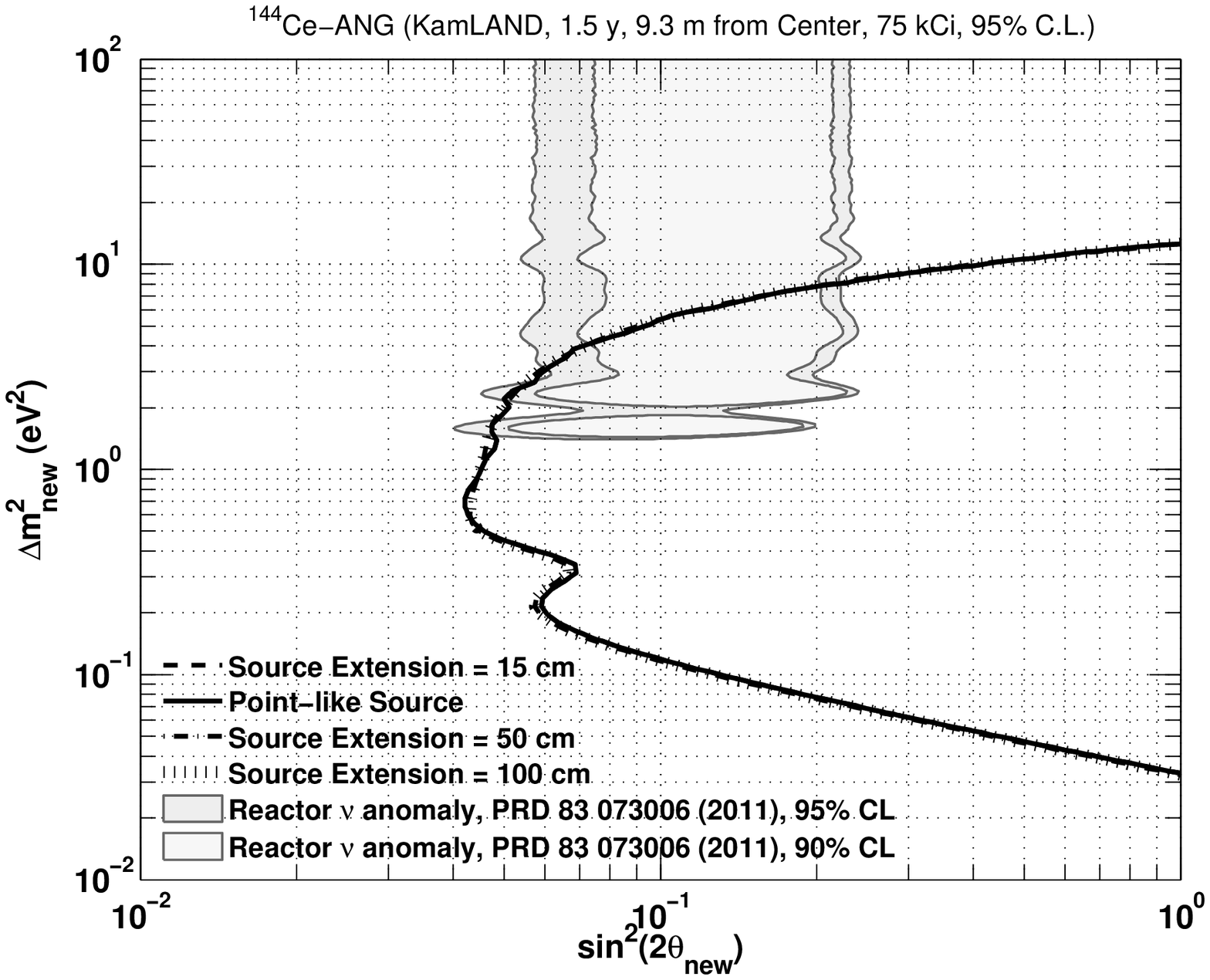}\caption{Impact of the extension of the source on the sensitivity. The source
extension is taken as a sphere ranging from 0 to of 100 cm diameter.
We observe that the sensitivity is not degraded at all by extensions
of the source up to 100 cm, for a rate+shape analysis (left) and a
shape only analysis (right). Exclusion contours are given at the 95\%
CL for 1.5 y of data taking with a source located 9.3 m away from
the detector center. The vertex resolution of the detector is taken
as 15 cm. Fiducial volume is taken as 0<R<6.5 m.\label{fig:ImpactSourceSize}}

\par\end{centering}

\centering{}
\end{figure}

\par\end{center}

\subsubsection{\noindent Energy resolution}

\noindent CeLAND sensitivity may depend on the energy resolution of
the KamLAND detector. As stated previously in this proposal, the energy
resolution of KamLAND detector is 6.5\%/(E\_vis (MeV))$^{0.5}$. The
impact on CeLAND sensitivity was evaluated for a 18 months long run
and the CeANG located 9.3 m away from the detector center. The energy
resolution was varied from 2.5\% to 15\% as a flat value independent
of energy. The results are shown in \figref{ImpactResol}. The sensitivity
is only marginally degraded for $\Delta$$m^{2}$ between 1 to 10
eV$^{2}$. Since KamLAND published energy resolution ranges from 4.4
\% to 6.5 \% at 1 MeV visible energy, corresponding conservatively
to the black lines on all plots since the positrons induced from the
source have an energy between 1 and 2.3 MeV. The energy resolution
is therefore not an issue for the CeLAND. Exclusion contours are given
at the 95\% CL for a fiducial volume taken as 0<R<6.5 m. Improvement
of the energy resolution would,slightly enhance the sensitivity of
the shape only analysis for $\Delta$$m^{2}$>2 eV$^{2}$.

\noindent \begin{center}
\begin{figure}[h]
\centering{}\includegraphics[scale=0.37]{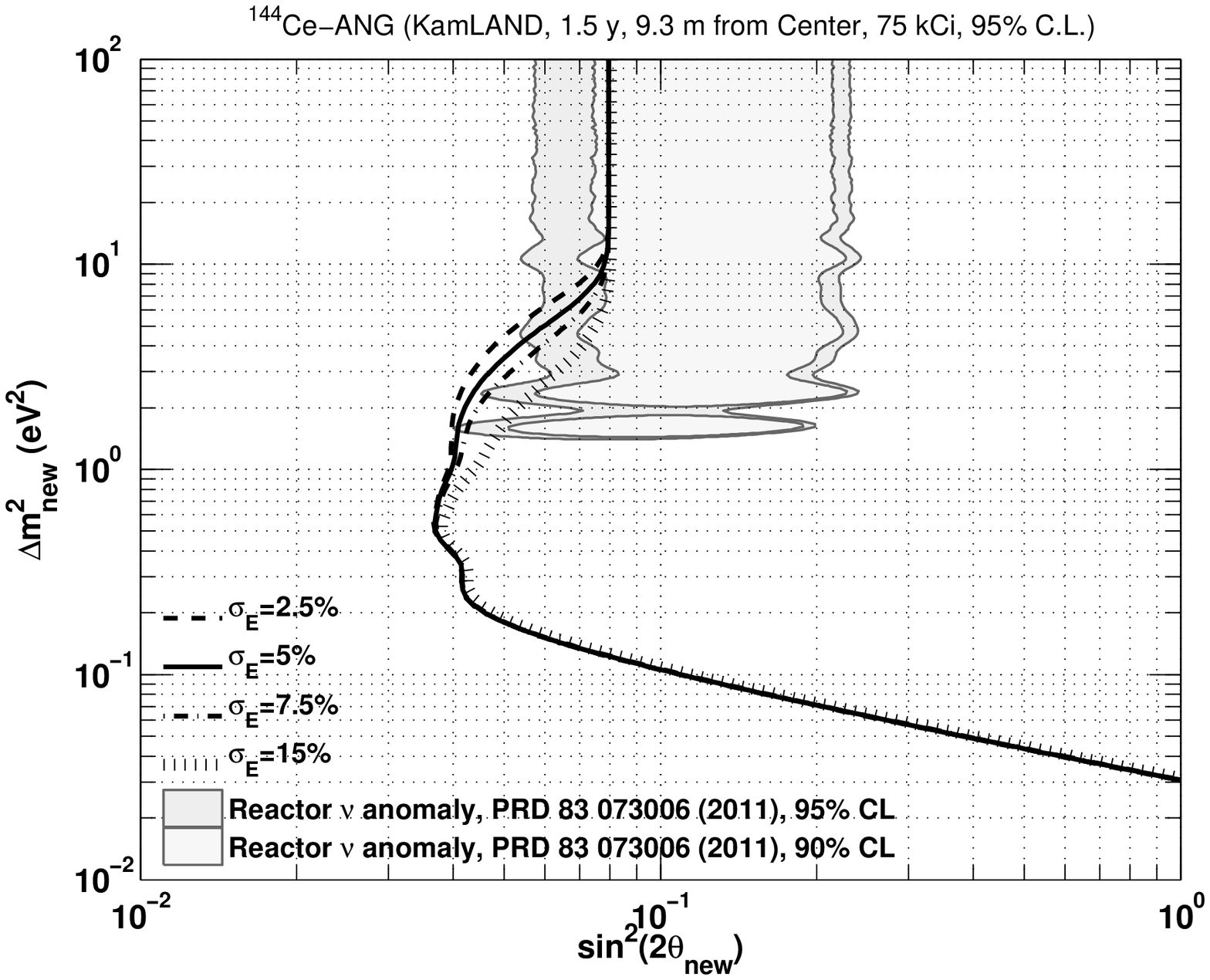}\includegraphics[scale=0.37]{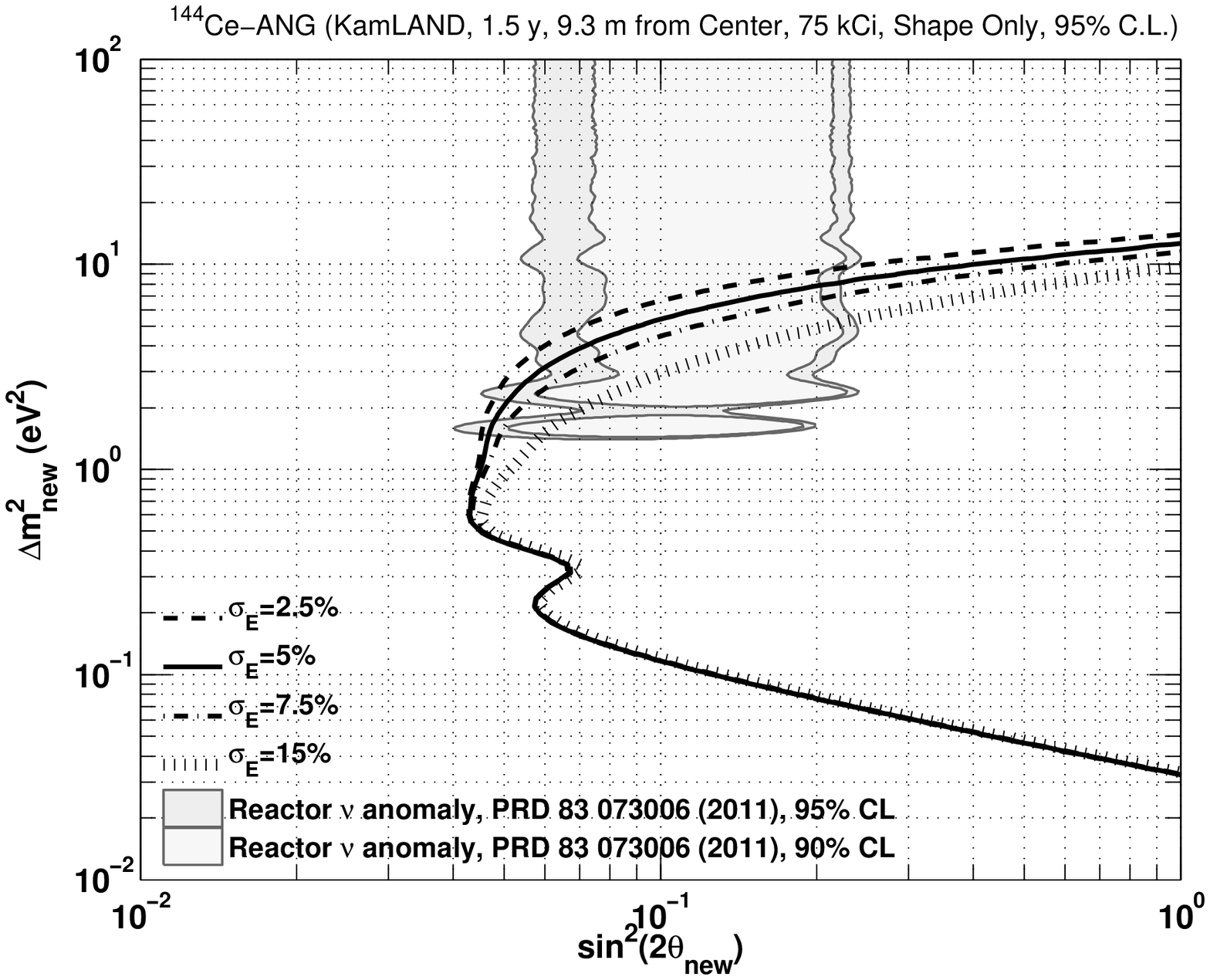}\caption{Impact of the KamLAND energy resolution on the sensitivity. We assume
a antineutrino generator located 9.3 m away from the detector center.
The sensitivity is computed for 18 months of data taking. We vary
the resolution from 2.5\% to 15\%, not depending on energy. The left
panel shows the rate+shape sensitivity assuming a normalization uncertainty
of 1.5\%, whereas the right panel shows the shape only sensitivity.
We observe that the sensitivity is not noticeably degraded for $\Delta m^{2}$
between 1 to 10 eV$^{2}$ in the case of the rate+shape analysis.
KamLAND published energy resolution is 5 \% at 1 MeV, corresponding
conservatively to the black lines on both plots since the positrons
induced from the source have an energy between 1 and 2.3 MeV. The
energy resolution is therefore not an issue for the CeLAND. Exclusion
contours are given at the 95\% CL for a fiducial volume is taken as
0<R<6.5 m.\label{fig:ImpactResol}}
\end{figure}

\par\end{center}

\subsubsection{\noindent Vertex resolution}

\noindent KamLAND's vertex resolution may impact sensitivity of CeLAND
and a simulation study was conducted to evaluate it. In this study,
the CeANG was located 9.3 m from the detector center for the duration
of 6 and 18 months. The vertex resolution was varied from 10 to 25
cm. The results are shown in \figref{ImpactVertexResol}. The sensitivity
is marginally degraded for $\Delta$$m^{2}$ between 1 to 10 eV2.
KamLAND's vertex resolution of 12 cm/(E\_vis{[}MeV{]})$^{0.5}$, or
8 cm to 12 cm in the CeLAND energy range of interest, corresponds
approximately to the black lines on all plots, and it is not expected
to be an issue, though reconstruction will have to be calibrated in
the outermost region of the target volume. In any case one expects
no concern induced by vertex resolution effect for the experiment.
Exclusion contours are given at the 95\% CL for a fiducial volume
is taken as 0<R<6.5 m.

\noindent \begin{center}
\begin{figure}[h]
\centering{}\includegraphics[scale=0.37]{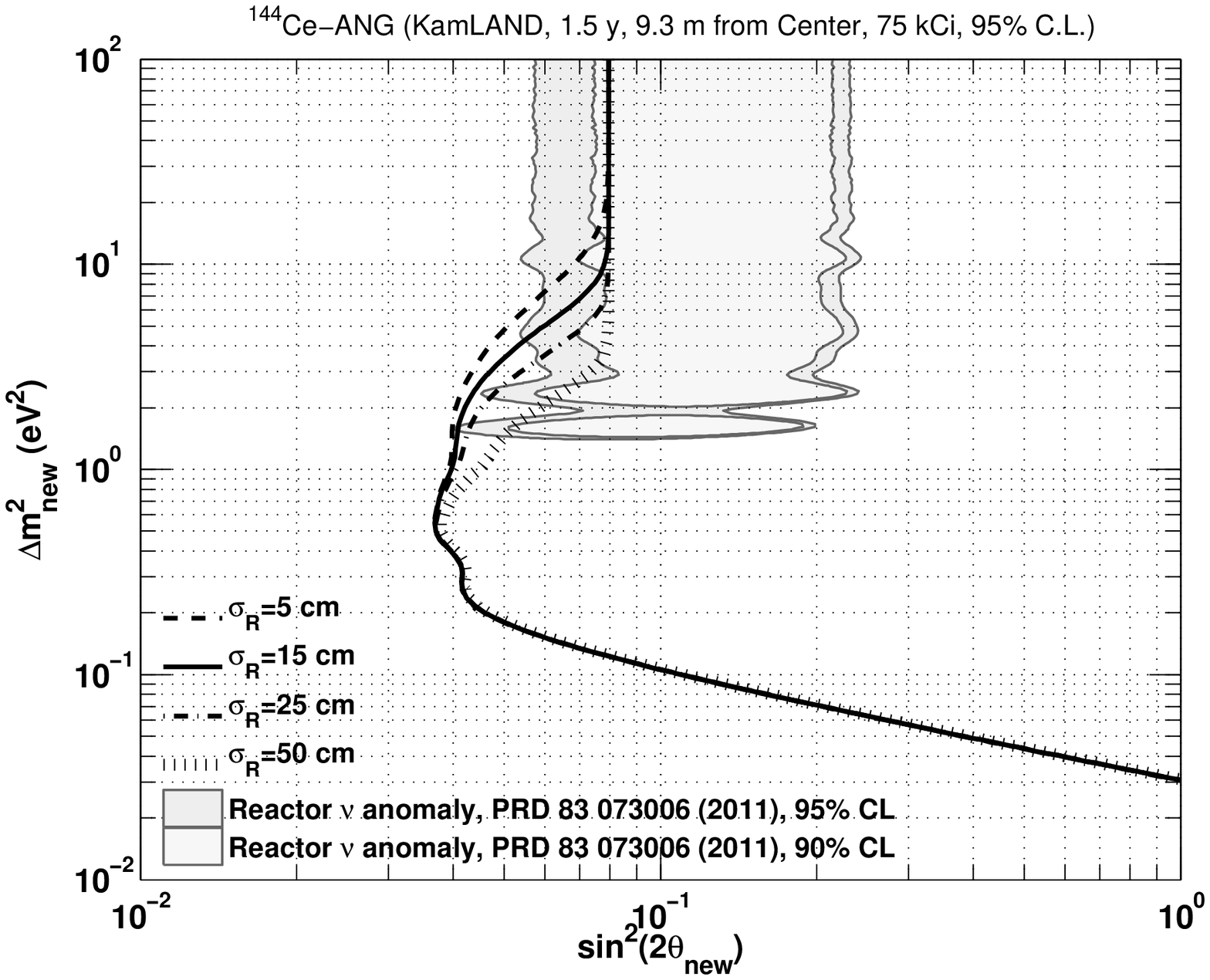}\includegraphics[scale=0.37]{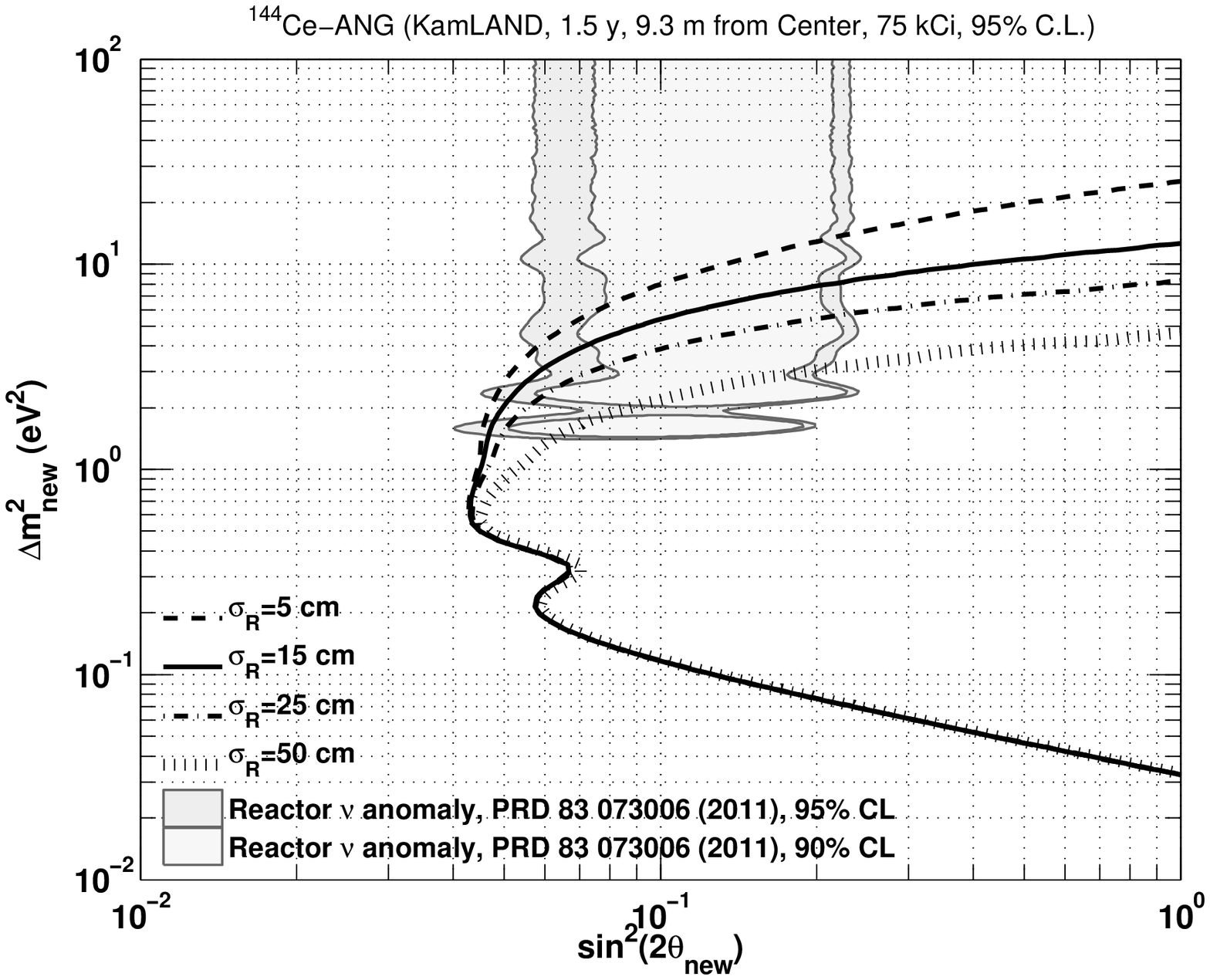}\caption{Impact of the KamLAND vertex resolution on the sensitivity. We assume
a antineutrino generator located 9.3 m away from the detector center.
The sensitivity is computed for 18 months of data taking. We vary
the resolution from 5 cm to 25 cm. The left panel shows the rate+shape
sensitivity assuming a normalization uncertainty of 1.5\%, whereas
the right panel shows the shape only sensitivity. The KamLAND published
vertex resolution is 15 cm, corresponding to the black lines on all
plots. Therefore one expects no concern induced by vertex resolution
effect for this phase of the experiment. Exclusion contours are given
at the 95\% CL for a fiducial volume is taken as 0<R<6.5 m.\label{fig:ImpactVertexResol}}
\end{figure}

\par\end{center}

\subsubsection{\noindent Uncertainty on the source activity}

\noindent Uncertainty in the knowledge of the CeANG activity directly
impacts CeLAND's rate measurement and more significantly the sensitivity
at delta $\Delta$$m^{2}$ above a couple of eV$^{2}$. The simulation
study was conducted to assess the sensitivity of CeLAND to the systematic
uncertainty in the rate of the CeANG. The CeANG's activity uncertainty
was varied in the range from 0.5\% to 3\%. The results are shown in
\figref{ImpactSourceActivity}. As can be seen in the plot, the impact
of the absolute rate uncertainty is significant especially for smaller
values of mixing angle and higher values of $\Delta$$m^{2}$. The
measurement of the absolute activity of the source will be performed
prior to deployment in the dedicated calorimeter described earlier.
Another measurement will be performed at the end of the deployment
run. The plan is to measure the CeANG activity with 1.5\% precision
or better. 

\noindent \begin{center}
\begin{figure}[h]
\begin{centering}
\includegraphics[scale=0.37]{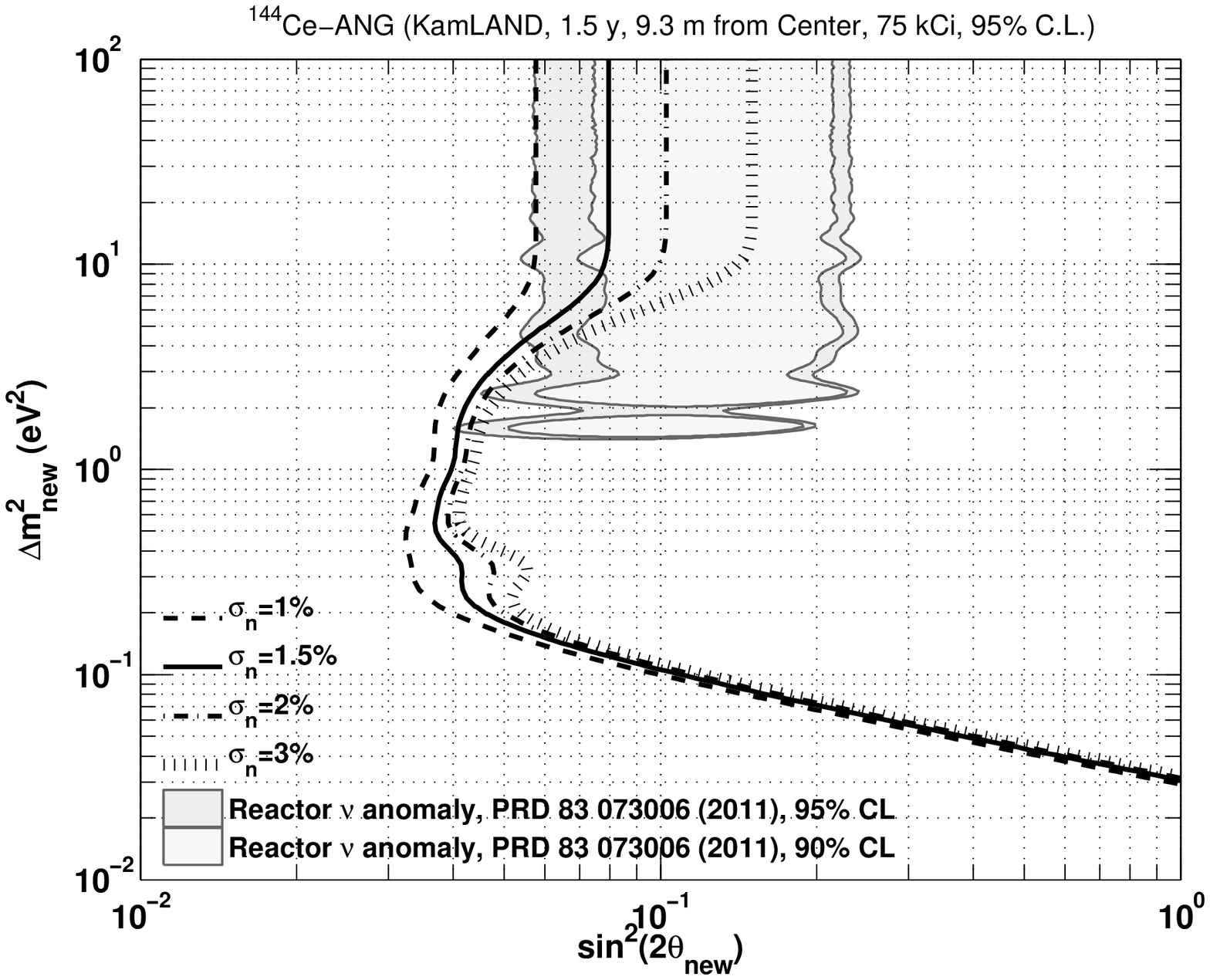}\caption{Impact on the knowledge of the source activity. We assume a antineutrino
generator located 9.3 m away from the detector center. The sensitivity
is computed for 18 months of data taking. The normalization error
is expected to be about 1.5\%, dominated by the uncertainty on the
measurement of the absolute activity of the antineutrino generator.
Exclusion contours are given at the 95\% CL for a fiducial volume
is taken as 0<R<6.5 m.\label{fig:ImpactSourceActivity}}

\par\end{centering}

\end{figure}

\par\end{center}

\section{\noindent $\mathrm{^{144}Ce-^{144}Pr}$ source disposal}

\noindent At the end of the experiment, the source activity will have
decreased by a factor of 6 or so. However, all radio-protection safety
rules will still apply for the removal of the source from the detector. 

\noindent The source disposal will be facilitated by the radioactive
decay of cerium. We provide below the evolution of the state of the
cerium ANG with time elapsed since its production (see also table
\ref{tab:Evolution-of-the-source}):
\begin{itemize}
\item \noindent at the end of the experiment, 2 years after the source delivery
(at 85 kCi), the source will only deliver 100 W at 15 kCi;
\item \noindent after 5 years, heat power will reach the watt scale (for
1 kCi) where heat issues can be neglected, and dose rate at 1 m of
the shielded source will be of the order of the natural dose rate
(some 0.1 \textmu{}Sv/h, depending on the site); 
\item \noindent after 20 years, the source will reach the 10 MBq scale (corresponding
to some 100 kBq in $\gamma$ activity) where manipulation with bare
hand is conceivable and the dose rate at 1 m of the naked source will
be of the order of the natural dose rate ;
\item \noindent after 25 years, the source will reach about 100 Bq/kg, which
correspond to natural activity of standard matter;
\item \noindent after 40 years, the source would reach the Bq scale, but
long-lived impurities will probably limit the decrease in the activity.
\end{itemize}
\noindent \begin{center}
\begin{table}[h]
\centering{}%
\begin{tabular}{cccccc}
\hline 
Source age & Heat & $\beta$ activity & 2.2 MeV $\gamma$ & Dose rate with  & Dose rate without \tabularnewline
 & (W) &  & activity & shielding (\textmu{}Sv/h) & shielding (\textmu{}Sv/h)\tabularnewline
\hline 
\hline 
1 day & 678 & 84.8 kCi & 594 Ci & 16.4 & $6.6\times10^{6}$\tabularnewline
1 week & 668 & 83.6 kCi & 585 Ci & 16.1 & $6.5\times10^{6}$\tabularnewline
1 month & 631 & 79.0 kCi & 553 Ci & 15.3 & $6.2\times10^{6}$\tabularnewline
3 month & 546 & 68.3 kCi & 478 Ci & 13.2 & $5.3\times10^{6}$\tabularnewline
$\nicefrac{1}{2}$ year & 438 & 54.9 kCi & 384 Ci & 10.6 & $4.3\times10^{6}$\tabularnewline
1 year & 279 & 35.0 kCi & 245 Ci & 6.8 & $2.7\times10^{6}$\tabularnewline
2 years & 115 & 14.4 kCi & 101 Ci & 2.8 & $1.1\times10^{6}$\tabularnewline
3 years & 47.3 & 5.92 kCi & 41 Ci & 1.1 & $4.6\times10^{5}$\tabularnewline
5 years & 8.0 & 1.00 kCi & 7.0 Ci & 0.19 & $7.8\times10^{4}$\tabularnewline
10 years & $9.5\times10^{-2}$ & 12 Ci & 3.1 GBq & $2.3\times10^{-3}$ & $9.2\times10^{2}$\tabularnewline
20 years & $1.3\times10^{-5}$ & 61 MBq & 430 kBq & $3.2\times10^{-7}$ & 0.13\tabularnewline
40 years & $2.5\times10^{-13}$ & 1.2 Bq & 8.3 mBq & $6.2\times10^{-15}$ & $2.5\times10^{-9}$\tabularnewline
\hline 
\end{tabular}\caption{\label{tab:Evolution-of-the-source}Evolution of the source, starting
with an initial $\beta$ activity of 85 kCi. The dose rate is calculated
with formula \ref{eq:dose-boe} at 1 m from the center of the source
($\sim80$ cm of the shielding if any).}
\end{table}

\par\end{center}

\noindent For disposal the antineutrino generator will return to PA
Mayak in Russia. The transport cost could be reduced by keeping the
source in Japan (Kamioka mine) until its activity is less than 16
kCi (approx. 2 years after production) such as a regular type B(U)
container could be used for air transportation without special arrangement.
This should implicitly happen after the data taking period and further
measurement of the source activity.

\section{\noindent Conclusion}

\noindent The CeLAND experiment offers to deploy a 75~kCi $\mathrm{^{144}Ce-^{144}Pr}$
antineutrino generator next or inside the KamLAND detector in order
to test both the reactor and the gallium anomalies. The concept of
the experiment is to measure the energy and position dependence of
the detected neutrino flux, characteristic of neutrino oscillation
into a 4th state. Consequently this will allow to search for eV-scale
light sterile neutrino oscillations with mass splittings of $\geq$0.1~eV$^{2}$
and sin$^{2}(2\theta_{new})>0.05$. With 1.5 year of data, the anomalous
allowed region of mixing parameters could be probed non ambiguously.
CeLAND plans to make use of the existing KamLAND detector with well-measured
backgrounds and systematics. Beside upgrading the KamLAND installation
to safely host the source, most of the new effort involves creating,
for the first time, an 75~kCi $\mathrm{^{144}Ce-^{144}Pr}$ antineutrino
and delivering it to the detector, where it could be placed in a first
phase outside the detector, 9.3 m from the center (this phase is the
main purpose of our current proposal), and possibly later inside the
target liquid scintillator, but only after the KamLAND-Zen 0\textgreek{nbb}
run is complete and if a hint of oscillation would be discovered during
the first phase. The cost-scale of the whole experiment is estimated
to be less than 5 M\$, and data taking could start within the next
1.5 years, depending of funding and logistic issues being currently
investigated.\newpage{}

\appendix

\section{\noindent Interaction length in Tungsten alloy}

\noindent In reference \cite{key-1} we found plots of the cross section
for photon on all elements taking from the evaluated data base ENDF/B-VI.
Examples on W, Fe, Ni,U are given in \figref{CrossSectionGamma}.
These data are used to precisely compute the interaction length in
tungsten alloys. 

\noindent \begin{center}
\begin{figure}[h]
\centering{}\includegraphics[scale=0.45]{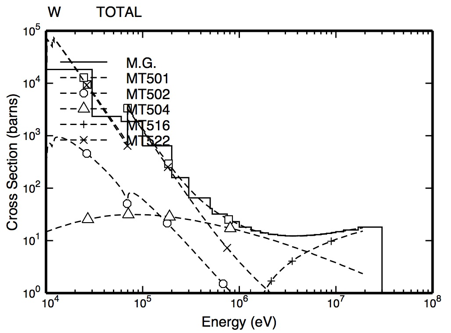}\includegraphics[scale=0.45]{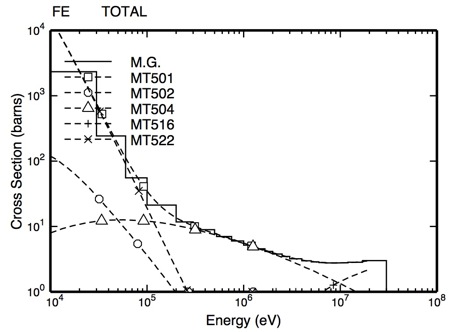}\\
\includegraphics[scale=0.45]{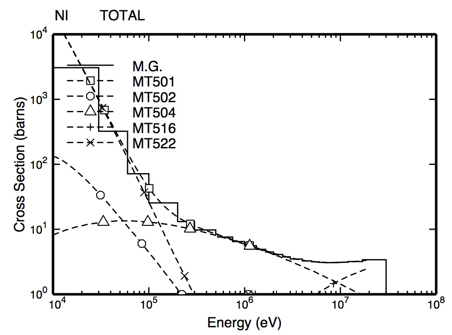}\includegraphics[scale=0.45]{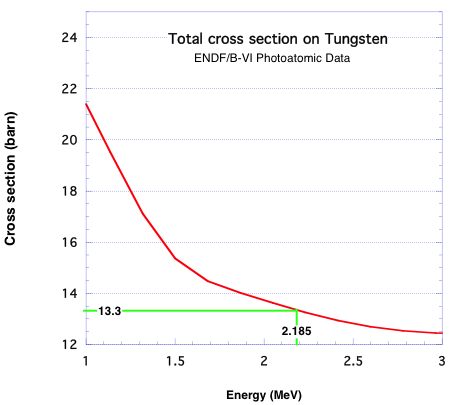}\caption{\label{fig:CrossSectionGamma}Cross sections of the interaction of
$\gamma's$ with tungsten, iron, nickel, and tungsten alloy (18.5
g/cm$^{3}$).}
\end{figure}

\par\end{center}

\noindent The total interaction length is 1.238 \textpm{} 0.14 cm
(see details in \tabref{DataAttLengthDensimet185}). For comparison,
the interaction lengths in pure tungsten and uranium metals are respectively
1.19 cm and 1.13 cm. For a tungsten alloy shielding having a density
of 18.5 \textpm{} 0.2 gm/cm$^{3}$the attenuation factor in a 35 cm
thick shielding is (5.25 \textpm{} 1.4 ) \texttimes{} $10$$^{\lyxmathsym{\textminus}13}$.

\noindent \begin{center}
\begin{table}[h]
\centering{}%
\begin{tabular}{cccccccc}
\hline 
Element & E (keV) & $\sigma$ (barn) & A & Z & \% in mass & \# nuclei/cm$^{3}$ & $\lambda_{i}$ \tabularnewline
\hline 
\hline 
W & 2179 & 13.3 \textpm{} 1 & 183.84 & 74 & 97 & 5.878 $10{}^{22}$ & 1.279 \textpm{} 0.096 cm\tabularnewline
Fe & 2189 & 5.09 \textpm{} .15 & 55.845 & 26 & 1.5  & 0.299 $10{}^{22}$ & 65.65 \textpm{} 0.19 cm\tabularnewline
Ni & 2188 & 5.28 \textpm{} .15 & 58.6934 & 28 & 1.5  & 0.284 $10{}^{22}$ & 66.52 \textpm{} 1.9 cm\tabularnewline
\hline 
\end{tabular}\caption{\label{tab:DataAttLengthDensimet185}Data used for the computation
of the attenuation length in tungsten alloy. $\lambda_{i}$ is the
partial interaction length. The total interaction length : 1.238 \textpm{}
0.14 cm.}
\end{table}

\par\end{center}

\noindent \clearpage\newpage{}

\end{document}